\documentstyle[preprint,graphicx,epsfig,tighten,eqsecnum,floats,aps,amsfonts,amssymb]{revtex}

\def\be{\begin{equation}}
\def\ee{\end{equation}}
\def\ba{\begin{eqnarray}}
\def\ea{\end{eqnarray}}

\def\A{{\cal A}}
\def\C{{\cal C}}
\def\CE{\C^{\rm Eucl}}
\def\D{{\cal D}}
\def\G{{\cal G}}
\def\Gauss{{\cal C}_{\rm G}}
\def\H{{\cal H}}

\def\I{{\cal I}}

\def\O{{\cal O}}
\def\P{{\cal P}}
\def\R{{\cal R}}
\def\S{{\cal S}}
\def\T{{\cal T}}

\def\Ab{{\bar \A}}
\def\bA{{\bar A}}
\def\bg{{\bar g}}
\def\Gb{{\bar \G}}

\def\h{{\hat h}}
\def\hA{{\hat A}}
\def\hC{{\hat \C}}
\def\hCE{{\hC}^{\rm Eucl}}

\def\hH{{\hat H}}
\def\hJ{{\hat J}}

\def\hL{{\hat L}}
\def\hP{{\hat P}}
\def\hR{{\hat R}}
\def\hT{{\hat T}}
\def\hq{{\hat q}}
\def\hV{{\hat V}}
\def\hG{{\hat \Gauss}}

\def\div{\rm div}
\def\const{{\rm const}}
\def\Cyl{{\rm Cyl}}
\def\cyl{\rm Cyl}

\def\su{{\rm su}}
\def\SU{{\rm SU}}
\def\SO{{\rm SO}}
\def\so{{\rm so}}
\def\u(1){{\rm u(1)}}
\def\U{{\rm U}}
\def\SL{{\rm SL}}
\def\diff{{\rm diff}}
\def\Diffb{{\rm Diff}}
\def\Diff{{\rm Diff}}
\def\TDiff{{\rm TDiff}}

\def\inv{{\rm inv}}

\def\Tr{{\rm Tr\,}}
\def\covps{{{\bf \Gamma}_{\!\rm cov}}}
\def\canps{{{\bf \Gamma}_{\!\rm can}}}

\def\a{\alpha}
\def\IH{\Delta}
\def\g{\gamma}

\def\lp{{\ell}_{\rm Pl}}
\def\Real{{\mathbb R}}
\def\Comp{{\mathbb C}}

\def\j{{\bf j}}
\def\l{{\bf l}}
\def\s{{\bf s}}
\def\M{M}
\def\f{\frac}
\def\man{{\cal M}}
\def\ub{\underbar}
\def\q{{}^o\!q}
\def\e{{}^o\!e}
\def\w{{}^o\!\omega}
\def\L{\ell}

\def\b{$\bullet\,\,$}

\def\aM{{\mathbb{A}}}
\def\eM{{\mathbb{P}}}
\def\heM{{{\hat{\eM}}}}
\def\bM{{\mathbb{B}}}

\def\CM{{\bf C}}
\def\hCM{\hat{\CM}}
\def\GM{\mathbf{G}}

\def\ggot{\mathfrak{g}}
\def\Nat{{\mathbb N}}

\begin{document}


\title{Background Independent Quantum Gravity: \\A Status Report}
\author{Abhay\ Ashtekar${}^{1,3,4}$, Jerzy Lewandowski${}^{2,1,3,4}$}
\address{1. Physics Department, 104 Davey, Penn State, University
Park, PA 16802, USA\\ 2. Institute of Theoretical Physics,
University of Warsaw, ul. Ho\.{z}a 69, 00-681 Warsaw, Poland\\ 3.
Max Planck Institut f\"ur Gravitationally, Albert Einstein
Institut, 14476 Golm, Germany\\
4. Erwin Schr\"odinger Institute, Boltzmanngasse 9, 1090 Vienna,
Austria}

\maketitle

\begin{abstract}

The goal of this article is to present an introduction to loop
quantum gravity ---a background independent, non-perturbative
approach to the problem of unification of general relativity and
quantum physics, based on a quantum theory of geometry. Our
presentation is pedagogical. Thus, in addition to providing a
bird's eye view of the present status of the subject, the article
should also serve as a vehicle to enter the field and explore it
in detail. To aid non-experts, very little is assumed beyond
elements of general relativity, gauge theories and quantum field
theory. While the article is essentially self-contained, the
emphasis is on communicating the underlying ideas and the
significance of results rather than on presenting systematic
derivations and detailed proofs. (These can be found in the listed
references.) The subject can be approached in different ways. We
have chosen one which is deeply rooted in well established physics
and also has sufficient mathematical precision to ensure that
there are no hidden infinities. In order to keep the article to a
reasonable size, and to avoid overwhelming non-experts, we have
had to leave out several interesting topics, results and
viewpoints; this is meant to be an introduction to the subject
rather than an exhaustive review of it.
\end{abstract}
\bigskip
\indent\emph{Pacs {04.60Pp, 04.60.Ds, 04.60.Nc, 03.65.Sq}}

\vfill\break

\tableofcontents\vfill\break


\section{Introduction}
\label{s1}

This section is divided into three parts. In the first, we outline
the general, conceptual viewpoint that underlies loop quantum
gravity; in the second, we recall some of the central physical
problems of quantum gravity; and in the third, we summarize the
progress that has been made in addressing these issues and sketch
the organization of the paper.

\subsection{The viewpoint}
\label{s1.1}

In this approach, one takes the central lesson of general
relativity seriously: gravity \textit{is} geometry whence, in a
fundamental theory, there should be no background metric. In
quantum gravity, geometry and matter should \textit{both} be `born
quantum mechanically'. Thus, in contrast to approaches developed
by particle physicists, one does not begin with quantum matter on
a background geometry and use perturbation theory to incorporate
quantum effects of gravity. There \textit{is} a manifold but no
metric, or indeed any other fields, in the background.%
\footnote{In 2+1 dimensions, although one begins in a completely
analogous fashion, in the final picture one can get rid of the
background manifold as well. Thus, the fundamental theory can be
formulated combinatorially \cite{aa1,ahrss}. To achieve this goal
in 3+1 dimensions, one needs a much better understanding of the
theory of (intersecting) knots in 3 dimensions.}

At the classical level, Riemannian geometry provides the
appropriate mathematical language to formulate the physical,
kinematical notions as well as the final dynamical equations of
modern gravitational theories. This role is now taken by
\textit{quantum} Riemannian geometry, discussed in sections
\ref{s4} and \ref{s5}. In the classical domain, general relativity
stands out as the best available theory of gravity, some of whose
predictions have been tested to an amazing accuracy, surpassing
even the legendary tests of quantum electrodynamics. Therefore, it
is natural to ask: \textit{Does quantum general relativity,
coupled to suitable matter} (or supergravity, its supersymmetric
generalization) \emph{exist as a consistent theory
non-perturbatively?}

In the particle physics circles, the answer is often assumed to be
in the negative, not because there is concrete evidence against
non-perturbative quantum gravity, but because of an analogy to the
theory of weak interactions. There, one first had a 4-point
interaction model due to Fermi which works quite well at low
energies but which fails to be renormalizable. Progress occurred
not by looking for non-perturbative formulations of the Fermi
model but by replacing the model with the Glashow-Salam-Weinberg
renormalizable theory of electro-weak interactions, in which the
4-point interaction is replaced by $W^\pm$ and $Z$ propagators. It
is often assumed that perturbative non-renormalizability of
quantum general relativity points in a similar direction. However
this argument overlooks the crucial fact that, in the case of
general relativity, there is a qualitatively new element.
Perturbative treatments pre-suppose that space-time can be assumed
to be a continuum \textit{at all scales} of interest to physics
under consideration. This appears to be a safe assumption in
theories of electro-weak and strong interactions. In the
gravitational case, on the other hand, the scale of interest is
given by the Planck length $\lp$ and there is no physical basis to
\textit{pre-suppose} that the continuum picture should be valid
down to that scale. The failure of the standard perturbative
treatments may be largely due to this grossly incorrect assumption
and a non-perturbative treatment which correctly incorporates the
physical micro-structure of geometry may well be free of these
inconsistencies.

Note that, even if quantum general relativity did exist as a
mathematically consistent theory, there is no a priori reason to
assume that it would be the `final' theory of all known physics.
In particular, as is the case with classical general relativity,
while requirements of background independence and general
covariance do restrict the form of interactions between gravity
and matter fields and among matter fields themselves, the theory
would not have a built-in principle which \textit{determines}
these interactions. Put differently, such a theory would not be a
satisfactory candidate for unification of all known forces.
However, just as general relativity has had powerful implications
in spite of this limitation in the classical domain, quantum
general relativity should have qualitatively new predictions,
pushing further the existing frontiers of physics. Indeed,
unification does not appear to be an essential criterion for
usefulness of a theory even in other interactions. QCD, for
example, is a powerful theory even though it does not unify strong
interactions with electro-weak ones. Furthermore, the fact that we
do not yet have a viable candidate for the grand unified theory
does not make QCD any less useful.

Finally, the quantum theory of geometry provides powerful tools to
do quantum physics in absence of a background space-time. Being
kinematical, it is not rigidly tied to general relativity (or
supergravity) and may well be useful also in other approaches to
quantum gravity.

\subsection{Physical questions of quantum gravity}
\label{s1.2}

Approaches to quantum gravity face two types of issues: Problems
that are `internal' to individual approaches and problems that any
approach must face. Examples of the former are: Incorporation of
physical ---rather than half flat--- gravitational fields in
twistor theory; mechanisms for breaking of supersymmetry and
dimensional reduction in string theory; and issues of space-time
covariance in the canonical approach. In this sub-section, we will
focus on the second type of issues by recalling some of the long
standing issues that \emph{any} satisfactory quantum theory of
gravity should address.

$\bullet$ \textit{Big-Bang and other singularities}: It is widely
believed that the prediction of a singularity, such as the
big-bang of classical general relativity, is primarily a signal
that the physical theory has been pushed beyond the domain of its
validity. A key question to any quantum gravity theory, then, is:
What replaces the big-bang? Are the classical geometry and the
continuum picture only approximations, analogous to the `mean
(magnetization) field' of ferro-magnets? If so, what are the
microscopic constituents? What is the space-time analog of a
Heisenberg quantum model of a ferro-magnet? When formulated in
terms of these fundamental constituents, is the evolution of the
\textit{quantum} state of the universe free of singularities?
General relativity predicts that the space-time curvature must
grow unboundedly as we approach the big-bang or the big-crunch but
we expect the quantum effects, ignored by general relativity, to
intervene, making quantum gravity indispensable before infinite
curvatures are reached. If so, what is the upper bound on
curvature? How close to the singularity can we `trust' classical
general relativity? What can we say about the `initial
conditions', i.e., the quantum state of geometry and matter that
correctly describes the big-bang? If they have to be imposed
externally, is there a \textit{physical} guiding principle?

$\bullet$ \textit{Black holes:} In the early seventies, using
imaginative thought experiments, Bekenstein \cite{jb} argued that
black holes must carry an entropy proportional to their area.
About the same time, Bardeen, Carter and Hawking (BCH) showed that
black holes in equilibrium obey two basic laws, which have the
same form as the zeroth and the first laws of thermodynamics,
provided one equates the black hole surface gravity $\kappa$ to
some multiple of the temperature $T$ in thermodynamics and the
horizon area $a_{\rm hor}$ to a corresponding multiple of the
entropy $S$ \cite{bch}. However, at first this similarity was
thought to be only a formal analogy because the BCH analysis was
based on \textit{classical} general relativity and simple
dimensional considerations show that the proportionality factors
must involve Planck's constant $\hbar$. Two years later, using
quantum field theory on a black hole background space-time,
Hawking \cite{swh} showed that black holes in fact radiate quantum
mechanically as though they are black bodies at temperature $T =
\hbar\kappa/2\pi$. Using the analogy with the first law, one can
then conclude that the black hole entropy should be given by
$S_{\rm BH} = a_{\rm hor}/4G\hbar$. This conclusion is striking
and deep because it brings together the three pillars of
fundamental physics ---general relativity, quantum theory and
statistical mechanics. However, the argument itself is a rather
hodge-podge mixture of classical and semi-classical ideas,
reminiscent of the Bohr theory of atom. A natural question then
is: what is the analog of the more fundamental,
Pauli-Schr\"odinger theory of the Hydrogen atom? More precisely,
what is the statistical mechanical origin of black hole entropy?
What is the nature of a quantum black hole and what is the
interplay between the quantum degrees of freedom responsible for
entropy and the exterior curved geometry? Can one derive the
Hawking effect from first principles of quantum gravity? Is there
an imprint of the classical singularity on the final quantum
description, e.g., through `information loss'?

$\bullet$ \textit{Planck scale physics and the low energy world:}
In general relativity, there is no background metric, no inert
stage on which dynamics unfolds. Geometry itself is dynamical.
Therefore, as indicated above, one expects that a fully
satisfactory quantum gravity theory would also be free of a
background space-time geometry. However, of necessity, a
background independent description must use physical concepts and
mathematical tools that are quite different from those of the
familiar, low energy physics. A major challenge then is to show
that this low energy description does arise from the pristine,
Planckian world in an appropriate sense, bridging the vast gap of
some 16 orders of magnitude in the energy scale. In this
`top-down' approach, does the fundamental theory admit a
`sufficient number' of semi-classical states? Do these
semi-classical sectors provide enough of a background geometry to
anchor low energy physics? Can one recover the familiar
description? Furthermore, can one pin point why the standard
`bottom-up' perturbative approach fails? That is, what is the
essential feature which makes the fundamental description
mathematically coherent but is absent in the standard perturbative
quantum gravity?

There are of course many more challenges: the issue of time, of
measurement theory and the associated questions of interpretation
of the quantum framework, the issue of diffeomorphism invariant
observables and practical methods of computing their properties,
practical methods of computing time evolution and S-matrices,
exploration of the role of topology and topology change, etc etc.
However, it is our view that the three issues discussed in detail
are more basic from a physical viewpoint because they are rooted
in general conceptual questions that are largely independent of
the specific approach being pursued; indeed they have been with us
longer than any of the current leading approaches.

\subsection{Organization}
\label{s1.3}

In recent years, a number of these fundamental physical issues
were addressed in loop quantum gravity. These include:  i) A
natural resolution of the big-bang singularity in homogeneous,
isotropic quantum cosmology [103-117]; ii) A statistical
mechanical derivation of the horizon entropy, encompassing
astrophysically interesting black holes as well as cosmological
horizons [122-141]; and, iii) The introduction of semi-classical
techniques to make contact between the background independent,
non-perturbative theory and the perturbative, low energy physics
in Minkowski space [142-160]. In addition, advances have been made
on the mathematical physics front. In particular, these include:
iv) A demonstration that all Riemannian geometric operators have
discrete eigenvalues, implying that the space-time continuum is
only an approximation [65-80]; v) A systematic formulation of
quantum Einstein equations in the canonical approach [85-102];
and, vi) The development of spin-foam models which provide
background independent path integral formulations of quantum
gravity [161-173]. These developments are also significant. For
example, in contrast to v), quantum Einstein's equations are yet
to be given a precise mathematical meaning in quantum
geometrodynamics ---a canonical approach that predates loop
quantum gravity by two decades or so--- because the products of
operators involved are divergent.

All these advances spring from a detailed quantum theory of
geometry that was systematically developed in the mid-nineties.
This theory is, in turn, an outgrowth of two developments: a)
Formulation of general relativity (and supergravity) as a
dynamical theory of connections, with the same phase space as in
Yang-Mills theories [12-16]; and, b) heuristic but highly
influential treatments of quantum theories of connections in terms
of loops [33-38].%
\footnote{This is the origin of the name `loop quantum gravity'.
Even though loops play no essential role in the theory now, for
historical reasons, this name is widely used. The current
framework is based on graphs introduced in \cite{al2,jb1,jl-sn}.}
In this review, we will first provide a brief but self-contained
and pedagogical introduction to quantum geometry and then discuss
its applications to problems mentioned above.

The article is organized as follows. In section \ref{s2} we recall
connection formulations of general relativity. (Readers who are
primarily interested in quantum geometry rather than dynamical
issues of general relativity may skip this section in the first
reading.) The next four sections present the basics of quantum
theory. In section \ref{s3} we summarize the overall strategy used
in the construction of quantum kinematics; in section \ref{s4}, we
discuss background independent formulations of general quantum
theories of connections; in section \ref{s5}, the basics of
quantum Riemannian geometry and in section \ref{s6} the basics of
quantum dynamics. These sections are self-contained and the reader
is referred to the original papers only for certain proofs,
technical subtleties and details that are interesting in their own
right but not essential to follow the general approach. Sections
\ref{s7}-\ref{s9} are devoted to applications of quantum geometry
and a summary of current directions, where the treatment is less
pedagogical: while the main ideas are spelled out, the reader will
have to go through at least some of the original papers to get a
thorough working knowledge. Section \ref{s10} contains a summary
and the outlook.

For simplicity, most of the discussion in the main body of the
review is focussed on the gravitational field. There is a large
body of work on coupling of gauge, fermionic and scalar fields to
gravity where the quantum nature of underlying geometry modifies
the physics of matter fields in important ways. Appendix \ref{a1}
illustrates these issues using the Einstein-Maxwell theory as an
example. Appendix \ref{a2} contains a list of symbols which are
frequently used in the review.

For a much more detailed review, at the level of a monograph, see,
\cite{ttrev}. Less pedagogical overviews, at the level of plenary
lectures in conferences, can be found in \cite{crrev,aarev}.

\section{Connection theories of gravity}
 \label{s2}

General relativity is usually presented as a theory of metrics.
However, it can also be recast as a dynamical theory of connections.%
\footnote{Indeed, in the late forties both Einstein and
Schr\"odinger had recast general relativity as a theory of
connections. However, the resulting theory was rather complicated
because they used the Levi-Civita connection. Theory simplifies if
one uses spin-connections instead.}
Such a reformulation brings general relativity closer to gauge
theories which describe the other three fundamental forces of
Nature in the sense that, in the Hamiltonian framework, all
theories now share the same kinematics. The difference, of course,
lies in dynamics. In particular, while dynamics of gauge theories
of other interactions requires a background geometry, that of
general relativity does not. Nonetheless, by a suitable
modification, one can adapt quantization techniques used in gauge
theories to general relativity. We will see in sections
\ref{s4},\ref{s5} and \ref{s6} that this strategy enables one to
resolve the functional analytic difficulties which have prevented
`geometrodynamical' approaches to quantum gravity, based on
metrics, to progress beyond a formal level.

In this section, we will present a self-contained introduction to
connection formulations of general relativity. However, we will
not follow a chronological approach but focus instead only on
those aspects which are needed in subsequent sections. For a
discussion of other issues, see [2, 12-23].

Our conventions are as follows. $\man$ will denote the
4-dimensional space-time manifold which we will assume to be
topologically $\M\times \Real$, equipped with a fixed orientation.
For simplicity, in this section we will assume that $\M$ is an
oriented, compact 3-manifold without boundary. (Modifications
required to incorporate asymptotic flatness can be found in
\cite{aa1} and those needed to allow an isolated horizon as an
inner boundary can be found in \cite{afk,abl2}.) For tensor fields
(possibly with internal indices), we will use Penrose's abstract
index notation. The space-time metric will be denoted by
$g_{\mu\nu}$ and will have signature -,+,+,+ (or, occasionally,
+,+,+,+. In the Lorentzian case, space-time will be assumed to be
time-orientable.) The torsion-free derivative operator compatible
with $g_{\mu\nu}$ will be denoted by $\nabla$ and its curvature
tensors will be defined via $R_{\alpha\beta\gamma}{}^\delta
K_\delta = 2\nabla_{[\alpha} \nabla_{\beta]} K_\gamma$;
$R_{\alpha\beta} = R_{\alpha\beta\gamma}{}^\beta$; and $R=
g^{\alpha\beta}R_{\alpha\beta}$. For the tetrad formalism, we fix
a 4-dimensional vector space $V$ equipped with a fixed metric
$\bar{\eta}_{IJ}$ of signature -,+,+,+ (or +,+,+,+), which will
serve as the `internal space'. Orthonormal co-tetrads will be
denoted by $e_\alpha^I$; thus $g_{\alpha\beta} =\bar{\eta}_{IJ}
e^I_\alpha e^J_\beta$. In the passage to the Hamiltonian theory,
the metric on a space-like Cauchy surface $M$ will be denoted by
$q_{ab}$ and the spatial co-triads  will be denoted by $e_a^i$.
Finally, we will often set $k = 8\pi G$ where $G$ is Newton's
constant. Due to space limitation, we will focus just on the
gravitational part of the action and phase space. For inclusion of
matter, see e.g., \cite{art,afk}; for extension to supergravity,
see, e.g., \cite{sugra}; and for ideas on extension to higher
dimensions, see \cite{fk}.

\subsection{Holst's modification of the Palatini action}
\label{s2.1}

In the Palatini framework, the basic gravitational variables
constitute a pair $(e_\mu{}^I{}, \omega_\mu{}^I{}_J)$ of 1-form
fields on $\man$ taking values, respectively, in $V$ and in the
Lie algebra $so(\bar\eta)$ of the group $SO(\bar\eta)$ of the
linear transformations of $V$ preserving $\bar\eta_{IJ}$. Because
of our topological assumptions, the co-frame fields $e_\mu{}^I$
are defined globally; they provide an isomorphism between
$T_x\man$ and $V$ at each $x\in \man$. The action is given by
\be {\bf S}_{(P)}(e,\omega)\ =\ \frac{1}{4k}\int_\man
\epsilon_{IJKL}\,e^I\wedge e^J\wedge\, \Omega^{KL}\ee
where $\epsilon_{IJKL}$ is an alternating tensor on $V$ compatible
with $\bar\eta_{IJ}$ such that the orientation of
$\epsilon_{\a\beta\gamma\delta} = \epsilon_{IJKL}\, e^I_\a
e^J_\beta e^K_\gamma e^L_\delta$ agrees with the one we fixed on
$\M$ and
\be \Omega\ :=\ d\omega + \omega\wedge\omega,\ee
is the curvature of the connection 1-form $\omega_\mu{}^I{}_J$.
The co-frame $e_\mu^I$ determines a space-time metric $g_{\mu\nu}
= \eta_{IJ}\, e^I_\mu e^J_\nu$. Thus, in contrast to the more
familiar Einstein-Hilbert action, ${\bf S}_{(P)}$ depends on an
additional variable, the connection $\omega_\mu^{IJ}$. However,
the equation of motion obtained by varying the action with respect
to the connection implies that $\omega_\mu^{IJ}$ is in fact
completely determined by the co-frame:
\be de + \omega\wedge e\ =\ 0.\label{De} \ee
If we now restrict ourselves to histories on which the connection
is so determined, ${\bf S}_{(P)}$ reduces to the familiar
Einstein-Hilbert action:
\be {\bf S}_{(P)}(e,\omega(e))\ =\ \frac{1}{2k}\, \int_\man d^4x\,
\sqrt{|\det g|} R\, . \ee
where $R$ is the scalar curvature of $g_{\mu\nu}$. Therefore, the
equation of motion for the metric is the same as that of the
Einstein-Hilbert action.

The action ${\bf S}_{(P)}$ is invariant under diffeomorphisms of
$\man$ as well as \emph{local} $SO(\bar\eta)$ transformations
\be (e,\omega) \mapsto (e',\,\omega')\ =\ (b^{-1}e, {b^{-1}}\omega
b + b^{-1}db)\, .\label{rotations}\ee
It is straightforward but rather tedious to perform a Legendre
transform of this action and pass to a Hamiltonian theory
\cite{abj}. It turns out that the theory has certain second class
constraints and, when they are solved, one is led to a triad
version of the standard Hamiltonian theory of geometrodynamics;
all reference to connection-dynamics is lost. This can be remedied
using the following observation: there exists another invariant,
constructed from the pair $(e, \omega)$, with the remarkable
property that its addition to the action does not change equations
of motion. The modified action, discussed by Holst \cite{holst},
is given by%
\footnote{This modification was strongly motivated by the very
considerable work on the framework based on(anti-)self-dual
connections in the preceding decade (see e.g. \cite{aa1})
particularly by the discovery of an action for general relativity
using these variables \cite{lagrangian}. Also, our presentation
contains several new elements which, to our knowledge, have not
appeared in the literature before.}:
\be {\bf S}_{(H)}(e,\omega)\ =\ {\bf S}_{(P)}(e,\omega) -
\frac{1}{2k\gamma}\int_\man e^I\wedge e^J\wedge\Omega_{IJ}
\label{holst}\ee
where $\gamma$ is an arbitrary but fixed number, called the
\emph{Barbero-Immirzi} parameter. For applications to quantum
theory, it is important to note that \emph{ $\gamma$ can not be
zero.} The purpose of this section is to analyze this action and
the Hamiltonian theory emerging from it. In sections \ref{s2.2}
and \ref{s2.3} we will show that the Hamiltonian theory can be
naturally interpreted as a background independent, dynamical
theory of connections.

Recall that, in Yang-Mills theories, one can also add a
`topological term' to the action which does not change the
classical equations of motion because its integrand can be
re-expressed as an exterior derivative of a three-form. In the
present case, while the extra term is not of topological origin,
because of the first Bianchi identity it vanishes identically on
histories on which (\ref{De}) holds. Therefore the situation is
similar in the two theories in some respects: in both cases, the
addition of the term does not change the classical equations of
motion and it induces a canonical transformation on the classical
phase space which fails to be unitarily implementable in the
quantum theory. Consequently, the parameter $\gamma$ is in many
ways analogous to the well-known $\theta$ parameter in the
Yang-Mills theory \cite{rt1,gop}. Just as the quantum theory has
\emph{inequivalent} $\theta$-sectors in the Yang-Mills case, it
has \emph{inequivalent} $\gamma$-sectors in the gravitational
case.

We will conclude this preliminary discussion by exhibiting the
symplectic structure in the covariant phase space formulation.
Here the phase space $\covps$ is taken to be the space of
(suitably regular) solutions to the field equations on $\man$. To
define the symplectic structure, one follows the following general
procedure. Denote by $\bar\delta \equiv (\bar\delta e, \bar\delta
\omega)$ tangent vectors in the space of histories. Since field
equations are satisfied on $\covps$, the change in the Lagrangian
4-form $L_4$ under a variation along $\bar\delta$ is for the form
$$(\bar\delta L_4)|_{\covps} = d L_3(\bar\delta)$$
for some 3-form $L_3$ on $\man$ which depends linearly on
$\bar\delta$. One can now define a 1-form ${\bf \Theta}$ on the
space of histories via: ${\bf \Theta} (\bar\delta) := \int_M\,
L_3(\bar\delta)$. The symplectic structure ${\bf \Omega}$ is
simply the pull-back to $\covps$ of the curl of ${\bf \Theta}$ on
the space of histories. In our case, $\bar\delta L_{(H)}$ is given
by
$$ \bar\delta L_{(H)}\ =\ - \frac{1}{2\gamma k
}\,\,
d\left[e^I\wedge e^J\wedge \bar\delta\left(
\omega_{IJ}-\frac{\gamma}{2}\epsilon_{IJKL}\omega^{KL}
\right)\right]. $$
whence the symplectic structure is given by:
\be {\bf \Omega}(\delta_1,\delta_2)\ =\
-\frac{1}{k\gamma}\int_{\M} \left[\delta_{[1}(e^I\wedge
e^J)\right]\wedge
\left[\delta_{2]}(\omega_{IJ}-\frac{\gamma}{2}\epsilon_{IJKL}\omega^{KL})
\right] \label{symp}\ee
for all tangent vectors $\delta_1$ and $\delta_2$ to $\covps$.
{}From general considerations, it follows that the value of the
integral is independent of the specific choice of a Cauchy surface
$\M$ made in its evaluation.

Using the fact that the tangent vectors to $\covps$ must, in
particular, satisfy the linearized version of (\ref{De}) it is
easy to verify that the $\gamma$-dependent term in (\ref{symp})
vanishes identically: Not only is $\covps$ independent of the
value of $\gamma$ (see footnote 5) but so is the symplectic
structure ${\bf \Omega}$ on it. As is clear from (\ref{symp}), the
momentum conjugate to $e^I\wedge e^J$, on the other hand, does
depend on the choice of $\gamma$. Thus, as noted above, like the
$\theta$ term in the Yang-Mills theory, the $\gamma$ term in
(\ref{holst}) only induces a canonical transformation on the phase
space.

In effect the canonical transformation is induced by the map
$$
 X_{IJ}\ \mapsto \frac{1}{2}
\left(X_{IJ}\ -\ \frac{\gamma}{2}\epsilon_{IJKL}X^{KL}\right) $$
on $so(\bar\eta)$. It is easy to verify that this map is a vector
space isomorphism on $so(\bar\eta)$ except when
$$ \gamma^2\ =\ \sigma := {\rm sgn}(\det\bar\eta{})\, , $$
the sign of the determinant of the metric tensor $\bar\eta_{IJ}$.%
\footnote{As a result, for generic values of $\gamma$, the
equation of motion for the connection resulting from variation of
${\bf S}_{(H)}$ with respect to $\omega_\mu^{IJ}$ is again
(\ref{De}). Hence the space of solutions obtained by varying ${\bf
S}_{(H)}$ is the same as that obtained by varying ${\bf S}_{(P)}$.
For the exceptional values, this equation says that the (anti-)
self dual part of $\omega_\mu^{IJ}$ equals the (anti-) self dual
part of the connection compatible with the co-frame $e_\mu^I$.
However, it is again true that the spaces of solutions obtained by
extremizing ${\bf S}_{(H)}$ and ${\bf S}_{(P)}$ are the same
\cite{aa1}.}
At these exceptional values of $\gamma$, the map is a projection
onto the subspace of $so(\eta)$ corresponding to the eigenvalue
$-\gamma\sigma$ of the Hodge-dual operator $\star: X_{IJ}\mapsto
\frac{1}{2}\epsilon_{IJ}{}^{KL}X_{KL}$. Furthermore, in this case
the map is a \emph{Lie algebra homomorphism}. In the Riemannian
case (when $\bar\eta$ has signature +,+,+,+) this occurs for
$\gamma= \pm 1$ while in the Lorentzian case (when $\bar\eta$ has
signature -,+,+,+) it occurs for $\gamma = \pm i$. In all these
exceptional cases the theory has a richer geometrical structure.
In particular, the combination $\textstyle{1\over 2}(\omega_{IJ} -
\gamma\,{}^\star \omega_{IJ})$ that occurs in the symplectic
structure is again a (half-flat) connection.

Chronologically, the background independent approach to quantum
gravity summarized in this review originated from a reformulation
of general relativity in terms of these half-flat connections
\cite{aa2,aa3}. It turns out that all equations in the classical
theory simplify considerably and underlying structures become more
transparent in these variables. They are also closely related to
Penrose's non-linear gravitons \cite{nlg} and Newman's H-space
constructions \cite{hspace}. In the Riemannian signature, one can
continue to use these variables also in the quantum theory. In the
Lorentzian case, on the other hand, the half-flat connections take
values in the Lie algebra of non-compact groups and functional
analysis on spaces of such connections is still not sufficiently
well-developed to carry out constructions required in the quantum
theory. Therefore, in the Lorentzian case, most progress has
occurred by working in sectors with real values of $\gamma$ where,
as we will see, one can work connections with compact structure
groups.

In section \ref{s2.2} we will summarize the situation with
half-flat connections in the Riemannian case and in section
\ref{s2.3} we will discuss the Lorentzian theory using real valued
$\gamma$ sectors.

\subsection{Riemannian signature and half flat connections}
\label{s2.2}

\subsubsection{Preliminaries}
\label{2.2.1}

Let us then assume that $\bar\eta_{IJ}$ is positive definite.
Since $\sigma = 1$, the half flat case corresponds to setting
$\gamma = \pm 1$. Let us set
\be \omega^{(+)}{}_{IJ}\ =\ \frac{1}{2}\left( \omega_{IJ} -
\frac{\gamma}{2}\epsilon_{IJ}{}^{KL}\omega_{KL} \right) \ee
so that $\omega^{(+)}$ is the anti-self dual part of $\omega$ if
$\gamma = 1$ and self dual, if $\gamma = -1$. In these cases, the
Holst action simplifies to:
\be {\bf S}_{(H)}(e, \omega^{(+)}) \ =\
-\frac{1}{k\gamma}\int_\man
\Sigma_{(+)}^{IJ}\wedge\Omega^{(+)}{}_{IJ} \ee
where  $\Sigma_{(+)}^{IJ}$ is the (anti-)self dual part of
$e^I\wedge e^J$,
$$ \Sigma_{(+)}^{IJ} = \frac{1}{2}\, \left (e^I\wedge e^J -
\frac{\gamma}{2}\epsilon^{IJ}{}_{KL}e^K\wedge e^L\right)\, , $$
and $\Omega^{(+)}_{IJ}$ is both the (anti-)self dual part of
$\Omega_{IJ}$ \emph{and} the curvature of $\omega^{(+)}_{IJ}$:
$$ \Omega^{(+)} = d\omega^{(+)}\ +\ \omega^{(+)}\wedge
\omega^{(+)}\, . $$
\emph{Note that the theory under consideration is full
(Riemannian) general relativity}; we are just describing it in
terms of the fields $(e^I, \omega^{(+)}_{IJ})$ where
$\omega^{(+)}_{IJ}$ is a half flat (i.e., self dual or anti-self
dual) connection.

The symplectic form (\ref{symp}) now simplifies to:
\ba \label{sympl2} {\bf \Omega}(\delta_1,\delta_2)\ &=&\
-\frac{2}{k\gamma}\int_{\M}
\left[\delta_{[1}\Sigma_{(+)}^{IJ}\right]\wedge
\left[\delta_{2]}\omega^{(+)}{}_{IJ}\right] \
\nonumber\\
&=& \ \int_\M d^3x\, \left[\delta_{1}P^a_{IJ}\,\delta_{2}A^{IJ}_a
       - \delta_{2}P^a_{IJ}\,\delta_{1}A^{IJ}_a
\right]\, ,\ea
where $A_{IJ}$, the pullback to $\M$ of $\omega^{(+)}_{IJ}$,
represents the configuration variable and
$$P^a_{IJ} := -\frac{1}{2k\gamma}\,\,{\eta}^{abc}\,
\Sigma^{(+)}_{bcIJ}\, ,$$
its canonically conjugate momentum. Here and in what follows
$\eta^{abc}$ will denote the metric independent Levi-Civita
density on $\M$ whose orientation is the same as that of the fixed
orientation on $\M$. Hence $P^a_{IJ}$ is a pseudo vector density
of the weight $1$ on $\M$.%
\footnote{In terms of coordinates, for any smooth field $V^{IJ}$
and 1-form $f_a$ on $\M$, the 3-form $V^{IJ} f_a P^a_{IJ}
dx^1\wedge dx^2\wedge dx^3$ is a volume element on $\M$ which is
independent of the choice of coordinates $(x^1,x^2,x^3)$.}

\subsubsection{The Legendre transform}
\label{s2.2.2}

Let us introduce on $\man$ a smooth (`time') function $t$ such
that $dt$ is everywhere non-zero and each $t= {\rm const}$ slice
is diffeomorphic with $\M$. Introduce a vector field $t^\a$ such
that $t^\a\nabla_a t = 1$. Thus, $t^\a$ is to be thought of as the
`time-evolution vector field'. Denote by $n^\a$ the unit normal to
the $t= {\rm const}$ slices $\M$ and decompose $t^\a$ as $t^\a = N
n^\a + N^\a$ with $N^{\a} n_\a =0$. The function $N$ is called the
\emph{lapse} and the vector field $N^\a$ the \emph{shift}. We will
denote by $q^a_\a$ and $q_a^\a$ the projection operator on to
vector and co-vector fields on $\M$. Finally, a tensor field
$T^{\a\ldots\beta}{}_{\g\ldots\delta}$ which is orthogonal in each
of its indices to $n^\mu$  will be identified with its projection
$T^{a\ldots b}{}_{c\ldots d} := q^a_\a\ldots q^b_\beta
q^\g_c\ldots q^\delta_d\, T^{\a\ldots\beta}{}_{\g\ldots\delta}$.

With these preliminaries out of the way, it is now straightforward
to perform the Legendre transform. The calculation is remarkably
short (especially when compared to the Legendre transform in the
metric variables; see, e.g. page 47 of \cite{aa1}). In terms of
fields $A_a^{IJ}$ and $P^a_{IJ}$ introduced above, one obtains%
\footnote{Here, and in the remainder of this paper, in the Lie
derivative of a field with internal indices will be treated simply
as scalars (i.e., ignored). Thus, ${\cal L}_t A_a^{IJ} = t^b
\partial_b A_a^{IJ} + A_b^{IJ} \partial_a t^b$.}:
\be \label{lt1} {\bf S}_{(H)} = \int dt\int_M d^3x\,
\left(P^a_{IJ}\, {\cal L}_t {A}_a^{IJ} -
h_{(+)}(A,P,N,N^a,\omega^{(+)}\!\cdot t)\right) \label{S+}\ee
where the Hamiltonian density $h_{(+)}$ is given by
 \be h_{(+)} = -(\omega^{(+)}_{IJ}\cdot t)\, G^{IJ}
 + N^a C^{(+)}_a + N C_{(+)}
\label{h+}\, , \ee
with
\ba \label{constraints1}
 G_{IJ} &:=&\D^{(+)}_aP^a_{IJ}:=\partial_aP^a_{IJ}
 + A_{aI}{}^{K} P^a_{KJ} + A_{aJ}{}^{K} P^a_{IK}\nonumber\\
C^{(+)}_{a}\ &:=&\ P^b_{IJ}F^{IJ}_{ab}\nonumber\\
C_{(+)}\ &:=&\ - \,\f{k}{\sqrt{|\det q|}}\,\,
P^a_{I}{}^{J}P^{b}_{J}{}^K\, F_{ab}{}_K{}^I\, .\ea
Here $F_{ab}^{IJ}$ is the curvature of $A_a^{IJ}$, $F = dA +
A\wedge A$ and $q$ is the determinant of the 3-metric
$$q_{ab}:= q_a^\a q_b^\beta\, g_{\a\beta} $$
on $\M$. The form of (\ref{h+}) confirms that, as suggested by
(\ref{sympl2}), we should regard $A_a^{IJ}$ as the configuration
variable and $P^a_{IJ}$ as its momentum. The momentum is related
in a simple way to the 3-metric:
$$ -\Tr P^aP^b\ =\ P^a_{IJ}P^{bIJ}\ =\ \frac{1}{k^2}(\det q)\,
q^{ab}\, , $$

Note that $\omega\cdot t, N$ and $N^a$ are Lagrange multipliers;
there are no equations governing them. The basic dynamical
variables are only $A_a^{IJ}$ and $P^a_{IJ}$; all other dynamical
fields are determined by them. Variation of ${\bf S}_{(H)}$ with
respect to these multipliers yields constraints:
\be G_{IJ}\ =0;\quad  C_a^{(+)}\ =0; \quad {\rm and} \quad
C^{(+)}\ = 0. \ee
As is always the case (in the spatially compact context) for
theories without background fields the Hamiltonian is a sum of
constraints. Variations of the action with respect to $A_a^{IJ}$
and $P^a_{IJ}$ yield the equations of motion for these basic
dynamical fields. The three constraints (\ref{constraints1}) and
these two evolution equations are equivalent to the full set of
Einstein's equations.

\subsubsection{The Hamiltonian framework}
\label{s2.2.3}

It follows from the Legendre transform (\ref{lt1}) that the
\emph{canonical} phase space $\canps$ consists of canonically
conjugate pairs of fields $(A_a^{IJ}, P^a_{IJ})$ of $\M$. The only
non-trivial Poisson bracket is:
\be \{A^{IJ}_a(x),P_{KL}^b(y)\} := \frac{1}{2}\left(
\delta^I_{[K}\delta^J_{L]} -
\frac{\gamma}{2}\delta^I_{[M}\delta^J_{N]}\epsilon^{MN}{}_{KL}
\right) \delta^b_a\delta(x,y)  \ee
A key point is that the configuration variable $A_a^{IJ}$ is again
a connection on the 3-manifold $\M$ but the structure group is now
the \emph{spin group} ${\rm SO}^{(+)}(\bar\eta)$ (which, in the
Riemannian case now under consideration, is isomorphic to
$SU(2)$).%
\footnote{The full group ${\rm SO}(\bar\eta)$ does admit an action
on the phase space, given by $(P,\, A)\mapsto \left(b^{-1}Pb,\
b^{-1}A b + (b^{-1}db)^{(+)}\right)$, where ${}^{(+)}$, stands for
the projection onto $so^{(+)}(\bar\eta)$ in ${\rm so}(\bar\eta)$.
However, because of the projection, $A$ does not transform as an
${\rm SO}(\bar\eta)$ connection.}
Thus, in the Hamiltonian framework, general relativity has been
cast as a dynamical theory of a spin connection.

The basic canonically conjugate variables are subject to three
sets of constraints, spelled out in (\ref{constraints1}). It is
easy to verify that the Poisson bracket between any two
constraints vanishes on the constraint surface; in Dirac's
terminology, they are of \emph{first class}. The first constraint,
$G_{IJ}^{(+)}$, generates internal gauge transformations in ${\rm
SO}^{(+)}(\bar\eta)$. Modulo these gauge rotations, the second,
$C_a^{(+)}$, generates diffeomorphisms on $\M$, and the third,
$C^{(+)}$, generates `evolutions' along $Nn^\a$. Using the
relation between $P^a_{IJ}$ and the 3-metrics $q_{ab}$ on $\M$,
one can show that these equations are equivalent to the full set
of Einstein's equations. However, one can work just with the
connections $A_a^{IJ}$ and their conjugate momenta, without any
direct reference to metrics, even when gravity is coupled to
matter \cite{art}. In this sense, gravity can be regarded as a
`gauge theory' which has the same phase space $\canps$ as that of
a ${\rm SO}^{(+)}(\bar\eta)$ Yang-Mills theory but a fully
constrained dynamics which does not refer to a background
space-time metric.

\subsection{Generic real value of $\gamma$}
\label{s2.3}

The formulation of general relativity as a dynamical theory of
half-flat connections, presented in section \ref{s2.2} has been
studied in detail also for Lorentzian signature [2, 12-23].
However, in that case, certain subtleties arise because the
connection is complex-valued and the structure group is
non-compact. We have chosen to bypass these issues because, as
explained in section \ref{s2.1}, for passage to quantum theory we
have in any case to use compact structure groups, i.e., real
values of $\gamma$. Therefore, in this sub-section we will let
$\gamma$ take any non-zero real value. Although we are now
primarily interested in the -,+,+,+ signature, our analysis will
apply also to the +,+,+,+ case.

\subsubsection{Preliminaries}
 \label{s2.3.1}

It is convenient to first carry out a partial gauge fixing. Let us
fix an internal vector field $n^I$ with $n^In_I = \sigma$ (the
signature of $\bar\eta_{IJ}$). We will require it to be constant
(in the sense that from now on, we will restrict ourselves to flat
derivative operators $\partial$ which annihilate $n^I$, in
addition to $\bar\eta_{IJ}$). Let $V_\bot$ be the 3-dimensional
subspace of $V$ orthogonal to $n^I$. Elements of $V_\bot$ will
carry lower case superscripts, $i,j,\ldots k$ and the projection
operator on to $V_\bot$ will be denoted by $q_I^i$. In particular,
then,
$$ \eta_{ij} = q_i^I q_j^J\, \bar\eta_{IJ} $$
is the induced metric on $V_\bot$. Because we have fixed $n^I$,
the group ${\rm SO}(\bar\eta)$ is now reduced to its subgroup
${\rm SO}(\eta)$ which leaves $n^I$ invariant. Finally, the
alternating tensor $\epsilon_{IJKL}$ on $V$ naturally induces an
alternating tensor $\epsilon_{ijk}$ on $V_\bot$ via:
$$ \epsilon_{ijk} = q_i^I q_j^J q_k^K n^L\, \epsilon_{LIJK}$$

Next, let us introduce a `time function' $t$ and the associated
structure as in the beginning of section \ref{s2.2.2}, with the
following additional provisos if the signature is Lorentzian: the
vector field $t^\a$ is future directed and $n^\a$ is the future
directed unit time-like normal to $\M$. We will now allow only
those co-frame fields $e_\a^I$ which are `compatible' with the
fixed $n^I$ in the sense that $n^\a := n^I e^\a_I$ is the unit
normal to the given foliation. (Note that every co-frame is gauge
related to one satisfying this condition; see (\ref{rotations}).)
Each of these co-frames $e_\a^I$ naturally defines an orthonormal
co-triad $e^i_a := e^I_\a q^i_I q^\a_a$: on each leaf $\M$ of the
foliation, the induced metric $q_{ab}$ is given by $q_{ab} = e_a^i
e_b^j \eta_{ij}$. Similarly, the connection 1-form
$\omega_\a^{IJ}$ naturally defines two ${\rm so(3)}$-valued
1-forms on $\M$:
\be \Gamma^i_a :=  \f{1}{2}\, q_a^\a q^i_I\, \epsilon^{IJ}{}_{KL}
n_J \,\omega_\a^{KL}  \quad {\rm and} \quad  K_a^i := q_I^i q_a^\a
\,\omega_\a^{IJ}n_J\, . \ee
These 1-forms have natural geometric interpretations. $\Gamma_a^i$
is a ${\rm so}(\eta)$-connection on $\M$ and it is compatible with
$e_a^i$ if $\omega_\a^{IJ}$ is compatible with $e_\a^{I}$. Thus,
if (\ref{De}) holds, we have:
\be de^i + \epsilon^i{}_{jk}\Gamma^j\wedge e^k = 0. \ee
$K_a^i$ is the extrinsic curvature of $\M$ if (\ref{De}) holds:
\be \label{K} K_a^i =  (q^\a_a q_\beta^b\, \nabla_\a
n^\beta)\,e_b^i \ee

In terms of these fields, the symplectic structure (\ref{symp})
can be re-expressed as:
\be {\bf \Omega}(\delta_1,\delta_2)\ =\ \int_\M d^3x\,
\left(\delta_{1}P^a_i\,\delta_{2}A^i_a - \delta_{2}P^a_i\,
\delta_{1}A^i_a\right) \ee
where
\be P^a_i := \frac{1}{2k\gamma}\, e^j_be^k_c\, \eta^{abc}\,
\epsilon_{ijk}, \quad {\rm and} \quad A^i_a := \Gamma^i_a - \sigma
\gamma K^i_a \label{PA}\ee
Note that $A_a^i$ is a connection 1-form on $\M$ which takes
values in ${\rm so}(\eta)$. $P^a_i$ is again a vector density of
weight 1 on $M$ which now takes values in (the dual of) ${\rm
so}(\eta)$. Geometrically, it represents an orthonormal triad
$\tilde{E}^a$ of density weight 1 on $\M$:
\be k\gamma P^a_i\ =\  \sqrt{|\det q|}\, e^a_i \equiv
\tilde{E}^a_i \quad {\rm whence} \quad |\det q|\, q^{ab} = k^2
\gamma^2 P^a_i P^b_j\, \eta^{ij} \label{P}\ee
where $\det q$ is the determinant of the 3-metric $q_{ab}$ on
$\M$.

Let us summarize. Through gauge fixing, we first reduced the
internal gauge group from $\SO(\bar{\eta})$ to $\SO(\eta)$. The
new configuration variable $A_a^i$ is a ${\rm so}(\eta)$-valued
connection on $M$, constructed from the spin-connection
$\Gamma_a^i$ compatible with the co-triad $e_a^i$ and the
extrinsic curvature $K_a^i$. Apart from a multiplicative factor $k
\gamma$, the conjugate momentum $P^a_i$ has the interpretation of
a triad with density weight 1. Note that the relation (\ref{PA})
between the canonical variables $A_a^i, P^a_i$ and the geometrical
variables $e_a^i$ and $K_a^i$ holds also in the half-flat case; it
is just that there is also an additional restriction, $\sigma^2
\gamma^2 =\pm 1$.

\subsubsection{The Legendre transform}
\label{s2.3.2}

Let us return to the Holst action (\ref{holst}) and perform the
Legendre transform as in section \ref{s2.2.2}. Again, the
calculations are simple but the full expression of the resulting
Hamiltonian density $h$ is now more complicated. As before one
obtains:
\be \label{lt2} {\bf S}_{(H)} = \int dt\int _\M d^3x\,
\left(P^a_{i}{\cal L}_t{A}_a^{i} -
h(A^i_a,P^a_i,N,N^a,\Gamma^\cdot t)\right) \ee
with $h$ given by
\be  h = (\omega^{i}\cdot t) G_{i} + N^a C_a + N C\label{h}
\ee
Again $\omega^{i}\cdot t := -\textstyle{1\over 2}\epsilon^{ijk}
\omega_{jk}\cdot t,  N^a$ and $N$ are Lagrange multipliers.
However, now the accompanying constraints acquire additional
terms:
\ba \label{constraints2}
 G_{i}\ =\, \D_aP^a_{i} &:=&\partial_aP^a_{i} + \epsilon_{ij}{}^kA_a^j P^a_{k}
\quad C_a\ =\ P^b_{i}F^{i}_{ab} -
\frac{\sigma-\gamma^2}{\sigma\gamma}K_a^i\,G_i\nonumber\\
C\ &=&\ \frac{k\gamma^2}{2\sqrt{|\det q|}}P^a_{i}P^{b}_j
\left[\epsilon^{ij}{}_k F^{k}_{ab} + (\sigma-\gamma^2)
2K^i_{[a}K^j_{b]}\right] \nonumber\\
&+& (\gamma^2-\sigma)k\,\partial_a\left(\frac{P^a_i}{\sqrt{|\det
q|}}\right)\, G^i \ea
Here, $F_{ab}^k$ is the curvature of the connection $A_a^i$ and
$|\det q|$ can be expressed directly in terms of $P^a_i$:
\be |\det q| \ =\ \frac{(k\gamma)^3}{\sqrt{|\det \eta|}} \det
P\label{detq}\ee
%
Thus, the overall structure of the constraints is very similar to
that in the half flat case. However there is a major new
complication in the detailed expressions of constraints: now they
involve also $K_a^i= (1/\sigma\gamma) (\Gamma_a^i-A_a^i) $ and
$\Gamma_a^i$ is a non-polynomial function of $P^a_i$.
\footnote{Although the possibility of using  real  $\gamma$ was
noted already in the mid-eighties, this choice was ignored in the
Lorentzian case because the term $K_a^i$ seemed unmanageable in
quantum theory. The viewpoint changed with Thiemann's discovery
that this difficulty can be overcome. See section \ref{s6}.}
(Since these terms are multiplied by $(\sigma -\gamma^2)$, they
disappear in the half-flat case.)

\subsubsection{Hamiltonian theory}
\label{s2.3.3}

Now the canonical phase space $\canps$ consists of pairs $(A_a^i,
P^a_i)$ of fields on the 3-manifold $M$, where $A_a^i$ is a
connection 1-form  which takes values in ${\rm so}(\eta)$ and
$P^a_i$ is a vector density of weight 1 which takes values in the
dual of ${\rm so}(\eta)$. The only non-vanishing Poisson bracket
is:
\be \{A^{i}_a(x),\, P_{j}^b(y)\} := \delta^i_j
\delta^b_a\delta(x,y)\label{bracket} \ee
Thus, the phase space is the same as that of a Yang-Mills theory
with $SO(\eta)$ as the structure group.  There is again a set of
three constraints, (\ref{constraints2}), which are again of first
class in Dirac's terminology. The basic canonical pair evolves via
Hamilton's equations:
$$ \dot{A}_a^i = \{A_a^i,\, H\}, \quad \dot{P^a_i} = \{P^a_i,\,
H\}$$
where the Hamiltonian is simply $H = \int_M d^3x\, h$. The set of
three constraints and these two evolution equations are completely
equivalent to Einstein's equations. Thus, general relativity is
again recast as a dynamical theory of connections.

Before analyzing the phase space structure in greater detail, we
wish to emphasize two important points. First, note that in the
Hamiltonian theory we simply begin with the fields $(A_a^i,
P^a_i)$; neither they nor their Poisson brackets depend on the
Barbero-Immerzi parameter $\gamma$. Thus, the canonical phase
space is manifestly $\gamma$ independent. $\gamma$ appears only
when we express geometrical fields ---the spatial triad $e^a_i$
and the extrinsic curvature $K_a^i$--- in terms of the basic
canonical variables (see (\ref{P}) and (\ref{K})). The second
point concerns a conceptual difference between the use of half
flat and general connections. The configuration variables in both
cases are connections on $M$. Furthermore, as noted in section
\ref{s2.3.1}, the relation (\ref{PA}) between these connections
and the fields $e_a^i, K_a^i$ is identical in form. However, while
the variable $A_a^{IJ}$ of section \ref{s2.2} is the pull-back to
$\M$ of a \emph{space-time} connection $A_\a^{IJ}$, the variable
$A_a^i$ now under consideration is not so obtained \cite{js}.
{}From the space-time geometry perspective, therefore, $A_a^i$ is
less natural. While this is a definite drawback from the
perspective of the classical theory, it is not a handicap for
canonical quantization. Indeed a space-time geometry is analogous
to a trajectory in particle mechanics and particle trajectories
play no essential role in quantum mechanics.

Finally, let us analyze the structure of constraints. As one would
expect, the first constraint, $G_i =0$ is simply the `Gauss law'
which ensures invariance under internal ${\rm SO}(\eta)$
rotations. Indeed, for any smooth field $\Lambda^i$ on $\M$ which
takes values in ${\rm so}(\eta)$, the function
\be \Gauss(\Lambda) := \int_\M d^3x\, \Lambda^i G_i \ee
on the phase space generates precisely the internal rotations
along $\Lambda^i$:
\be \{A_a^{i},\, \Gauss(\Lambda)\} =  - \D_a\Lambda^{i},\quad {\rm
and}\quad  \{P^a_{i},\Gauss(\Lambda)\}\ =
\epsilon_{ij}{}^k\Lambda^jP^a_{k}.\ee
To display the meaning of the second constraint $C_a$ of
(\ref{constraints2}), it is convenient to remove from it the part
which generates internal rotations which we have already analyzed.
Therefore, For each smooth vector field ${\vec N}$ on $\M$ let us
define
\be C_{\rm Diff}(\vec{N}) := \int_M d^3x\, \left( N^a P^b_i
F_{ab}^i - (N^aA_a^i) G_i\right)  \ee
This constraint function generates diffeomorphisms along ${\vec
N}$:
\be \{A^i_a,\, \C_{\rm Diff}(\vec{N})\} = {\cal L}_{\vec{N}}A^i_a,
\quad {\rm and} \quad \{P_i^a,\, \C_{\rm Diff}(\vec{N})\}= {\cal
L}_{\vec{N}}P_i^a, \ee
Finally, let us consider the third constraint in
(\ref{constraints2}). For quantization purposes, it is again
convenient to remove a suitable multiple of the Gauss constraint
from it. Following Barbero and Thiemann, we will set:
\be \C(N) = \frac{k\gamma^2}{2} \int_\M d^3x\,  N\,
\frac{P^a_{i}P^{b}_j}{\sqrt{|\det q|}} \left[\epsilon^{ij}{}_k
F^{k}_{ab} + 2(\sigma-\gamma^2) K^i_{[a}K^j_{b]}\right]\,
.\label{BT} \ee
As one might expect, this constraint generates time evolution,
`off' $\M$. The Poisson brackets between these specific
constraints are:\\
\be \{\Gauss(\Lambda),\, \Gauss(\Lambda')\}\, =\, \{\Gauss
([\Lambda, \,\Lambda'])\}\, ;\quad\quad \{\Gauss(\Lambda),\,
\C_{\rm Diff}(\vec{N})\}\, =\, -\Gauss({\cal L}_N \Lambda)\, ; \ee
\be \{\C_{\rm Diff}(\vec{N}),\, \C_{\rm Diff}(\vec{N}')\}\, =\,
\C_{\rm Diff}([\vec{N}, \vec{N}'])\, ; \ee
\be \{\Gauss(\Lambda),\, \C{}(N)\} \, =\, 0\, ;  \quad\quad
\{\C_{\rm Diff}(\vec{N}),\, \C{}(M)\} \, =\, -\C{}({\cal L}_N M)\,
; \ee
and
\be \{\C{}(N),\, \C{}(M)\}\, =\, k^2\gamma^2 \sigma \left(\C_{\rm
Diff}(\vec{S}) + \Gauss(S^aA_a)\right) +
(\sigma-\gamma^2)\,\Gauss\left(\frac{[P^a\partial_a N,\,
P^b\partial_b M]} {|\det q|}\right)\, .\ee
In the last equation, the vector field $S^a$ is given by
\be S^a\, =\, (N\partial_b M- M\partial_b N)\,
\frac{P_i^bP^{ai}}{|\det q|}\ee
As in geometrodynamics, the smearing fields in the last Poisson
bracket depend on dynamical fields themselves. Therefore, the
constraint algebra is open in the BRST sense; we have structure
functions rather than structure constants. We will return to this
point in section \ref{s6}.

To summarize, both the Euclidean and Lorentzian general relativity
can be cast as a dynamical theory of (real-valued) connections
with compact structure groups. The price in the Lorentzian sector
is that we have to work with a real value of the Barbero-Immerzi
parameter, for which the expressions of constraints and their
Poisson algebra are more complicated.

\textsl{Remarks}:\\
\indent 1. For simplicity, in this section we focussed just on the
gravitational field. Matter couplings have been discussed in
detail in the literature using half-flat gravitational connections
in the framework of general relativity as well as supergravity
(see, e.g., \cite{art,sugra}). In the matter sector, modifications
required to deal with generic $\g$ values of the Barbero-Immirzi
parameter are minimal.

2. In the purely gravitational sector considered here, the
internal group for general real values of $\gamma$ is ${\rm
SO}(\eta)$. For cases we focussed on, $\eta_{ij}$ is positive
definite whence ${\rm SO}(\eta) = SO(3)$. However, since we also
wish to incorporate spinors, in the remainder of the paper we will
take the internal group to be $\SU(2)$. This will also make the
structure group the same in the generic and half-flat cases.

3. Throughout this section we have assumed that the frames,
co-frames and metrics under consideration are non-degenerate.
However, the final Hamiltonian framework can be naturally extended
to allow degenerate situations. Specifically, by replacing the
scalar lapse function $N$ with one of density weight $-1$, one can
allow for the possibility that the fields $P^a_i$ become
degenerate, i.e., have $\det P = 0$. Somewhat surprisingly,
dynamics continues to be well-defined and one obtains an extension
of general relativity with degenerate metrics. For details, see,
e.g., \cite{deg1,deg2,deg3,deg4,deg5}.

\section{Quantization strategy}
\label{s3}

In sections \ref{s4} and \ref{s5} we will provide a systematic,
step by step construction of background independent quantum
theories of connections (including general relativity) and a
quantum theory of geometry. Since that treatment is mathematically
self contained, the procedure involved is rather long. Although
individual steps in the construction are straightforward, the
motivation, the goals, and the relation to procedures used in
standard quantum field theories may not always be transparent to
an uninitiated reader. Therefore, in this section, we will provide
the motivation behind our constructions, a summary of the
underlying ideas and a global picture that will aid the reader to
see where one is headed.

\subsection{Scalar field theories}
\label{s3.1}

To anchor the discussion in well-established physics, we will
begin by briefly recalling the construction of the Hilbert space
of states and basic operators for a free massive scalar field in
Minkowski space-time, within the canonical approach. (For further
details, see, e.g., \cite{charap}.) The Classical configuration
space ${\cal C}$ is generally taken to be the space of smooth
functions $\phi$ which decay rapidly at infinity on a $t={\rm
const}$ slice, $M$. {}From one's experience in non-relativistic
quantum mechanics, one would expect quantum states to be
`square-integrable functions' $\Psi$ on ${\cal C}$. However, Since
the system now has an infinite number of degrees of freedom, the
integration theory is now more involved and the intuitive
expectation has to be suitably modified.

The key idea, which goes back to Kolmogorov, is to build the
infinite dimensional integration theory from the finite
dimensional one. One begins by introducing a space ${\cal S}$ of
`probes', typically taken to be real test functions $e$ on the
spatial slice $M$. Elements of ${\cal S}$ probe the structure of
the scalar field $\phi \in {\cal C}$ through linear functions
$h_e$ on ${\cal C}$:
\be h_e(\phi) = \int_M d^3x\, e(x) \phi(x) \ee
which capture a small part of the information in the field $\phi$,
namely `its component along $e$'. Given a set $\a$ of probes,
$h_{e_1}, \ldots ,h_{e_n}$ and a (suitably regular) complex-valued
function $\psi$ of $n$ real variables, we can now define a more
general function $\Psi$ on ${\cal C}$,
\be \Psi(\phi) := \psi(h_{e_1}(\phi), \ldots, h_{e_n}(\phi)) ,\ee
which depends only on the $n$ `components' of $\phi$ singled out
by the chosen probes. (Strictly, $\Psi$ should be written
$\Psi_\a$ but we will omit the suffix for notational simplicity.)
Such functions are said to be \emph{cylindrical}. We will denote
by $\cyl_\a$ the linear space they span. Given a measure
$\mu_{(n)}$ on $\Real^n$, we define an Hermitian inner product on
$\cyl_\a$  in an obvious fashion:
\be \langle \Psi_1,\, \Psi_2 \rangle := \int_{\Real^n}
d\mu_{(n)}\, (\bar{\psi_1}\, \psi_2)
(h_{e_1}(\phi),\ldots,h_{e_n})\, \ee
The idea is to extend this inner product to the space $\cyl$ of
\emph{all} cylindrical functions, i.e., the space of all functions
on ${\cal C}$ which are cylindrical with respect to \emph{some}
set of probes. However, there is an important caveat which arises
because a given function $\Psi$ on ${\cal C}$ may be cylindrical
with respect to \emph{two different} sets of probes. (For example,
every $\Psi\in \cyl_\a$ is also in $\cyl_\beta$ where $\beta$ is
obtained simply by enlarging $\a$ by adding new probes.) The inner
product will be well-defined only if the value of the integral
does not depend on the specific set $\a$ of probes we use to
represent the function. This requirement imposes consistency
conditions on the family of measures $\mu_{(n)}$. These conditions
are non-trivial. But they \emph{can} be met. The simplest example
is provided by setting $\mu_{(n)}$ to be normalized Gaussian
measures on $\Real^n$. Every family $\{\mu_{(n)}\}$ of measures
satisfying these consistency conditions enables us to integrate
general cylindrical functions and is therefore said to define a
\emph{cylindrical measure} $\mu$ on ${\cal C}$. The Cauchy
completion $\H$ of $(\cyl, \langle\, ,\,\rangle)$ is then taken to
be the space of quantum states.

If this construction were restricted to any one set $\a$ of
probes, the resulting Hilbert space would be (infinite dimensional
but) rather small because it would correspond to the space of
quantum states of a system with only a finite number of degrees of
freedom. The huge enlargement, accommodating the infinite number
of degrees of freedom, comes about because we allow
\emph{arbitrary} sets $\alpha$ of probes which provide a `chart'
on all of ${\cal C}$, enabling us to incorporate the infinite
number of degrees of freedom in the field $\phi$.

Let us examine this issue further. Any one cylindrical function is
a `fake' infinite dimensional function in the sense that its
`true' dependence is only on a finite number of variables.
However, in the Cauchy completion, we obtain states which
`genuinely' depend on an infinite number of degrees of freedom.
However, in general, these states can not be realized as functions
on ${\cal C}$. In the case of free fields, the appropriate
measures are Gaussians (with zero mean and variance determined by
the operator $\Delta - \mu^2$) and all quantum states can be
realized as functions on the space ${\cal S}^\prime$ of tempered
distributions, the topological dual of the space ${\cal S}$ of
probes. In fact, the cylindrical measure can be extended to a
regular Borel measure $\mu$ on ${\cal S}^\prime$ and the Hilbert
space is given by $\H = L^2({\cal S}^\prime, d\mu)$. ${\cal
S}^\prime$ is referred to as the \emph{quantum configuration
space}. Finally, as in Schr\"odinger quantum mechanics, the
configuration operators $\hat\phi(f)$ are represented by
multiplication and momentum operators $\hat\pi(f)$ by derivation
(plus a multiple of the `divergence of the vector field $\int d^3x
\delta/\delta\phi(x)$ with respect to the Gaussian measure'
\cite{als}).%
\footnote{For interacting field theories rigorous constructions
are available only in low space-time dimensions. The $\lambda
\phi^4$ theory, for example, is known to exist in 2 space-time
dimensions but now the construction involves non-Gaussian
measures. For a brief summary, see \cite{charap}.}
This `Schr\"odinger representation' of the free field is entirely
equivalent to the more familiar Fock representation.

Thus, the overall situation is rather similar to that in quantum
mechanics. The presence of an infinite number of degrees of
freedom causes only one major modification: the classical
configuration space ${\cal C}$ of smooth fields is enlarged to the
quantum configuration space ${\cal S}^\prime$ of distributions.
Quantum field theoretic difficulties associated with defining
products of operators can be directly traced back to this
enlargement.

\subsection{Theories of Connections}
\label{s3.2}

We saw in section \ref{s2} that general relativity can be recast
in such a way that the configuration variables are $\SU(2)$
connections on a `spatial' manifold $\M$. In this section we will
indicate how the quantization strategy of section \ref{s3.1} can
be modified to incorporate such background independent theories of
connections. We will let the structure group to be an arbitrary
compact group $G$ and denote by $\A$ the space of all suitably
regular connections on $M$. $\A$ is the classical configuration
space of the theory.%
\footnote{Since the goal of this section is only to sketch the
general strategy, for simplicity we will assume that the bundle is
trivial and regard connections as globally defined 1-forms which
take values in the Lie algebra of $G$.}

The idea again is to decompose the problem in to a set of finite
dimensional ones. Hence, our first task is to introduce a set of
probes to extract a finite number of degrees of freedom from the
connection field. The new element is gauge invariance: now the
probes have to be well adapted to extracting gauge invariant
information from connections. Therefore, it is natural to define
cylindrical functions through holonomies $h_e$ along edges $e$ in
$M$. This suggests that we use edges as our probes. Unlike in the
case of a scalar field, holonomies are not linear functions of the
classical field $A$; in gauge theories, the duality between the
probes and classical fields becomes non-linear.

Denote by $\alpha$ graphs on $M$ with a finite number of edges
$e$. Then, given a connection $A$ on $M$, holonomies $h_e(A)$
along the edges $e$ of $\a$ contain gauge invariant information in
the restriction to the graph $\a$ of the connection $A$. While
these capture only a finite number of degrees of freedom, the full
gauge invariant information in $A$ can be captured by considering
\emph{all possible} graphs $\a$.

The strategy, as in section \ref{s3.1}, is to first develop the
integration theory using single graphs $\a$. If the graph $\a$ has
$n$ edges, the holonomies $h_{e_1}, \ldots, h_{e_n}$ associate
with every connection $A$ an n-tuple $(g_1, \ldots, g_n)$ of
elements of $G$. Therefore, given a (suitably regular) function
$\psi$ on $G^n$, we can define a function $\Psi$ on the classical
configuration space $\A$ as follows:
\be \Psi(A) := \psi(h_{e_1}(A), \ldots, h_{e_n}(A)\ee
These functions will be said to be cylindrical with respect to the
graph $\a$ and their space will be denoted by $\cyl_\a$. To define
a scalar product on $\cyl_\a$, it is natural to choose a measure
$\mu_{(n)}$ on $G^n$ and set
\be\label{ip0} \langle \Psi_1, \Psi_2  \rangle := \int_{G^n}
d\mu_{(n)}\,\bar\psi_1 \psi_2\, \ee
This endows $\cyl_\a$ with a Hermitian inner product. This
analysis is completely analogous to that used in lattice gauge
theories, the role of the lattice being played by the graph $\a$.

However, as in section \ref{s3.1}, elements of $\cyl_\a$ are
`fake' infinite dimensional functions because they depend only on
a finite number of `coordinates', $h_{e_1},\ldots, h_{e_n}$, on
the infinite dimensional space $\A$. To capture the full
information contained in $\A$, we have to allow \emph{all
possible} graphs in $M$.%
\footnote{Note that this strategy is quite different from the
standard continuum limit used in lattice approaches to Minkowskian
field theories. Our strategy is well-suited to background
independent theories where there is no kinematic metric to provide
scales. Technically, it involves a `projective limit'
\cite{mm,al4}.}
Denote by $\cyl$ functions on $\A$ which are cylindrical with
respect to \emph{some} graph $\a$. The main challenge lies in
extending the integration theory from $\cyl_\a$ to $\cyl$. Again
the key subtlety arises because $\Psi_1$ and $\Psi_2$  in $\cyl$
may be cylindrical with respect to many graphs and there is no a
priori guarantee that the value of the inner product is
independent of which of these graphs are used to perform the
integral on the right side of (\ref{ip0}). The requirement that
the inner product be well-defined imposes severe restrictions on
the choice of measures $\mu_{(n)}$ on $G^n$. However, as discussed
in section \ref{s5}, there is a natural choice compatible with the
requirement that the theory be diffeomorphism covariant, imposed
by our goal of constructing a background independent quantum
theory \cite{al2,jb1,charap,mm,al3,ol,st1,lost}.

As in section \ref{s3.1}, a consistent set of measures $\mu_{(n)}$
on $G^n$ provides a cylindrical measure on $\A$ and a general
result ensures that such a measure can be naturally extended to a
regular Borel measure on an extension $\Ab$ of $\A$ \cite{charap}.
The space $\Ab$ is called the \emph{quantum configuration space}.
It contains `generalized connections' which can not be expressed
as continuous fields on $M$ but nonetheless assign well-defined
holonomies to edges in $M$. These are referred to as \emph{quantum
connections}. Conceptually, the enlargement from $\A$ to $\Ab$
which occurs in the passage to quantum theory is very similar to
the enlargement from ${\cal C}$ to ${\cal S}^\prime$ in the case
of scalar fields. This enlargement plays a key role in quantum
theory (especially in the discussion of surface states of a
quantum horizon discussed in section \ref{s8}). It is an imprint
of the fact that, unlike in lattice theories, here we are dealing
with a genuine field theory with an infinite number of degrees of
freedom.

By now, the structure of the quantum configuration space $\Ab$ is
well understood \cite{ai,al2,al4,cf2,cf1}. In particular, using an
algebraic approach (which has been used so successfully in
non-commutative geometry), differential geometry has been
developed on $\Ab$ \cite{al4}. It enables the introduction of
physically interesting operators discussed in sections
\ref{s4.3},\ref{s5} and \ref{s6}.

Ideas sketched in this section are developed systematically in the
next two sections. We begin in section \ref{s4.1} by discussing
quantum mechanics on a compact Lie group $G$ and use it to
introduce the quantum theory of connections on a graph in section
\ref{s4.2}. The quantum theory of connections in the continuum is
discussed in section \ref{s4.3}. This structure is then used in
section \ref{s5} to introduce quantum geometry.

\section{Quantum theories of connections: background independent
kinematics} \label{s4}

In this section,  we will construct a kinematical framework for
background independent, quantum theories of connections in the
abstract, without direct reference to section \ref{s2}. To bring
out the generality of these constructions, we will work with gauge
fields for which the structure group is any compact Lie group $G$.
This discussion of theories of connections is divided in to three
parts. In the first, we provide a gentle introduction to the
subject via quantum mechanics of a `particle' on the group
manifold of a compact Lie group $G$; in the second, we consider
the quantum kinematics of a (background independent) lattice gauge
theory with structure group $G$ on an arbitrary graph; and, in the
third, we consider connections in the continuum with structure
group $G$.

Constructions based on a general compact Lie group are important,
e.g., in the discussion of the Einstein-Yang-Mills theory.
However, for quantum geometry and for formulation of quantum
Einstein's equations, as we saw in section \ref{s2}, the relevant
group is $G= \SU(2)$. Therefore, we will often spell out the
situation for this case in greater detail. We will use the
following conventions. The dimension of the $G$ will be $d$ and
its Lie-algebra will be denoted by $\ggot$. Occasionally we will
use a basis $\tau^i$ in $\ggot$. In the case $G = \SU(2)$, the
Lie-algebra $\ggot = \su(2)$ will be identified with the Lie
algebra of all the complex, traceless, anti-self adjoint $2$ by
$2$ matrices. Then the Cartan-Killing metric $\eta_{ij}$ is given
by:
\be \label{ck} \eta(\xi,\zeta)= -2\Tr(\xi\zeta)\, ,\ee
for all $\xi,\zeta\in \su(2)$. In this case our $\tau_i$ will
constitute an ortho-normal basis satisfying
\be \label{tau} [\tau_i,\tau_j]= \epsilon^k{}_{ij}\tau_k\, . \ee

\subsection{Quantum mechanics on a compact Lie group $G$}
\label{s4.1}

Let us consider a `free' particle on the group manifold of a
compact Lie group $G$. In this sub-section, we will discuss
(classical and) quantum mechanics of this particle. The quantum
Hilbert space and operators will be directly useful to quantum
kinematics of theories of connections discussed in the next two
sub-sections. The theory described in this section also has some
direct physical applications. For example in the case $G=\SO(3)$,
it describes `a free spherical top' while if $G = \SU(2)$, it
plays an important role in the description of hadrons in the
Skyrme model.

\subsubsection{Phase space}
\label{s4.1.1}

The configuration space of the particle is the group manifold of
$G$ and the phase space is its cotangent bundle $T^\star(G)$. The
natural Poisson bracket between functions on $T^\star(G)$ is given
by:
 \be \{f_1,\, f_2\}\ =\ \frac{\partial f_1}{\partial q^i}
 \frac{\partial {f_2} }{\partial p_i} - \frac{\partial
 {f_2}}{\partial q^i}\frac{\partial f_1}{\partial p_i} \ee
where $q^i$ are coordinates on $G$ and $(q^i,p_i)$ are the
corresponding coordinates on $T^\star(G)$.

Every smooth function $f$ on $G$ defines a configuration variable
and every smooth vector field $X^i$, a momentum variable $P_X :=
X^ip_i$ on $T^\star(G)$. As on any cotangent bundle, (non-trivial)
Poisson brackets between them mirrors the action of vector fields
on functions and the Lie bracket between vector fields:
 \be \label{pb1} \{P_x\ ,\, f\} = - {\cal L}_X f; \quad {\rm and}
 \quad \{ P_X,\,P_Y \}= - P_{[X,Y]}.\ee
These configuration and momentum observables will be said to be
\textit{elementary} in the sense that they admit unambiguous
quantum analogs.

Being a Lie group, $G$  admits two natural Lie algebras of vector
fields, each of which is isomorphic with the Lie algebra $\ggot$
of $G$. Given any $\xi\in \ggot$, we can define a left
(respectively, right) invariant vector field $L^{(\xi)}$
(respectively, $R^{(\xi)}$) on $G$ such that
 \be L^{(\xi)} f(g)\, =\, \frac{d}{dt}f(ge^{t\xi}),\quad {\rm and}
 \quad R^{(\xi)} f(g)\, =\, \frac{d}{dt}f(e^{-t\xi}g). \ee
(The sign convention is such that $L^{(\xi)}\mapsto R^{(\xi)}$
under $g\mapsto g^{-1}$.)

The corresponding momentum functions on $T^\star(G)$ will be
denoted by $J^{(L,\xi)}, J^{(R,\xi)}$. These are generalization of
the familiar `angular momentum functions' on $T^\star\SO(3)$. Each
set forms a $d$ dimensional vector space which is closed under the
Poisson bracket. Since any vector field $X$ on $G$ can be
expressed as a (functional) linear combination of $L^{(\xi)}$
($R^{(\xi)}$), it suffices to restrict oneself only this $2d$-
dimensional space of momentum observables.

Since the particle is `free', the Hamiltonian is given just by the
kinetic term:
\be H(p,q)\ =\  \eta^{ij} p_i p_j,\ee
where $\eta_{ij}$ is a metric tensor defined on $G$ and invariant
with respect to the left and right action of $G$ on itself. Given
an orthonormal basis $\tau_i$, $i=1,\ldots, d$, in $\ggot$, and we
denote $J^{(L, \tau_i)}$ by $J^{(L)}_i$ and $J^{(R, \tau_i)}$ by
$J^{(R)}_i$, then the Hamiltonian can be rewritten as:
 \be H(p,q)\ =\ J^{(L)}_iJ^{(L)}_j\eta^{ij}\
 =\ J^{(R)}_iJ^{(R)}_j\eta^{ij}. \ee
We will see that all these basic observables naturally define
operators in the quantum theory.

\subsubsection{Quantization}
\label{s4.1.2}

Since $G$ is equipped with the normalized Haar measure $\mu_H$,
the Hilbert space of quantum states can be taken to be the space
$L^2(G,d\mu_H)$ of square integrable functions on $G$ with respect
to the Haar measure. (For a detailed discussion, see
\cite{creutz,jbbook}.) The configuration and momentum operators
can be introduced as follows. To every smooth function $f$ on $G$,
we can associate a configuration operator $\hat{f}$ in the obvious
fashion
\be (\hat{f}\psi)(g)\ =\ f(g)\psi(g), \ee
and to every momentum function $X^ip_i$, a momentum operator
$\hat{J}_{(X)}$ via:
\be \label{mom1} (\hat{J}^{(X)} \psi) (g) = i \,\, [{\cal L}_X\,
\psi  + \frac{1}{2}(\div X)\,\psi ](g)\, , \ee
where $\div X$ is the divergence of the vector field $X$ with
respect to the invariant volume form on $G$ (and, for later
convenience, we have left out the factor of $\hbar$.) It is
straightforward to check that the commutators of these
configuration and momentum operators mirror the Poisson brackets
between their classical counterparts. Of particular interest are
the operators associated with the left (and right) invariant
vector fields associated with an orthonormal basis $\tau_i$ of
$\ggot$. We will set
\be \hL_i= \hJ^{(L)}_i \quad {\rm and} \quad \hR_i = \hJ^{(R)}_i
\ee
Since the divergence of right and left invariant vector fields
vanishes, the action of operators is given just by the
Lie-derivative term, i.e., formally, by the Poisson bracket
between the momentum functions and $\psi$. In terms of these
operators, the quantum Hamiltonian is given by
\be \hH\ =\ \hL_i \hL_j \eta^{ij}\ =\ \hR_i \hR_j \eta^{ij}  \ = -
\Delta, \ee
where $\Delta$ is the Laplace operator on $G$.

\subsubsection{Spin states.}
\label{s4.1.3}

In theories of connections developed in the next two subsections,
a `generalized spin-network decomposition' of the Hilbert space of
states will play an important role. As a prelude that
construction, we will now introduce an orthogonal decomposition of
the Hilbert space $L^2(G, d\mu_H)$ into \textit{finite}
dimensional subspaces. Let $j$ label inequivalent irreducible
representations of $G$, let $V_j$ denote the carrier space of the
$j$-representation and let $V^\star_j$ be its dual. Then, the
Peter Weyl theorem provides the decomposition we are seeking:
\be\label{dec1gen} L^2(G, d\mu_H)\, =\, \oplus_j {\cal S}_j, \quad
{\rm with} \quad {\cal S}_j\ =\ V_j\otimes V^\star_j \ee

In the case $G= \SU(2)$ we can make this decomposition more
explicit. As is well-known from quantum mechanics of angular
momentum, in this case the eigenvalues of the operator $\hJ^2=
-\Delta$ are given by $j(j+1)$, where $j$ runs through all the
non-negative half-integers and labels the irreducible
representations. Each carrier space $V_j$ is now $2j+1$
dimensional. We can further decompose each ${\cal S}_j= V_j\otimes
V^\star_j$ into orthogonal 1-dimensional subspaces. Fix an element
$\xi\in\su(2)$ and consider the pair of commuting operators,
$\hL_{(\xi)}$ and $\hR_{(\xi)}$. Given $j$, every pair of
eigenvalues, $j_{(L, \xi)}, j_{(R, \xi)}$, of these operators,
each in $ -j, -j+1,...,j,$, defines a 1-dimensional eigensubspace
${\cal S}_{j, j_{(L,\xi)},j_{(R,\xi)}}$. Thus, we have
\be\label{dec1} L^2(\SU(2), d\mu_H)\ =\ \oplus_j {\cal S}_j\ =\
\oplus_{j, j_{(L, \xi)}, j_{(R, \xi)}}\,\, {\cal S}_{j, j_{(L,
\xi)}, j_{(R, \xi)}}\, . \ee
This fact will lead us to spin network decomposition in the next
two subsections.

%
%
%
%
%
%

\subsection{Connections on a graph}
\label{s4.2}

Before considering (field) theories of connections, let us
consider an intermediate quantum mechanical system, that of
connections on a fixed graph $\a$ with a finite number of edges.
This system is equivalent to lattice gauge theory on $\a$
\cite{creutz}. In the next sub-section, we will see that field
theories of connections in the continuum can be obtained by
appropriately `gluing' theories associated with all possible
graphs on the given manifold, in the manner sketched in section
\ref{s3.2}.

A graph may be thought of as a collection of edges and vertices
and will serve as a `floating' lattice.%
\footnote{ More precisely, a \textit{graph} $\a$ is a finite set
of compact 1-dimensional sub-manifolds of $\M$ called
\textit{edges} of $\a$, such that: i) every edge is either an
embedded interval with boundary (an open edge with end-points);
or, an embedded circle with a marked point (a closed edge with an
`end point'); or an embedded circle (a loop); and, ii) if an edge
intersects any other edge of $\a$ it does so only at one or two of
its endpoints. The end points of an edge are called vertices. This
precise definition is needed to ensure that our Hilbert space $\H$
of \ref{s4.3} is sufficiently large and admits a generalized
`spin-network decomposition.}
(`Floating', because the edges need not be rectangular. Indeed
since we do not have a background metric, terms like `rectangular'
have no invariant meaning.) A graph $\a'$ will be said to be
\textit{larger} than another graph $\a$ (or \textit{contain}
$\a$), $\a\ge \a'$, if every edge $e$ of $\a$ can be written as
$e= e'_1\circ \ldots \circ e'_k$ for some edges $e'_1, \ldots
e'_k$ of $\a'$.

\subsubsection{Spaces of connections on a graph}
\label{s4.2.1}

A $G$ \textit{connection $A_\a$ on a graph $\a$} is the set of
$\ggot$ valued 1-forms $A_e$ defined on each edge $e$ of $\a$. For
concreteness we will suppose that each $A_\a$ is given by the
pullback to $\a$ of a smooth $\ggot$-valued 1-form on $\M$.%
\footnote{Throughout this paper, we will work with a fixed
trivialization. \emph{In this section, lower case Greek letters
always refer to graphs and not to indices on space-time fields};
indeed, we do not use space-time fields in most of the remainder
of this review.}
Thus, one can think of a connection on $\a$ simply as an
equivalence class of smooth connections on $\M$ where two are
equivalent if their restrictions to each edge of $\a$ agree. (This
concrete representation of $A_\a$ will make the passage to section
\ref{s4.2} more transparent but is not essential in this section.)

Denote the space of $G$ connections on $\a$ by $\A_\a$. This space
is infinite dimensional because of the trivial redundancy of
performing local gauge transformations along the edges of $\a$. As
in lattice gauge theories it is convenient to remove this
redundancy to arrive at a finite dimensional space $\Ab_\a$, which
can be taken to be the relevant configuration space for any
(background independent) theory of connections associated with the
graph $\a$.

A gauge transformation $g_\a$ in $\G_\a$ is a map $g_\a:
x_\a\rightarrow G$ from all points $x_\a$ on $\a$. Thus, $g_\a$
can be thought of as the restriction to $\a$ of a $G$-valued
function defined on $\M$. Under $g_\a$, connections $A_\a$
transform as:
\be A_\a \mapsto\ g_\a^{-1} A_\a g_\a + g_\a^{-1}\, d_\a g_\a, \ee
where $d_\a$ is the exterior derivative along the edges of $\a$.
Let us now consider the quotient spaces
\be \Ab_\a\ :=\ \A_\a/\G^0_\a, \quad {\rm and} \quad
 \Gb_\a\ :=\ \G_\a/\G^0_\a, \ee
where $\G^0_\a$ is the subgroup given by all local gauge
transformations  $g_\a$ which are identity on the vertices of
$\a$. Let us choose an arbitrary but fixed orientation of each
edge of $\a$. Then, every element $\bA_\a\in \Ab_\a$ can be
identified with the $G$ values $\bA_\a(e)$ of the parallel
transport (i.e., holonomy) defined by any connection $A_\a$ in the
equivalence class $\bA_\a$.%
\footnote{For simplicity, in the main body of the paper we will
discuss the case where every edge of $\a$ has two vertices. If a
graph $\a$ admits a closed edge $e'$ without vertices, then
$\G^0_\a$ contains \textit{all} gauge transformations at points of
$e'$ (since $e'$ has no vertex). Hence, $\bA_\a(e') \in G/{\rm
Ad}$, where the quotient is by the adjoint action on $G$. Thus,
for a general graph, if $n_o$ denotes the number of closed edges
without vertices and $n_1$ the remaining edges (so that $n = n_0 +
n_1$), the image of the map $\I_E$ defined below is [$G/Ad]^{n_o}
\times G^{n_1}$. All our constructions and results can be extended
to general graphs in a straightforward manner.}
Thus, we have natural 1-1 maps between $\Ab_\a$ and $G^n$ and
between $\Gb_\a$ and $G^m$
 \ba
 \I_E &:& \Ab_\a \longrightarrow G^n;\quad
 \I_E(\bA_\a) = (\bA_\a(e_1),...,\bA_\a(e_1)),
 \label{iso}\\
 \I_V &:& \Gb_\a \longrightarrow G^m; \quad
 \I_V(\bg_\a) = (\bg(v_1),...,\bg(v_m))
 \ea
where $e_1,...,e_n$ are the edges of $\a$ and $v_1,...,v_m$ the
vertices. Note that $\I_E$ depends on the orientation of edges. In
the next section, this map will play a key role and we will ensure
that our final results are insensitive to the choice of the
orientation.

Following lattice gauge theory, we will refer to $\bA_\a$ as a
\textit{configuration variable} of theories of connections on $\a$
and $\bg_\a$ as a (residual) \textit{gauge transformation} on the
configuration variables. Since $\Gb_\a$, the group of the
(residual) gauge transformations, has a non-trivial action on
$\Ab_\a$, \textit{physical configuration space} is given by the
quotient $\Ab_\a/\Gb_\a$.

\emph{Remark}: The quotient $\Ab_\a/\Gb_\a$ can be characterized
in the following way \cite{al2}. Fix a vertex $v_0$ in $\a$. Let
$\alpha_1,...,\alpha_h$ be free generators of the first homotopy
group of $\a$, based at $v_0$ (that is every loop in $\a$
beginning in $v_0$ is a  product of the generators and their
inverses, and this decomposition is unique). The map
\be \bA\ \mapsto\ (\bA(\alpha_1),...,\bA(\alpha_h)) \ee
from $\Ab$ to $G^h$ defines a 1-1 correspondence between
$\Ab_\a/\Gb_\a$ and $G^m/G$ where the quotient is with respect to
the residual gauge action $(U_1,...,U_h)g :=
(g^{-1}U_1g,...,g^{-1}U_hg)$.

\subsubsection{Quantum theory}
\label{s4.2.2}

Since $\Ab_\a$ is the configuration space, it is natural to
represent quantum states as square-integrable functions on
$\Ab_\a$. This requires that we define a measure on $\A_\a$. An
obvious strategy is to use the map $\I_E$ of (\ref{iso}) to
represent $\Ab_\a$ by $G^n$ and use the Haar measure on $G$. This
endows $\Ab_\a$ with a natural measure which we denote by
$\mu_\a^o$. Thus, the space of quantum states can be taken to be
the Hilbert space $\H_\a = L^2(\Ab_\a, d\mu_\a^o)$. Let us denote
the pull-backs of functions $\psi$ on $G^n$ to functions on
$\Ab_\a$ by $\Psi$:
\be \label{rep} \Psi = \I_E^\star\,\, \psi .\ee
Since $\I_E$ is a \textit{bijection}, every function $\Psi$ on
$\Ab_\a$ can be so represented, enabling us to think of quantum
states $\Psi$ in $\H_\a$ as functions $\psi$ on $G^n$. Then, the
inner-product can be written as:
 \be \label{ip1}
 \langle \Psi_1,\, \Psi_2 \rangle \, =\,  \int_{G^n}\, d\mu_H^o\,
 \bar \psi_1 \, \psi_2\,\,  , \ee
where $\mu_H^o$ is the Haar measure on $G^n$. Since the Haar
measure is invariant under $g \mapsto g^{-1}$, the inner product
does not depend on the choice of the orientation of edges of $\a$,
made in the definition of $\I_E$. It is easy to verify that the
inner product is also invariant under the induced action on
$\H_\a$ of the residual group $\Gb_\a$ of the gauge
transformations.

We will now introduce a number of interesting operators on
$\H_\a$, which will turn out to be useful throughout this paper.
Clearly, $\H_\a$ is the tensor product of the spaces $L^2(G,
d\mu_H)$, each associated with an edge of $\a$. Using the
operators $\hL_i$ and $\hR_i$ on $L^2(G, d\mu_H)$ and the fact
that the correspondence (\ref{iso}) associates a copy of $G$ to
each edge in $\a$, we define certain operators $\hJ_i^{(v,e)}$ on
$\H_\a$. Given a vertex $v$ of $\a$, an edge $e$ with $v$ as an
end-point, and a basis $\tau_i$ in $\ggot$, we set:
\be \hat{J}^{(v,e)}_i \, \Psi = \I_E^\star\, \left[(1\otimes ...
\otimes 1\otimes \hJ_i \otimes 1 \otimes...\otimes 1 )\psi \right]
\ee
where the non-trivial action is only on the copy of $G$ associated
with the edge $e$, and where $\hJ_i = \hL_i$ if the vertex $v$ is
the origin of the edge $e$ and $\hJ_i = \hR_i$ if $v$ is the
target of $e$. Thus, the edge $e$ dictates the copy of $G$ on
which $\hJ^{(v,e)}_i$ has non-trivial action while the vertex $v$
determines if the action is through the left or right invariant
vector field.

\subsubsection{Generalized spin network decomposition}
\label{s4.2.3}

The product structure $\H_\a\sim [L^2(G, d\mu_H)]^{\otimes n}$
enables us to import results of the last sub-section on quantum
mechanics on $G$. In particular, using (\ref{dec1gen}), $\H_\a$
can be decomposed in to finite-dimensional sub-spaces $\H_{\a,
\j}$ where $\j = \{ j_1,\ldots j_n\}$ assigns to each edge of $\a$
an irreducible representations of $G$. The individual sub-spaces
$\H_{\a,\j}$ can be further decomposed into irreducible
representations of the action of the group of residual gauge
transformations. Let $\l = \{ l_1,\ldots, l_v\}$ assign to each
vertex of $\a$ an irreducible representation of $G$. Then, each
$\H_{\a,\j}$ can be further decomposed in to subspaces $\H_{\a,
\j,\l}$ consisting of all vectors which belong to the irreducible
representation $\l$ of the group of residual gauge transformations
at every vertex $v$. Then, we have:
 \be
 \H_\a = \oplus_{\j}\, \H_{\a,\j}  =  \oplus_{\j, \l}\,
\, \H_{\a,\j, \l},\label{dec2}
 \ee
The gauge invariant subspace of $\H_\a$ corresponds to the
labelling of vertices
\be \l = {\vec 0} \quad {\hbox{\rm i.e., }}\quad  \ell_{v} =
\text{the\ trivial\ representation}\quad\hbox{\rm for all $v$ in
}\,\, \a . \ee

For applications to quantum geometry, let us make this
decomposition more explicit in the case when $G = \SU(2)$. This
discussion will also serve to make the somewhat abstract
construction given above by  providing a more detailed description
of the labels $\j, \l$.

\emph{Example:} With each edge $e$ of $\a$, we associate an
operator $\hJ_e^2$:
\be \label{j} (\hJ_{e})^2\ :=\ \eta^{ij}\, \hJ_i^{(v,
e)}\hJ_j^{(v, e)} . \ee
where, $\eta_{ij}$ is again the Cartan-Killing metric (\ref{ck})
on $\su(2)$, and $v$ is the source or target of $e$. Since they
act on different copies of $\SU(2)$, all these operators commute
with each other. Since each of these operators has eigenvalues
$j_e (j_e +1)$ where $j_e$ is a non-negative half-integer, each
simultaneous eigenspace $\H_{\j}$ of this set of operators is
labelled by $\j= (j_{e_1},..., j_{e_n})$. Thus, we have a
decomposition of the total Hilbert space:
\be \H_\a =  \oplus_{\j} \,\, \H_{\a,\,\j} \ee
The individual subspaces $\H_{\a,\,\j}$ are the natural extensions
of spaces $\S_j$ introduced in sub-section \ref{s4.1.3} in the
case of a single copy of $\SU(2)$. They have several interesting
properties: i) Each $\H_{\j}$ is a finite dimensional sub-space of
$\H_\a$; and, ii) it is preserved by the action of every
$\hJ_i^{(v,e)}$; and, iii) it is preserved by the (induced) action
of gauge transformations in $\Gb_\a$ (which act non-trivially at
vertices of $\a$).

Finally, we can carry out a further decomposition by introducing
additional commuting operators. Of particular importance are
vertex operators $[\hJ^v]^2$, associated with each vertex $v$ of
$\a$. These are defined by
\be \label{l} [\hJ^{v}]^2\ :=\ \eta^{ij}\, \hJ^{v}_i\hJ^{v}_j \, ,
\quad {\rm where}\quad \hJ^{v}_i\ :=\ \sum_{e'\ {\rm at}\ v}
\hJ_i^{(v,e')}, \ee
where the sum extends over all edges $e'$ intersecting at $v$.
Heuristically, $\hJ^e_i$ can be regarded as angular momentum
operators `living on the edge $e$' and $\hJ^v_i$, as the
\emph{total} angular momentum operators `arriving' at the vertex
$v$. It is easy to check that the operators $[\hJ^v]^2$ commute
with the operators $[\hJ^e]^2$. Hence, if we denote eigenvalues of
$[\hJ^v]^2$ by $l_v(l_v +1)$ the subspaces $\H_\j$ can be further
decomposed and we arrive at a finer decomposition of the total
Hilbert space:
 \be
 \H_\a = \oplus_{\j}\, \H_{\a,\j}  =  \oplus_{\j, \l}\,
\, \H_{\a,\j, \l},
 \ee
where $\H_{\j,\l}$ is a simultaneous eigenspace of operators
$[\hJ^e]^2$ and $[\hJ^v]^2$.

\textsl{Remark}: One can enlarge the set of commuting operators
and further refine the decomposition of $\H_\a$. We illustrate the
procedure for $G=SU(2)$. At each vertex $v$, let us first order
the intersecting edges, $(e'_1,...,e'_k)$ say. Then, introduce the
following (rather large) set of operators:
\ba (\hJ_i^{(v, e'_1)}+\hJ_i^{(v,e'_2)})&\eta^{ij} &(\hJ_j^{(v,
e'_1)}+\hJ_j^{(v,e'_2)}),
\ldots,\nonumber\\
(\hJ_i^{(v, e'_1)}+ \hJ_i^{(v,e'_2)} + \hJ_i^{(v,e'_3)}
)&\eta^{ij} & (\hJ_j^{(v, e'_1)}+ \hJ_j^{(v,e'_2)} +
\hJ_j^{(v,e'_3)} ),
\ldots,\nonumber\\
\ldots\quad \ldots \nonumber \\
(\hJ_i^{(v,e'_1)}+\ldots+ \hJ_i^{(v,e'_{k-1})})& \eta^{ij} &
(\hJ_j^{(v,e'_1)}+\ldots+\hJ_j^{(v,e'_{k-1}i)}), \label{s} \ea
where each bracket contains sum only over operators defined at a
given vertex $v$. These operators commute with each other and with
our earlier operators $[\hJ^e]^2$ and $[\hJ^v]^2$. If we label the
eigenvalues of these operators by $\s$, each simultaneous
eigenspace $\H_{(\j,\l,\s)}$ can be labelled by the triplet
$(\j,\s,\l)$. Each $\H_{(\j,\l,\s)}$ is the irreducible
representation of the group of the gauge transformations $\Gb_\a$
corresponding to the half-integer values  $l_v$.

Note that, given arbitrary labelling $\j$ of the edges of $\a$,
the remaining two labellings  $\l,\s$ are restricted by some
inequalities: at each vertex $v$, we must have:
\ba
 s_{1,2} &\in &\{|j_{e'_1}-j_{e'_2}|, |j_{e'_1}-j_{e'_2}|
 +1,...,j_{e'_1}+j_{e'_2}\}, \label{ineq1}\\
 s_{1,2,3}&\in &\{|s_{1,2}-j_{e'_3}|,..., s_{1,2}+j_{e'_3}\}
 \label{ineq2}\\
\ldots \nonumber\\
 l_v &\in &\{|s_{1,2,...,k-1}-j_{e'_k}|,...,|s_{1,2,...,k-1}
 +j_{e'_k}|\}\label{ineq3}.
 \ea
When these conditions are met, we obtain the following orthogonal
decomposition of $\H_\j$:
 \be \H_\a = \oplus_{j}\, \H_{\j},\ \quad
  \H_{\j}= \oplus_{\l}\, \H_{(\j,\l)}\quad {\rm and}\quad
  \H_{(\j,\l)} \oplus_{\s}\H_{\j,\l,\s} \ee
where the labellings $\j$, $\l$ and $\s$ take positive,
half-integer values, subject to the inequalities (\ref{ineq1}) --
(\ref{ineq3}).  Gauge invariant states subspaces are labelled by
trivial $\l$.

\subsection{Connections on $\M$}
\label{s4.3}

Let us now turn to field theories of $G$-connections, such as
general relativity, discussed in section \ref{s2}. Given any graph
$\a$ on $\M$, each connection $A$ on $\M$ defines, just by
restriction, a connection $A\!\mid_\a$ on $\a$. Furthermore, $A$
is completely determined by the collection $\{A\!\mid_\a\}$
defined by considering all possible graphs $\a$ on $\M$.
Therefore, we will be able to construct a background independent
quantum kinematics for theories of connections on $\M$ by weaving
together quantum theories of connections on graphs (constructed in
section \ref{s4.2}).

\subsubsection{The classical phase space}
\label{s4.3.1}

For simplicity, we will consider $G$-connections on a
\emph{trivial} bundle over $M$. This restriction is motivated by
the fact that the main application of this framework will be to
quantum geometry where $G=\SU(2)$ and all $\SU(2)$ bundles over a
3-manifold are trivial. Since all the structures we introduce are
gauge covariant, it is convenient to fix a global trivialization
once and for all and regard smooth $\ggot$-valued 1-forms $A$ on
$\M$ as connections.%
\footnote{For background material see, e.g., \cite{jbbook}. It is
quite straightforward to generalize the framework introduced in
this subsection to non-trivial bundles on n-dimensional manifold
$M$ \cite{jb1,lost}.}
The space of all such 1-forms will be the classical configuration
space and denoted by $\A$. The phase space will consist of pairs
$(A_a^i,P^a_i)$, where $A\in\A$ and $P$ is an $\ggot$-valued
vector density defined on $\M$. Following the standard terminology
from Yang-Mills theory, we will refer to $A_a^i$ as
\textit{connections} and $P_a^i$ as the analogs of Yang-Mills
\textit{electric fields}. As we saw in section \ref{s2}, in the
gravitational case, $k P^a_i = (8\pi G\g)  P^a_i$ also has the
interpretation of an orthonormal triad (of density weight one),
where $\g$ is the Barbero-Immirzi parameter. While this fact plays
no role in this section, it will be used crucially in section
\ref{s5} to introduce quantum Riemannian geometry.

 The Poisson bracket between any two smooth functions on the phase
 space is given by:
 \be \label{pb2}
 \{f_1(A,P), {f_2}(A,P)\} = \int_\M d^3x\, \left(
 \frac{\delta f_1}{\delta A^i_a} \frac{\delta
 {f_2}}{\delta P_i^a} -
 \frac{\delta {f_2} }{\delta A^i_a} \frac{\delta f_1}
 {\delta P_i^{a}}\right)\, .
 \ee
The gauge group $\G$ is the group of $G$-valued functions $g$ on
$\M$. This group has a natural action on the phase space, given
by:
\be (A\cdot g,\, P\cdot g)\ =\ (g^{-1}Ag + g^{-1}dg,\, g^{-1}Pg).
\ee
The \textit{elementary} classical observables that will have
direct quantum analogs are (complex-valued functions of)
holonomies $A(e)$ along paths $e$ in $\M$ and fluxes $P(S,f)$ of
electric fields (smeared by $\ggot$-valued functions $f$) across
2-surfaces $S$ in $\M$. In this sub-section, we will fix our
conventions, introduce precise definitions of these phase space
functions, and explore some of their properties.

In the main body of this paper, for technical simplicity we will
restrict ourselves to oriented, analytic 3-manifolds $\M$ and use
only \textit{closed-piecewise analytic edges $e$ and
closed-piecewise analytic sub-manifolds} $S$ in $\M$.%
\footnote{More precisely, we assume that for each edge $e: [t_n,\,
t_1]\mapsto \M$, the interval $[t_n,\, t_1]$ admits a covering by
closed intervals $[t_n,\, t_{n-1}], \ldots [t_{2},\, t_1]$ such
that the image of each of these closed intervals in $\M$ is
analytic. Each surface $S$ is an topological sub-manifold of $M$
such that its closure $\bar{S}$ is of the form $\bar{S} = \cup_I\,
\bar{S}_I$ where each $\bar{S}_I$ is a compact, analytic
sub-manifold of $\M$ (possibly) with boundary. (These assumptions
can be relaxed and one can work with just smooth structures; see
\cite{bs1,bs2,cf4,lt}.)}
Given an edge $e: [t_2,\, t_1]\rightarrow \M$ on $\M$ and a
connection $A$, the parallel transport from $e(t_1)$ to $e(t)$
along $e$ is defined by the following differential equation and
initial condition,
\be \frac{d}{dt}U_e(t,t_1;\,A)\ =\ -A_a(e(t))\,
\dot{e}^a(t)\,U(t,t_1;A),\quad {\rm and}\quad  \ U(t_1,t_1; A)=I.
\ee
Given $A\in\A$, the parallel transport along entire $e$ will be
denoted by $A(e)$:
\be A(e)\ :=\ U_p(t_2, t_1 ;A). \ee
Thus, $A(e) \in G$, it is unchanged under orientation-preserving
re-parameterizations of $e$, and has two key properties which will
play an important role in the next sub-section.
\be\label{A} A(e_2\circ e_1)= A(e_2)A(e_1),\ \ A(e^{-1})=
A(e)^{-1}, \ee
where $e^{-1}$ is obtained from $e$ by simply reversing the
orientation.

The `electric flux' is defined using our surfaces $S$. Fix on $S$
a smooth function $f$ with values in the dual $\ggot^\star$ of the
Lie algebra $\ggot$ and define the (smeared) flux of $P$ through
$S$ as:
\be \label{E1}
 P(S,f) := \int_S f_i\, \Sigma^i\, ,
\ee
where $\Sigma_{ab}^i = \eta_{abc} P^{ci}$ is the 2-form dual of
the electric field.

As a prelude to quantization, let us calculate Poisson brackets
between these observables. Since the phase space is a cotangent
bundle, the configuration observables have vanishing Poisson
brackets among themselves. (As in sections \ref{s4.1.2} and
\ref{s4.2.2}, this will make it possible to introduce a
configuration representation in which quantum states are functions
of connections.) The Poisson bracket between configuration
observables $A(e)$ and momentum observables $P(S,f)$ can be easily
calculated and has a simple, geometrical structure. Any edge $e$
with $e\cap S \not= \emptyset$ can be trivially written as the
union of `elementary' edges which either lie in $S$, or intersect
$S$ in exactly one of their end-points. (This can be achieved
simply by introducing suitable new vertices on $e$.) Then for each
of these `elementary' edges $e$ which intersect $S$ at a point
$p$, we have:
\be\label{AE} \{ A(e),P{(S, f)} \}\ =\ -\left[\f{\kappa{(S,e)}}{2}
\right]\,\, \times\, \cases{ A(e)\tau^i f_i(p) & if $p$ is the
source of $e$\cr
 - f_i(p) \tau^i A(e) & if $p$ is the target of $e$\cr}\ee
where $\tau^i$ is any orthonormal basis in $\ggot$ and
$\kappa(S,e)$ is $0$ or $\pm 1$:
 \be\label{kappa} \kappa{(S,e)}\, =\ \cases{0, &if $e\cap S =
 \emptyset$, or $e\cap S =e$ modulo the end points\cr
 +1, &if $e$ lies above $S$\cr
 -1, &if $e$ lies below $S$\cr}\ee
Thus, the bracket vanishes if $e$ and $S$ don't intersect or $e$
lies within (the closure of) $S$ and, if they have a `simple'
intersection, is given by a linear combination of the
configuration observables $A(e)$, where the coefficients are
determined by the value of the smearing field $f$ at the
intersection point.

The bracket between the momentum observables, by contrast, is not
as straightforward because of the following technical complication
\cite{acz}. The configuration (holonomy) observables $A(e)$ are
obtained by smearing $A$ along 1-dimensional curves $e$ while the
momentum (electric flux) observables are obtained by smearing the
electric field 2-forms $e$ on 2-dimensional surfaces. Since
connections $A$ are 1-forms and dual-electric fields $\Sigma$
2-forms, this smearing is geometrically most natural, particularly
when there is no background metric. However, in contrast to the
standard practice in field theories where smearing is done in 3
dimensions, our smearing fields are themselves `distributional'
from the full 3-dimensional perspective. Therefore one has to
exercise due care in evaluating Poisson brackets. The 1 and
2-dimensional smearings in the definitions of $A_e$ and $P(S,f)$
are `just right' for the calculation of the Poisson bracket
(\ref{AE}) to go through. However technical subtleties arise in
the evaluation of the Poisson brackets between smeared electric
fields. If one naively uses Poisson brackets (\ref{pb2}) to
conclude that the Lie-bracket between the momentum operators
$P(S,f)$ must vanish, then (\ref{AE}) implies that the Jacobi
identity between $A(e), P(S,f), P(\tilde{S}, \tilde{f})$ fails to
be satisfied. The correct procedure is to use the fact that the
momentum variables $P(S,f)$ are of the form `$X_{(S,f)}\cdot P$'
for some vector fields $X_{(S,f)}$ on the configuration space $\A$,%
\footnote{More precisely, $X_{(S,f)}$ are derivations on the ring
of functions of holonomies (i.e., on the space $\cyl$ defined
below). {}From the perspective of textbook treatments of field
theories, these functions are `singular', being supported on one
dimensional edges rather than three dimensional open sets of $\M$.
This is the origin of the counter-intuitive result on Jacobi
identity.}
whence the (non-trivial) Lie-bracket between $P(\tilde{S},
\tilde{f})$ is dictated by the action of the vector fields
$X_{S,f}$ on the ring of functions of holonomies. Now, in general
these vector fields fail to commute. Hence (as on a general
cotangent bundle, see (\ref{pb1})) the Poisson bracket between
momentum variables fails to vanish in general. As in section
\ref{s4.1.1}, the correct Lie algebra between our elementary
configuration and momentum observables is given by the geometric
Lie algebra of functions and vector fields on the configuration
space ${\cal A}$. This Lie algebra naturally incorporates
(\ref{AE}) and provides non-trivial Poisson brackets between the
momentum observables. It will be mirrored in the commutators of
the corresponding elementary \textit{quantum} operators.

\subsubsection{Quantum configuration space $\Ab$ and Hilbert
space $\H$} \label{s4.3.2}

In quantum mechanics of systems with a finite number of degrees of
freedom, states are represented by functions on the classical
configuration space. By contrast, as explained in section
\ref{s3}, in field theories quantum states are functions on a
larger space ---the quantum configuration space. Only certain
`nice' functions on the classical configuration space admit an
extension to the larger space. In our case, these are the
so-called `cylindrical functions' on $\A$. Fix a graph $\a$ with
$n$ edges. Then, given a $C^\infty$ complex-valued function $\phi$
on $G^n$, we can define a function $\Phi_{\a}$ on $\A$ via:
\be\label{cyl} \Phi_{\a} (A)\ =\ \phi(A(e_1),...,A(e_n)). \ee
(Strictly, $\Phi$ should carry a subscript $(\a, \phi)$. However,
for notational simplicity, we will drop $\phi$.) The space of such
functions will be denoted by $\cyl_\a$.%
\footnote{Occasionally, we will need to let $\phi$ be only $C^n$.
The space of resulting cylindrical functions $\Phi$ will be
denoted by $\cyl^{(n)}_\a$.}
A function $\Phi$ on $\A$ will be said to be \textit{cylindrical}
if it arises from this construction for \textit{some} graph $\a$.
Note that: i) there is a natural isomorphism between $\cyl_\a$ and
the space of functions on $\Ab_\a$ (see Eq. (\ref{iso})); and, ii)
every function which is cylindrical with respect to a given graph
$\a$ is automatically cylindrical with respect to a larger graph
$\a'$; $\cyl_\a \subseteq \cyl_{\a'}$. These facts are used
repeatedly in this section. We will denote the space of all
cylindrical functions by $\cyl$; thus
$$\cyl = \cup_\a \, \cyl_\a \, .$$

Given any one graph $\a$, using (\ref{ip1}), we can introduce a
natural inner-product on $\cyl_\a$
 \be\label{ip2}
 \langle \Phi_{\a},\, \Psi_{\a} \rangle\, =  \, \int_{G^n} d\mu_H^o\,
 \bar \phi \, \psi\, . \ee
The Cauchy completion of this space provides a Hilbert space which
is naturally isomorphic with $\H_\a = L^2(\Ab_\a, d\mu^o_\a)$ of
section \ref{s4.2.2} and for notational simplicity, we will denote
it also by $\H_\a$.

The idea is to introduce an inner product on the space of
\textit{all} cylindrical functions via (\ref{ip2}). Suppose we are
given two cylindrical functions $\Phi_{\a_1}$ and $\Psi_{\a_2}$
based on two distinct graphs $\a_1$ and $\a_2$. Then, we can
introduce a third graph $\a_3$ which contains all the edges and
vertices of $\a_1$ and $\a_2$, regard the two functions as
cylindrical with respect to $\a_3$, and attempt define the inner
product between $\Phi_{\a_1}$ and $\Psi_{\a_2}$ using (\ref{ip2})
with $\a = \a_3$. The key question now is whether the resulting
number is independent of the specific $\a_3$ used in this
construction. Fortunately, the right and left invariance
properties of the Haar measure and the fact that we have chosen it
to be normalized (so that $\int_G d\mu_H = 1$) imply that the
answer is in the affirmative \cite{al2,jb1}. Thus, thanks to the
Haar measure on $G$, $\cyl$ has a natural Hermitian inner product.
Denote its Cauchy completion by $\H$. This is our Hilbert space
for quantum kinematics of background independent theories of
connections.

Since the Cauchy completion $\H_\a$ of $\cyl_\a$ is simply
$L^2(\Ab_\a, d\mu^o_\a)$, every element of $\H_\a$ can be
represented as a function on $\Ab_\a$ (more precisely, an
equivalence class of functions on $\Ab_a$, where two are
equivalent if they differ on a set of measure zero).
Unfortunately, this is \textit{not} true of $\H$: while every
element of $\cyl$ is a function (indeed a very simple one!) on
$\A$, in the Cauchy completion, one picks up limit points which
\textit{can not} be represented as functions on $\A$. As noted in
section \ref{s3}, this is a standard occurrence in systems with an
infinite number of degrees of freedom (in particular, field
theories). It is then natural to ask: Is there an enlargement
$\Ab$ of $\A$ such that $\H$ is isomorphic with the space of all
square-integrable functions on $\Ab$ (with respect to some regular
Borel measure)? The answer is in the affirmative
\cite{ai,al2,jb1}. This $\Ab$ is called the \textit{quantum
configuration space.}

Surprisingly, one can give a rather simple characterization of
$\Ab$, which will turn out to be extremely useful \cite{mm,al4}.
Let us denote an element of $\Ab$ by $\bA$ and call it a
\textit{quantum connection}. $\bA$ assigns to each edge $e$ in
$\M$ an element $\bA(e)$ of $G$ such that:
\be\label{bA} \bA(e_2\circ e_1)= \bA(e_2) \bA(e_1), \quad {\rm
and} \quad \bA(e^{-1})= (\bA(e))^{-1}\, .\ee
Thus, every smooth connection $A$ automatically defines a
generalized connection (see (\ref{A})); in this case $\bA(e)$ is
just the standard holonomy. However, a general $\bA$ can be
\textit{arbitrarily discontinuous}; there are no requirements on
it other than (\ref{bA}). This is why $\Ab$ is much larger than
$\A$. Nonetheless, in a natural topology (due to Gel'fand), $\A$
is \textit{densely} embedded in $\Ab$ and $\Ab$ is compact
\cite{al2}. Thus, the quantum configuration space $\Ab$ can be
naturally thought of as a completion of the classical
configuration space $\A$. Finally, we note an important fact that
will be used often in regularization procedures in quantum theory:
Given any quantum connection $\bA$ and any graph $\a$, there
exists a smooth connection $A$ such that $\bA(e) = A(e)$ for all
edges $e$ of the graph.

Let us now consider quantum states. Now, one can show that the
family of induced Haar measures $\mu_\a^o$ on $\Ab_\a$ defines a
regular, Borel measure on $\Ab$ \cite{al2,jb1,mm,al3,al4}. We will
denote it by $\mu^o$. As discussed in \ref{s4.3.5},
diffeomorphisms on $\M$ have a natural induced action on $\Ab$.
The measure $\mu^o$ is \textit{invariant}
under this action.%
\footnote{Initially, this came as a surprise because, in the
mathematical community, there was a widespread expectation that
non-trivial diffeomorphism invariant measures do not exist. Note,
however, that our $\mu_o$ is defined on $\Ab$ rather than $\A$,
the space used in those heuristic arguments. We have included this
brief discussion of measure theory to highlight the fact that our
constructions are on a sound mathematical footing; in contrast to
the habitual situation in the physics literature, our functional
integrals are not formal but well-defined and finite. The measure
$d\mu_o$ was introduced in \cite{al2,jb1} and discussed from
different perspectives in \cite{al3,mm,al4}. }
This is why $\H \equiv L^2(\Ab, d\mu_0)$ is an appropriate Hilbert
space of states for background independent theories of
connections.

Finally, let us highlight the essential steps that lead to $\mu^0$
as this series of steps will be used repeatedly to introduce other
structures, such as operators, on $\H$. We begin with spaces
$\Ab_\a$. Each $\Ab_\a$ admits a measure $\mu^o_\a$. This family
of measures is \textit{consistent} in the sense that, given a
function $f$ on $\Ab$ which is cylindrical with respect to a
graphs $\a$ and  $\a'$,
\be \int_{\Ab_\a}d \mu_{\a}^o \, f_\a \,   = \, \int_{\Ab_{\a'}}
 d\mu_{\a'}^o \,f_{\a'}\, .\ee
 It is this consistency that ensures the existence of
$\mu^o$ on $\Ab$. More generally, regular Borel measures (a well
as geometrical structures) on $\Ab$ are defined by `gluing
together' consistent structures on finite spaces $\Ab_a$ (via the
so-called `projective techniques' \cite{mm,al3,al4}). Several
families of such measures have been constructed, $\mu^o$ being the
simplest and, because of certain uniqueness results
\cite{hs,ol,st1,lost} (discussed at the end of section
\ref{s4.3.4}), the most useful of them.

As emphasized in section \ref{s3}, the passage from $\A$ to $\Ab$
is highly non-trivial and comes about \emph{because we are going
beyond lattice gauge theories and incorporating the infinite
number of degrees of freedom in the connection $A$}. As in
Minkowskian field theories, while the classical configuration
space $\A$ is densely embedded in the quantum configuration space
$\Ab$ (in the natural Gel'fand topology on $\Ab$), measure
theoretically, $\A$ is sparse: $\A$ is contained in a set of zero
$\mu_o$ measure \cite{mm}. The fact that general quantum states
have support on `non-classical' connection is not a `mere
mathematical technicality': In the Hilbert space language, this is
the origin of field theoretic infinities, e.g., the reason why we
can not naively multiply field operators. Hence, to ensure that
there are no hidden infinities, it is necessary to pay due
attention to the quantum configuration space $\Ab$.

\subsubsection{Generalized spin networks}
\label{s4.3.3}

It seems natural to attempt to decompose $\H$ as a direct sum of
the Hilbert spaces $\H_\a$ associated with various graphs and then
use the constructions introduced in section \ref{s4.2.2} to carry
out further orthogonal decompositions into \textit{finite}
dimensional Hilbert spaces. However, this idea encounters an
elementary obstruction. Recall that each function on $\Ab$ which
is cylindrical with respect to a graph $\a$ is also cylindrical
with respect to every larger graph. Thus, regarded as subspaces of
$\H$,  the Hilbert spaces $\H_\a$ can not be mutually orthogonal.
To get around this obstacle, let us introduce new Hilbert spaces:
Given a graph $\a$, let $\H'_\a$ be the subspace of $\H_\a$ which
is \textit{orthogonal} to the subspace $\H_{\tilde\a}$ associated
with \textit{every} graph $\tilde\a$ which is \textit{strictly}
contained in $\a$. Through the introduction of these $\H'_\a$, we
remove the undesired redundancy; if $f \in \H'_\a$, it cannot
belong to $\H'_{\beta}$ for any graph $\beta$ distinct from $\a$.
While the definition of $\H'_\a$ may seem unwieldy at first, using
(\ref{dec2}) it is easy to provide an explicit description of
$\H'_\a$ which we now describe.

Consider assignments $\j^\prime = \{j^\prime_1,\ldots,
j^\prime_{n})$ of irreducible representations of $G$ to edges of
$e$ such that each representation is \emph{non-trivial}. Next, let
$\l^\prime = \{l^\prime_1, \ldots, l^\prime_{n_v} \}$ denote
assignments of irreducible representations to vertices of $\a$
which are non-trivial at each \emph{spurious} vertex of $\a$,
where a vertex $v$ is spurious if it is bivalent, and if the edges
$e_i$ and $e_{i+1}$ which meet at $v$ are such that $e_{i}\circ
e_{i+1}$ is itself an analytic edge (so that $v$ `just serves to
split an edge'). Then, $\H'_\a$ is given by:
\be \H'_{\a} = \oplus_{\j',\l'}\, \H'_{\a,\, \j',\l'}\, , \ee
Condition on $\j^\prime$ in the definition of $\H'_{\a,\,\j'}$ is
necessary because functions in $\cyl$ which result from allowing
any of the $(j_1, \ldots , j_n) $ to vanish belong to
$\H'_{\tilde\a}$ where $\tilde{\a}$ is a smaller graph obtained by
`removing those edges for which $j_i$ vanished'. Condition on
$\l^\prime$ removes the redundancy that would otherwise arise
because a function in $\H_{\a_1,\, \j'}$ also defines a function
in $\H_{\a_2, \,\j'}$ where $\a_2$ is obtained merely by splitting
one or more edges of $\a_1$ just by insertion of new vertices. We
can now write the desired decomposition of $\H$:
\be \label{snd} \H = \oplus_{\a}\, \H'_\a = \oplus_{\a,\,
\j'}\,\,\H'_{\a, \,\j'}\, . \ee

\emph{Example:} Again, for $G=\SU(2)$, we can make this procedure
more concrete. Given a  graph $\a$, the space $\H'_{\a}$  is the
subspace of the space $\H_{\a}$ spanned by all the simultaneous
eigenvectors of the operators $[\hJ^v]^2$ and $[\hJ^e]^2$ such
that $i)$ the eigenvalue $j'_e(j'_e+1)$ of $[J^e]^2$ is non-zero
for \textit{each} edge $e$ in $\a$; and $ii)$ the eigenvalue
$\ell'_v$ of $[\hJ^v]^2$ is non-zero at each spurious vertex $v$.

When $G= \SU(2)$, the $\j$ and the $\l$ are sets of half-integers,
or spins,  and the decomposition (\ref{snd}) is referred to as the
\emph{spin-network decomposition} of $\H$. For a general gauge
group, it is called the \emph{generalized spin-network
decomposition} of $\H$. The subspaces $\H_{\a, \,\j'}$ are
\textit{finite} dimensional and their elements are referred to as
\textit{generalized spin-network states} of quantum theories of
connection%
\footnote{Sometimes, the term is used to refer to an orthonormal
basis of states in $\H_{\a, \j'}$, chosen in a specific
calculation. However, introduction of such a basis requires
additional structure. What is naturally available on $\H$ is only
the decomposition (\ref{snd}) rather than a generalized
spin-network basis.}
on $\M$. As we will see, spin network sub-spaces $\H_{\a,\j'}$ are
left invariant by interesting geometric operators. In this sense,
the decomposition (\ref{snd}) has a direct physical significance.
Because of the sum over all graphs in (\ref{snd}), the Hilbert
space $\H$ is very large. Indeed, when it was first constructed,
it seemed to be `too large to be controllable'. However, the later
introduction of projective techniques \cite{mm,al3,al4}, spin
networks \cite{rs3,jb2} and the orthogonal decomposition
\cite{almmt1} of the Hilbert space showed that quantum theory can
be in fact developed relatively easily by importing techniques
from lattice gauge theories and quantum mechanics of spin systems.

\textit{Remark}: Tri-valent spin networks were introduced by Roger
Penrose already in '71 in a completely different approach to
quantum gravity \cite{rp}. He expressed his general view of that
construction as follows: \textsl{``I certainly do not want to
suggest that the universe `is' this picture \ldots But it is not
unlikely that essential features of the model I am describing
could still have relevance in a more complete theory applicable to
more realistic situations''}. We will see in section \ref{s4},
trivalent graphs are indeed `too simple' for semi-classical
considerations but Penrose's overall vision \textit{is} realized
in a specific and precise way in quantum geometry.

\subsubsection{Elementary quantum operators}
\label{s4.3.4}

Recall that in non-relativistic quantum mechanics, typically, one
first defines operators on the space $\S$ of smooth functions with
rapid decay at infinity and then extends them to self-adjoint
operators on the full Hilbert space, $L^2(R^3)$. We will follow a
similar strategy; now the role of $\S$ will be played by $\cyl$.

Let us begin with the configuration operators. Classical
configuration variables are represented by complex-valued,
cylindrical function $f$ on $\Ab$. We define corresponding quantum
operators $\hat{f}$ which also acts by multiplication:
\be (\hat{f}\, \Psi )(\bA)\, = \, f(\bA) \, \Psi (\bA) \ee

Next, let us define momentum operators $\hP_{(S,f)}$, labelled by
a 2-surface $S$ and $\ggot$-valued smearing fields $f^i$ on $S$.
As with operators $\hL_i$ and $\hR_i$ in section \ref{s4.1.2},
this action is given just by the Poisson brackets between the
classical momentum and configuration observables: For all $\Psi
\in \cyl$, we have:
\be (\hP_{(S,f)} \Psi)(\bA)\, =\,  i\hbar \{P{(S,f)},\,
\Psi\}(\bA) \ee

For later use, let us make the action of the momentum operators
explicit. If $\Psi \in \cyl_\a$, we have
\be \label{ESf1}\hP_{(S,f)}\, \Psi = \f{\hbar}{2}\, \sum_v\,
f^i(v) \left[ \sum_{e\ {\rm at}\ v}\kappa{(S,e)} \hJ_i^{(v,e)}\,
\Psi \right]\ee
in terms of $\kappa{(S,e)}$ of (\ref{kappa}). With domain
$\cyl^{(2)}$, consisting of twice differentiable cylindrical
functions, these operators are essentially self-adjoint (i.e.,
admit a unique self-adjoint extension) on $\H$. An alternate
expression, which brings out the interpretation of $\hP_{(S,f)}$
as the `flux of the electric field through $S$', can be given in
terms of operators $\hJ_{i(u)}^{S,v}$ and $\hJ_{i (d)}^{S,v}$ on
$\cyl_\a$, associated with a 2-surface $S$ and vertices $v$ of
$\a$ at which $\a$ intersects $S$ (where $u$ stands for `up' and
$d$ for `down'). If the edges $e_1,\ldots , e_u$ of $\a$ lie
`above' $S$ and $e_{u+1},\ldots ,e_{u+d}$ lie `below' $S$, then we
set
\ba \label{ESf2}
\hJ_{i(u)}^{S,v}\ =\ \hJ_i^{(v,e_1)}+ ... + \hJ_i^{(v,e_u)}\, ,
\nonumber\\
\hJ_{i(d)}^{S,v}\ =\  \hJ_i^{(v,e_{u+1})}+
...+\hJ_i^{(v,e_{u+d})}. \label{d} \ea
In terms of these operators, we have:
\be \hP_{(S,f)}\ =\ \f{\hbar}{2} \, \sum_{v\in S} f^i(v)
\,(\hJ_{i(u)}^{S,v} - \hJ_{i(d)}^{S,v}), \ee
where the sum is over all points in $S$. (The operator is well
defined on $\cyl$ because, when acting on a cylindrical function,
only a finite number of terms in the uncountable sum are
non-zero.)

\subsubsection{Gauge and Diffeomorphism symmetries}
\label{s4.3.5}

In the classical domain, automorphisms of the bundle on which
connections are defined are symmetries of the theory. The group of
these symmetries is the semi-direct product of the group of smooth
local gauge transformations with the group of smooth
diffeomorphisms on $\M$. In this sub-section we will examine these
symmetries from the quantum perspectives. Modifications arise
because on the one hand quantum connections can be arbitrarily
discontinuous and on the other hand they are associated only with
closed-piecewise analytic edges (see section \ref{s4.3.1}).

Let us begin with gauge transformations. Given a local $G$
rotation $\bg:\M\rightarrow G$ there is an active mapping on $\Ab$
defined by
\be\label{gb} \bg \cdot \bA(e) = g(v_+)\,\bA (e)\,(g(v_-))^{-1},
\ee
for all edges $e$ in $\M$ with source $v_-$ and target $v_+$. Note
that $\bg$ can be an arbitrarily discontinuous $G$-valued function
on $\M$. We will denote the group of these gauge transformations
by $\Gb$. The natural measure $\mu_o$ on $\Ab$ is invariant under
$\Gb$, whence the corresponding action of $\Gb$ on $\H$ is
unitary:
\be (U_{\bg} \Psi_\a )(\bA) = \Psi(\bg\cdot\bA)\, .\ee
Each of the subspaces $\H'_{\a}$, $\H'_{\a, \j'}$ and $\H'_{\a,
\j', \l'}$ is left invariant by this action. Furthermore, each
quantum state in $\H'_{\a, \j', \l'= {\vec 0}}$ is gauge
invariant, i.e., mapped to itself by all $U_{\bg}$. This
observation will be useful in section \ref{s6.1} to obtain a
characterization of the Hilbert space of solutions to the quantum
Gauss constraint.

Let us now turn to diffeomorphims on $\M$. Since we have
restricted ourselves to closed-piecewise analytic edges, analytic
diffeomorphisms on $\M$ have a natural action on $\Ab$. However
there is a larger group of maps ${\varphi}: \M\rightarrow \M$
which has a natural action on $\Cyl$ \cite{lost}. Let ${\varphi}$
be a $C^n$ diffeomorphism of $\M$ such that every permissible
graph on $\M$ is mapped to a permissible graph.%
\footnote{Recall that each edge $e$ of a permissible graph is
closed-piecewise analytic. Therefore, the requirement is that for
every analytic embedding $e:[0,1]\rightarrow M$ of the interval,
the image ${\varphi}(e([0,1]))$ is a finite sum
$\cup_Ie_I([0,1])$, where each $e_I:[0,1]\rightarrow M$ is again
an analytic embedding.}
Then we can define the action of ${\varphi}$ in the space $\Ab$ of
the quantum connections, namely
\be\label{bp} {\varphi}\cdot\bA(e) := \bA ({\varphi}(e))\ee
for all paths $e$ in $\M$. Denote the group of such
diffeomorphisms by $\Diffb$.  Each element ${\varphi}$ of this
group naturally defines an isomorphism in $\Cyl$. Moreover, the
measure $\mu_o$ is $\Diffb$ invariant, therefore an operator
$U_{\bar{\varphi}}$ defined  in $\H$ by each $\bar{\varphi}$,
namely
\be U_{\bar{\varphi}}\Psi(\bA)\ :=\ \Psi(\bar{\varphi}\cdot(\bA))
\ee
is unitary. However, under that induced action of $\Diffb$, none
of the subspaces $\H'_{\a}$, $\H'_{\a, \j'}$ and $\H'_{\a, \j',
\l'}$ is left invariant; they transform covariantly.

The group $\Diffb$  is a sub-group of all $C^n$ diffeomorphisms
but it is considerably larger than the group of the entire
analytic diffeomorphisms. The crucial difference is the local
character of $\Diffb$: for every point $x\in M$ and every open
neighborhood ${\cal U}_x$ containing $x$, there is a ${\varphi}\in
\Diffb$ which moves $x$ nontrivially within ${\cal U}_x$ but is
trivial outside ${\cal U}_x$ \cite{lost}. A generic element of
$\Diffb$ is fails to be analytic. Roughly, $\Diffb$ can be thought
of as the group of piecewise analytic, $C^n$ diffeomorphisms. This
group will play an important role in the imposition of the
diffeomorphism constraint in section \ref{s6.2}.

\emph{Remark:} In the kinematic description constructed so far,
$\Psi(\bA)= 1$ is the only gauge and diffeomorphism invariant
state in $\H$.%
\footnote{Since diffeomorphisms on $M$ are generated by a first
class constraint, one would expect that \emph{all} physical states
should be diffeomorphism invariant. We will see in section
\ref{s6} that this expectation is indeed borne out but the
physical states belong to $\cyl^\star$ which is considerably
larger than the kinematical Hilbert space $\H$.}
{}From this symmetry considerations, one can regard it as the
`ground state'. It is annihilated by all the momentum (or triad)
operators. Elements of $\cyl_\a$ represent `excited states', where
the geometry is excited only along the edges of $\a$; the smeared
triad $\hat{E}_{S,f}$ has a non-trivial action on these states
only if $S$ intersects at least one edge of $\a$. Since these
basic excitations are 1-dimensional, the quantum geometry is said
to be \emph{polymer-like}. If the graph has just a few edges, we
have a highly quantum mechanical state ---the analog of a state of
the quantum Maxwell field with just a few photons. To approximate
a classical geometry, one needs a \emph{highly} excited state,
with a huge number of edges, cris-crossing $\M$ `very densely'.

Let us summarize our discussion of quantum kinematics for
background independent theories of connections. In section
\ref{s4.3.1}, we introduced a Lie algebra of holonomy and flux
functions on the classical phase space \cite{acz}. In the
subsequent sub-sections, we constructed a natural, diffeomorphism
covariant representation of the quantum analog of this
holonomy-flux algebra. For pedagogical reasons, we chose a
constructive approach and developed the theory step by step
starting from quantum mechanics on a compact Lie algebra $G$ and
passing through the quantum theory of connections on graphs. The
actual development of the subject, on the other hand, began with a
broader perspective and first principles \cite{ai,al2,jb1}). The
main problem is that of finding the physically appropriate
representation of the holonomy-flux algebra. The starting point
was the observation \cite{ai} that $\cyl$, which serves as the
algebra of configuration variables, has the structure of an
Abelian $\star$-algebra. By completing it in the sup-norm one
obtains the $C^\star$ algebra $\overline{\cyl}$ of quantum
configuration operators. The strategy was to first seek its
representations and then represent the momentum operators on the
resulting Hilbert spaces. A general theorem due to Gel'fand
ensures that \emph{every} representation of $\overline{\cyl}$ is
of the following type: The Hilbert space is the space of
square-integrable functions on a compact Hausdorff space ---called
the Gel'fand spectrum of the $C^\star$-algebra--- with respect to
a regular Borel measure, and the configuration operators act on it
by multiplication. The non-trivial fact is that the structure of
$\cyl$ is such that the spectrum is easy to exhibit: it is
precisely our space $\Ab$ \cite{al2}. Thus, the representation of
the algebra of elementary variables we constructed step by step is
in fact rooted in the general Gel'fand representation theory.

Even though this procedure is quite general and well-motivated,
one can nonetheless ask why we did not adopt the more general
algebraic approach but focused instead on a specific
representation. Interestingly, several partial uniqueness theorems
have been established indicating that the requirement of general
covariance suffices to select a unique cyclic representation of
the kinematic quantum algebra \cite{hs,ol,st1,lost}. This is the
quantum geometry analog to the seminal results by Segal and others
that characterized the Fock vacuum in Minkowskian field theories.
However, while that result assumed not only Poincar\'e invariance
but also specific (namely free) dynamics, it is striking that the
present uniqueness theorems make no such restriction on dynamics.
Thus, the quantum geometry framework is surprisingly tight. These
results seem to suggest that, for background independent theories,
the full generality of the algebraic approach may be unnecessary:
if there is a unique diffeomorphism invariant representation, one
might as well restrict oneself to it. For non-trivially
constrained systems such as general relativity, this is fortunate
because a satisfactory and manageable algebraic treatment of
theories with such constraints is yet to become available.

\section{Quantum Riemannian geometry}
\label{s5}

In this section, we will introduce simple geometric operators on
$\H$. Recall from section \ref{s2} that the internal group for the
phase space of general relativity is $\SU(2)$ and the Riemannian
geometry is coded in the triad field $\tilde{E}^a_i = k\g P^a_i
\equiv 8\pi G\gamma P^a_i$ (of density weight one), where $\g > 0$
is the Barbero-Immirzi parameter (see (\ref{P}). Therefore, in the
quantum theory we set $G= \SU(2)$ and our geometric operators are
built from the (smeared) triad operators $\hat{P}_{(S,f)}$.
Because of space limitation, we will only discuss the area and the
volume operators [65-78] which have had direct applications, e.g.,
in the entropy calculations and quantum dynamics. For the length
operator, see \cite{tt2}.

\subsection{Area operators}
\label{s5.1}

Let $S$ be either a closed 2-dimensional sub-manifold of $\M$ or
an open 2-dimensional sub-manifold without boundary. In the
classical theory, its area is a function on the phase space given
by $A(S) = \int_S d^2x\, \sqrt{h}$, where $h$ is the determinant
of the intrinsic 2-metric $h_{ab}$ on $S$. Our task is to
construct the quantum operator corresponding to this phase space
function and analyze its properties. (For further details, see
\cite{al5}).

\subsubsection{Regularization} \label{s5.1.1}

A natural strategy is to first re-express $A_S$ in terms of the
`elementary' observables $P(S, f)$, and then replace each $P(S,
f)$ by its unambiguous quantum analog. This strategy naturally
leads to a regularization procedure which we now summarize.

Let us divide $S$ in to a large number of elementary cells, $S_I$,
with $I = 1,2,\ldots N$. On each cell, introduce an internal triad
$\tau^i$ and, using its elements as test fields $f^i$, set
$P{(S_I,\tau^i)} = P^i(S_I)$. Next, recall from (\ref{P}) that the
orthonormal triad $\tilde{E^a_i}$ of density weight one is related
to the momentum field $P^a_i$ via $\tilde{E}^a_i = 8\pi G\g
P^a_i$. Set
\be [A_S]_N\, =\, 8\pi G\g \sum_{I=1}^N \sqrt{P^i(S_I)
P^j(S_I)\eta_{ij}}\, . \ee
where $\eta_{ij}$ is again the Cartan-Killing metric on $\su(2)$.
Then, $[A_S]_N$ is an approximate expression of the area $A(S)$ in
the following sense: as the number of cells goes to infinity such
that the coordinate size of the cells $S_I$ goes to zero uniformly
in $I$, we have
\be \lim_{N\to \infty}\, \,[A_S]_N\ = A_S\, . \ee
Since each $P^i(S_I)$ gives rise to an unambiguously defined
quantum operator, $[A_S]_N$ represents a suitable `regularized
area function' and the limit $N \to \infty$ corresponds to the
operation of removing the regulator. In the quantum theory, then,
we first define an approximate area operator by first noting that,
for each $I$, $\hat{P}^i(S_I) \hat{P}^j(S_I)\eta_{ij}$ is a
positive definite self-adjoint operator on $\H$ with a
well-defined (positive) square-root, and setting
\be \label{approxAhat} [\hat{A}_S]_N \, :=\, 8\pi G\g
\,\sum_{I=1}^N \sqrt{\hat{P}^i(S_I) \hat{P}^j(S_I)\eta_{ij}}\, .
\ee
To obtain an explicit expression of this operator, let us restrict
its action to Hilbert space $\H_\a$ associated with any one graph
$\a$. Let us first refine the partition  sufficiently so that
every elementary cell $S_I$ intersects $\a$ transversely at most
at one point (or contains a segment of $\a$). Then, using the
expression (\ref{ESf1}) of the smeared triad operators, we
conclude that a non-zero contribution to the sum in
(\ref{approxAhat}) comes only from those $S_I$ which intersect
$\a$ and, furthermore, a subsequent refinement of the partition
does not change the result. Thus, for any given $\a$, the limit
$N\rightarrow \infty$ is reached already at a finite step;
somewhat surprisingly, the removal of the regulator can be
achieved rather easily in the quantum theory. It is
straightforward to verify that the resulting operator $\hat{A}_{S,
\a}$ is given by:
\be \label{hatA}\hA_{S,\a}\, =\, 4\pi\g \lp^2\,
\sum_{v}\sqrt{-\Delta_{S,v,\a}}\, . \ee
where $v$ ranges through all the vertices of $\a$ which lie on $S$
and the `vertex Laplace operator' $\Delta_{S,v,\a}$ is defined on
$\cyl_\a$ as
\be \Delta_{S,v,\a}\ =\ -(\hJ_{i(u)}^{S,v} - \hJ_{i(d)}^{S,v})
(\hJ_{j(u)}^{S,v} - \hJ_{j(d)}^{S,v})\, \eta^{ij}. \ee
(The `up' and `down' operators $\hJ_{i(u)}^{S,v}$ and
$\hJ_{i(d)}^{S,v}$ are defined in (\ref{ESf2}).) Thus, on each
$\H_\a$ we have obtained a non-negative, self-adjoint area
operator $\hat{A}_{S,\a}$.

The question is if these operators can be glued together to obtain
a well-defined area operator on the full Hilbert space $\H$. As in
the definition of measures on $\Ab$, this is a question about
\textit{consistency} of the family. More precisely, suppose an
element $\Psi$ of $\cyl$ belongs both to $\cyl_{\a_1}$ and
$\cyl_{\a_2}$, for two different graphs $\a_1$ and $\a_2$. The
question is whether $\hA_{S,\a_1} \Psi$ equals $\hA_{S, \a_2}
\Psi$ as an element of $\cyl$. The answer is in the affirmative.
Thus, there is a non-negative, self-adjoint operator $\hat{A}_S$
on $\H$ whose restriction to $\cyl_\a$ is given by (\ref{hatA})
for any graph $\a$.

In fact, we can also define an \textit{area element operator}
corresponding to $\sqrt{h(x)}$ whose integral over $S$ gives the
total area operator $\hA_S$. Fix a point $x\in S$, and consider a
refinement such that $x$ is contained in the interior of a cell
$S_I$. Introduce an approximate area element $[\sqrt{h(x)}]_N$ via
\be \sqrt{[h(x)]_N}\, =\, \frac{8\pi G \g}{\epsilon^2}\,
\sqrt{P^i(S_I)P^j(S_I)\, \eta_{ij}} \ee
where $\epsilon^2$ is the coordinate area of the cell $S_I$. As we
let $N$ tend to infinity, shrinking the coordinate size of the
cells uniformly, $\sqrt{{h(x)}_N}$ tends to $\sqrt{h(x)}$. As
before, we can pass to a regularized quantum operator
\be \widehat{[\sqrt{h(x)]_N}} \, =\, \frac{8\pi G\g}{\epsilon^2}\,
\sqrt{\hat{P}^i(S_I)\hat{P}^j(S_I)\, \eta_{ij}} \ee
simply by replacing the smeared $P$'s with corresponding
operators. Finally, we remove the regulator. The result is a
well-defined operator-valued distribution on $\H$, whose action on
$\cyl_\a$ is given by:
\be \widehat{\sqrt{h_S(x)}}  =\, 4\pi\g \lp^2\, \sum_{v}
\delta^2_{(S)}(x,v)\, \sqrt{-\Delta_{S,v,\a}}. \ee
where $\delta^2_{(S)}(x,v) $ is the 2-dimensional Dirac
distribution on $S$ and the sum is over intersections $v$ of $\a$
and $S$. Again, this family of operators is consistent and
therefore defines an operator $\widehat{\sqrt{h_S(x)}}$ on $\H$.

\emph{Remark}: In the definition of the momentum operators
$\hat{P}(S,f)$ and area operators $\hat{A}_S$,  we only considered
2-manifolds $S$ without boundary. Now, if we sub-divide $S$ as
$S=S'\cup I\cup S''$, where $S'$ and $S''$ are two 2-dimensional
sub-manifolds without the boundaries and $I$ is a 1-dimensional
sub-manifold without boundary, then while classically $A_S -
(A_{S'} + A_{S''}) = 0 $, because of the distributional nature of
quantum geometry, $\hA_S - (\hA_{S'} + \hA_{S''})$ is non-zero
since its action on graphs with edges passing through $I$ is
non-trivial. To obtain additivity of areas, it is then natural to
regard this operator as defining the quantum area $\hat{A}_{S,I}$
of $I$, although $I$ is a 1-manifold. Proceeding in this manner,
one is led to assign a quantum area operator $\hat{A}_{V,S}$ also
to a point $v$ of $S$. Detailed examination shows that this
operator is just $4\pi\g\lp^2\, \sqrt{-\Delta_{S, v}}$. {}From
this perspective, then, $\hA_S = \sum_v \hat{A}_{S,v}$; the
quantum area of a surface is obtained by summing up the `areas
associated with all points' in it!

\subsubsection{Properties of area operators} \label{s5.1.2}

In each $\cyl_\a$, the area operator is defined  by
\be\label{hA} \hA_{S,\a}\ =\ \int_S\, d^2x \widehat{\sqrt{h_{S,
\a}}\,(x)} =\ 4\pi\g \lp^2\,\sum_v \sqrt{-\Delta_{S,v,\a}}. \ee
With domain $\cyl_\a^{(2)}$, consisting of twice differentiable
functions on $\Ab_\a$, this operator is essentially self-adjoint
on $\H_\a$. Since this family of operators is consistent, the
resulting area operator, with domain $\cyl^{(2)}$, is also
essentially self-adjoint on $\H$. By inspection, the operator is
gauge invariant (i.e. commutes with the vertex operators
$\hat{J}_i^v$ generating $\SU(2)$ gauge rotations at vertices
$v$). Since its definition does not require a background
structure, it is diffeomorphism covariant.

The eigenvalues of the operator are given by finite sums
\be\label{as} a_S\ =\  4\pi\g \lp^2 \, \sum_I \sqrt{-\lambda_I},
\ee
where $\lambda_I$ are arbitrary eigenvalues of the operators
$\Delta_{S,v_I}$. Now, this operator can be cast in a convenient
form as a sum of three commuting operators,
\be -\Delta_{S,v}\ =\ 2(\hJ^{(d)}_{S,v})^2 + 2(\hJ^{(u)}_{S,v})^2
- (\hJ^{(u)}_{S,v} + \hJ^{(d)}_{S,v})^2, \ee
which makes its eigenvalues transparent. These are given by
\be\label{lambda} -\lambda\ =\ 2j^{(u)}(j^{(u)}+1) +
2j^{(d)}(j^{(d)}+1) - j^{(u+d)}(j^{(u+d)}+1), \ee
where $j^{(u)}, j^{(d)}$ and $j^{(u+d)}$ are arbitrary
half-integers subject to the standard condition
\be \label{jconstraint} j^{(u+d)}\ \in \{|j^{(u)}- j^{(d)}|,
|j^{(u)}- j^{(d)}|+1, ... , j^{(u)}+j^{(d)}\}. \ee
Thus, the general eigenvalues of the area operator are given
finite sums:
\be \label{spectrum} a_S \ =\  4\pi\g \lp^2 \, \sum_I
\sqrt{2j^{(u)}(j^{(u)}+1) + 2j^{(d)}(j^{(d)}+1) -
j^{(u+d)}(j^{(u+d)}+1)}\ee
where the $j$s are subject to the constraint (\ref{jconstraint}).
Thus, all eigenvalues are discrete and the area gap  ---the
smallest non-zero eigenvalue $a_S$--- is given by
\be \Delta a_S\ =\ 4\pi\g\, \lp^2 \,\,
\frac{\sqrt{3}}{2}.\label{gap} \ee
The level spacing between consecutive eigenvalues is \emph{not}
uniform but decreases \emph{exponentially} for large eigenvalues.
This implies that, although the eigenvalues \emph{are}
fundamentally discrete, the continuum approximation becomes
excellent \emph{very rapidly}. On the full kinematic Hilbert space
$\H$ ---as opposed to the gauge invariant sub-space considered
below--- all these properties are insensitive to the topology of
$S$.

\subsubsection{ The gauge invariant subspace} \label{s5.1.3}

Let us now restrict ourselves to gauge invariant subspace $\H_{\rm
inv}$ of $\H$. This is spanned by elements of $\cyl$ which have
zero eigenvalue for every vertex operator $\hat{J}_i^v$ (i.e., in
the terminology of section \ref{s4.2}, states in the subspaces
$l_v =0$ for all vertices $v$). Now, the spectrum of the area
operator $\hA_S$ \emph{depends on some global properties of} $S$.
If the closure of $S$ is a manifold with a non-trivial boundary,
then the spectrum is the same as in (\ref{spectrum}). However, if
$\partial S=\emptyset$, then gauge invariance imposes certain
additional conditions on the total spin `coming in' $S$.

Let us focus this case. Suppose first, that $S$ divides $M$ in to
two disjoint open sets (as would happen if $\M$ were $R^3$ and $S$
a 2-sphere in it). Then the spins $j^{(u)}_I$, $j^{(d)}_I$ in
(\ref{lambda}) have to satisfy the following condition
\be \sum_I j^{(u)}_I\ \in \Nat, \ \ \sum_I j^{(d)}_I\ \in \Nat.
\ee
where $\Nat$ is the set of natural numbers. In the case when $S$
has no boundary  but $\M\setminus S$ is connected (as can happen
if $M$ is a 3-torus and $S$ a 2-torus in it) the condition is
milder,
\be \sum_I j^{(u)}_I + \sum_I j^{(d)}_I\ \in \Nat. \ee
In particular, in these cases, the area gap increases. In the
first case, it is given by $4\pi \g\, \lp^2\,\, (\sqrt{2})$ while
in the second case, by $4\pi \g\, \lp^2$. Thus, there is an
interesting interplay between topology of $(M,S)$ and the area
gap.

If there are no fermionic fields, then all physically relevant
states lie in the gauge invariant sub-space $\H_{\rm inv}$ of $\H$
now under consideration.  However, in presence of fermions, the
gravitational part of the state by itself will not be gauge
invariant at vertices where fermions are located. In particular,
then, if there are fermions in the interior of $S$ --say when $S$
is a 2-sphere-- the area eigenvalues of $S$ are less restricted
and we can `detect' presence of these fermions from these
eigenvalues!

\emph{Remarks}:\\
(i) Fix a surface $S$ and consider only those states in $\cyl$ for
which the graph has no edge which lies within $S$ and which are
gauge invariant at each vertex where $S$ intersects the graph.
(This is in particular the case if all intersections of $S$ with
the graph are at simple bi-valent vertices.) In this case,
$j_I^{(u+d)}=0$ and $j_I^{(u)}=j_I^{(d)}$, and the area spectrum
simplifies considerably to
\be\label{rev} a_S\ =\ 8\pi\g \lp^2\, \sum_I \sqrt{j_I(j_I+1)},
\ee
It was first believed, incorrectly, that these are all the area
eigenvalues. However, in the case of an isolated horizon, only
these eigenvalues are relevant and hence, even now, one often sees
only this expression in the literature in place of the complete
spectrum (\ref{spectrum}).\\
(ii) It follows from the definition (\ref{hatA}) of area operators
that $\hat{A}_S$ and $\hat{A}_{S'}$ fail to commute if the
surfaces $S$ and $S'$ intersect. This is a striking property
because it implies that the Riemannian geometry operators can not
all be diagonalized simultaneously.%
\footnote{Thus, the assertion that `the spin-network basis
diagonalizes all geometrical operators' that one sometimes finds
in the literature is incorrect. As we saw in section \ref{s4.3.3},
while there is a natural spin network decomposition (\ref{snd}) of
$\H$, there is no natural spin network basis. Given a surface $S$,
we \emph{can} find a spin-network basis which diagonalizes
$\hat{A}_S$ but there is no basis which diagonalizes area
operators associated with all surfaces.}
At one level this is not surprising because, even in quantum
mechanics, if the configuration space is a non-trivial manifold,
in general the momentum representation does not exist. However,
this result brings out a fundamental tension between
connection-dynamics and geometrodynamics. As we saw, quantum
connection-dynamics is very `tight'; once we choose the holonomies
$A(e)$ and the `electric fluxes' $P(S,f)$ as basic variables,
there is essentially no freedom in the background independent
quantization. Thus, under these seemingly mild assumptions, one is
led to conclude that the metric representation does not exist (at
least in the obvious sense). Although this non-commutativity is of
considerable conceptual interest, in semi-classical states the
expectation value of the commutator would be extremely small; the
non-commutativity appears to have no observable effects except at
the Planck scale \cite{acz}.

\subsection{Volume operators}
\label{s5.2}

Let $R$ be an open subset of $\M$. In the classical theory, its
volume is a function on the phase space given by  $V_R\ =\
(\sqrt{8\pi G\g})^3 \int_R d^3x \sqrt{|\det P|}$ (see
(\ref{detq})). Our task is to construct the quantum operator
corresponding to this phase space function and analyze its
properties. (For further details, see \cite{al6,jl1,rdp,tt1}).

\subsubsection{Regularization}
\label{s5.2.1}

As in the case of the area operator, we will first recast the
classical expression of $V_R$ in terms of the `elementary'
observables $P(S, f)$, and then replace each $P(S, f)$ by its
unambiguous quantum analog. This will provide the regularized
volume operator. However the final step, in which the regulator is
removed, turns out to be technically more subtle than that in
section \ref{s5.1.1} and will require an additional construction.

\begin{figure}
\centerline{\hbox{\epsfig{figure=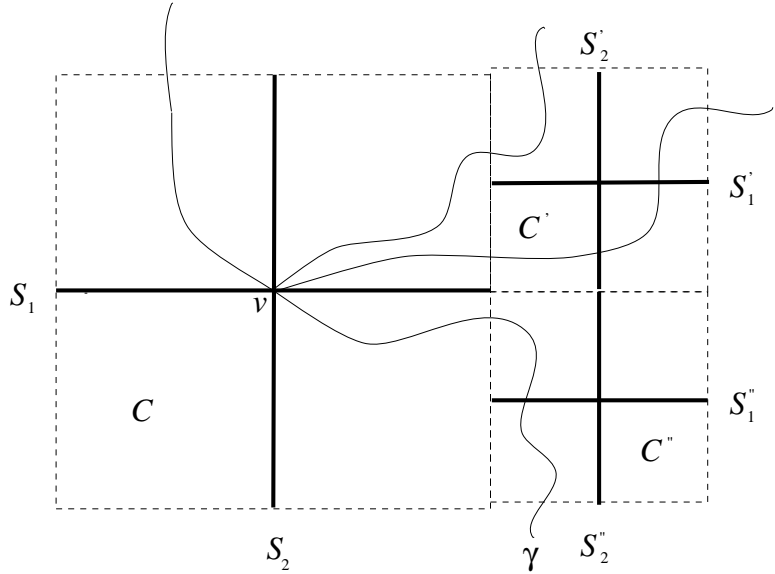,width=3in,angle=270}}}
\caption{The figure illustrates a partition $\P_\epsilon$ with
cells $C, C', C''$ (the dashed lines) and 2-surfaces $S_a, {S'}_a,
{S''}_a$ (the bold faced lines). $v$ is a vertex of  a graph $\g$.
For simplicity, one dimension has been dropped.} \label{volfig}
\end{figure}

Let us fix a coordinate system $(x^a)$ in $R$ and a positive
number $\epsilon$. We then define a partition $\P_\epsilon$ of $R$
as follows. Divide $R$ in to a family $\C$ of cells $C$ such that
each $C$ is a cube with volume less than $\epsilon$ in the given
coordinate system and two different cells share only points on
their boundaries. In each cell $C$, introduce three 2-surfaces
$s=(S^1, S^2, S^3)$, such that each of the surfaces splits $C$
into two disjoint parts, and $x^a|_{S^a}=\const$ for $a=1,2,3$.
The family of pairs $(C,s)$ defines $\P_\epsilon$ (see Fig
\ref{volfig}). Given a partition $\P_\epsilon$, we can introduce
an approximate expression of the volume $V_R$:
\ba V_\R^{\P_{\epsilon}}\ &=&\ \sum_{C\subset \R}\sqrt{|q_s|}
\quad {\rm where} \nonumber \\
q_{C,s}\ &=& \ \frac{(8\pi G \g)^3}{3!}\epsilon^{ijk}\,
\eta_{abc}\, P^i(S^a)P^j(S^b)P^k(S^c), \label{q} \ea
It is easy to verify that
$$ \lim_{\epsilon \to 0} V_R^{\P_\epsilon} = V_R \, ;$$
the dependence on the coordinate system and the partition
disappears in the limit.

To pass to the quantum theory, we need to define a consistent
family of volume operators $\hat{V}_{R,\a}$, one for each graph
$\a$. Let us then fix a graph $\a$ and consider a partition
$\P_{\epsilon}$ such that each vertex $v$ of $\a$ is the
intersection point of the triplet of 2-surfaces $S^1,S^2,S^2$ in
some cell $C_V$. Then, we can easily promote the approximate
volume function $V_R^{\P_\epsilon}$ to a quantum operator
$\hV_{R,\a}^{\P_\epsilon}$:
\ba \hV_{R, \a}^{\P_{\epsilon}}\ &=& \ \sum_{C\subset
\R}\sqrt{|\hq_{s, C}|}, \quad {\rm where}\nonumber \\
\hq_s\ &=& \  \frac{(8\pi G \g)^3}{3!}\, \epsilon_{ijk}\,
\eta_{abc}\, \hP^i (S^a)\hP^j (S^b) \hP^k (S^c),\label{hq} \ea
This operator is well-defined on $\cyl_\a^{(3)}$, the space of
thrice differentiable functions on $\Ab_\a$. Furthermore, the
limit $\epsilon \rightarrow 0$, of the operator exists. However,
unlike in the classical theory, it carries a memory of the
partition $\P_\epsilon$ used in the regularization process. This
is a new complication which did not occur in the case of the
simpler area operators. But one can handle it simply by averaging
with respect to the essential background structures, used in the
construction of the partition $\P_\epsilon$, prior to the removal
of the regulator. This extra step can be carried out in detail
\cite{al6}. The resulting operator is given by:
\ba\label{vol1} \hV_{R,\a} &=&\ \kappa_o \, \sum_{v}\,
\sqrt{|\hat{q}_{v,\a}|}, \quad {\rm where} \nonumber \\
\hq_{v, \a} \ &=& \  (8\pi \g \lp^2)^3 \, \frac{1}{48}\,\,
\sum_{e,e',e''}\, \epsilon_{ijk}\, \epsilon(e,e',e'')\,
\hJ_i^{(v,e)}\hJ_j^{(v,e')}\hJ_k^{(v,e'')} \ea
Here $\kappa_o$ is an undetermined overall constant resulting from
the averaging (usually set equal to 1); $v$ runs over the set of
vertices of $\a$; each of $e$, $e'$, and $e''$, over the set of
edges of $\a$ meeting at $v$; and $\epsilon(e,e',e'')$ is the
orientation factor. (Thus, $\epsilon(e,e',e'') = 0$ if the tangent
vectors to the three edges are planar, i.e., lie in a 2-plane, at
$v$,  and $\pm 1$ if the orientation they define is the same as or
opposite to the fiducial orientation of $\M$.) It is
straightforward to verify that this family of operators is
consistent and hence defines a single densely defined operator
$\hat{V}_\R$ on $\H$ with domain $\cyl^{(3)}$.

Again, it is also meaningful to introduce in each $\cyl^{(3)}_\a$
a \emph{volume element operator}
\be \widehat{\sqrt{q(x)}}_\a\ =\ \kappa_o
\sum_{v}\delta^{(3)}(x,v)\, \sqrt{|\hq_v,\a|}. \ee
The family of these operators is consistent and we thus have a
densely defined operator $\widehat{\sqrt{q(x)}}$ on $\H$
satisfying: $\hV_R = \int_R\, d^3x\, \widehat{\sqrt{q(x)}}$.

Finally, in the classical the Poisson bracket $\{A_a^i (x), V_R\}$
between the connection at a point $x$ and the volume of a region
containing that point is proportional to the co-triad $e_a^i(x)$.
This fact has been exploited to introduce co-triad operators
which, in turn, have led to a definition of the length operator
\cite{tt2} and features prominently in the discussion of quantum
dynamics of \ref{s6.3} \cite{tt3,tt4,tt5,tt6}.

\subsubsection{Properties of volume operators}
\label{s5.2.2}

By inspection, $\hV_\R$ is gauge invariant and covariant with
respect to diffeomorphisms of $\M$: under $\varphi: \M \rightarrow
\M$, we have $\hV_R \rightarrow \hV_{\varphi\cdot R}$. The
\emph{total} volume operator $\hV_{\M}$ is diffeomorphism
invariant. Hence, its action is well-defined on the diffeomorphism
invariant sub-space of $\cyl^\star$. This property plays an
important role in the analysis of the Hamiltonian constraint.

Because of the presence of $\epsilon(e,e',e'')$ in (\ref{vol1}),
it is clear that $\hq_{v,\a} = 0$ if all edges meeting at $v$ are
planar. In particular, then, $\hq_{v,\a} = 0$ if $v$ is a
bi-valent vertex. \emph{More surprisingly, $\hq_{v, \a}\, \Psi=0$
also for every tri-valent vertex $v$ provided the state $\Psi$ is
invariant with respect to the gauge transformations at $v$}
\cite{rl1}. Indeed, let $e,e',e''$ be the edges of $\a$ which meet
at $v$, and, for definiteness, suppose that
$\epsilon(e,e',e'')=1$. Then, using gauge invariance at $v$, i.e.,
$\hJ_i^{(v,e)} + \hJ_i^{(v,e')} + \hJ_i^{(v, e'')} = 0$, we
obtain:
\ba \f{48}{(8\pi \g \lp^2)^3}\, \hq_{v,\a}\, \Psi &=&\
\epsilon^{ijk}\, \hJ_i^{(v,e)} \hJ_j^{(v,e')}\hJ_k^{(v, e'')}\,
\Psi\
\nonumber\\
&=& \ -\epsilon^{ijk} \hJ_i^{(v,e)}(\hJ_j^{(v,e)}+\hJ_j^{(v,e'')})
\hJ_k^{(v,e'')}\,\Psi\ \nonumber\\
&=& -2i(\hJ^{(v,e)}_j\hJ^{(v,e'')}_k \eta^{jk}-
\hJ^{(v,e)}_i\hJ^{(v,e'')}_j\eta^{ij})\,\Psi\nonumber\\
&=&\ 0. \ea

As with the area operator, it is easy to show that all eigenvalues
of $\hq_{v, \a}$, $\hV_{R, \a}$ and $\hV_R$ are real and discrete.
The spectrum ---i.e., the set of all eigenvalues--- of $\hV_R$ is
the same irrespective of the specific open region $R$. Given a
point $x \in \M$, the spectrum --i.e., the complete set of
eigenvalues--- of $\widehat{\sqrt{q(x)}}$ is simply given by the
union of the spectra of the restrictions of $\hq_x$ to each of the
(finite dimensional) spin-network spaces $\S_{\a,\j, \l}$ in the
orthogonal decomposition of $\H$. Because of this  property, many
eigenvalues and eigenvectors of the volume operators $\hat{V}_R$
operators have been calculated in a number of special cases
\cite{rdp,rl2,dp1,dp2}. However the complete spectrum, or even an
estimate of how the number of eigenvalues grows for large volumes,
is not yet known.

On the space of gauge invariant states, the simplest eigenvectors
arise from 4-valent vertices. Even in this case, the full set of
eigenvalues is not known. However, a technical simplification
enables one to calculate the matrix element of the volume
operator, which have been useful in the analysis of the quantum
Hamiltonian constraint. Let $v$ be a 4-valent vertex of $\alpha$
at which edges $e,e',e'',e'''$ meet and consider the action of
$\hq_{v, \a}$ on the subspace $S_{\a,\j,\l}$ with $l_v =0$. Then,
\be \hq_{v, \a}\ =\ (8\pi\g\lp^2)^3\,\,
\frac{1}{8}\,\,\kappa(e,e',e'',e''')\,\epsilon^{ijk}
\hJ_i^{(v,e)}\hJ_j^{(v,e')}\hJ_k^{(v,e'')} \ee
where $\kappa(e,e',e'',e''')\, \in \{-2,-1,0,2,3,4\}$ depending on
the diffeomorphism class of the four tangent vectors at $v$. Using
gauge invariance at $v$, the expression can be cast in the form:
\be \hq_{v, \a}\ =\ (8\pi \g \lp^2)^3\, \frac{1}{32i}\,
\kappa(e,e',e'',e''')\, [(\hJ^{(v,e')}+\hJ^{(v,e'')})^2,
(\hJ^{(v,e)}+\hJ^{(v,e'')})^2], \ee
which simplifies the task of calculating its matrix elements in
the subspace $\H_{\a, \j,\l}$ \cite{tt1}.

Finally, we note a property of the volume operator which plays an
important role in quantum dynamics. Let $R(x,\epsilon)$ be a
family of neighborhoods of a point $x\in \M$. Then, given any
element $\Psi$ of $\cyl^{(3)}_\a$,
$$ \lim_{\epsilon \to 0} \hV_{R(x,\epsilon)}\, \Psi $$
exists but is not necessarily zero. This is a reflection of the
`distributional' nature of quantum geometry.

\subsubsection{`External' regularization}
\label{s5.2.3}

Since the basic momentum variables are smeared on 2-surfaces, in
the regularization procedure for defining geometric operators one
invariably begins by re-expressing the geometric functions on the
classical phase space in terms of $P(S,f)$. However, there is
considerable freedom in achieving this and, while different
expressions may yield the same function on the classical phase
space when the regulator is removed, their quantum analogs need
not share this property. This is the standard `factor ordering
problem' of quantum theory. In particular, in the procedure
summarized in section \ref{s5.2.1}, we expressed the volume of
each elementary cell $C$ in terms of three 2-surfaces $S^a$
($a=1,2,3$) which lie \emph{inside} that cell. This strategy goes
under the name `internal regularization'. A natural alternative is
to use the six 2-surfaces $\tilde{S}^\alpha$, $\alpha = 1,\ldots
6$ which \emph{bound} the cell. This `external regularization'
strategy was first introduced by Rovelli and Smolin \cite{rs4} for
gauge invariant states on tri-valent graphs. Although it was later
realized that this volume operator is identically zero on these
states \cite{rl1}, Rovelli and de Pietri \cite{rdp} showed that
the method extends also to non-trivial situations.

A detailed analysis \cite{al6} shows that this strategy is equally
viable, once due attention is paid to the convergence issues (that
arise while removing the regulator) by carefully constructing the
partition of $R$. Then, the final volume operator is again of the
form (\ref{vol1}) but given by:
\ba\label{extvol} \hV_{R,\a}^{\rm Ext} &=&\ \kappa_o \, \sum_{v}\,
\sqrt{|\hat{q}_{v,\a}^{\rm Ext}|}, \quad {\rm where} \nonumber \\
\hq_{v, \a}^{\rm Ext} \ &=& \  (8\pi \g \lp^2)^3 \,
\frac{1}{48}\,\, \sum_{e\not=e'\not=e''\not=e}\, \epsilon^{ijk}\,
\hJ_i^{(v,e)}\hJ_j^{(v,e')}\hJ_k^{(v,e'')}\ea
%
A state $\Phi$ which is cylindrical with respect to a graph $\a$
and gauge invariant at a tri-valent vertex $v$ is again
annihilated by the new $\hq_{v, \a}^{\rm Ext}$. Furthermore, at
gauge invariant 4-valent vertices, $\hq_{v, \a}^{\rm Ext}$ agrees
with $\hq_{v, \a}$, \emph{modulo} a multiplicative factor which
depends on the diffeomorphism class of the tangent vectors at $v$.
In spite of this close relation on simple states, the two
operators are fundamentally different because of the absence of
the orientation factor $\epsilon (e,e', e'')$ in (\ref{extvol}).
In particular, because of this factor, the operators constructed
in section \ref{s5.2.1} know about the differential structure at
vertices of graphs. By contrast, the action of (\ref{extvol}) is
`topological'.

In the literature, internally regulated operators of section
\ref{s5.2.1} are used more often. For example, Thiemann's analysis
of properties volume operators in the continuum \cite{tt1} and
Loll's analysis on lattices \cite{rl2,rl3} refer to (\ref{vol1}).
The same is true of the volume operators used by Thiemann and
others in the discussion of the Hamiltonian constraint.

\section{Quantum dynamics}
\label{s6}

The quantum geometry framework provides the appropriate arena for
a precise formulation of quantum Einstein's equations. As
indicated in section \ref{s2}, because of the difficult problems
of background independent regularization of products of operator
valued distributions, quantum Einstein's equations still remain
formal in geometrodynamics. In connection-dynamics, by contrast,
we have a well-defined Hilbert space $\H$ of kinematical states
and it is natural to attempt to represent left sides of quantum
Einstein's equations by well-defined operators on $\H$. Now, in
interacting (low dimensional) quantum field theories, there is a
delicate relation between quantum kinematics and dynamics: unless
the representation of the basic operator algebra is chosen
appropriately, typically, the Hamiltonian fails to be well-defined
on the Hilbert space. For a complicated system such as general
relativity, then, one would imagine that the problem of choosing
the `correct' kinematic representation would be extremely
difficult (see, e.g. \cite{kk}). However, a major simplification
arises from the striking uniqueness result discussed at the end of
section \ref{s4}: the requirement of general covariance picks out
a unique representation of the algebra generated by holonomies and
electric fluxes \cite{hs,ol,st1,lost}. Therefore we have a
\emph{single arena} for background independent theories of
connections and a natural strategy for implementing dynamics
provided, of course, this mathematically natural, kinematical
algebra is also `physically correct'. (This proviso exists also
for the quantum field theories referred to above.) As we will
summarize in this section, this strategy has led to well-defined
candidates for quantum Einstein's equations.

Recall from section \ref{s2} that because general relativity has
no background fields, the theory is fully constrained in its phase
space formulation. To pass to the quantum theory, one can use one
of the two standard approaches: i) find the reduced phase space of
the theory representing `true degrees of freedom' thereby
eliminating the constraints classically and then construct a
quantum version of the resulting unconstrained theory; or ii)
first construct quantum kinematics for the full phase space
ignoring the constraints, then find quantum operators
corresponding to constraints and finally solve quantum constraints
to obtain the physical states. Loop quantum gravity follows the
second avenue, which was initiated by Dirac.%
\footnote{Thus, in the canonical approach, the entire quantum
dynamics is fully incorporated by solving quantum constraints.
This may seem surprising because in the classical theory we have
both the constraint and the evolution equations. However, because
the evolution is generated by constraints in the Hamiltonian
framework, in the quantum theory dynamics is encoded in the
operator constraints. A simple example is provided by a free
particle in Minkowski space, where the constraint $g^{ab}p_a p_b +
m^2=0$ on the classical phase space becomes $\Box \phi - m^2 \phi
=0$, governing dynamics in the quantum theory }
This program has been carried out to completion in many simpler
systems, such as 2+1 dimensional gravity \cite{ahrss,aa1,sc1} and
a number of mini-superspaces in 3+1 dimensions \cite{atu}, where
one can explicitly see that the procedure incorporates all of
quantum Einstein equations. Readers who are not familiar with
quantization of constrained systems should first familiarize
themselves with the subject through simple examples (see, e.g.
\cite{at}). To adequately handle conceptual and technical
intricacies encountered in general relativity, Dirac's original
program has to be modified and extended suitably. We will use the
resulting framework, called \emph{refined algebraic quantization}.
For further details, see, e.g., \cite{dm,almmt1,gm}.

\subsection{The Gauss constraint}
\label{s6.1}

Recall from section \ref{s2} that the Gauss constraint, $G_i
\equiv \D_aP^a_i =0$, generates internal $\SU(2)$ rotations on the
phase space of general relativity. More precisely, given an ${\rm
su(2)}$-valued function $\xi$ on $\M$, we can use it as a smearing
field to obtain a phase space function
\be \Gauss(\xi)\ =\  -\int_\M d^3x\, P_i^a(x)\D_a\xi^i(x)\, , \ee
which generates infinitesimal canonical transformations $(A, P)
\rightarrow (A - {\cal D} \xi, P+[\xi,\, P])$. Using the heuristic
ans\"atz $P \rightarrow -i\hbar\delta/\delta A$, it is
straightforward to promote $\Gauss(\xi)$ to a well-defined
operator on $\H$ \cite{almmt1}. For any $\Psi_\a \in
\cyl_\a^{(1)}$ we have:
\be \hG(\xi)\,\Psi_\a =  \hbar\sum_v \,\sum_e\,\, (\xi^i(v)
J_i^{(v,e)})\,\Psi_\a \ee
where the first sum extends over all vertices $v$ of $\a$ and the
second over all edges $e$ meeting at $v$. Apart from the factor of
$\hbar$, this action coincides with that of the generator of gauge
transformations on $\H_\a$ discussed in section \ref{s4.2.2}. This
family of operators on $\H_\a$ is consistent and defines a
self-adjoint operator on $\H$ which we will also denote by
$\hG(\xi)$. Finite gauge transformations are generated by the
1-parameter unitary groups generated by these operators.

Physical states belong to the kernel $\H^{\rm G}_{\rm inv}$ of
$\hG(\xi)$ for all $\xi \in {\rm su(2)}$. Because the action of
$\hG$ is familiar, the kernel is easy to find: In terms of the
Hilbert space decompositions discussed in section \ref{s4.3.3},
$$ \H^{\rm G}_{\rm inv} =\bigoplus_{\a, \j} \,\,\H^\prime_{\alpha,\j,
\l = {\bf 0}}\, .$$
Note that these states are automatically invariant under
\emph{generalized} gauge transformations in $\Gb$ and can be
regarded as functions on the reduced quantum configuration space
$\Ab/\Gb$. $\H^{\rm G}_{\rm inv}$ is a sub-space of $\H$ because
zero is in the discrete part of the spectrum of the constraint
operator $\hG(\xi)$. In particular, $\H^{\rm G}_{\rm inv}$
inherits a Hilbert space structure from $\H$ and $\H^{\rm G}_{\rm
inv} = L^2(\Ab/\Gb, d\mu_o^{\rm G})$, where $d\mu_o^{\rm G}$ is
the natural measure on $\Ab/\Gb$, the push-forward of $d\mu_o$
under the natural projection map from $\Ab$ to $\Ab/\Gb$. Every
gauge invariant operator ---such as areas $\hA_S$ and volumes
$\hV_R$ of section \ref{s5}--- has a well defined action on
$\H^{\rm G}_{\rm inv}$.

The fact that the Gauss constraint could be imposed so easily and
that the structure of $\H^{\rm G}_{\rm inv}$ is so simple hides
the non-triviality of the procedure. For example if, in place of
$\Ab$, one uses one of the standard distribution spaces as the
quantum configuration space, the imposition of the Gauss
constraint and construction of the Hilbert space of physical
states becomes complicated and it is not obvious that these
difficulties can be surmounted.

\subsection{The diffeomorphism constraint}
\label{s6.2}

Let us now consider the diffeomorphism constraint. We will find
that the imposition of this constraint is more complicated because
of a key difference: While there is an infinite dimensional
subspace $\H^{\rm G}_{\rm inv}$ of $\H$ that is invariant under
the $\SU(2)$ gauge rotations, since diffeomorphisms move graphs,
the only element of $\H$ left invariant by the action of all
diffeomorphisms is the constant function on $\Ab$\,! As a result,
solutions to the quantum constraints lie not in the kinematical
Hilbert space $\H$ but in a larger space, the dual $\cyl^\star$ of
$\cyl$. This is not unusual. Even in simple quantum mechanical
systems, such as a particle in $R^3$ with a constraint $p_x = 0$,
solutions to the constraint fail to have finite norm in the
kinematic Hilbert space $L^2(R^3)$ and belong to a larger space,
e.g., the space of distributions in $R^3$. In a similar fashion,
we will be able to construct a systematic framework and obtain the
general solutions to the diffeomorphism constraint.

\subsubsection{Strategy}
\label{s6.2.1}

Recall from section \ref{s2.3.3} that each vector field $N^a$ on
$\M$ defines a constraint function $\C_{\Diff}(\vec{N})$ on the
gravitational phase space:
\be\label{diff}  \C_{\Diff}(\vec{N})\ =\ \int_\M d^3x \,
\left(N^aF^i_{ab}P^b_i - P^a {\cal D}_a (N^bA_b^i)\right) .\ee
Under infinitesimal canonical transformations generated by
$\C_{\Diff}(\vec{N})$, we have: $(A, P) \mapsto (A+ {\cal L}_{\vec
N} A ,\, P + {\cal L}_{\vec N} P)$. In the mathematically precise
literature on constrained systems it is the \emph{finite} gauge
transformations generated by constraints that are of primary
interest in the quantum theory. Therefore, in our case, it is
appropriate to impose the diffeomorphism constraint by demanding
that the physical states be left invariant under finite
diffeomorphisms $\varphi$ generated by $\vec{N}$. Since the
measure $d\mu_o$ on $\Ab$ is diffeomorphism invariant, the induced
action of $\varphi$ on $\H$ is unitary. Thus, given the vector
field $N^a$, we obtain a 1-parameter family $\varphi(\lambda)$ of
diffeomorphisms on $\M$ and a corresponding family
$\hat\varphi(\lambda)$ of unitary operators  on $\H$. But this
family fails to be weakly continuous in $\lambda$ because
$\cyl_\a$ is \emph{orthogonal} to $\cyl_{\varphi\cdot \a}$ if
$\varphi$ moves $\a$. Hence, the infinitesimal generator of
$\hat\varphi(\lambda)$ fails to exist. (For details, see Appendix
C in \cite{almmt1}.) However, this creates no obstacle because,
for the quantum implementation of the constraint, we can work
directly with finite diffeomorphisms: Physical states are to be
invariant under the induced action $\hat\varphi$ of appropriate
diffeomorphisms $\varphi$ on $\M$.

To solve the constraint, we will use the `group averaging
procedure', generally available for such constraints%
\footnote{The quantum Gauss constraint can be rigorously
implemented also through group averaging over $\Gb$. The final
result is the same as that obtained in section\ref{s6.1}. For
pedagogical purposes, in section \ref{s6.1} we adopted a procedure
which is closer to the one followed for the scalar constraint in
section \ref{s6.3}.}
(see, e.g., \cite{dm,almmt1}): Physical states will be obtained by
averaging elements of $\cyl$ with respect to the induced action of
the diffeomorphism group. It is intuitively obvious that the
result of group averaging will be diffeomorphism invariant.
However, although one begins with states in $\cyl$, the result
naturally belongs to $\cyl^\star$, the algebraic dual of $\cyl$.%
\footnote{In the end, one would have to introduce a suitable
topology on $\cyl$ which is finer than the Hilbert space topology
and let $\cyl^\star$ be the topological dual of $\cyl$. The
program is yet to reach this degree of sophistication and, for the
moment, the much bigger algebraic dual is used.}
In finite dimensional constrained systems, one generally uses a
triplet, $\S \subset L^2(R^n)= H_{\rm kin} \subset \S^\star$,
where $\S$ is typically the space of smooth functions with rapid
decay at infinity, and $\S^\star$, the space of distributions. The
solutions to constraints are obtained by averaging elements of
$\S$ with respect to the group generated by constraints and they
typically belong to $\S^\star$ rather than to the kinematical
Hilbert space $H_{\rm kin}$ \cite{dm,gm}. In the present case, we
have a completely analogous situation and now the triplet is $\cyl
\subset \H \subset \cyl^\star$.

Finally there is an important technical subtlety coming from the
fact that, graphs $\a$ are required to have {closed-piecewise
analytic} edges. The classical phase space, on the other hand,
consists of smooth (i.e. $C^n$) fields $(A, P)$. Smooth
diffeomorphisms $\varphi$ correspond to finite canonical
transformations generated by the constraint (\ref{diff}) and have
a well-defined action on the phase space. It is just that the
action does not extend to our full algebra of `elementary
variables' since their definition involves closed-piecewise
analytic edges and surfaces. A natural strategy to impose the
diffeomorphism constraint, therefore, is to enlarge the framework
and allow smooth edges and surfaces. This is possible
\cite{bs1,bs2,lt,cf4} but then the technical discussion becomes
much more complicated because, e.g., two smooth curves can
intersect one another at an infinite number of points. Here we
will adopt an `in-between' approach and use the sub-group $\Diff$
of all $C^n$ diffeomorphisms of $\M$, introduced in section
\ref{s4.3.5}, which has a well-defined action on our elementary
variables and the Hilbert space $\H$. {}From a physical
perspective, this is more appropriate than averaging with respect
to just the analytical diffeomorphisms and from a mathematical
perspective it enables us to bypass the complications associated
with non-analytical edges and surfaces.

\subsubsection{Physical states}
\label{s6.2.2}

Our task now is to construct the general solution to the
diffeomorphism constraint. For this, we will use the spin-network
decomposition (\ref{snd}): $\H = \oplus_\a \, \H^\prime_\a$. Let
us begin by introducing some notation. Given a graph $\a$, denote
by $\Diff_\a$ the sub-group of $\Diff$ which maps $\a$ to itself
and by $\TDiff_\a$ its subgroup which has trivial action on $\a$,
i.e., which preserves every edge of $\a$ and its orientation. The
induced action, $\widehat{\TDiff}_\a$, is \emph{trivial} on
$\cyl_\a$. Next, let ${\rm Diff}_\a$ be the group of all the
diffeomorphisms that preserve $\a$. Then, the quotient
\be {\rm GS}_\a\ =\ {{\Diff}_\a}/{\TDiff_\a}, \ee
is the group of \emph{graph symmetries} of $\a$. It is a finite
group  and it has a non-trivial induced action $\widehat{\rm
GS}_\a$ on $\cyl_\a$. In the group averaging procedure,
consistency requires that one must divide by the `volume' of the
orbits of these groups \cite{almmt1}.

To construct the general solution to the diffeomorphism
constraint, we proceed in two steps. First, given any $\Psi_\a \in
\H^\prime_\a$, we average it using \emph{only the group of graph
symmetries} and obtain a projection map $\hP_{\diff,\a}$ from
$\H_\a^\prime$ to its subspace which is invariant under
$\widehat{\rm GS}_\a$:
\be \hP_{\diff,\a} \Psi_\a\ :=\ \frac{1}{N_\a}\, \sum_{\varphi\in
{GS}_\a}\, \varphi\star\Psi_\a , \ee
where $N_\a$ is the number of the elements of ${\rm GS}_\a$ (the
volume of the obit of ${\rm GS}$) and $\varphi\star \Psi_a$
denotes the pull-back of $\Psi_\a$ under $\varphi$. The map
extends naturally  to a projection $\hP_\diff$ from $\H= L^2(\Ab,
d\mu_o)$ to its subspace which is invariant under $\widehat{\rm
GS}_\a$ for all $\a$.

In the second step, we wish to average with respect to the
remaining diffeomorphisms \emph{which move the graph $\a$}. This
is a very large group and the result of averaging now belongs to
$\cyl^\star$ rather than $\H$. Thus, to each $\Psi_\a\in
\H_\a^\prime$, we now associate an element $(\eta(\Psi_\a)|\, \in
\cyl^\star$, defined by its (linear) action on arbitrary
cylindrical functions $|\Phi_\beta \rangle \in \cyl$:
\be\label{av} (\eta(\Psi_{\a})\,|\,\Phi_\beta \rangle \ =
\sum_{\varphi\in {\rm Diff}/{\rm Diff}_\a}\, \,\langle
\varphi\star \hP_{\diff,\a} \Psi_\a ,\,\, \Phi_\beta \rangle, \ee
where the bracket on the right side denotes the inner product
between elements of $\H$. Although $\varphi\in {\rm Diff}/{\rm
Diff}_\a$ contains an infinite number of elements $\varphi$, for
any given $\beta$ only a finite number of terms are non-zero,
whence $\eta(\Psi_{\a})$ is well-defined. However, there is no
vector $\eta$ in $\H$ such that $\langle \eta,\, \Phi_\beta
\rangle$ equals the right side of (\ref{av}) for all
$\Phi_\beta\in\cyl$. Thus, $(\eta(\Psi_{\a})|$ is a `genuine
distribution' on $\Ab$ rather than a function. Because of the
diffeomorphism invariance of the scalar product on $\H$,
$\eta(\Psi_\a)$ is invariant under the action of ${\rm Diff}(\M)$:
\be (\eta(\Psi_{\a})\,|\, \varphi\star\Phi_\beta \rangle\ =\
(\eta(\Psi_{\a})\, |\, \Phi_\beta\rangle \ee
for all $\varphi \in {\rm Diff}(\M)$. We will denote the space of
these solutions to the diffeomorphism constraint by
$\cyl^\star_\diff$.  Finally, since $\Psi_{\a}$ was an arbitrary
element of $\cyl$, we have constructed a map:
\be\eta: \cyl \rightarrow \cyl^\star_\diff.  \ee
Thus, \emph{every} element of $\cyl$ gives rise, upon group
averaging, to a solution to the diffeomorphism constraint. In this
sense, we have obtained the general solution to the diffeomorphism
constraint. The map $\eta$ is the analog of the projection from
$\H$ to its gauge invariant sub-space $\H^{\rm G}_{\rm inv}$ in
the case of the Gauss constraint. However, because of the
differences between the two constraints discussed above, $\eta$ is
\emph{not} a projection since it maps $\cyl$ onto a
\emph{different} space $\cyl^\star_\diff$. Nonetheless, the group
averaging procedure naturally endows the solution space with a
Hermitian inner product,
\be \label{ip3}(\eta(\Psi)\, |\, \eta(\Phi)) := (\eta(\Psi)\,
|\,\Phi \rangle\, , \ee
since one can show that the right side is independent of the
specific choice of $\Psi$ and $\Phi$ made in the averaging
\cite{dm,almmt1}. (For subtleties, see \cite{almmt1,gm}). We will
denote by $\H_{\diff}$ the Cauchy completion of
$\cyl^\star_\diff$. Finally, we can obtain the \emph{general
solution to both the Gauss and the diffeomorphism constraints} by
simply restricting the initial $\Psi \in \cyl$ to be gauge
invariant, i.e., to belong to $\cyl\cap \H^{\rm G}_\inv$. We will
denote this space of solutions by $\cyl^\star_\inv$:
\be \cyl^\star_\inv = \eta(\cyl\cap \H^{\rm G}_\inv)\ee

What is the situation with respect to operators?  Note first that
there do exist non-trivial (gauge and) diffeomorphism invariant
operators on $\cyl$; an example is the total volume operator
$\hV_{\M}$. Let $\O$ be such an operator. Its dual, $\O^\star$, is
well defined in $\cyl^\star_\inv$:
\be (\O^\star \eta(\Psi)\, |\, \Phi \rangle := (\eta(\Psi)\, |\,
\O\Phi \rangle\, . \ee
(Furthermore, one can show that $\O^\star$ preserves the image of
$\H^\prime_\a$ for very $\a$.) The operator $\O^\star$ is self
adjoint with respect to the natural scalar product (\ref{ip3}) on
$\H_{\diff}$ if and only if $\O$ is self adjoint in $\H$. This
property shows that the scalar product on $\H$ is not only
mathematically natural but also `physically correct'.

Let us summarize. The basic idea of the procedure used to solve
the diffeomorphism constraint is rather simple: One averages the
kinematical states with the action of the diffeomorphism group to
obtain physical states. But the fact that this procedure can be
implemented in detail is quite non-trivial. For example, a
mathematically precise implementation still eludes the
geometrodynamics program. Furthermore, even the final answer
contains certain subtleties. We will conclude by pointing them
out.

\emph{Remarks:}  i) Note that while $(\eta(\Psi)|$ is a solution
to the diffeomorphism constraint for any $\Psi \in \cyl$, it is
\emph{not} true that there is a 1-1 correspondence between
elements of $\cyl$ and solutions to the diffeomorphism constraint.
This is because the map $\eta$ has a non-trivial kernel. In
particular, the projection map $\hat{P}_{\diff}$ itself has a
non-trivial kernel which, by (\ref{av}) is also in the kernel of
$\eta$. (In addition, elements of $\cyl$ of the form $ a_0 \Psi_\a
+ a_1{\varphi_1}\star\Psi_\a +...+ a_n {\varphi_n}\star\Psi_\a,
\,\, {\rm with} \quad  a_0+...+a_n\ =\ 0 $ are also in the kernel
of $\eta$.) Therefore statements such as ``solutions to the
diffeomorphism constraint are diffeomorphism classes of spin
network states'' that one often finds in the literature are
only heuristic.\\
ii) One also finds claims to the effect that the diffeomorphism
constraint can be imposed simply by replacing spin networks
embedded in a manifold $M$ by abstract, non-embedded spin
networks. Within the systematic approach summarized in this
section, these claims are simply incorrect (for a detailed
discussion in the context of 2+1 gravity on a lattice, see, e.g.,
\cite{cz}). In particular a graph, one of whose edges is knotted,
can not be mapped by a diffeomorphism to one in which all edges
are unknotted, whence the mapping $\eta$ sends spin-network states
associated with the two graphs to \emph{distinct} solutions to the
diffeomorphism constraint. As abstract, non-embedded graphs, on
the other hand, they can be equivalent and define the same
spin-network functions. One can imagine a new approach in which
one simply declares that the diffeomorphism constraint is to be
incorporated by replacing embedded spin-networks by abstract ones.
But since the original diffeomorphism constraint acts on the basic
canonical variables $(A,E)$ on $\M$ and the action can be
transferred to graphs only if they are embedded, it would be
difficult to justify such an approach from first principles.\\
iii) Note that there are continuous families of 4 or higher valent
graphs which can not be mapped to one another by $C^n$
diffeomorphisms with $n>0$. Consequently, states in $\H_{\diff}$
based on two of these graphs are mutually orthogonal. Thus, even
though we have `factored out' by a very large group $\Diff$, the
Hilbert space $\H_{\diff}$ is still non-separable. However, if we
were to let $n=0$, i.e. consider homeomorphisms of $\M$ which
preserve the family of graphs under consideration, then these
`problematic' continuous families of graphs would all be
identified in the group averaging procedure and the Hilbert space
of solutions to the diffeomorphism constraint \emph{would be}
separable. However, since the classical constraints do not
generate homeomorphims, and furthermore homeomorphisms do not even
have a well-defined action on the phase space, it is difficult to
`justify' this enlargement of $\Diff$ from direct physical
considerations.\\
iv)  Note that $\cyl^\star_\diff$ is a proper subset of the space
$\cyl^\star_{\rm Diff}$ of \emph{all} elements of $\cyl^\star$
invariant under $\Diff$. However, every $(\Psi| \in
\cyl^\star_{\rm Diff}$ can be uniquely decomposed as
\be\label{decomp} (\Psi| \ =\ \sum_{[\a]}\, (\Psi_{[\a]}| \, ,
\quad {\rm with}\,\, (\Psi_{[\a]}|\,\in\, \eta(\cyl_\a), \ee
where $[\a]$ runs through the diffeomorphism classes of graphs.
The sum on the right side is uncountable but the result is a
well-defined element of $\cyl^\star$ because, in its action on any
cylindrical function, only finite number of the terms fail to
vanish.

\subsection{The Scalar Constraint}
\label{s6.3}

The canonical transformations generated by the Gauss and the
diffeomorphism constraints are \emph{kinematical} gauge symmetries
of the classical theory in the sense that, in the space-time
picture, they operate at a `fixed time'. The crux of quantum
dynamics lies in the scalar constraint. One can imagine
implementing it in the quantum theory also by a group averaging
procedure. However, this strategy is difficult to adopt because
the \emph{finite} canonical transformations generated by this
constraint are not well-understood even at the classical level.
Therefore, one follows the procedure used for the Gauss
constraint: construct a quantum operator corresponding to the
classical, smeared constraint function and then seek its kernel.
Because the form of this constraint is so intricate, its
implementation is still rather far from being as clean and
complete as that of the other two constraints. In particular,
genuine ambiguities exist in the regularization procedure and
distinct avenues have been pursued [95-102]. What is non-trivial
at this stage is the \emph{existence} of well-defined strategies.
Whether any of them is fully viable from a physical perspective is
still an open issue. In this summary, we will essentially follow
the most developed of these approaches, introduced by Thiemann
\cite{tt3,tt4,tt5}. However, to bring out quantization ambiguities
we have generalized the method, emphasizing points at which there
is freedom to modify the original procedure and still arrive at a
well-defined constraint operator. Our emphasis is more on
clarifying the underlying conceptual structure than on providing
efficient calculational tools.

\subsubsection{Regulated classical expression}
\label{s6.3.1}
As with area and volume operators, our first task is to re-express
the classical expression of the scalar constraint as a Riemann sum
involving only those phase space functions which have direct
quantum analogs. Recall from section \ref{s2} that in terms of
real connection variables, the scalar constraint (\ref{BT})
smeared with a lapse $N$ can be written as a sum of two terms:
\be \C(N) = (\frac{\g}{4k})^{\f{1}{2}}\, \int_\M d^3x\, N\,
\frac{P^a_{i}P^{b}_j}{\sqrt{\det P}}\left[\epsilon^{ij}{}_k
F^{k}_{ab} + 2(\sigma-\gamma^2) K^i_{[a}K^j_{b]}\right] \ee
where, as before, $k = 8\pi G$ and we have used the relation $\det
q = (\kappa \g)^3 \det P$. Had we worked in the +,+,+,+ signature
and in the half-flat sector $\g^2 = \sigma$, the second term would
have been zero. Thus, the first term has the interpretation of the
the scalar constraint of Euclidean general relativity. Therefore,
the full Lorentzian constraint can be written as
\be \C(N) = \sqrt{\gamma}\CE(N) - 2(1+\gamma^2) \T(N)\, \label{C}
\ee
where we have used $\sigma = -1$ corresponding to the Lorentzian
signature.

\begin{figure}
\begin{center}
\includegraphics[height=3in]{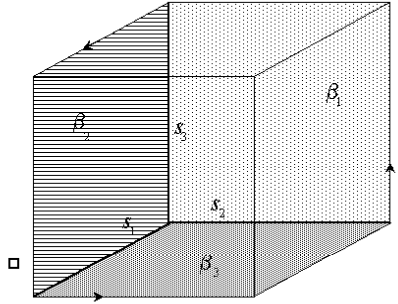}
\caption{An elementary cell $\Box$ in a cubic partition.
$s_1,s_2,s_3$ are the edges of the cell and $\beta_1,
\beta_2,\beta_3$ the three oriented loops which are boundaries of
faces orthogonal to these edges.}\label{hamreg1}
\end{center}
\end{figure}

Let us begin by exploring the first term. In comparison with
geometric operators discussed in section \ref{s5}, we now have
three sets of complications. First, the expression of $\CE(N)$
involves not only triads $P^a_i$ but also curvature $F_{ab}^i$ of
the connection $A_a^i$. However, following the standard procedure
in gauge theories, it is straightforward to express curvature in
terms of holonomies which can be directly promoted to operators.
The second complication arises from the fact that the expression
of $\T(N)$ involves extrinsic curvature terms. Fortunately, we
will see that these can be expressed using the Poisson bracket
between $\CE$ and the total volume, both of which have
well-defined operator analogs. The final complication is the
presence of the volume element $\sqrt{\det P}$ in the denominator.
At first, this seems to be a fatal drawback. A key insight of
Thiemann's \cite{tt3,tt4} was that this is not the case (see also
\cite{glmp} for a further discussion). For, the combination
\be e^i_a \ :=\ \frac{\sqrt{k\g}}{2}\, \eta_{abc} \epsilon^{ijk}\,
\frac{P^b_j P^c_k}{\sqrt{\det P}}. \ee
representing the co-triad $e^i_a$ can be expressed as a manageable
Poisson bracket:
\be e^i_a(x)\ =\ \frac{2}{k\g}\{A^i_a(x),V\}.\label{e} \ee
Using this fact, the Euclidean scalar constraint part $\CE(N)$ is
written as
\be \CE(N)\ =\ -\frac{2}{k^2\g^{\f{3}{2}}}\,\int_M d^3x\, N(x)\,
\eta^{abc}\, \Tr\Big(F_{ab}(x)\{A_c(x), V\} \Big)
 \label{Eu}\,\ee
We will see that this this expression is well-suited for
quantization.

 The second term $\T(N)$ in the expression
(\ref{C}) of the constraint is given by:
\be \T(N)\ =\ \frac{\sqrt{\gamma}}{2{\sqrt k}}\int_M d^3x\, N\,
\left(K_{[a}^iK_{b]}^j\, \f{P^a_i P^b_j}{\sqrt{\det P}}\right) \ee
To cast this term in the desired form, we first note that $K_a^i$
can be expressed as a Poisson bracket bracket,
\be K_a^i = \f{1}{k\g}\,\{A_a^i,\, \bar{K}\} \ee
where $\bar{K}$ is the integral of the trace of the extrinsic
curvature
\be \bar{K}\ =\ k\gamma\int_M d^3x\, K^i_a\, P_i^a \, . \ee
Now $\bar{K}$ itself can be expressed as a Poisson bracket
\be \bar{K}\ =\ \gamma^{-\frac{3}{2}}\{\CE(1),V\}\, . \ee
Hence $\T(N)$, can be expressed as:
\be \T(N) = -\frac{2}{k^4\g^3}\, \int_M d^3x\,N(x) \, \eta^{abc}\,
\Tr \Big(\{A_a(x), \bar{K}\} \{A_b(x), \bar{K}\}\{A_c(x), V
\}\Big)\, . \ee

Thus, to express the constraint in terms of variables adapted to
quantum theory, it only remains to re-express the curvature and
connection terms appropriately. Now, if $s$ is a line segment of
coordinate length $\varepsilon$ and if a loop $\beta$ is the
boundary of a coordinate plane $P$ of area $\varepsilon^2$  we
have:
\ba \{\textstyle{\int}_s A,\, V\} &=& -[\bA(s)]^{-1}\, \{\bA(s),\,
V\} + o(\varepsilon)\nonumber\\
 \{\textstyle{\int}_s A,\, \bar{K}\} &=&
 -[\bA(s)]^{-1}\, \{\bA(s),\, \bar{K}\} +o(\varepsilon)\ea
and,
\be \textstyle{\int}_P F\ =\ \frac{1}{2}(\bA(\beta^{-1}) -
\bA(\beta)) + o(\varepsilon^2)\, . \ee
These formulas provide a concrete strategy to replace the
connection and curvature terms in terms of holonomies. For
example, if $M$ is topologically $R^3$, it is simplest to
introduce a cubic partition where the coordinate length of edges
of elementary cells is $\varepsilon$. Denote by ${s}_1$, ${s}_2$,
${s}_3$, the edges of an elementary cell $\Box$ based at a vertex
$v_\Box$ and by $\beta_1$, $\beta_2$, and $\beta_3$ the three
oriented loops based at $v_\Box$ which are boundaries of faces
orthogonal to these edges (see Figure 2). Then, the
$\sum_{\Box}\CE_\Box(N)$, where
\be \CE_\Box(N)\ =\ -\frac{2 N(v_\Box)}{k^2\gamma^{\frac{3}{2}}}
\sum_{I} \Tr\Big((\bA(\beta_I) - \bA(\beta_I^{-1}))
\bA(s_I)^{-1}\{\bA(s_I), V\}\Big)\, ,\ee
is a Riemann sum which converges to $\CE(N)$ as the cell size
tends to zero (and the number of cells tends to infinity).
Similarly, the sum $\sum_{\Box}\T_\Box(N)$
\be \T_{\Box}(N)\, =\, \frac{2N(v_\Box)}{k^4\gamma^3}
\epsilon^{IJL} \Tr\Big(\bA(s_I^{-1})\{\bA(s_I),\bar{K} \}
\bA({s_J}^{-1}) \{\bA(s_J),\bar{K} \}\bA({s_L}^{-1})\{\bA(s_L),V
\}\Big)\,  \ee
is a Riemann sum which converges to $\T(N)$ as the cell size tends
to zero. These Riemann sums can therefore be regarded as providing
a  `regularization' of the classical constraint. As in the
discussion of the geometric operators of section \ref{s5}, the
idea is to first replace classical quantities in the `regularized
expression' by their quantum counterparts and \emph{then} remove
the regulator. \emph{A remarkable feature of this regularization,
first pointed out by Rovelli and Smolin \cite{rs5}, is that the
regulating parameter $\varepsilon$ has disappeared from the
expression}. Hence it is not necessary to multiply the constraint
by a suitable power of $\varepsilon$ before removing the
regulator; no renormalization is involved.

\begin{figure}
\begin{center}
\includegraphics[height=3in]{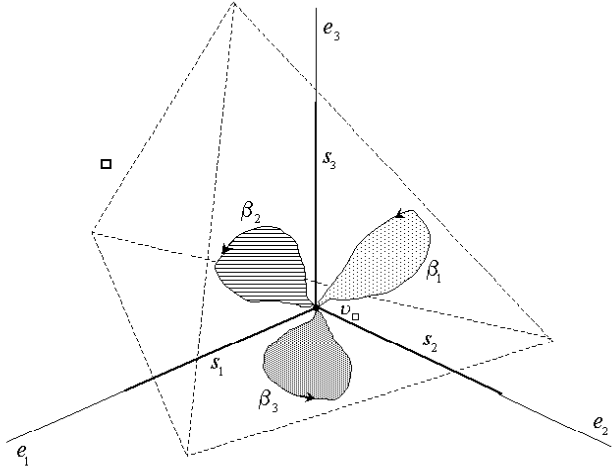}
\caption{An elementary cell $\Box$ in a general partition.
Segments $s_i$ now lie along the edges of the given graph which
has a vertex $v_\Box$ in the interior of $\Box$. Each of the loops
$\beta_i$ originates and ends at $v_\Box$ and lies in a
co-ordinate plane spanned by two edges.}\label{hamreg2}
\end{center}
\end{figure}

The cubic partition is the simplest example of a more general
classical regularization. The available freedom can be summarized
as follows. To every value $\varepsilon\in [0,\varepsilon_0]$,
assign a partition of $\Sigma$ into cells $\Box$ of possibly
arbitrary shape. In every cell $\Box$ of the partition we define
edges $s_J$, $J=1,...,n_s$ and loops $\beta_i$, $i=1,...,n_\beta$,
where $n_s,n_\beta$ may be different for different cells (see
Figure 3). Finally, fix an arbitrarily chosen representation
$\rho$ of $SU(2)$. This entire structure will be denoted by
$R_\varepsilon$ and called a \emph{permissible classical
regulator} if the following property holds:
\be\label{CTlim} \lim_{\varepsilon\rightarrow
0}\CE_{R_{\varepsilon}}(A, E)\ =\ \CE(A, E) \quad {\rm and}\quad
\lim_{\varepsilon\rightarrow 0}\T_{R_{\varepsilon}}(A, E)\ =\
\T(A, E)\, , \ee
where
\ba \CE_{R_{\varepsilon}}\ &=&\
\sum_{\Box}\CE_{R_\varepsilon\Box},\\
 \CE_{R_\varepsilon}\ &=&\ \frac{N(v_\Box)}{k^2\gamma^{\frac{3}{2}}}
 \sum_{iJ}C^{iJ}\Tr \Big(
\left(\rho[\bA(\beta_i)]-\rho[\bA(\beta^{-1}_{i})]\right)\rho[\bA({s_J}^{-1})]\{
\rho[\bA(s_J)],\, V \}\Big), \label{CE}\\
T_{R_\varepsilon}\ &=&\ \sum_\Box T_{R_\varepsilon\Box}\\
T_{R_\varepsilon\Box}\ &=&\ \frac{N(v_\Box)}{k^4\gamma^3}
\sum_{I,J,K}T^{IJK} \Tr \Big(
\rho[\bA(s_I^{-1})]\{\rho[\bA({s_I})],\bar{K} \}
\rho[\bA({s_J}^{-1})] \{ \rho[\bA(s_J)],\, \bar{K}\}\nonumber\\
&{}& \times\,\, \rho[\bA({s_K}^{-1})] \{ \rho[\bA(s_K)],\,
V\}\Big)\, ;\label{T}\, \ea
and,  $C^{iJ}$, $T^{IJK}$ are fixed constants, \emph{independent}
of the scale parameter $\varepsilon$. A large family of the
classical regulators can be constructed by modifications of the
cubic example, changing the shape of the cells, loops and edges,
and their relative positions suitably.

\subsubsection{The quantum scalar constraint}
\label{s6.3.2}

Our task is to first promote the regulated classical constraint to
a quantum operator and then remove the regulator. In the detailed
implementation of this procedure, one encounters three non-trivial
issues.

As in the case of geometric operators, the first step is rather
straightforward because the regulated expressions involve only
those phase functions which have direct quantum analogs. However,
while the `obvious' quantum operator would be well-defined on
states which are cylindrical with respect to any one graph, at the
end we have to ensure that the resulting family of operators is
consistent. This is the first non-trivial issue. The simplest way
to address it is to use the decomposition
$\H=\oplus_{\a}\H^\prime_\a$ of the Hilbert space, introduced in
Section \ref{s4.3.3}, and define the quantum constraint on each
$\H^\prime_\a$ separately. Because of the orthogonality of any two
$\H^\prime_\a$, the resulting family of operators would then be
automatically consistent.

Let us begin with $\CE_{R_\varepsilon}$. Fix a subspace
$\H^\prime_\a$ of $\H$. The quantum operator can be obtained
simply by promoting the holonomies and volume functions to
operators and replacing the Poisson brackets by $1/i\hbar$-times
commutators. Thus, for any given graph $\a$ and  $\varepsilon >0$,
\be \hCE_{R_{\varepsilon, \alpha}}(N)\ :=\ \sum_\Box\hC^E_\Box(N)
\label{hCE1}\ee
with
\be {\hCE}_{\Box}(N) \ :=\
-\frac{iN(v_\Box)}{k^2\gamma^{\frac{3}{2}}\hbar}
\sum_{iJ}C^{iJ}\Tr \Big(\left(
\rho[\bA(\beta_i)]-\rho[\bA(\beta_{i}^{-1})]\right)\rho[\bA({s_J}^{-1})]
\Big[\rho[\bA(s_J)],\, \hat{V} \Big]\Big), \label{hCE2}\ee
is a densely defined operator on $\H'_\a$ with domain $\D_\a =
\H^\prime_\a \cap\cyl_\a$ for any classical regulator
$R_\varepsilon$.  We now encounter the second non-trivial issue:
we have to ensure that the final operator is diffeomorphism
covariant. To address it, we need to use regulators which are not
fixed but transform covariantly as we move from a graph $\a$ to
any of its images under diffeomorphisms. Therefore, we will
restrict our regulators appropriately.

A \emph{diffeomorphism covariant quantum regulator}
$R_{\varepsilon,\alpha}$ is a family of permissible classical
regulators, one for each choice of the graph $\a$, satisfying the
following properties:
\begin{itemize}
\item[a)] the partition is sufficiently refined in the sense that
every vertex $v$ of $\a$ is contained in exactly one cell of
$R_{\varepsilon,\a}$; and,%
\item[b)] if $(\a,v)$ is diffeomorphic to $(\alpha',v')$ then, for
every $\varepsilon$ and $\varepsilon^\prime$, the quintuple
$(\alpha, v, \Box, (s_I),(\beta_J))$ is diffeomorphic to the
quintuple $(\alpha', v', \Box', (s'_I),(\beta'_J))$ where $\Box$
and $\Box'$ are the cells of $R_\varepsilon$ and $R_\varepsilon'$
respectively, containing $v$ and $v'$ respectively.
\footnote{We need the restriction only on cells which contain
vertices because the properties of the volume operator imply that
the action of $\hCE_{\Box}$ is non-trivial only if one of the
segments $s_I$ of the regulator intersects a vertex of $\a$.}
\end{itemize}
Such diffeomorphism compatible quantum regulators exist; an
explicit example is given in \cite{tt4}. Given such a
$R_{\varepsilon,\a}$, for every value of $\varepsilon$, the
operators ${\hCE}_{R_{\varepsilon,\alpha}}(N)$ are densely defined
on $\H^\prime_\a$ with a common domain $\D_\a$. This family of
operators determines a densely defined operator
$\hCE_{R_\varepsilon}$ on the full Hilbert space $\H$ with domain
$\cyl$, independently of the value of $\varepsilon$. Furthermore,
for any value of $\varepsilon$, this domain is mapped to itself by
the operator $\hCE_{R_\varepsilon}$.

Thus, it only remains to remove the regulator. Here we encounter
the third non-trivial issue. Typically
$\hCE_{R_{\varepsilon,\a}}|\Psi_\a \rangle$ is orthogonal to
$\hCE_{R_{\varepsilon^\prime,\a}}|\Psi_\a \rangle$ if
$\varepsilon\not=\varepsilon^\prime$, whence the operator does not
converge (even in the weak topology) on $\H$. This is a rather
general problem associated with the topology of $\H$; we
encountered it also while defining the operator analog of the
diffeomorphism constraint $\C_{\rm Diff}(\vec{N})$. Recall,
however, that solutions to the diffeomorphism constraint also fail
to lie in the kinematical Hilbert space $\H$; they belong to
$\cyl^\star$, the algebraic dual of $\cyl$. Therefore, for the
consistency of the whole picture, what we need is the action of
the scalar constraint only on a sufficiently large subspace of
$\cyl^\star$ and not on $\H$. \emph{And this action is
well-defined and non-trivial.} More precisely, for each $(\Psi|
\in \cyl^\star$, the action of the regulated constraint operator
is naturally given by:
\be \left[(\Psi|\hCE_{R_\varepsilon}\right]|\Phi \rangle\,\,
:=\,\, (\Psi| \left[\hCE_{R_\varepsilon}|\Phi \rangle\right] \ee
for all $|\Phi \rangle \, \in \cyl^\star$. We can now remove the
regulator in the obvious fashion. Define
\be \hCE(N)\ =\ \lim_{\varepsilon\rightarrow 0}\,
\hCE_{R_\varepsilon}(N) \ee
via
\be \left[(\Psi|\hCE(N)\right]|\Phi \rangle\ =\
\lim_{\varepsilon\rightarrow 0}\,
(\Psi|\left[\hCE_{R_\varepsilon}(N)|\Phi \rangle\right].
\label{CElim}\ee
Note that the limit has to exist only pointwise, i.e., for each
$|\Phi \rangle\, \in \cyl$ separately. As a consequence the domain
of the operator, the set of $(\Psi|$ in $\cyl^\star$ for which the
limit exists, is quite large. In particular, as discussed below,
it includes a large class of solutions to the diffeomorphism
constraints. The term $\T(N)$ in the scalar constraint can be
handled in a completely parallel fashion. Specifically, we can
first define the operator $\hat{\bar{K}}$ through
\be \hat{\bar{K}}\ :=\ \frac{i}{\hbar\gamma^{\frac{3}{2}}}\,
[\hV,\hC^E(1)] \ee
and use $\hat{\bar{K}}$ and the quantum regulator
$R_{\varepsilon,\alpha}$ to define the regulated operator
$\hat\T_\Box(N)$:
\ba (\Psi|\hat\T_\Box(N)\ = \ \frac{iN(v_\Box)}{k^4\gamma^3
\hbar^3}T^{IJK}\,\, (\Psi| \Tr\Big(&& \bA(s_I^{-1})[\bA_{s_I},\,
\hat{\bar{K}}] \bA({s_J}^{-1})[\bA(s_J),\,\hat{\bar{K}}]
\nonumber\\
&&\times \bA({s_K}^{-1}) [\bA(s_K),\,\hV]\Big),\label{TE} \ea
on the domain $\cyl^\star$. Collecting these definitions, we now
have a regulated scalar constraint operator:
\be (\Psi|\,\hat{\C}_{R_\varepsilon}(N) := (\Psi|
\Big(\sqrt{\gamma}\hCE_{R_\varepsilon}(N)\ -  2(1 + \gamma^2)
\sum_{\Box}\hT_\Box(N)\Big)\label{Clim} \ee
for all $(\Psi| \in \cyl^\star$. Again, we can remove the
regulator by taking the limit as $\varepsilon \rightarrow 0$ as in
(\ref{CElim}). By construction, the action of this operator is
diffeomorphism covariant. Thus, each diffeomorphism covariant
quantum regulator defines a scalar constraint operator. Since
there is a great deal of freedom in choosing these regulators,
there is considerable quantization ambiguity. Nonetheless, all
these constructions exhibit some very non-trivial properties. We
will conclude this section by providing two illustrations.

First, as mentioned in the beginning of this section, it is
significant that well defined prescriptions exist to give precise
meaning to quantum Einstein equations in a background independent
setting. In geometrodynamics, for example, the Wheeler-DeWitt
equation still remains only formal. Secondly, these constructions
match surprisingly well with the solutions of the diffeomorphism
constraint. To see this note first that, irrespective of the
choice of the diffeomorphism covariant quantum regulator, \emph{up
to diffeomorphisms}, the operator $\hC_{R_\varepsilon}(N)$ is
independent of $\varepsilon$: for every $\varepsilon,\,
\varepsilon'$ and $|\Psi \rangle\, \in \cyl$, there is a
diffeomorphism $\varphi$ such that
\be \hC_{R_{\varepsilon'}}(N)\, |\Psi \rangle\ =\
U_\varphi\hC_{R_{\varepsilon'}}(N) |\Psi \rangle\, ,
 \ee
for every $N$. Next, suppose that $(\Psi| \in\cyl^\star$ is
diffeomorphism invariant. Then, for every lapse function $N$, the
result $(\Psi| (\hC_{R_\varepsilon}(N))$, is in fact
\emph{independent} of $\varepsilon$, and so is the expression
under the limit on the right hand side of (\ref{Clim}). Hence the
regulator can be removed trivially \cite{rs5}. Thus because of the
form of the regulated operators, diffeomorphism invariant states
in $\cyl^\star$ (constructed by group averaging elements of
$\H^\prime_\a\cap \cyl_\a$) are automatically in the domain of the
scalar constraint operator. This tight matching between the way in
which the two constraints are handled is quite non-trivial.

\textsl{Remark:} In the original construction by Thiemann
\cite{tt4}, the Hamiltonian constraint operator was defined on
$\cyl^\star_{diff}$. Now, as we saw in section \ref{s2}, the
Poisson bracket of any two scalar constraints is given by a
diffeomorphism constraint in the classical theory. Therefore, on
diffeomorphism invariant states, one would expect the quantum
scalar constraint operators to commute. Irrespective of the choice
of the regulator $R_\varepsilon$, they do. To obtain a more
stringent test, the domain of the Thiemann operator was extended
slightly in \cite{glmp,lm}. The extended domain, called the
`habitat' \cite{lm}, also includes certain elements of
$\cyl^\star$ which are not diffeomorphism invariant. Nonetheless,
it turned out that the commutator between scalar constraints
continues to vanish on the habitat. This may seem alarming at
first. However, it turns out that the quantum operator
corresponding to the classical Poisson bracket also annihilates
every state in the habitat \cite{glmp}. Thus, there is no
inconsistency; the habitat just turned out to be too small to
provide a non-trivial viability criterion of this quantization
procedure. The domain of the operator introduced in this section
includes the habitat and the same result continues to hold. More
importantly, it is likely that this domain is significantly larger
and may contain semi-classical states. If this turns out to be the
case, stronger viability criteria to test this quantization
procedure will become available. In particular, in addition to the
relation to the classical Poisson algebra of constraints, one may
be able to analyze the relation between the classical evolution
and the action of the constraint on semi-classical states.

\subsubsection{Solutions to the scalar constraint}
\label{s6.3.3}

In this section, we will illustrate how the difficult problem of
finding solutions to the quantum constraints can be systematically
reduced to a series of simpler problems. For this, we will need to
make the quantum regulator $R_{\varepsilon,\a}$ more specific.

The most convenient class of regulators requires some
modifications in the original construction due to Thiemann.
Restrictions defining this class can be summarized as follows. Fix
a graph $\a$ and consider a cell $\Box$ containing a vertex $v$ in
the partition of $M$ defined by the regulator. The first
restriction is that every edge $s_k$ assigned to $\Box$ must be a
proper segment of an edge incident at $v$, oriented to be outgoing
at $v$. The next restriction is on the closed loops $\beta$.  To
every pair of edges $e_I, e_J$ assign a triangular closed loop
$\beta_{IJ}$ such that: i) the loop contains $v$ but no other
point in the graph $\a$; ii) it lies `between' the 2-edges, in a
2-plane containing the edges, where the plane is defined up to
diffeomorphisms preserving $\a$; and iii) it is oriented clockwise
with respect to the orientation defined in the plane by the
ordered pair of segments $s_I, s_J$. (The double index $IJ$
labelling the loop corresponds to the single index $i$ before,
e.g. in (\ref{CE}).) Finally, the constants $C^{IJK}$ and
$T^{IJK}$ of the regulator are, respectively, $\pm\kappa_1,\, 0$
and $\pm\kappa_2,\, 0$ depending on the orientation of a triad of
vectors tangent to the segments $s_I,s_J,s_K$ at $v$ relative to
the background orientation on $M$, where $\kappa_1$ and $\kappa_2$
are fixed constants.

Given such a regulator, the action of the resulting operators
$\hCE(N)$ and $\hC(N)$ on diffeomorphism invariant elements of
$\cyl^\star$ has a rather simple geometric structure which can be
roughly summarized as follows. Suppose $(\Psi_\a|\in \cyl^\star$
is obtained by group averaging a state in $\H^\prime_\a$. Then, if
$\a$ contains no closed loops of the type introduced by the
regulator at any of its vertices, it is annihilated by both the
operators. If $\a$ does contain such closed loops, $\hCE(N)$
removes one loop, $\hC(N)$ removes two loops, and in each case
there is also a possible change in the intertwiners at the vertex.
Following a terminology introduced by Thiemann in his
regularization, closed loops of the type introduced by the
regulator will be called \emph{extraordinary}.

More precisely, constraint operators act as follows. Consider a
labelled graph $(\a_0,j_0)$ such that no labelled graph belonging
to the same diffeomorphism class contains an extraordinary loop
labelled by $j(\rho)$ where $\rho$ is the representation used in
the regularization. Call a labelled graph with this property
\emph{simple}. Given a simple graph $\a$, all states $(\Psi_\a|$
in $\cyl^\star_{\rm diff}$ obtained by group averaging elements of
$\H^\prime_a$ are in the kernel of $\hC(N)$. This is a large class
of solutions. However, these states are annihilated by each of the
two terms, $\hCE(N)$ and $\h\T(N)$, of $\hC(N)$ separately whence
they solve both the Euclidean and the Lorentzian scalar
constraint. In this sense, they are the analogs of time-symmetric
solutions to the classical Hamiltonian constraint and will at best
capture very special physical situations.

More interesting solutions can be obtained starting from graphs
which do admit extraordinary edges. We begin by introducing some
notation. Consider the set of all the labelled graphs $(\a',j')$
that can be obtained from a given $(\a_0,j_0)$ by creation of $n$
extraordinary loops labelled by $j(\rho)$, and by the
diffeomorphisms. Denote this set by $\Gamma^{(n)}_{[(\a_0,j_0])]}$
and denote by $\D^{(n)}_{[(\a_0,j_0)]}$ the linear span of the
corresponding diffeomorphism averaged  spin-network states. The
resulting spaces are \emph{finite} dimensional and have trivial
intersection with one another:
\be ([(\a_0,j_0)],n)\not= ([(\a'_0,j'_0)],n')\ \Rightarrow\
\D^{(n)}_{[(\a_0,j_0)]} \cap \D^{(m)}_{[(\a'_0,j'_0)]}=\{ 0\}. \ee
As a consequence, one can show that they have the following very
useful property: Every $(\Psi| \in \cyl^\star_{\diff}$ can be
uniquely decomposed as
\be \label{dec} (\Psi| \ =\ \sum_{\a,j,n} (\Psi|_{(n)[(\a,j)]}\, ,
\quad {\rm where} \quad (\Psi|_{(n)[(\a,j)]}\in
\D^{(n)}_{[(\a,j)]}. \ee
The availability of this decomposition systematizes the task of
finding solutions to the scalar constraint.

Let us begin with $\hCE(N)$. For the Euclidean theory, one can
obtain the following surprising result:
\be (\Psi|\hCE(N) \ =\ 0 \ \Leftrightarrow\ \
(\Psi|_{(n)[(\a,j)]}\hCE\ =\ 0, \ \ {\rm for}\ \ {\rm every}\,
[(\a,j)], n.
 \ee
Thus, \emph{$(\Psi|$ is a solution of the Euclidean part of the
constraint if and only if each of its components with respect to
the decomposition (\ref{dec}) is a also solution.} This is a very
useful property because the problem of finding a general
diffeomorphism invariant solution to the Euclidean constraint is
reduced to that of finding solutions in \emph{finite} dimensional
subspaces. On each of these sub-spaces, one has just to find the
kernel of certain matrices, a problem that can be readily put on a
computer. Reciprocally, given any diffeomorphism invariant
solution to the constraint  ---e.g. the state supported just on
flat connections---  the decomposition provides a family of new
solutions. In 2+1 dimensions this property implies that any
semi-classical state can be obtained by a superposition of these
`elementary' solutions.

Finally, let us turn to the full (Lorentzian) scalar constraint
operator, $\hC(N)$. In the present scheme, the problem of
obtaining diffeomorphism invariant solutions is reduced to a
hierarchy of steps. More precisely, the equation
\be (\Psi|\hC(N)\ =\ 0, \ee
is equivalent to the following hierarchy of equations
\ba (\Psi_{(1)[\a,j]}|\, \hat{\T}(N) &=&  0\, ,\nonumber\\
(\Psi_{(2)[\a,j]}|\, \hat{\T}(N) &=&  (\Psi_{(1)[\a,j]}|\,
\hCE(N)\, ,
\nonumber\\
                      .&.&.\\
(\Psi_{(n+1)[\a,j]}|\, \hat{\T}(N) &=& \sum_n\,
(\Psi_{(n)[\a,j]}|\,
\hCE(N)\, , \nonumber\\
.&.&. \ea
In general, the procedure involves infinitely many steps. However,
it gives a partial control on the solutions and suggests new
ans\"atze, e.g. requiring that the series terminate after a finite
number of steps.

\textsl{Remark:} Since the procedure outlined above is a variation
on Thiemann's original strategy, it is worth comparing the
relative merits. Thiemann's procedure is simpler in that two of
the three edges of triangles $\beta_i$ (holonomy around which
captures the curvature term in $\CE(N)$) are along edges of the
graph under consideration. However, now the analogs of our spaces
$\D^{(n)}_{(\alpha,j)}$ overlap, making the procedure for solving
the constraint more complicated.

To summarize, in this section we have presented a general
framework for defining the Hamiltonian constraint and for finding
its solutions. This procedure provides a good handle on the
problem and also brings out the ambiguities involved.
Specifically, each choice of a diffeomorphism covariant quantum
regulator $R_{\varepsilon,\a}$ gives rise to a quantum constraint
operator $\hC$ on $\D$. For each choice of the regulator, there is
also a certain factor ordering freedom which was ignored for
brevity. In general these operators will differ from each other,
defining distinct quantum dynamics and one has to invoke
\emph{physical} criteria to test their viability. Quantum
cosmology results discussed in section \ref{s7} favor the factor
ordering used here. There have also been attempts at restricting
the freedom in the choice of the quantum regulator by imposing
heuristically motivated conditions. However, a canonical choice
has not emerged. Thus, there is still a great deal of ambiguity
and it is not clear if any of the candidates are fully viable. A
key criterion is that the the solution set to the constraints
should be rich enough to admit a large number of semi-classical
states. This issue will not be systematically resolved until one
has a greater control on the semi-classical sector of the theory.
As discussed above there is now a general strategy to find
solutions, whence one can hope to address this issue. Partial
support for this strategy comes from 2+1 dimensional Euclidean
general relativity. As mentioned above, in this theory, all
semi-classical states can be recovered by superposing the
`elementary solutions' to quantum constraints, obtained via our
systematic procedure. This result is encouraging because the 2+1
theory has all the conceptual problems associated with the absence
of a background geometry. However, it can not be taken as a strong
indication because the 2+1 dimensional theory has only a finite
number of degrees of freedom.

\bigskip
This concludes the general framework for quantum kinematics and
dynamics. In the next three sections we will discuss various
applications.

\section{Applications of quantum geometry: Quantum cosmology}
\label{s7}

In cosmology, one generally freezes all but a finite number of
degrees of freedom by imposing spatial homogeneity (and sometimes
also isotropy). Because of the resulting mathematical
simplifications, the framework provides a simple arena to test
ideas and constructions introduced in the full theory both at the
classical and quantum levels. Moreover, in the classical regime,
the symmetry reduction captures the large scale dynamics of the
universe as a whole quite well. Therefore, in the quantum theory,
it provides a fertile test bed for analyzing the important issues
related to the fate of the initial singularity (highlighted in
section \ref{s1}). Over the last three years, Bojowald and his
collaborators have made striking advances in this area by
exploiting the quantum nature of geometry [104-117]. In this
section we will provide a self-contained summary of the core
developments using constructions which mimic the ones introduced
in sections \ref{s4}--\ref{s6}. (For subtleties and details, see
especially \cite{abl:qc}.)

Loop quantum cosmology also provides a number of lessons for the
full theory. However, to fully understand their implications, it
is important to keep track of the differences between the symmetry
reduced and the full theories. The most obvious difference is the
tremendous simplification resulting from the reduction of a field
theory to a mechanical system. However, there are also two other
differences which make it conceptually and technically \emph{more}
complicated, at least when one tries to directly apply the methods
developed for the full theory in section \ref{s6}. First, the
reduced theory is usually treated by gauge fixing and therefore
fails to be diffeomorphism invariant. As a result, key
simplifications that occur in the treatment of full quantum
dynamics do not carry over and, in a certain sense, dynamics now
acquires \emph{new} ambiguities in the reduced theory! The second
complication arises from the fact that spatial homogeneity
introduces distant correlations. Consequently, in contrast to
section \ref{s4}, quantum states associated with distinct edges
and electric flux operators associated with distinct 2-surfaces
are no longer independent. Both these features give rise to
certain complications which are \emph{not} shared by the full
theory. Once these differences are taken in to account, loop
quantum cosmology can be used to gain valuable insights about
certain qualitative features of the methods introduced in section
\ref{s6.3} to formulate and solve the Hamiltonian constraint in
the full theory.

\subsection{Phase space}
 \label{s7.1}

For simplicity, we will restrict ourselves to spatially
homogeneous, isotropic cosmologies. (For non-isotropic models, see
\cite{mb9,bdv}. Specifically, we will focus only on the case where
the isometry group $\S$ is the Euclidean group. Then the
3-dimensional group ${T}$ of translations (ensuring homogeneity)
acts simply and transitively on the 3-manifold $\M$. Therefore,
$\M$ is topologically $\Real^3$. It is convenient to fix on $M$ a
fiducial flat metric $\q_{ab}$, an associated constant orthonormal
triad $\e^a_i$ and the dual co-triad $\w_a^i$.

Let us now turn to the gravitational phase space in the connection
variables. As we saw in section \ref{s2}, in the full theory, the
phase space consists of pairs $(A_a^i,\, P^a_i)$ of fields on a
3-manifold $\M$, where $A_a^i$ is an $\SU(2)$ connection and
$P^a_i$ a triplet of vector fields with density weight 1. A pair
$(A^\prime_a{}^i,\, P^{\prime a}_i)$ on $\M$ will be said to be
spatially homogeneous and isotropic or, for brevity,
\emph{symmetric} if for every $s\in \S$ there exists a local gauge
transformation $g: \M\rightarrow SU(2)$, such that
\be \label{ge} (s^\star A^\prime, \, s^\star P^\prime)\ =\
(g^{-1}A'g\, + \, g^{-1}dg, g^{-1}P'g). \ee
As is usual in cosmology, we will fix the local diffeomorphism and
gauge freedom. To do so, note first that for every symmetric
$(A^\prime,\, P^\prime)$ (satisfying the Gauss and diffeomorphism
constraints) there exists an equivalent pair $(A,\, P)$ (under
(\ref{ge})) such that
\be\label{ss} A\ =\ \underbar{c}\,\, \w^{i} \tau_i,\quad {P}\ =\
\underbar{p}\, \sqrt{\q}\,\,\, \e_{i}\tau^i \ee
where $\underbar{c}$ and $\underbar{p}$ are constants, carrying
the only non-trivial information contained in the pair $(A^\prime,
E^\prime)$, and the density weight of ${P}$ has been absorbed in
the determinant of the fiducial metric. Denote by $\A_S$ and ${\bf
\Gamma}^S_{\rm grav}$ the subspace of the gravitational
configuration space $\A$ and of the gravitational phase space
${\bf \Gamma}$ defined by (\ref{ss}). Tangent vectors $\delta$ to
${\bf \Gamma}^S_{\rm grav}$ are of the form:
\be \label{tv} \delta = (\delta A,\, \delta P), \quad {\rm with}
\quad \delta A \equiv (\delta \underbar{c})\,\,  \w_a^i, \,\,\,
\delta{P} \equiv (\delta \underbar{p})\,\, \e^a_i . \ee
Thus, $\A_S$ is 1-dimensional and  ${\bf \Gamma}^S_{\rm grav}$ is
2-dimensional: we made a restriction to \emph{symmetric} fields
and solved and gauge-fixed the gauge and the diffeomorphism
constraints, thereby reducing the infinite, local, gravitational
degrees of freedom to just one.

Because $\M$ is non-compact and fields are spatially homogeneous,
various integrals featuring in the Hamiltonian framework of
section \ref{s2} diverge. This is in particular the case for the
symplectic structure of the full theory. However, one can bypass
this problem in a natural fashion: Fix a `cell' ${\cal V}$ adapted
to the fiducial triad and restrict all integrations to this cell.
The volume $V_o$ of this cell (with respect to the fiducial metric
$\q_{ab}$) can also be used to absorb the dependence of the basic
variables $\underbar{c}, \underbar{p}$ on the fiducial $\q_{ab}$.
Let us rescale the basic variables to remove this dependence,
\be  c := V_o^{\f{1}{3}} \underbar{c} \quad {\rm and} \quad
     p := 8\pi G \g V_o^{\f{2}{3}} \underbar{p}\, , \ee
and express the gravitational symplectic structure ${\bf \Omega}$
on ${\bf \Gamma}$ in terms of them:
\ba {\bf \Omega}(\delta_1, \, \delta_2)\,  &=& \, \int_{\cal V}
d^3x \left(\delta_1 A^i_a(x)\delta_2 P_i^a(x) - \delta_2
A^i_a(x)\delta_1 P_i^a(x)\right) \nonumber\\
&=& \, {3} \,\, dc \wedge dp\, . \ea
This expression also makes no reference to the fiducial metric (or
the volume $V_o$ of the cell ${\cal V}$). We will work with this
phase space description. Note that now the configuration variable
$c$ is dimensionless while the momentum variable $p$ has
dimensions $({\rm length})^2$. (While comparing results in the
full theory, it is important to bear in mind that these dimensions
are different from those of the gravitational connection and the
triad there.) In terms of $p$, the physical triad and co-triad are
given by:
\be \label{e1} e^a_i = ({\rm sgn}\,p) |p|^{-\frac{1}{2}}\,\,\,
(V_o^{\frac{1}{3}}\,  \e^a_i), \quad {\rm and} \quad e_a^i = ({\rm
sgn}\, {p}) |p|^{\frac{1}{2}}\,\,\, (V_o^{- \frac{1}{3}}\, \w_a^o)
\ee

We have specified the gravitational part of the reduced phase
space. We will not need to specify matter fields explicitly but
only note that, upon similar restriction to symmetric fields and
fixing of gauge and diffeomorphism freedom, one is led to a finite
dimensional phase space also for matter fields. Finally, let us
turn to constraints. Since the Gauss and the diffeomorphism
constraints are already satisfied, there is a single non-trivial
Scalar/Hamiltonian constraint (corresponding to a constant lapse):
\be \label{scalar} -\frac{6}{\gamma^2}\,\, c^2\, {\rm sgn}p\,
\sqrt{|p|} \, + \, C_{\rm matter}\ =\ 0\, . \ee

\subsection{Quantization: kinematics}
\label{s7.2}

We will now adapt the general procedure of sections \ref{s4} and
\ref{s5} to the symmetry reduced phase space and emphasize how it
leads to interesting departures from the `standard' quantum
cosmology in geometrodynamic variables.

\subsubsection{Elementary variables}
\label{s7.2.1}

Let us begin by singling out `elementary functions' on the
classical phase space which are to have unambiguous quantum
analogs. In the full theory, the configuration variables were
constructed from holonomies $A(e)$ associated with edges $e$ and
momentum variables, from $E(S,f)$, triads $E$ smeared with test
fields $f$ on 2-surfaces. But now, because of homogeneity and
isotropy, we do not need all edges $e$ and surfaces $S$. Symmetric
connections $A$ in $\A_S$ can be recovered knowing holonomies
$h(e)$ along straight lines in $\M$. Similarly, it is now
appropriate to smear triads only by constant fields, $f_i =
\tau_i$, and across squares to which the fiducial triads
$\e^a_i$ are tangent.%
\footnote{Indeed, we could just consider edges lying in a single
straight line and a single rectangle bounded by $\e^a_i$. We chose
not to break the symmetry artificially and consider instead all
lines and all rectangles.}

The $\SU(2)$ holonomy along an edge $e$ is given by:
\be A(e) = \cos \f{\L c}{2} + 2 [\sin \f{\L c}{2}]\,\,
(\dot{e}^a\w_a^i)\, \tau^i \ee
where $\L V_o^{1/3}$ is the oriented length of the edge.
Therefore, a typical element of the algebra generated by sums of
products of matrix elements of these holonomies can be written as
\be \label{f} F(A) = \sum_j \xi_j \,\, e^{i\L_j c} \, \ee
where $j$ runs over a finite number of integers (labelling edges),
$\L_j \in \Real$ and $\xi \in \Comp$. These are precisely the
\emph{almost periodic functions} which have been studied in the
mathematical literature in detail. One can regard a finite number
of edges as providing us with a graph (since, because of
homogeneity, the edges need not actually meet in vertices now) and
the function $F(A)$ as a cylindrical function with respect to that
graph. The vector space of these almost periodic functions is the
space of cylindrical functions of symmetric connections and will
be denoted by $\cyl_S$.

To define the momentum functions, we are now led to smear the
triads with constant test functions and integrate them on a square
(with respect to the fiducial metric). The resulting phase space
function is then just $p$ multiplied by a kinematic factor. We
will therefore regard $p$ itself as the momentum function. In
terms of classical geometry, $p$ is related to the physical volume
of the elementary cell ${\cal V}$ via $V = V_o |p|^{3/2}$.
Finally, the only non-vanishing Poisson bracket between elementary
functions is:
\be \{F(A),\, p\} \,=\, \frac{8\pi \gamma G}{6} \sum_j (i\L_j
\xi_j)\, e^{i\L_j c}\, . \ee
Since the right side is again in $\cyl_S$, the space of elementary
variables is closed under the Poisson bracket. Note that, in
contrast with the full theory, now there is only one momentum
variable whence non-commutativity of triads is no longer an issue.
Therefore, the triad representation also exists in quantum theory.
In fact it turns out to be convenient in making the quantum
dynamics explicit.

\subsubsection{Representation of the algebra of elementary
variables} \label{s7.2.2}

To construct quantum kinematics, let us seek a representation of
this algebra of elementary variables. We will find that the
quantum theory is quite different from the `standard'
geometrodynamical quantum cosmology. This differences arises from
the fact that the configuration variables are not smooth functions
of compact support on $\A_s$, but rather, almost periodic
functions. As we saw in section \ref{s7.2.1}, this choice can be
directly traced back to the full theory where holonomies play a
primary role. Because we are repeating the procedure used in the
full theory as closely as possible, the fundamental discreteness
underlying polymer geometry will trickle down to quantum cosmology
and lead to results which are \emph{qualitatively different} from
those in standard quantum cosmology.

Recall from section \ref{s4.3.4} that one can construct the
representation of the algebra of elementary variables using
Gel'fand theory. In the reduced model under consideration, the
theory implies that the Hilbert space must be the space of square
integrable functions on a suitable completion $\Ab_S$ of the
classical configuration space $\A_S$. Now, $\A_S = \Real$ and
$\Ab_S$ is the Gel'fand spectrum of the $C^\star$ algebra of
almost periodic functions (see (\ref{f})) on $\A_S$. This is a
well-understood space, called the \emph{Bohr compactification} of
the real line (discovered by the mathematician Harold Bohr, Neils'
brother). This is now the quantum configuration space. It is an
Abelian group and carries a canonical, normalized Haar measure
$\mu_o^S$. Following the procedure used in the full theory, it is
natural to set $\H^S_{\rm grav} = L^2(\Ab_S, d\mu_o^S)$ and use it
as the Kinematical Hilbert space for the gravitational sector of
the theory. The general theory can also be used to represent the
algebra of elementary variables by operators on $\H^S_{\rm grav}$.

As in section \ref{s4}, one can make this representation concrete.
Set
\be {\cal N}_\L (\bA) = e^{i\L c}\ee
and introduce on $\cyl$ the following Hermitian inner-product:
\be \langle{{\cal N}_{\L_1}|{\cal N}_{\L_2}} \rangle =
\delta_{\L_1,\L_2} \ee
where the right side is the Kronecker delta, rather than Dirac.
Then $\H^S_{\rm grav}$ is the Cauchy completion of this space.
Thus the almost periodic functions ${\cal N}_{\L}$ constitute an
orthonormal basis in $\H_{\rm grav}^S$; they play the role of spin
networks in the reduced theory. $\cyl_S$ is dense in $\H_{\rm
grav}^S$, and serves as a common domain for all elementary
operators. The configuration and momentum operators have expected
actions:
\ba (\hat{F} {\cal N}_\L )({\bA}) &=& F({\bA})\, {\cal N}_\L
({\bA})
 \nonumber\\
(\hat{p} {\cal N}_\L )({\bA}) &=&  \f{8\pi\L\gamma\lp^2}{6}\,\,
{\cal N}_\L ({A})\, \, . \ea
As in the full theory, the configuration operators are bounded,
whence their action can be extended to the full Hilbert space
$\H^S_{\rm grav}$, while the momentum operators are unbounded but
essentially self-adjoint. The basis vectors ${\cal N}_\L$ are
normalized eigenstates of $\hat{p}$. As in quantum mechanics, let
us use the bra-ket notation and write ${\cal N}_\L (c) = \langle
c|\L\rangle$. Then,
\be \hat{p}\, |\L\rangle \, = \, \f{8\pi\L\gamma\lp^2}{6}\,\, |\L
\rangle \, \equiv \, p_\L\, |\L\rangle \, . \ee
Using the relation $V = |p|^{3/2}$ between $p$ and physical volume
of the cell ${\cal V}$ we have:
\be \label{vol2} \hat{V}\, |\L\rangle =  \left( \frac{8\pi \g
|\L|}{6}\right)^{\f{3}{2}}\,\, \lp^3\, |\L\rangle\, \equiv V_\L\,
|\L\rangle . \ee
This provides us with a physical meaning of $\L$: when the
universe is in the quantum state $|\L\rangle$, (modulo a fixed
constant) $|\L|^{3/2}$ is the \emph{physical} volume of the cell
${\cal V}$ \emph{in Planck units}. Thus, in particular, while the
volume $V_o $ of the cell ${\cal V}$ with respect to the fiducial
metric $\q_{ab}$ may be `large', its physical volume in the
quantum state $|\L =1\rangle$ is just $(\g/6)^{3/2} \lp^3$.

As our notation makes it clear, the construction of the Hilbert
space and the representation of the algebra is entirely parallel
to that in the full theory. In the full theory, holonomy operators
are well-defined but there is no operator representing the
connection itself. Similarly, $\hat{\cal N}_\L$ are well defined
unitary operators on $\H^S_{\rm grav}$ but they fail to be
continuous with respect to $\L$, whence there is no operator
corresponding to $c$ on $\H_{\rm grav}^S$. Thus, as in section
\ref{s4.3}, to obtain physically interesting operators, one has to
\emph{first express them in terms of the elementary variables
${\cal N}_\L$ and $p$ and then promote those expressions to the
quantum theory.}

There is, however, one important difference between the full and
the reduced theories: while eigenvalues of the momentum (and other
geometric) operators in the full theory span only a discrete
subset of the real line, now every real number is a permissible
eigenvalue of $\hat{p}$. This difference can be directly
attributed to the high degree of symmetry. In the full theory,
eigenvectors are labelled by a pair $(e, j)$ consisting of
continuous label $e$ (denoting an edge) and a discrete label $j$
(denoting the `spin' on that edge), and the eigenvalue is dictated
by $j$. Because of homogeneity and isotropy, the pair $(e,j)$ has
now collapsed to a single continuous label $\L$. Note however that
there \emph{is} a weaker sense in which the spectrum is discrete:
all eigenvectors are \emph{normalizable}. Hence the Hilbert space
can be expanded out as a direct \emph{sum} ---rather than a direct
integral--- of the 1-dimensional eigenspaces of $\hat{p}$; i.e.,
the decomposition of identity on $\H_S$ is given by a (continuous)
sum
\be I = \sum_{\L}\, |\L\rangle \langle \L| \ee
rather than an integral. Consequently, the natural topology on the
spectrum of $\hat{p}$ is \emph{discrete}. Although weaker, this
discreteness plays a critical role both technically and
conceptually.

\subsection{Triad operator}
\label{s7.3}

In the reduced classical theory, curvature is simply a multiple of
the \emph{inverse} of the square of the scale factor
$a=\sqrt{|p|}$. Similarly, the matter Hamiltonian invariably
involves a term corresponding to an \emph{inverse} power of $a$.
Therefore, one needs to obtain an operator corresponding to the
inverse scale factor, or the triad (with density weight zero) of
(\ref{e1}). In the classical theory, the triad coefficient
diverges at the big bang and a key question is whether quantum
effects `tame' the big bang sufficiently to make the triad
operator (and hence the curvature and the matter Hamiltonian) well
behaved there.

Now, given a self adjoint operator $\hat{A}$ on a Hilbert space,
the function $f(\hat{A})$ is well-defined if and only if $f$ is a
measurable function on the spectrum of $A$. Thus, for example, in
non-relativistic quantum mechanics, the spectrum of the operator
$\hat{r}$ is the positive half of the real line, equipped with the
standard Lesbegue measure, whence the operator $1/\hat{r}$ is a
well-defined, self-adjoint operator. By contrast, the spectrum of
$\hat{p}$ has discrete topology and $[\hat{p}]^{-1}$ is not a
measurable function of $\hat{p}$. More explicitly, since $\hat{p}$
admits a \emph{normalized} eigenvector $|\L=0\rangle$ with zero
eigenvalue, the naive expression of the triad operator fails to be
densely defined on $\H^S_{\rm grav}$. One could circumvent this
problem in the reduced model in an ad-hoc manner by just making up
a definition for the action of the triad operator on
$|\L=0\rangle$. But then the result would have to be considered as
an artifact of a procedure expressly invented for the model and
one would not have any confidence in its implications for the big
bang. However, we saw in section \ref{s6.3} that a similar problem
arises in the full theory and can be resolved using a strategy due
to Thiemann. It is natural to use the same procedure also in
quantum cosmology. As in the general theory, therefore, we will
proceed in two steps. In the first, we note that, on the reduced
phase space ${\bf \Gamma}_{\rm grav}^S$, the triad coefficient
${\rm sgn}\, p\, |p|^{-\frac{1}{2}}$ can be expressed as the
Poisson bracket $\{c,\, V^{{1}/{3}} \}$ which can be replaced by
$i\hbar$ times the commutator in quantum theory. However, as in
the full theory, a second step is necessary because there is no
operator $\hat{c}$ on $\H^S_{\rm grav}$ corresponding to $c$: one
has to re-express the Poisson bracket in terms of
\emph{holonomies} which do have unambiguous quantum analogs. The
resulting triad (coefficient) operator is given by:
\be \label{qtriad1} \widehat{\left[{\frac{{\rm
sgn}(p)}{\sqrt{|p|}}}\right]}\, = \, -\frac{12i}{8\pi \g\lp^2}
\left(\sin\frac{c}{2} \hat{V}^{\frac{1}{3}} \cos\frac{c}{2}-
\cos\frac{c}{2} \hat{V}^{\frac{1}{3}} \sin\frac{c}{2}\right) \ee
%
%
%

where $V_\L$ is the eigenvalue of the volume operator given in
(\ref{vol2}). Although the triad involves both configuration and
momentum operators, it commutes with $\hat{p}$, whence its
eigenvectors are again $|\L\rangle$. The eigenvalues are given by:
\be \widehat{\left[{\frac{{\rm sgn}(p)}{\sqrt{|p|}}}\right]}\,\,
|\L\rangle = \f{6}{8\pi\gamma\lp^2}\, (V_{\L+1}^{1/3} -
 V^{1/3}_{\L-1}) \, |\L\rangle\, .\ee
where $V_\L$ is the eigenvalue of the volume operator (see
(\ref{vol2})).

A key property of the triad operator follows immediately: It is
bounded above! The upper bound is obtained at the value $\L= 1$:
\be
 |p|^{-\frac{1}{2}}_{\rm max}\, =\, \sqrt{\frac{12}{8\pi \gamma}}\,
 \lp^{-1}\,.
\ee
Since in the classical theory the curvature is proportional to
$p^{-1}$, in quantum theory, it is bounded above by $(12/\g)
\lp^{-2}$. Note that $\hbar$ is essential for the existence of
this upper bound; as $\hbar$ tends to zero, the bound goes to
infinity just as one would expect from classical considerations.
This is rather reminiscent of the situation with the ground state
energy of the hydrogen atom in non-relativistic quantum mechanics,
$E_o = -(m_{\rm e} e^4/2) (1/\hbar)$, which is bounded from below
because $\hbar$ is non-zero.

While this boundedness is physically appealing, at first it also
seems puzzling because the triad coefficient and the momentum are
algebraically related in the classical theory via $p \cdot ({\rm
sgn}p/{|p|^{1/2}})^2 = 1$ and $\hat{p}$ admits a \emph{normalized}
eigenvector with zero eigenvalue. A key criterion of viability of
the triad operator is that the classical relation should be
respected in an appropriate sense. More precisely, one can
tolerate violations of this condition on states only in the Planck
regime; the equality must be satisfied to an excellent
approximation on states with large $\L$ (i.e., with large volume).
Is this the case? We have:
\ba \label{approximate}
 \f{6}{\gamma\lp^2}\, (V_{\L+1}^{1/3} - V^{1/3}_{\L-1}) &=&
 \sqrt{\frac{6|\L|}{8\pi\gamma\lp^2}}\,\,\left(\sqrt{1+ 1/|\L|}-
 \sqrt{1- 1/|\L|}\right)\nonumber\\
 &=& \sqrt{\frac{6}{8\pi \gamma|\L|\lp^2}}\,\,(1+O(\L^{-2}))
\ea
%
Thus, up to order $O(\L^{-2})$, the eigenvalue of the triad
operator is precisely $1/\sqrt{|p_\L|}$, where $p_\L$ is the
eigenvalue of $\hat{p}$ (see (\ref{vol2})). On states representing
a large universe ($\L \gg 1$), the classical algebraic relation
between the triad coefficient and $p$ is indeed preserved to an
excellent approximation. Violations are significant only on the
eigen-subspace of the volume operator with eigenvalues of the
order of $\lp^3$ or less, i.e., in the fully quantum regime.

Since the classical triad diverges at the big bang, it is perhaps
not surprising that quantum effects usher-in the Planck scale.
However, the mechanism by which this came about is new and
conceptually important. For, the procedure did not call for a
cut-off or a regulator. The classical expression of the triad
coefficient we began with is \emph{exact} whence the issue of
removing the regulator does not arise. It \emph{is} true that the
quantization procedure is somewhat `indirect'. However, this was
necessary because \emph{the spectrum of the momentum operator
$\hat{p}$} (or of the `scale factor operator' corresponding to
$a$) \emph{is discrete} in the sense detailed in section
\ref{s7.2}. Had the Hilbert space $\H^{S}_{\rm grav}$ been a
direct integral of the eigenspaces of $\hat{p}$
---rather than a direct sum---  the triad operator could then have
been defined directly using the spectral decomposition of
$\hat{p}$ and would have been unbounded above. Indeed, this is
precisely what happens in geometrodynamics. Thus, the key
differences between the mathematical structures of the present
quantum theory and quantum geometrodynamics are responsible for
the boundedness of the triad coefficient on the entire Hilbert
space \cite{abl:qc}.

A natural question then is: how can there be such inequivalent
quantizations? After all, here we are dealing with a system with a
finite number of degrees of freedom. Doesn't the von Neumann
uniqueness theorem ensure that there is a unique representation of
the exponentiated Heisenberg relations? The answer is no: von
Neumann's theorem requires that the unitary operators $U(\L),
V(\lambda)$ corresponding respectively to the classical functions
$\exp\, i\L c$ and $\exp i\lambda p$ be \emph{weakly continuous}
in the parameters $\L$ and $\lambda$. As we noted in section
\ref{s7.2.2}, this assumption is violated by $U(\L)$ in our
representation ---this in fact is the reason why there is no
operator corresponding to $c$ and only the holonomies are well
defined. A priori, it may seem that by dropping the continuity
requirement, one opens up a Pandora's box. Which of the possible
representations is one to use? The most important aspect of this
construction is that it came directly from the full theory where,
as discussed at the end of section \ref{s4.3.4}, diffeomorphism
invariance severely constrains the choice representation.

\subsection{Quantum dynamics: The Hamiltonian constraint}
\label{s7.4}

Since the curvature is bounded above on the entire kinematical
Hilbert space $\H^S_{\rm grav}$, one might expect that the
classical singularity at the big bang would be naturally resolved
in the quantum theory. This turns out to be the case.

\subsubsection{The constraint operator} \label{s7.4.1}

Rather than starting from the reduced Hamiltonian constraint
(\ref{scalar}), to bring out the relation to the full theory, we
will return to the full constraint and use the procedure spelled
out in section \ref{s6.3}. Because of spatial homogeneity and
flatness, two simplifications arise: i) the two terms in the
expression (\ref{C}) of the full Hamiltonian constraint are now
proportional; and, ii) without loss of generality one can restrict
ourself to a constant lapse function $N$, and we will just set it
to one. Then, the gravitational part of the constraint can be
written as:
\be \C_{\rm grav} = - \f{V_o}{\sqrt{8\pi G} \gamma^{2}}\,
\epsilon_{ijk} F_{ab}^i\, \f{P^{aj} P^{bk}}{\sqrt{\det{P}}} \ee
As in section \ref{s6.3}, we have to `regulate' this classical
expression by writing it in terms of phase space functions which
can be directly promoted to quantum operators. As in the full
theory, we can express the curvature components $F_{ab}^i$ in
terms of holonomies. Consider a square $\a_{ij}$ in the $i$-$j$
plane spanned by two of the triad vectors $\e^a_i$, each of whose
sides has length $\L_o V_o^{1/3}$ with respect to the fiducial
metric $\q_{ab}$. Then, `the $ab$ component' of the curvature is
given by
\be \label{F} F_{ab}^i\tau_i = \w^i_a\,\, \w^j_b\,\,
\left(\f{A(\a_{ij})-1 }{\L_o^2V_o^{2/3}} + O(c^3\L_o)\right) \ee
The holonomy $A({\a_{ij}})$  around the square $\a_{ij}$ can be
expressed as a produce
\be
 A(\a_{ij}) = A (e_i)\, A(e_j)\, A(e_i^{-1})\, A(e_j^{-1})
\ee
where, holonomies along individual edges are given by
\be A(e_i) :=\cos \f{\L_o c}{2} + 2 \sin \f{\L_o c}{2}\, \tau_i \,
.\ee

Next, let us consider the triad term $\epsilon_{ijk} E^{aj}
E^{bk}/\sqrt{\det E}$. As in the full theory, this can be handled
through the Thiemann regularization. Thus, let us begin with the
identity on the symmetry reduced phase space ${\bf \Gamma}^S_{\rm
grav}$:
\be \label{cotriad}
 \epsilon_{ijk}\tau^i\, \f{P^{aj}P^{bk}}{\sqrt{\det{P}}} \,
=\, -2 (\gamma \L_o (8\pi G)^{1/2}\, V_o^{1/3})^{-1} \,
\epsilon^{abc}\, \w^k_c\, A(e_k)\, \{A(e_k^{-1}),\, V\} \ee
where $A(e_k)$ is the holonomy along the edge $e_k$ parallel to
the $k$th basis vector of length $\L_o V_o^{1/3}$ with respect to
$\q_{ab}$. Note that, unlike the expression (\ref{F}) for
$F_{ab}^i$, (\ref{cotriad}) is exact, i.e. does not depend on the
choice of $\L_o$.

Collecting terms, we can now express the gravitational part of the
`regulated' constraint as:
\be \label{reg}
 \C_{\rm grav}^{\L_o} = -4(8\pi G\gamma^3\L_o^3 )^{-1} \sum_{ijk}
\epsilon^{ijk}\,{\rm tr} (A(e_i) A(e_j) A(e_i^{-1}) A(e_j^{-1})\,
A(e_k)\{A(e_k^{-1}),\, V\}) \ee
where, the term proportional to identity in the leading
contribution to $F_{ab}^i$ in (\ref{F}) drops out because of the
trace operation and where we used
$\epsilon^{abc}\,\w^i_a\w^j_b\w^k_c = \sqrt{\q}\,\epsilon^{ijk}$.
In the limit $\L_o \rightarrow 0$, the right side of $\C_{\rm
grav}^{\L_o}$ reproduces the classical expression (\ref{scalar})
of the constraint. Thus,  $\L_o$ ---or the length of the edge used
while expressing $F_{ab}$ in terms of the holonomy around the
square $\alpha_{ij}$--- plays the role of a regulator in
(\ref{reg}). Because of the presence of the curvature term, there
is no natural way to express the constraint \emph{exactly} in
terms of our elementary variables; a limiting procedure is
essential. This faithfully mirrors the situation in the full
theory: there, again, the curvature term is recovered by
introducing small loops at vertices of graphs and the classical
expression of the constraint is recovered only in the limit in
which the loop shrinks to zero.

It is now straightforward to pass to quantum theory. The regulated
quantum constraint is:
\ba \label{qreg}
 \hat{C}_{\rm grav}^{(\L_o)} &=& 4i(8\pi\gamma^3\L_o^3\lp^2)^{-1}
 \sum_{ijk}\epsilon^{ijk}\, {\rm tr} (\bA(e_i) \bA(e_j)
 \bA(e_i^{-1}) \bA(e_j^{-1})\, \bA(e_k) [\bA(e_k^{-1}),\,
 \hat{V}])\nonumber\\
  &=& 96i(8\pi\gamma^3\L_o^3\lp^2)^{-1} \sin^2\frac{\L_oc}{2}
  \cos^2\frac{\L_oc}{2} \left(\sin\frac{\L_oc}{2}\hat{V}
  \cos\frac{\L_oc}{2}- \cos\frac{\L_oc}{2}\hat{V}
  \sin \frac{\L_oc}{2}\right) \ea
Its action on the eigenstates of $\hat{p}$ is
\be
 \hat{C}_{\rm grav}^{(\L_o)}\, |\L\rangle=
 3(8\pi\gamma^3\L_o^3\lp^2)^{-1}
 (V_{\L+\L_o}-V_{\L-\L_o}) (|\L+4\L_o\rangle-
 2|\L\rangle+ |\L-4\L_o\rangle)\, .
\ee
On physical states, this action must equal that of the matter
Hamiltonian $-8\pi G\hat{C}_{\rm matter}$.

In the full theory, one could remove the regulator to obtain a
well-defined operator on (suitable) diffeomorphism invariant
states in $\cyl^\star$. The reduced model, on the other hand, does
not have diffeomorphism invariance. Therefore, one would expect
that the obvious $\L_0 \rightarrow 0$ limit would run in to
problems. This is indeed what happens. In this limit, the
classical regulated expression (\ref{reg}) equals the Hamiltonian
constraint (\ref{scalar}) which, however, contains $c^2$.
Consequently, the naive limit of the operator $\hat{C}_{\rm
grav}^{(\L_o)}$ also contains $\hat{c}^2$. However, since
$\hat{c}^2$ is not well-defined on $\H^S_{\rm grav}$, now the
limit as $\L_o \rightarrow 0$ fails to exist. Thus, one can not
remove the regulator in the quantum theory of the reduced model.
This feature can be traced back directly to the symmetry reduction
\cite{abl:qc}.

A detailed analysis shows that the presence of $\L_o$ in the
quantum Hamiltonian constraint should be regarded as a
quantization ambiguity. Indeed, as we discussed in section
\ref{s6.3}, even in the full theory, there is a similar ambiguity
associated with the choice of the $j$ label used on the new edges
introduced to define the operator corresponding to $F_{ab}$
\cite{gr}. More precisely, in the full theory, the quantization
procedure involves the introduction of a pair of labels $(e,j)$
where $e$ is a continuous label denoting the new edge and $j$ is a
discrete label denoting the spin on that edge. Diffeomorphism
invariance ensures that the quantum constraint is insensitive to
the choice of $e$ but the dependence on $j$ remains as a
quantization ambiguity. In the reduced model, diffeomorphism
invariance is lost and the pair $(e, j)$ of the full theory
collapses into a single continuous label $\L_o$ denoting the
length of the edge introduced to define $F_{ab}$. The dependence
on $\L_o$ persists ---there is again a quantization ambiguity but
it is now labelled by a \emph{continuous} label $\L_o$.

If one works in the strict confines of the reduced model, there
does not appear to exist a natural way to remove this ambiguity.
In the full theory, on the other hand, one can fix the ambiguity
by assigning the \emph{lowest} non-trivial $j$ value, $j= 1/2$, to
each extra loop introduced to determine the operator analog of
$F_{ab}$. This procedure can be motivated by the following
heuristics. In the classical theory, one can use a loop enclosing
an arbitrarily small area in the $a$-$b$ plane to determine
$F_{ab}$ locally. In quantum geometry, on the other hand, the area
operator (of an open surface) has a lowest eigenvalue $a_o =
(\sqrt{3}\pi\gamma)\, \lp^2$, suggesting that it is
\emph{physically} inappropriate to try to localize $F_{ab}$ on
arbitrarily small surfaces. The best one could do is to consider a
loop spanning an area $a_o$, consider the holonomy around the loop
to determine the integral of $F_{ab}$ on a surface of area $a_o$,
and then extract an effective, local $F_{ab}$ by setting the
integral equal to $a_o F_{ab}$. It appears natural to use the same
physical considerations to remove the quantization ambiguity also
in the reduced model. Then, we are led to set the area of the
smallest square spanned by $\alpha_{ij}$ to $a_o$, i.e. to set
$(\gamma \L_o)\, \lp^2 = a_o$, or $\L_o = \sqrt{3}\pi$. Thus,
while in the reduced model itself, area eigenvalues can assume
arbitrarily small values, if one `imports' from the full theory
the value of the smallest non-zero area eigenvalue, one is
naturally led to set $\L_o = \sqrt{3}\pi$.

To summarize, in loop quantum cosmology, one adopts the viewpoint
that (\ref{qreg}), with $\L_o = \sqrt{3}\pi $, is the
`fundamental' Hamiltonian constraint operator which `correctly'
incorporates the underlying discreteness of quantum geometry and
the classical expression (\ref{scalar}) is an approximation which
is valid only in regimes where this discreteness can be ignored
and the continuum picture is valid. This viewpoint is borne out by
detailed calculations \cite{abl:qc}: the expectation values of
$\hat{C}_{\rm grav}^{\L_o}$ in semi-classical states reproduce the
classical constraint. Furthermore, one can calculate corrections
to the classical expression arising from the fundamental
discreteness and quantum fluctuations inherent in the
semi-classical quantum states \cite{abw}.

\subsubsection{Physical states} \label{s7.4.2}

Let us now solve the quantum constraint and obtain physical
states. For simplicity, we assume that the matter is only
minimally coupled to gravity (i.e., there are no curvature
couplings). As in non-trivially constrained systems, one expects
that the physical states would fail to be normalizable in the
kinematical Hilbert space $\H^S = \H^S_{\rm grav}\otimes \H^S_{\rm
matter}$ (see, e.g., \cite{dm,almmt1}). However, as in the full
theory, they do have a natural `home'. We again have a triplet
$$ \cyl_S \subset \H^S \subset \cyl^\star_S $$
of spaces and physical states will belong to $\cyl_S^\star$, the
algebraic dual of $\cyl_S$. Since elements of $\cyl_S^\star$ need
not be normalizable, as in section \ref{s6.3}, we will denote them
by $(\Psi|$. (The usual, normalizable bras will be denoted by
$\langle \Psi|$.)

It is convenient to exploit the existence of a triad
representation. Then, every element $(\Psi|$ of $\cyl^\star_S$ can
be expanded as
\be (\Psi| \, = \, \sum_{\L}\, \psi(\phi, \L) \langle\L|\ee
where $\phi$ denotes the matter field and $\langle \L|$ are the
(normalized) eigenbras of $\hat{p}$. Note that the sum is over a
continuous variable $\L$ whence $(\Psi|$ need not be normalizable.
Now, the constraint equation
\be (\Psi|\,\left(\hat{C}_{\rm grav}^{(\L_o)}\, +\,  8\pi
G\hat{C}_{\rm matter}^{(\L_o)} \right)^{\dagger}\, =0\ee
turns into the equation
\ba \label{de1}
 &&(V_{\L+5\L_o}-V_{\L+3\L_o})\psi(\phi, {\L+4\L_o}) -
 2(V_{\L+\L_o}-V_{\L-\L_o})\psi(\phi, \L) \nonumber\\
 &+&(V_{\L-3\L_o}-V_{\L-5\L_o})\psi (\phi, \L-4\L_o)
 = -\frac{1}{3}\,8\pi G\gamma^3\L_o^3\lp^2\,\,
 \hat{C}_{\rm matter}^{(\L_o)}(\L) \psi(\phi, \L)
\ea
for the coefficients $\psi(\phi, \L)$, where $\hat{C}_{\rm
matter}^{(\L_o)}(\L)$ only acts on the matter fields (and depends
on $\L$ via metric components in the matter Hamiltonian). Note
that, even though $\L$ is a continuous variable, the quantum
constraint is a \emph{difference} equation rather than a
differential equation. Strictly, (\ref{de1}) just constrains the
coefficients $\psi(\phi, \L)$ to ensure that $(\Psi|$ is a
physical state. As in the full quantum theory, we do not have a
background space-time, hence no natural notion of `time' or
`evolution'. However, since each $\langle \L|$ is an eigenbra of
the volume operator, it tells us how the matter wave function is
correlated with volume, i.e., geometry. Now, if one wishes, one
can regard $p$ as providing a heuristic `notion of time', and then
think of (\ref{de1}) as an evolution equation for the quantum
state of matter with respect to this time. (Note that $p$ goes
from $-\infty$ to $\infty$, negative values corresponding to
triads which are oppositely oriented to the fiducial one. The
classical big-bang corresponds to $p=0$.) This heuristic
interpretation often provides physical intuition for (\ref{de1});
one can regard it as a discrete `evolution' equation. However, it
is \emph{not} essential for what follows; one can forego this
interpretation entirely and regard (\ref{de1}) only as a
constraint equation.

What is the fate of the classical singularity? At the big bang,
the scale factor goes to zero. Hence it corresponds to the state
$|\L =0\rangle$ in $\H^S_{\rm grav}$. So, the key question is
whether the quantum `evolution' breaks down at $\L= 0$. Let us
examine this issue. Starting at $\L = -4N \L_o$ for some large
positive $N$, and fixing values of $\psi(\phi, -4N\L_o)$ and
$\psi(\phi, (-4N+4)\L_o)$, one can use the equation to determine
the coefficients $\psi(\phi, (-4N+4n)\L_0)$ for all $n > 1$,
\emph{provided} the coefficient of the highest order term in
(\ref{de1}) continues to remain non-zero. Now, it is easy to
verify that the coefficient vanishes if and only if $n= N$. Thus,
the coefficient $\psi(\phi, \L\! =\! 0)$ remains undetermined. In
its place, one just obtains a consistency condition constraining
the coefficients $\psi(\phi, \L\!=\! -4)$ and $\psi(\phi, \L\!=\!
-8)$. Now, since  $\psi(\phi, \L\!=\! 0)$ remains undetermined, at
first sight, it may appear that one can not `evolve' past the
singularity, i.e. that the quantum evolution also breaks down at
the big-bang. However, the main point is that this is \emph{not}
the case. For, the structure of the quantum scalar constraint is
such that \emph{the coefficient $\psi(\phi, \L\! =\!0)$ just
decouples from the rest}. This comes about because of two facts:
i) the minimally coupled matter Hamiltonians annihilate $\psi
(\phi, \L\!=\! 0) =0$ \cite{mb1,mb7}; and ii) $V_{\L_o} =
V_{-\L_o}$. Thus, unlike in the classical theory, evolution does
not stop at the singularity; the difference equation (\ref{de1})
lets us `evolve' right through it. In this analysis, we started at
$\L= -4N\L_o$ because we wanted to test what happens if one
encounters the singularity `head on'. If one begins at a generic
$\L$, the `discrete evolution' determined by (\ref{de1}) just
`jumps' over the classical singularity without encountering any
subtleties.

Next, let us consider the space of solutions. An examination of
the classical degrees of freedom suggests that the freedom in
physical quantum states should correspond to two functions
\emph{just} of matter fields $\phi$. The space of solutions to the
Hamiltonian constraint, on the other hand is much larger: there
are as many solutions as there are functions $\psi(\phi, \L)$ on
an interval $[\L^\prime- 4\L_o, \L^\prime+ 4\L_o)$, where
$\L^\prime$ is any fixed number. Thus suggests that a large number
of these solutions may be redundant. Indeed, to complete the
quantization procedure, one needs to introduce an appropriate
inner product on the space of solutions to the Hamiltonian
constraint. The physical Hilbert space is then spanned by just
those solutions to the quantum constraint which have finite norm.
In simple examples one generally finds that, while the space of
solutions to all constraints can be very large, the requirement of
finiteness of norm suffices to produce a Hilbert space of the
physically expected size.

For the reduced system considered here, one have a quantum
mechanical system and with a single constraint in quantum
cosmology. Hence it should be possible to extract physical states
using the group averaging technique of the `refined algebraic
framework' \cite{dm,almmt1}.  However, this analysis is yet to be
carried out explicitly and therefore one does not yet have a good
control on how large the physical Hilbert space really is. This
issue is being investigated.

To summarize, two factors were key to the resolution of the big
bang singularity: i) as a direct consequence of quantum geometry,
the Hamiltonian constraint is now a difference equation rather
than a differential equation as in geometrodynamics; and ii) the
coefficients in the difference equation are such that one can
evolve unambiguously `through' the singularity even though the
coefficient $\psi(\phi, \L=0)$ is undetermined. Both these
features are robust: they are largely insensitive to factor
ordering ambiguities and persist in more complicated cosmological
models \cite{mbh,mb9,bdv}. The qualitative changes introduced by
quantum geometry in kinematics and dynamics are significant only
in the Planck regime. A careful analysis shows that the discrete
evolution is extremely well approximated by the Wheeler DeWitt
differential equation at scales larger than $\L_o$ in a precise
sense \cite{abl:qc}. Thus the fundamental discreteness,
characteristic of loop quantum gravity, intervenes in a subtle way
precisely in the Planck regime where geometrodynamics fails to
resolve singularities. Since loop quantum cosmology mimics the
full theory as closely as possible, within the limitations
discussed at the very beginning of section \ref{s7}, these results
provide support for the approach to quantum dynamics in the full
theory.

\section{Applications: Quantum geometry of isolated horizons
and black hole entropy} \label{s8}

Loop quantum cosmology illuminates dynamical ramifications of
quantum geometry but within the limited context of
mini-superspaces where all but a finite number of degrees of
freedom are frozen. In this section, we will discuss a
complementary application where one considers the full theory but
probes consequences of quantum geometry which are not sensitive to
the details of how the Hamiltonian constraint is imposed. (For
further details, see \cite{abck,ack,abk,afk,abl2,abl1,acs,ac,aa4}.
For early work, see
\cite{earlyentropy1,earlyentropy2,earlyentropy3}.)

As was explained in section \ref{s1}, since mid-seventies, a key
question in the subject has been: What is the statistical
mechanical origin of the black hole entropy $S_{\rm BH} = ({a_{\rm
hor}/ 4\lp^2})$? What are the microscopic degrees of freedom that
account for this entropy? This relation implies that a solar mass
black hole must have $(\exp 10^{77})$ quantum states, a number
that is \textit{huge} even by the standards of statistical
mechanics. Where do all these states reside? To answer these
questions, in the early nineties Wheeler \cite{jw} had suggested
the following heuristic picture, which he christened `\emph{It}
from \emph{Bit}'. Divide the black hole horizon in to elementary
cells, each with one Planck unit of area, $\lp^2$ and assign to
each cell two microstates. Then the total number of states ${\cal
N}$ is given by ${\cal N} = 2^n$ where $n = ({a_{\rm hor}/
\lp^2})$ is the number of elementary cells, whence entropy is
given by $S = \ln {\cal N} \sim a_{\rm hor}$. Thus, apart from a
numerical coefficient, the entropy (`\emph{It}') is accounted for
by assigning two states (`\emph{Bit}') to each elementary cell.
This qualitative picture is simple and attractive. Thus, one is
led to ask: can these heuristic ideas be supported by a systematic
analysis from first principles? What is the origin of the
`elementary cells'? Why is each cell endowed with precisely two
states? And most importantly: \emph{what has all this to do with a
black hole?} The `It from Bit' considerations seem to apply to any
2-surface! Quantum geometry enables one to address all these
issues in detail.

The precise picture is much more involved: because the area
spectrum is quite complicated in quantum geometry, `elementary
cells' need not all carry the same area and the number of
`internal states' of each cell is also not restricted to two.
Nonetheless, it does turn out that a dominant contribution in the
entropy calculation comes from states envisaged by Wheeler. The
purpose of this section is to summarize the overall situation. Our
discussion is divided in to three parts. In the first, we recall
the `isolated horizon framework' in classical general relativity
which serves as a point of departure for quantum theory. In the
second we discuss the quantum geometry of the simplest
(undistorted and non-rotating) horizons and present the entropy
calculation allowing for the presence of minimally coupled matter.
In the third, we discuss extensions that include non-minimal
couplings, distortions and rotation.

\subsection{Isolated horizons}
\label{s8.1}

A systematic approach to the problem of entropy requires that we
first specify the class of horizons of interest. Since the entropy
formula is expected to hold unambiguously for black holes in
equilibrium, most analyses were confined to \textit{stationary},
eternal black holes (i.e., to the Kerr-Newman family in
4-dimensional general relativity). {}From a physical viewpoint
however, this assumption seems overly restrictive. After all, in
statistical mechanical calculations of entropy of ordinary
systems, one only has to assume that the given system is in
equilibrium, not the whole world. Therefore, it should suffice for
us to assume that the black hole itself is in equilibrium; the
exterior geometry should not be forced to be time-independent.
Furthermore, the analysis should also account for entropy of black
holes which may be distorted or carry (Yang-Mills and other) hair.
Finally, it has been known since the mid-seventies that the
thermodynamical considerations apply not only to black holes but
also to cosmological horizons \cite{gh}. A natural question
arises: Can these diverse situations be treated in a single
stroke? In classical general relativity, the isolated horizon
framework provides a natural avenue by encompassing all these
situations. It also provides a Hamiltonian framework which serves
as a natural point of departure for quantization
\cite{ack,afk,abl2}

Let us begin with the basic definitions \cite{afk}. In this
discussion, we need to begin with a 4-dimensional space-time
manifold $\man$ although we will return to 3-manifolds $\M$ once
we have a Hamiltonian framework.

A \emph{non-expanding horizon} $\IH$ is a null, 3-dimensional
sub-manifold of the 4-dimensional space-time $(\man,
g_{\a\beta})$,
with topology $S^2\times R$, such that:\\
i) the expansion $\theta_\ell$ of its null normal $\ell$ vanishes;
and,\\
ii) Field equations hold on $\IH$ with stress energy,
$T_{\a\beta}$, satisfying the very weak requirement that
$-T^\a{}_{\beta}\ell^\beta$ is a future-directed, causal vector.
(Throughout, $\ell^\a$ will be assumed to be future pointing.)\\
Note that: i) if the expansion vanishes for one null normal, it
vanishes for all; and, ii) the condition on stress energy is
satisfied by all the standard matter fields \emph{provided they
are minimally coupled to gravity}.

The definition ensures that the area of any 2-sphere cross-section
of the horizon is constant and matter flux across $\IH$ vanishes.
It also implies that the space-time derivative operator $\nabla$
naturally induces a unique derivative operator $\D$ on $\IH$.
Since $\IH$ is a null 3-surface, it has a degenerate intrinsic
`metric' $q_{ab}$ of signature 0,+,+. The pair $(q_{ab}, \D)$ is
referred to as \emph{the geometry} of $\IH$. The notion that the
black hole itself is in equilibrium is captured by requiring that
this geometry is time independent:

An \emph{isolated horizon} $(\IH, \ell)$ is a non-expanding
horizon $\IH$ equipped with a null normal $\ell$ such that $\ell$
is a symmetry of the geometry; i.e., ${\cal L}_\ell q_{ab} = 0$
and $[{\cal L}_\ell, \, D] = 0$ on $\IH$. \\
One can show that, generically, the null normal $\ell$ satisfying
these conditions is unique up to a constant rescaling. For
simplicity, we will assume that we are in the generic case
although the main results go through in all cases.

The isolated horizon definition extracts from the notion of the
Killing horizon just that `tiny' part which turns out to be
essential for black hole mechanics and, more generally, to capture
the notion that the horizon is in equilibrium, allowing for
dynamical processes and radiation in the exterior region. Indeed,
Einstein's equations admit solutions with isolated horizons in
which there is radiation arbitrarily close to the horizons
\cite{abf,jl2}. Note that in the definition uses conditions which
are \emph{local} to $\IH$. Thus, unlike event horizons, this
notion is local, not `teleological'. For our purposes, the two
important considerations are:\\
i) The definition is satisfied not only by event horizons of
stationary black holes but also by the standard cosmological
horizons. Thus, all situations in which thermodynamical
considerations apply are treated in one stroke. \\
ii) If one restricts oneself to space-times which admit an
internal boundary which is an isolated horizon, \emph{the action
principle and the Hamiltonian description is well-defined} and the
resulting phase space has an infinite number of degrees of
freedom. This would not be the case if one used general event
horizons or Killing horizons instead.

Next, let us examine symmetry groups of isolated horizons. A
\emph{symmetry} of $(\IH, \ell, q, \D)$ is a space-time
diffeomorphism which maps $\IH$ to itself; at most rescales $\ell$
by a constant, and preserves $q$ and $\D$. It is clear the
diffeomorphisms generated by any smooth extension of $\ell^\a$ are
symmetries. So, the symmetry group $G_\IH$ is at least
1-dimensional. The question is: Are there any other symmetries? At
infinity, we generally have a universal symmetry group (such as
the Poincar\'e or the anti-de Sitter) because all metrics under
consideration approach a fixed metric (Minkowskian or anti-de
Sitter) there. In the case of the isolated horizons, generically
we are in the strong field regime and space-time metrics do not
approach a universal metric. Therefore, the symmetry group is not
universal. However, there are only three universality classes:\\
i) Type I: the isolated horizon geometry is spherical; in this
case, $G_\IH$ is four dimensional;\\
ii) Type II: the isolated horizon geometry is axi-symmetric; in
this case, $G_\IH$ is two dimensional;\\
iii) Type III: the diffeomorphisms generated by $\ell^\a$ are the
only symmetries; $G_\IH$ is one dimensional.

Note that these symmetries refer \emph{only to} the horizon
geometry. The full space-time metric need not admit any isometries
even in a neighborhood of the horizon. Physically, type II
horizons are the most interesting ones. They include the
Kerr-Newman horizons as well as their generalizations
incorporating distortions (to due exterior matter of other black
holes) and hair. The zeroth and the first laws of black hole
mechanics can be naturally extended to type II isolated horizons
\cite{afk,abl2}. In particular, for the Einstein-Maxwell theory,
one can define the mass $M_\IH$ and angular momentum $J_\IH$ of
the horizon using only the intrinsic geometry of the isolated
horizon and show that the first law holds:
\be {\bf d} M_\IH = \frac{\kappa}{8\pi G}{\bf d} a_{\IH} + \Omega
{\bf d} J_\IH + \Phi {\bf d} Q_\IH \ee
where $\kappa, \Omega, \Phi$ are, respectively, the surface
gravity, angular velocity and the electric potential at the
horizon and ${\bf d}$ denotes the exterior derivative on the
(infinite dimensional) phase space. This law of isolated horizon
mechanics encompasses all black holes and cosmological horizons in
equilibrium, including the ones which arbitrary distortion and
rotation.

\subsection{Type I isolated horizons: Quantum theory}
\label{s8.2}

Let us first discuss type I isolated horizons in detail and then
generalize the results to include non-minimally coupled matter and
type II horizons. We will divide the type I discussion in to three
parts. In the first, we introduce the Hamiltonian formulation; in
the second, we describe the quantum horizon geometry; and in the
third, we summarize the entropy calculation.

\subsubsection{Hamiltonian framework}
\label{s8.2.1}

Consider the sector of general relativity consisting of
gravitational and matter fields for which the underlying
space-time admits an internal boundary which is a type I isolated
horizon $\IH$ \emph{with a fixed area} $a_o$. We will focus on
geometrical structures near $\IH$ and on the modifications of the
Hamiltonian framework of section \ref{s2} caused by the presence
of an internal boundary.

Denote by $S$ the 2-sphere intersection of the (partial) Cauchy
surface $\M$ with the isolated horizon $\IH$. Introduce on $\IH$
an internal vector field $r^i$, i.e. any isomorphism from the unit
2-sphere in the Lie algebra of $\SU(2)$ to $S$ and partially gauge
fix the internal $\SU(2)$ freedom to ${\rm U(1)}$ by requiring
that $r^i P^a_i = \sqrt{|\det q|}\, r^a$, where $r^a$ is the unit
normal to $S$. Then it turns out that \emph{the intrinsic geometry
of $\IH$ is completely determined by the pull-back $\ub{A}^ir_i
=:2W$ to $S$ of the connection $A^i$ on $M$} \cite{abl1}.
Furthermore, $W$ is in fact a spin-connection intrinsic to the
2-sphere $S$: $W = \textstyle{ 1\over 2} \underbar{A}^i r_i =
\textstyle{1 \over 2}\underline\Gamma^i r_i$ on $S$ (see
(\ref{PA}). Thus, if we consider orthonormal dyads $(m, \bar{m})$
on $S$ with internal rotation freedom in $SO(2)$, $W$ is a
connection on the corresponding $U(1)$ bundle. Now, this $U(1)$
bundle on $S$ is non-trivial and $\oint_S dW$ equals $ -2\pi$,
rather than zero. (But since the Chern class of any spin
connection is the same, $\oint_S \delta W =0$; tangent vectors
$\delta W$ to the phase space are genuine 1-forms, globally
defined on $S$. This fact will be useful in the discussion of the
symplectic structure.) Finally, the fact that $S$ is (the
intersection of $M$ with) a type I isolated horizon is captured in
a relation between the two canonically conjugate fields:
\be \label{bc1} F := dW=  - \f{2\pi}{a_o}\,{8\pi G\g}\,\,
\underline\Sigma^i\, r_i .\ee
where $\underline{\Sigma}^i$ is the pull-back to $S$ of the
2-forms $\Sigma^i_{ab} = \eta_{abc}P^a_j \eta^{ij}$ on $\M$, dual
to the momentum $P^a_i$. Thus, because of the isolated horizon
boundary conditions, fields which would otherwise be independent
are now related. As one would expect, the boundary conditions
\emph{reduce} the number of independent fields; in particular, the
pull-backs to $S$ of the canonically conjugate fields $A_a^i,
\Sigma^i_{ab}$ are completely determined by the $U(1)$ connection
$W$.

The main modification in the Hamiltonian framework of section
\ref{s2} is that the gravitational symplectic structure now
acquires a surface term:
\be \label{sym2} {\bf \Omega}(\delta_1, \delta_2) =  -\int_M \,
{\rm Tr}\, (\delta_1 A \wedge \delta_2 \Sigma - \delta_2 A \wedge
\delta_1 \Sigma) \, +\, \f{1}{2\pi}\,\f{a_o}{4\pi G\g } \oint_S \,
\delta_1 W \wedge \delta_2 W\,\, , \ee
where, as in section \ref{s2}, $\delta \equiv (\delta A, \delta
\Sigma)$ denote tangent vectors to the phase space ${\bf \Gamma}$.
Since $W$ is essentially the only `free data' on the horizon, it
is not surprising that the surface term of the symplectic
structure is expressible entirely in terms of $W$. However, it is
interesting and somewhat surprising that the new surface term is
precisely the symplectic structure of the $U(1)$-Chern Simons
theory. \emph{The symplectic structures of the Maxwell,
Yang-Mills, scalar and dilatonic fields do not acquire surface
terms.} Conceptually, this is an important point: this, in
essence, is the reason why (for minimally coupled matter) the
black hole entropy depends just on the area and not, in addition,
on the matter charges.

\subsubsection{Quantum horizon geometry}
\label{s8.2.2}

In the classical theory, the bulk fields determine the surface
fields just by continuity; there are no \emph{independent} degrees
of freedom on the surface in the classical phase space. In the
quantum theory, on the other hand the fields are distributional
and arbitrarily discontinuous whence the surface and the bulk
fields effectively decouple. It is this phenomenon that is
responsible for creating `independent surface states' in the
quantum theory.

The main task is to extend the `bulk' quantum geometry of sections
\ref{s4} and \ref{s5} to allow for the presence of an internal
boundary $S$. Now, the space of generalized connections $\Ab$ is a
product $\Ab = \Ab_V \times \Ab_S$, where a volume generalized
connection $\bA_V$ assigns an $\SU(2)$ element to any
(closed-piecewise analytic) edge lying in the bulk while $\bA_S$
assigns an $\U(1)$ element to each (closed-piecewise analytic)
edge lying in the surface $S$. Therefore, it is natural to begin
with a total Hilbert space $\H = \H_V \otimes \H_S$ where $\H_V$
is built from suitable functions of generalized connections in the
bulk and $\H_S$ from suitable functions of generalized surface
connections. The volume Hilbert space $\H_V$ is the one that comes
from bulk quantum geometry of section \ref{s4}: $\H_V = L^2(\Ab,
\mu_o)$. The question is: What would be the surface Hilbert space?
The answer is suggested by the structure of the surface term in
the symplectic structure: It should be the Hilbert space of
Chern-Simons theory on the horizon. Furthermore, the coefficient
in front of this surface term tells us that the quantum
Chern-Simons theory must have a (dimensionless) coupling
constant/level $k$ given by:
\be \label{level} k = \f{a_{o}}{4\pi\gamma\lp^2} \ee
But we also have to incorporate the boundary condition (\ref{bc1})
\emph{which ensures that $S$ is not any old 2-surface but an
isolated horizon}. The \emph{key} idea is to impose it quantum
mechanically, \emph{as an operator equation} on $\H$, i.e. via
\be \label{qbc}(1\otimes \hat{F})\, \Psi = - \frac{2\pi}{a_{o}}\,
8\pi G\g\,\,(\hat{\underline{\Sigma}}\cdot r)\otimes 1)\, \Psi \,
, \ee
where the notation emphasizes that $\hat{F}$ is a \emph{surface}
operator while $\hat{\underline\Sigma}$ is an operator on the
\emph{volume} Hilbert space. It is easy to show that a basis of
solutions is given by states of the type $\Psi = \Psi_V \otimes
\Psi_S$ where $\Psi_V$ is an eigenstate of the volume operator,
$\Psi_S$ an eigenstate of the surface operator with \emph{same}
eigenvalues. Now, all the eigenvalues of the bulk operator
%
%
on the right side of (\ref{qbc}) are known from bulk quantum
geometry of section \ref{s4.3.4} (see (\ref{ESf1}). They are given
by:
\be \label{triadevs} -\,(\f{2\pi}{a_{o}})\, \left(8\pi \lp^2 \,
\sum_I\, m_I\, \delta^3(x, p_I)\, \eta_{ab}\right)\, , \ee
where $m_I$ are half integers, the sum ranges over a finite set of
points ---called punctures--- on $S$ and where $\eta_{ab}$ is the
metric independent Levi-civita density on $S$. Therefore, the
quantum boundary condition (\ref{qbc}) tells us that $\H_S$ should
be the Hilbert space of $U(1)$-Chern-Simons theory on a punctured
2-sphere $S$ where the curvature $F$ has the form of a
$\delta$-distribution concentrated at a finite number of
punctures.

\begin{figure}
\begin{center}
\includegraphics[height=2in]{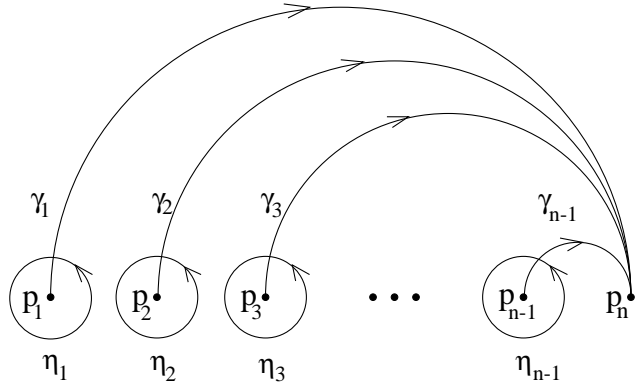}
\caption{Co-ordinatization of the surface phase space}
\label{entropyfig1}
\end{center}
\end{figure}

Let us then begin by fixing a set $\P$ of punctures on $S$ and
consider $\U(1)$-Chern-Simons theory on this punctured sphere. The
phase space of this theory is ${\bf \Gamma}_S^{\P} =
(\Ab_S^o)/(\Gb^{\P}\rtimes {\cal D}^{\P})$  where $\Ab_S^o$ is the
space of connections which is flat everywhere except at the
punctures; $\Gb^{\P}$ is the space of local $\U(1)$ gauge
transformations which are identity except at the punctures; ${\cal
D}^{\P}$ is the space of diffeomorphisms of $S$ which fix the
punctures and certain structure at the punctures; and $\rtimes$
stands for semi-direct product.%
\footnote{The extra structure one needs to fix at the punctures is
listed at the end of section 4.3.1 of \cite{abk}. {}From the
physics perspective, this is the most delicate of the technical
subtleties in the subject, although this procedure is `standard
practice' in the mathematics literature. It plays an important
role in the imposition of the diffeomorphism constraint and state
counting.}
This phase space is isomorphic with the torus
${\mathbb{T}}^{2(n-1)}$ if there are $n$ punctures in the set
$\P$, equipped with the natural symplectic structure on
${\mathbb{T}}^{2(n-1)}$. A convenient set of canonically conjugate
coordinates can be introduced as follows. Let us fix the $n$th
puncture as `origin' and, as in Fig \ref{entropyfig1}, denote by
$\g_I$, a family of curves joining the $I$th puncture to the
$n$th, (with $I = 1,2,\ldots, n-1)$) and by $\eta_I$, `small'
closed loops surrounding each of the first $n-1$ punctures. Then,
for each $I$, two holonomies around $\g_I$ and $\eta_I$ are
canonically conjugate. This phase space is often referred to as
the \emph{non-commutative torus}.

The Hilbert space $\H_S^{\P}$ of surface states results from
geometric quantization of this torus \cite{abk}. This is the space
of quantum states of the $U(1)$ Chern-Simons theory on $(S, \P)$.
The total surface Hilbert space $\H_S$ is the inductive limit of
these Hilbert spaces as the set $\P$ becomes larger and larger. As
discussed in section \ref{s4}, the volume Hilbert space $\H_V =
L^2(\Ab_v, d\mu_o)$ can also be obtained as the inductive limit of
the Hilbert spaces $(\H_V)_\a$ associated with graphs $\a$
\cite{mm,al3,al4}.

Next one has to impose the quantum boundary condition (\ref{qbc}).
\emph{This introduces a highly non-trivial test of the whole
framework.} Construction of the surface Hilbert space was strongly
motivated by (\ref{qbc}). However, now that it is complete, there
is no more freedom.  In the Chern-Simons Hilbert space, one can
compute the eigenvalues of the surface operator $\hat{F}$. This
calculation is \emph{completely independent} of the volume Hilbert
space; it has never heard of the quantum geometry in the bulk. The
\emph{key} question on which everything hinges is: Are the
eigenvalues of $\hat{F}$ the same as the eigenvalues
(\ref{triadevs}) of the bulk operator in (\ref{qbc})? If not,
there will be no solutions to the quantum boundary conditions! The
remarkable fact is that \emph{the infinite set of eigenvalues of
the two operators match}, even though the two calculations are
completely distinct. This comes about because the level of the
Chern-Simons theory is related to the Barbero-Immirzi parameter
$\gamma$ and the area $a_{o}$ in a very specific way, which in
turn is a consequence of the isolated horizon boundary conditions.
Thus there is a seamless matching between three completely
independent theories: the isolated horizon framework in classical
general relativity; the bulk quantum geometry; and, the
Chern-Simons theory on the punctured horizon. And this matching
provides a coherent mathematical description of the quantum
geometry of the horizon.

\begin{figure}
\begin{center}
\includegraphics[height=6cm]{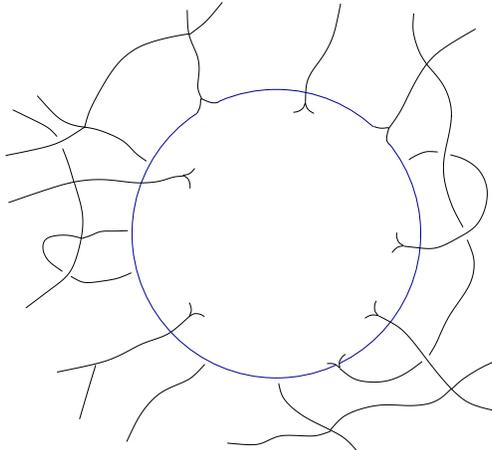}
\caption{Quantum Horizon. Polymer excitations in the bulk puncture
the horizon, endowing it with quantized area. Intrinsically, the
horizon is flat except at punctures where it acquires a quantized
deficit angle. These angles add up to endow the horizon with a
2-sphere topology.} \label{entropyfig2}
\end{center}
\end{figure}

Finally, one has to impose quantum Einstein's equations. The Gauss
constraint asks that the total state $\Psi_V\otimes \Psi_S$ should
be gauge invariant. The diffeomorphism constraint asks that
diffeomorphisms on $S$ should be regarded as gauge. Again there
are important mathematical subtleties. But the final picture is
simple. While each of the bulk and the surface states transforms
non-trivially under the remaining gauge freedom ($\Gb/\Gb^\P$),
the total state is gauge invariant as needed. Implementation of
the remaining diffeomorphism constraint (corresponding to ${\cal
D}/{\cal D}^\P$) asks that what matters is only the number of
punctures; their location is irrelevant.%
\footnote{The subtleties are: i) In the Chern-Simons theory, only
the exponentiated operator $\exp i \hat{F}$ is well-defined;
$\hat{F}$ itself is not. Therefore, the mathematically meaningful
quantum boundary condition is the exponentiated version of
(\ref{qbc}). ii) the $U(1)$ gauge group at the punctures is
replaced by the quantum $U(1)$ group. The deformation parameter is
supplied by the level $k$ of the Chern-Simons theory which is
required to be an integer because of the pe-quantization
requirements. This also implies that the deficit angles at each
puncture are quantized. iii) Recall that one has to fix certain
structure at the punctures in the construction of the surface
Hilbert space. Under ${\cal D}/{\cal D}^\P$, this structure
changes. Therefore, strictly, to begin with one has infinitely
many copies of the surface Hilbert spaces, one for each choice of
the extra structure and the fact that ${\cal D}/{\cal D}^\P$ is
gauge implies that only one of these copies is physically
relevant. That is, the fact that diffeomorphisms in ${\cal
D}/{\cal D}^\P$ are gauge is incorporated by `gauge fixing'.}

The Hamiltonian constraint, by contrast does not restrict the
surface states, i.e., the quantum geometry of the horizon. This is
because in the classical theory, the constraint is functionally
differentiable (i.e. generates gauge) only when the smearing
function (the lapse) goes to zero on the isolated horizon boundary
(and, as usual, at infinity). The time-evolution along the
isolated horizon is generated by a \emph{true Hamiltonian}, not
just the constraint.

Let us summarize. The physical surface Hilbert space $\H_S^{\rm
Phys}$ is given by $\H_S^{\rm Phys} = \oplus_n \H_s^n$, with
$\H_S^n$, the Hilbert space of the $\U(1)$ Chern-Simons theory on
the sphere $S$ with $n$ punctures, where the polymer excitations
of the bulk geometry intersect $S$. (See Fig \ref{entropyfig2}).
Let us focus on $\H_S^n$. Since $W$ is the intrinsic
spin-connection on $S$ and since $F$ vanishes except at the
punctures, the intrinsic geometry of the quantum horizon is flat
except at the $n$ punctures. The $\eta_i$-holonomies around these
punctures are non-trivial, whence the punctures carry deficit
angles which, furthermore, are quantized. They add up to $4\pi$,
providing a quantum analog of the Gauss-Bonnet theorem.

 \emph{Remarks}: i) Note that the above analysis makes a crucial use
of the horizon boundary condition; it is not applicable to a
general 2-surface. Thus the most important limitation of Wheeler's
`It from Bit' considerations is overcome. The strategy of
incorporating the boundary condition through an \emph{operator
equation} (\ref{qbc}) allows both the connection $W$ and the triad
$\underline{\Sigma}^i r_i$ to fluctuate and requires only that
they do so in tandem. This equation provides the first step in the
answer to the question: what is a quantum black hole?

ii) For extensions of this framework to include non-minimally
coupled fields and type II horizons (discussed in section
\ref{s8.3}), let us note that there are just three essential
mathematical ingredients which serve as the input for this
construction of the surface Hilbert space: a) the form of the
surface term in (\ref{sym2}) which shows that the surface
symplectic structure is that of the $\U(1)$-Chern-Simons theory
with level $k$ of (\ref{level});  b) the horizon boundary
condition (\ref{bc1}); and c) The spectrum (\ref{triadevs}) of the
triad operator $\underline{\Sigma}^ir_i$.

\subsubsection{Entropy: Counting surface states}
\label{s8.2.3}

In the classical theory, $a_{o}$ in the expression of the surface
term of the symplectic structure (\ref{sym2}) and in the boundary
condition (\ref{bc1}) is the horizon area. However in the
\emph{quantum theory}, $a_{o}$ has simply been a parameter so far;
we have not tied it to the \emph{physical area of the horizon}. To
calculate entropy, one has to construct a suitable
`micro-canonical' ensemble by relating $a_o$ to the physical area.

It follows from the definition of volume connections that, as
depicted in Fig \ref{entropyfig2}, the polymer excitations of the
bulk geometry puncture the horizon \emph{transversely} from the
`exterior'. Hence, the relevant area eigenvalues are those given
in (\ref{rev}):
$$ 8\pi \gamma\lp^2\, \sum_I \sqrt{j_I(j_I +1)}\, . $$
Therefore, one is led to construct the micro-canonical ensemble by
considering only that sub-space of the bulk theory which, at the
horizon, satisfies:
\be \label{micro1} a_{o} -\epsilon \le 8\pi\gamma \lp^2\, \sum_I\,
\sqrt{j_I(j_{I}+1)} \le a_{o} + \epsilon \ee
where $I$ ranges over the number of punctures, $j_I$ is the spin
label (the eigenvalue of the vertex operator $\hat{J}_{p_I}$
associated with the puncture $p_I$), and $\lp^2 < \epsilon \ll
a_0)$. In presence of matter fields carrying charges, one fixes
values of horizon charges $Q_{o}^{(\alpha)}$ (labelled by
$\alpha$) and restrict the matter configurations so that
\be \label{micro2} Q_{o}^{(\alpha)} -\epsilon^{(\alpha)} \le
Q_{o}^{(\alpha)} \le Q_{o}^{(\alpha)} + \epsilon^{(\alpha)}\ee
for suitably chosen $\epsilon^{(\alpha)}$. (As is usual in
statistical mechanics, the leading contribution to the entropy is
independent of the precise choice of these small intervals.) Now,
the physical states belonging to this ensemble contain information
also about gravitational and electromagnetic radiation far away
from the horizon which is obviously irrelevant to the calculation
of black hole entropy. What is relevant are the states directly
associated with the horizon of a given area $a_o$, and charges
$Q^{(\alpha)}_o$. One is therefore led to trace over the volume
degrees of freedom and construct a density matrix $\rho_\IH$
describing a maximum entropy mixture of surface satisfying
(\ref{micro1}) and (\ref{micro2}). The statistical mechanical
entropy is then given by $S_\IH = -{\rm Tr}(\rho_\IH\,
\ln\rho_\IH$).  As usual, this number can be calculated simply by
counting states:
\be S_\IH = \ln {\cal N}_\IH \ee
where ${\cal N}_\IH$ is the number of Chern-Simons surface states
consistent with the area and charge constraints. A detailed
analysis \cite{abk}, estimates this number and leads to the
expression of entropy of large black holes:
\be \label{entropy} S_{\IH} := \ln {\cal N}_{\IH} =
\frac{\gamma_o}{\gamma}\, \frac{a_{o}}{4\lp^2} +
o(\frac{\lp^2}{a_{o}}) ,\quad {\rm where} \quad \gamma_o =
\frac{\ln 2}{\sqrt{3} \pi } \ee
Thus, ignoring terms $o(\frac{\lp^2}{a_{o}})$, entropy is indeed
proportional to the horizon area. However, even for large black
holes, one obtains agreement with the Hawking-Bekenstein formula
\emph{only} in the sector of quantum geometry in which the
Barbero-Immirzi parameter $\gamma$ takes the value $\gamma =
\gamma_o$. Thus, while all $\gamma$ sectors are equivalent
classically, the standard quantum field theory in curved
space-times is recovered in the semi-classical theory only in the
$\gamma_o$ sector of quantum geometry. It is quite remarkable that
thermodynamical considerations involving \textit{large} black
holes can be used to fix the quantization ambiguity which dictates
such Planck scale properties as eigenvalues of geometric
operators.

Now, the value of $\gamma$ can be fixed by demanding agreement
with the semi-classical result just in one case ---e.g., a
spherical horizon with zero charge, or a cosmological horizon in
the de Sitter space-time, or, \ldots. Once the value of $\gamma$
is fixed, the theory is completely fixed and one can ask: Does
this theory yield the Hawking-Bekenstein value of entropy of
\textit{all} isolated horizons, irrespective of the values of
charges, angular momentum, and cosmological constant, the amount
of distortion, or hair. The non-trivial fact is that the answer is
in the affirmative. Thus, the agreement with quantum field theory
in curved space-times holds in \textit{all} these diverse cases.
Physical interpretation of $S_\IH$ is that it represents the
entropy that observers in the `external region' (used in the
construction of the phase space) associate to the horizon.

Why does $\gamma_o$ not depend on other quantities such as
charges? As noted in section \ref{s8.2.2}, only the gravitational
part of the symplectic structure develops a surface term at the
horizon; the matter symplectic structures have only volume terms.
(Furthermore, the gravitational surface term is insensitive to the
value of the cosmological constant.) Consequently, there are no
independent surface quantum states associated with matter. This
provides a natural `explanation' of the fact that the
Hawking-Bekenstein entropy depends only on the horizon geometry
and is independent of electro-magnetic (or other) charges (of
minimally coupled matter fields).

Finally, let us return to Wheeler's `It from Bit'. One can ask:
what are the states that dominate the counting? Perhaps not
surprisingly, they turn out to be the ones which assign to each
puncture the smallest quantum of area (i.e., spin value $j =
\textstyle{1\over 2}$), thereby maximizing the number of
punctures. In these states, each puncture defines Wheeler's
`elementary cell' and his two states correspond to whether the
deficit angle is positive or negative.

To summarize, quantum geometry naturally provides the micro-states
responsible for the huge entropy associated with horizons. In this
analysis, all black holes and cosmological horizons are treated in
an unified fashion; there is no restriction, e.g., to
near-extremal black holes. The sub-leading term has also been
calculated and shown to be proportional to $\ln a_{\rm hor}$
\cite{km,gm}. Finally, in this analysis quantum Einstein's
equations \textit{are} used. In particular, had the quantum Gauss
and co-vector/diffeomorphism constraints not been imposed on
surface states, the spurious gauge degrees of freedom would have
given an infinite entropy. However, because of the isolated
horizon boundary conditions, the scalar/Hamiltonian constraint has
to be imposed just in the bulk. Since in the entropy calculation
one traces over bulk states, the final result is insensitive to
the details of how this (or any other bulk) equation is imposed.
Thus, as in other approaches to black hole entropy, the
calculation does not require a complete knowledge of quantum
dynamics.

\subsection{Non-minimal couplings and type II horizons}
\label{s8.3}

We will now show that, while the introduction of non-minimal
couplings \cite{acs,ac} and distortion and rotation
\cite{aepv,aa4} does introduce interesting modifications, the
qualitative picture of section \ref{s8.2} remains unaltered.

\subsubsection{Non-minimal couplings}
\label{s8.3.1}

Consider a scalar field $\phi$ non-minimally coupled to gravity
through the action
\be
 {\bf S}[ g_{ab}, \phi]\, = \, \int d^4x\, \sqrt{- g}\left[
\frac{1}{16\pi G} f(\phi) R  - \frac{1}{2}  g^{ab} \nabla_a \phi
\nabla_b \phi - V(\phi)\right]
 \label{Action1}
 \ee
where $f$ is a nowhere vanishing function (minimal coupling
results if $f(\phi) =1$) and $V(\phi)$ is the potential. Now the
stress energy tensor does not satisfy even the weak energy
requirement ii) in the definition of a non-expanding horizon.
However, one can replace it by: \\
ii)' Field equations hold on $\IH$ and the scalar field $\phi$
satisfies ${\cal L}_\ell \phi =0$ on $\IH$.\\
to incorporate the idea that the scalar field is time independent
on $\IH$, reflecting the fact that the horizon is in equilibrium.
(For minimal couplings, time independence of matter fields on
$\IH$ is ensured by ii).) The isolated horizon framework then
leads to the zeroth and first laws \cite{acs}. However, now the
form of the first law is modified:
\be {\bf d} M = \frac{\kappa}{8\pi G}{\bf d}\, [\oint_S f(\phi)
d^2V_S] + \Omega {\bf d} J_\IH \ee
suggesting that the entropy should be given by:
\be S_\IH = \f{1}{4\lp^2}\, \oint_S f(\phi) d^2V . \ee
(The same conclusion is reached using the general framework of
\cite{jkm,iw} which deals with a broad class of theories but which
requires a globally defined Killing field with a bifurcate Killing
horizon.) The question now is: Can the statistical mechanical
derivation of entropy of \ref{s8.2}, based on quantum geometry, go
through also in this case? This is a non-trivial check on the
robustness of that framework, first because the seamless matching
between Chern-Simons theory and bulk geometry required for a
coherent description of the quantum horizon geometry is very
delicate, and second, because the entropy now depends not only on
geometry but also on the scalar field. In spite of these
non-trivialities, the framework does turn out to be robust.

For type I horizons, as one would expect, $\phi$ is constant on
$S$. Let us consider the sector of the phase space consisting of
fields for which $\IH$ is a type I horizon with fixed area $a_o$
and scalar field $\phi_o$. Then, the main modifications of the
discussion of \ref{s8.2}, caused by the non-minimal coupling, are
the following:\\
i) Denote by $\Pi^a_i$ the momentum conjugate to the gravitational
connection $A_a^i$. If we only have minimally coupled matter,
$\Pi^a_i = P^a_i$ and the geometrical triad $\tilde{E}^a_i$ is
given by $\tilde{E}^a_i = 8\pi G \g {P}^a_i$ (see (\ref{P})). With
non-minimal coupling, the geometrical triad involves \emph{both}
$\Pi^a_i$ \emph{and} the scalar field: $\tilde{E}^a_i = (8\pi G
\g/f(\phi))\, \Pi^a_i$. Conceptually, this is an important change
because quantum geometry is now dictated not just by the
gravitational variables $(A_a^i, \Pi^a_i)$ but also involves the
matter variable
$\phi$.\\
ii) The geometrical relation between the pulled back triad and the
gravitational connection on the horizon remains unaltered.
Therefore, in terms of the phase space variables, the boundary
condition (\ref{bc1}) is now replaced by:
\be \label{bc2} F := dW=  - \f{2\pi}{f(\phi_o)\, a_o}\, 8\pi G \g
\,\, {\underline{\Sigma}^i\, r_i} .\ee
where $\underline{\Sigma}^i_{ab}$ is now the pull-back to $S$ of
the dual $\eta_{abc} \Pi^{c}_j \eta^{ij}$ of the gravitational
momentum.\\
iii) The surface term in the symplectic structure is again given
by that of the $\U(1)$ Chern-Simons theory but the level is
modified:

\be \label{level2} k = \f{f(\phi_o)\,a_{o}}{4\pi\gamma\lp^2} \ee
iv) Since the description of the bulk Hilbert space in terms of
the gravitational momentum variables remains unaltered, the
eigenvalues of the gravitational momentum operator
$\underline{\Sigma}^i_{ab} r_i$ continue to given by
(\ref{triadevs}):
\be \label{triadevs2} -\,(\f{2\pi}{a_{o}})\, \left(8\pi \lp^2 \,
\sum_I\, m_I\, \delta^3(x, p_I)\, \eta_{ab}\right)\, , \ee
v) Finally, for the 2-surface $S$, the area eigenvalues are now
given by:
\be \label{rev2} \f{8\pi \gamma\lp^2}{f(\phi_o)}\, \sum_I
\sqrt{j_I(j_I +1)}\, . \ee
in place of (\ref{rev}).

Thus, in the key equations, $a_o$ is just replaced by
$f(\phi_o)a_o$ everywhere. One can now repeat the analysis of
section \ref{s8.2} (using the `polymer representation also of the
scalar field \cite{acs}.) Equations (\ref{bc2}), (\ref{level2})
and (\ref{rev2}) now imply that the quantum boundary condition
does have `enough' solutions: although the level of the
Chern-Simons theory and the boundary condition are both modified,
the delicate interplay between the surface and the volume sectors
required for a coherent theory of the geometry of quantum horizons
survives in tact. The state counting procedure can be repeated and
(\ref{rev2}) now implies that the entropy is given by:
\be \label{entropy2} S_{\IH} := \ln {\cal N} =
\frac{\gamma_o}{\gamma}\, \frac{f(\phi_o) a_{o}}{4\lp^2} +
o(\frac{\lp^2}{a_{o}}) ,\quad {\rm where} \quad \gamma_o =
\frac{\ln 2}{\sqrt{3} \pi }\, . \ee
Thus, if $\g = \g_o$, one obtains the answer suggested by the
classical analysis. Note that the value $\g_o$ of the
Barbero-Immirzi parameter is the \emph{same} as it was for
minimally coupled matter fields.

\subsubsection{Inclusion of distortion and rotation}
\label{s8.3.2}

Let us now extend the framework of section \ref{s8.1} to type II
horizons \cite{aepv,aa4}. It turns out that the \emph{type II
problem can be mapped on to the type I problem} already at the
classical level. Thus, in the quantum theory, the underlying
mathematics will be the same as that in the type I case. However,
the physical meaning of the Chern-Simons connection on the
boundary will be different. As is usual when one maps a given,
complicated problem to a mathematically simpler one, the physical
non-trivialities are contained in the map. In the present case it
is the map that carries all the information about distortion and
angular momentum.

For brevity, let us focus just on the gravitational sector and
ignore other fields on $\IH$. The  `free data' on type II isolated
horizon is again determined by an $\U(1)$ connection $V :=
\textstyle{\f{1}{2}}\ub{A}^ir_i$ on $S$. However, now the
connection also has the information about distortion and angular
momentum : $\ub{A}^i\, r_i = \underline{\Gamma}^ir_i + \g
\underbar{K}^i r_i$, where $\underline{\Gamma}^i$ is a real
$\U(1)$ connection on the spin-bundle over $S$ which now carries
information about distortion and $\underbar{K}^ir_i$ is a globally
defined, real-valued 1-form on $S$ which carries information about
angular momentum (see(\ref{PA}). The gauge and diffeomorphism
invariant characterization of the free data $V$ can be coded in a
pair, $M_n, J_n, n =0,1,2,\dots,\infty$ of mass and angular
momentum multipoles \cite{aepv}. In the type I case, $M_0$ is the
mass and all other horizon multipoles vanish. In the type II case,
$M_0$ is again the mass, $J_0$ continues to vanish, $J_1$ is the
angular momentum and higher multipoles represent departures from
sphericity. The $M_n$ and $J_n$ have the interpretation of `source
multipoles' of the black hole.

Recall from section \ref{s8.1} that two ingredients from the
classical theory play the key role in quantization: the isolated
horizon boundary condition (\ref{bc1}) and the surface term in the
symplectic structure (\ref{sym2}). Now, (\ref{bc1}) is replaced
by:
\be \label{bc3}
 F := dV = f\, 8\pi G\g \,\,\underline{\Sigma}^i r_i \ee
The major difference from the type I case is that $f$ is \emph{not
a constant but a genuine function on} $S$ (determined by the
Newman-Penrose component $\Psi_2$ of) the Weyl curvature.
Similarly, while the symplectic structure does have a surface term
which is fully determined by the surface connection $V$, it is
\emph{not} the Chern-Simons symplectic structure for $V$. So, at
first, the situation appears to be quite different from that in
the type I case.

However, one can in fact map the present problem to the type I
problem which was already solved. To see this, note first that if
one is interested in any one macroscopic black hole, one has to
fix its macroscopic parameters. In the globally vacuum context of
classical general relativity, for example, one would fix the mass
(or the horizon area) and the angular momentum. In the present,
very general discussion of isolated horizons, we have allowed
arbitrary distortions. Therefore, to fix the macroscopic black
hole, one has to fix all multipole moments $M_n,J_n$. The type I
phase space can also be constructed my fixing all multipoles (now
$M_0 \not=0$ and all other multipoles zero). Hence, one would
intuitively expect that this sector is `of the same size' as the
type I sector. However, is this really the case? More importantly,
do the arguments that one should be considering a
\emph{Chern-Simons theory on a punctured 2-sphere} go through?

The answer to both questions turns out to be in the affirmative:
One can explicitly coordinatize the type II surface phase space
${\bf \Gamma}_S$ with a new $\U(1)$ connection $W$ on the
spin-bundle over $S$ such that the surface symplectic structure
${\bf \Omega}_S$ is given by:
\be \label{sym5} {\bf \Omega}_S(\delta_1, \delta_2) = \f{1}{8\pi
G}\, \f{a_o}{\gamma \pi} \oint_S \, \delta_1 W \wedge \delta_2 W\,
. \ee
and the curvature of $W$ is given by:
\be \label{W} dW = - \f{2\pi \g}{a_o}\, 8\pi G\g\,\,
\underline{\Sigma}^ir_i \ee
Thus, the sector of the surface phase space corresponding to any
fixed set of multipoles on a type II horizon is isomorphic with
the phase space of a type I horizon. Moreover, the horizon
boundary condition in the type II case (when expressed in terms of
$W$) are identical to those in the type I case. Therefore in terms
of the surface connection $W$, one can proceed with quantization
as we did before in the type I case. All mathematics underlying
the quantum horizon geometry and the state counting is the same!
However, the physical meaning of symbols and constructions is
different. In particular, in the type I case, $W$ was the natural
spin connection which directly described the horizon geometry and
therefore the punctures where its curvature is concentrated could
be directly associated with deficit angles. In the present case,
it is the connection $V$ that determines the physical horizon
geometry and not $W$ and the relation between the two involves
distortion and rotation. Classically, this non-trivial information
is coded in multipole moments of $V$. On the quantum Hilbert
space, one can introduce the multipole moment operators and their
eigenvalues distinguish the physical situation of interest, coded
in $V$, from the physics of the fiducial connection $W$. Thus,
there are non-trivial differences on issues related to
interpretation. However, the counting argument of \ref{s8.3} is
unaffected by these.

To summarize, one can treat generic type II isolated horizons via
following steps: i) construct parameters (the multipoles) which
characterize these horizons in an invariant fashion
macroscopically (i.e. in the classical theory); and, ii) Introduce
an isomorphism from the phase space of horizons of interest to
that of type I horizons, which maps the physical isolated horizon
condition to the horizon condition in the type I case. Together,
these properties enable us to construct the quantum theory of
horizon geometry and count the horizon states. \emph{The procedure
guarantees that the value of the Barbero-Immirzi parameter that
reproduces the Hawking Bekenstein formula for large black holes is
the same as that used in the type I case, independent of the
values of the mass and angular momentum multipoles.} The value is
thus robust. Finally, note that because this analysis incorporates
arbitrary distortions, we are going well beyond the Kerr-Newman
family. The method encompasses a vast class of astrophysically
realistic black holes.

\section{Current Directions}
\label{s9}

In the last six sections, we presented a self-contained summary of
the quantum geometry framework and its physical applications which
have been worked out in detail. In this section, we turn to
current research. There are two major thrusts: i) recovery of low
energy physics through semi-classical quantum geometries; and, ii)
spin foam models, which provide a `sum over histories' approach
based on quantum geometry. Since these are frontier areas, a
finished physical picture is yet to emerge. Therefore, our
discussion will be briefer.

\subsection{Low energy physics}
\label{s9.1}

A basic premise of loop quantum gravity is that there should be no
background fields; everything, including space-time geometry is
dynamical and treated quantum mechanically from the start.
However, of necessity, a background independent description must
use physical concepts and mathematical tools that are quite
different from those normally used in low energy quantum physics
which is rooted in classical, Minkowskian geometry. A major
challenge, then, is to show that this low energy description does
arise from the pristine, Planckian world in an appropriate sense.
This challenge is now being met step by step, although one is
still far from reaching the final goal.

Let us begin by listing some of the main issues and questions.
Loop quantum gravity is based on \textit{quantum geometry}, the
essential discreteness of which permeates all constructions and
results. The fundamental excitations are 1-dimensional and
polymer-like. A convenient basis of states is provided by spin
networks. Low energy physics, on the other hand, is based on
quantum field theories which are rooted in a flat space continuum.
The fundamental excitations of these fields are 3-dimensional,
typically representing wavy undulations on the background
Minkowskian geometry. The convenient Fock-basis is given by
specifying the occupation number in one particle states labelled
by momenta and helicities. At first sight, the two frameworks seem
disparate. \textit{What then is the precise sense in which the
Fock states are to arise in the low energy limit of the full
theory?}

{}From a mathematical physics perspective, the basic variables of
quantum geometry are holonomies (or Wilson loops) of the
gravitational connection $A$ along 1-dimensional curves and fluxes
of the conjugate momenta (the triads) $E$ across 2-surfaces. In
the final quantum theory, the connection $A$ fails to be a
well-defined operator(-valued distribution); only the holonomies
are well-defined. In perturbative quantum field theories, by
contrast, the vector potential operators are distributions,
whence, a priori, their holonomies fail to be well-defined
operators. Similarly, fluxes of electric field operators across
2-surfaces fail to be well-defined on the Fock space of photons.
Heuristically, then, it would appear that, even at a kinematic
level, loop quantum gravity describes a `phase' of gauge theories
which is distinct from the one used in electrodynamics. Since it
is generally believed that distinct phases carry distinct physics,
it is natural to ask: \textit{Is the well-tested, macroscopic
`Coulomb phase' of low energy gravity compatible at all with the
Planck scale discreteness of quantum geometry?} If so, in what
sense? How does it emerge from loop quantum gravity?

So far these issues have been analyzed through simple examples,
where the focus is on constructing mathematical and conceptual
tools that will be ultimately necessary for the systematic
analysis of quantum fields on semi-classical states of quantum
geometry \cite{st2} .

\subsubsection{Quantum mechanics of particles}
\label{9.1.1}

In non-relativistic quantum mechanics, one generally begins with
the Weyl algebra generated by operators $U(\lambda) = \exp
i\lambda X$ and $V(\mu) = \exp i(\mu/\hbar) P$ and seeks
representations in which $U(\lambda)$ and $V(\mu)$ are represented
by 1-parameter family of unitary operators which are weakly
continuous in $\lambda$ and $\mu$. The von Neumann uniqueness
theorem tells us that every irreducible representation of this
algebra is isomorphic with the Schr\"odinger representation.
Therefore, one typically develops the theory using just this
representation. However, if one drops the requirement of weak
continuity, say in $\mu$, new representations become available.

Specifically, there is one in which states $\Psi(k)$ are almost
periodic functions of $k= p/hbar$. In this representation,
operators $U(\lambda), V(\mu)$ are unitary as desired but the
self-adjoint generator of $V(\mu)$, which provides the momentum
operator in the Schr\"odinger representation, fails to exist. The
position operator $X$, on the other hand, does exist and is
self-adjoint. Furthermore, its spectrum is discrete in the sense
that all its eigenvectors are normalizable. This representation is
referred to as `polymer particle'  because of its close
mathematical similarities with the `polymer' representation of the
algebra generated by holonomies and electric fluxes introduced in
section \ref{s4}. (Indeed, the underlying mathematical framework
is the same as that used in quantum cosmology in section \ref{s7},
but the physical interpretations are very different.) $X$ is
analogous to the electric flux operators and its eigenstates
provide us with analogs of spin network states. $V(\mu)$ is
analogous to the holonomies. Just as the connection operator does
not exist in quantum geometry, the generator of space translations
----the momentum operator of the Schr\"odinger theory---  does not
exist on the Hilbert space $\H_{\rm poly}$ of the polymer particle
representation. While the absence of the standard momentum
operator is alarming from the perspective of non-relativistic
quantum mechanics, \emph{heuristically} it can be thought of as
arising from a fundamental discreteness of spatial geometry.
However, this motivation can not be taken too literally:
non-relativistic quantum mechanics has limitations which become
manifest \emph{much} before quantum gravity discreteness can
become significant.

Rather, the primary motivation in this study is mathematical: we
have a simple toy model to probe the questions raised in the
beginning of this section. In this analogy, Schr\"odinger quantum
mechanics plays the role of quantum theories used in low energy
physics and the main question is: Can the polymer particle
framework reproduce the results of Schr\"odinger quantum
mechanics, in spite of the fact that the two descriptions are
fundamentally so different? The answer is in the affirmative and
the analysis has provided some conceptual and technical insight to
recover low energy physics from the Planck scale framework based
on polymer geometry.

Main results \cite{afw} can be summarized as follows:

\b Although the standard creation and annihilation operators fail
to be well-defined in $\H_{\rm poly}$, their exponentials
\emph{are} well-defined and can be used to construct coherent
states purely in the polymer framework. As one might expect of
semi-classical states, they belong to $\cyl_{\rm poly}^\star$, the
analog of $\cyl^\star$ of quantum geometry. A key question is
whether they can be regarded as semi-classical states. At first
this appears to be difficult because $\cyl_{\rm poly}^\star$ does
not carry a Hermitian inner product. However, one \emph{can}
provide a meaningful criterion of semi-classicality through a
notion of \emph{shadow states} (explained in section \ref{s9.1.2})
and verify that these coherent states satisfy the criterion.

\b As in quantum cosmology, by introducing a length scale $\mu_o$
which is thought of as arising from the fundamental discreteness
of spatial geometry, one can define a momentum operator and the
kinetic energy term in the Hamiltonian using $V(\mu)$:
$$ \f{P^2}{2m} = \f{\hbar^2}{2m}\, \f{1}{\mu_o^2}\, [2- V(\mu_o)
- V(-\mu_o) ]\, . $$
with $\mu_o \ll d$, where $d$ is the smallest length scale in the
problem.%
\footnote{For a harmonic oscillator, $d = \sqrt{\hbar/m\omega}$,
which , for the vibrational modes of a CO molecule is $10^{-10}
{\rm cm}$. Since laboratory experiments show no signature of
discreteness at the $10^{-17}{\rm cm}$ scale, it is safe to take
$\mu_o \le 10^{-19}{\rm cm}$}
The Schr\"odinger equation then reduces to a difference equation.
For the case of the harmonic oscillator, one can transform it to
the well-known Mathieu equation and, using the rather large body
of results on this equation, show that all energy eigenstates are
non-degenerate and eigenvalues discrete, given by:
$$ E_n \sim  (2n+1) \f{\hbar\omega}{2} -
\f{2n^2+2+1}{16}\,\left(\f{\mu_o}{d}\right)^2\, \f{\hbar\omega}{2}
+ O\left(\f{\mu_o}{d}\right)^4 $$
Thus the `polymer corrections' to the Schr\"odinger eigenvalues
become significant only when $n \sim 10^7$! Using the notion of
shadow states, one can also show that that there is a precise
sense in which the eigenvectors are `close' to the Schr\"odinger
eigenvectors. Since in the final picture one is in effect using a
discrete approximation to the Schr\"odinger equation, it may seem
`obvious' that a close agreement with Schr\"odinger quantum
mechanics must occur. However, the detailed analysis contains a
number of subtleties and the agreement emerges only when these
subtleties are handled appropriately \cite{afw}. More importantly,
one does not simply begin with the Schr\"odinger equation and
discretize it `by hand'. Rather, one follows procedures that are
natural from the `polymer' perspective and arrive at the discrete
substitute of the Schr\"odinger equation.

Thus, polymer particle has turned out to be a simple toy model to
illustrate how the gap between inequivalent mathematical
frameworks can be bridged and how they can lead to physically
equivalent results in the `low energy regime' in spite of the deep
conceptual and structural differences at a fundamental level.

\subsubsection{The Maxwell field and linearized gravity}
\label{s9.1.2}

The next two models that have been studied in detail are the
Maxwell field and linearized gravity in Minkowski space-time
\cite{mv1,al7,jv,afg,mv2,agv}. Since our goal is to only to
provide a bird's eye view, we will focus the main discussion on
the Maxwell case and return to linearized gravity at the end.

Following the general procedures outlined in section \ref{s4} to
the case when the gauge group $G$ is $\U(1)$, one can construct a
`polymer representation' of the Maxwell field (see Appendix
\ref{a1}. Polymer representations were first introduced by
Buchholz and Fredenhagen in a different context.) The goal is to
understand its relation to the standard Fock representation of
photons. Again, the viewpoint is \emph{not} that the polymer
representation provides a better physical description of photons
in Minkowski space. Rather, the primary goal is to develop
mathematical and conceptual tools to compare the disparate
descriptions, tools which will be finally useful in understanding
the relation between quantum field theories on semi-classical
quantum geometries representing classical space-times and
continuum quantum field theories in these space-times.

The first major difference between the polymer and the Fock
representations lies in their algebras of elementary observables.
In the polymer representation, these are given by holonomies
$\aM(e)$ of the Maxwell connection $\aM$ along edges $e$ in $M=
{\mathbb{R}}^3$, and electric fields $\eM(g)$ smeared by smooth
1-forms $g$ of compact support in $M$. In the Fock representation,
by contrast, the configuration variables $\aM(f)$ are vector
potentials smeared with smooth vector densities $f$ of compact
support; $\aM(e)$ fail to be well-defined \cite{ai2}. To resolve
this tension, one can proceed as follows \cite{mv1,al7}. Introduce
a test bi-tensor field $r_a^{a'}(x,x')$ which is a 1-form in its
$x$ dependence and a vector density in its $x'$ dependence:
\be \aM^{(r)}_a(x):= \int_M d^3x'\,
r_a^{a'}(x,x')\aM_{a'}(x'),\quad {\rm and} \quad
\eM^{a'}_{(r)}(x'):= \int_M d^3x\, r_a^{a'}(x,x')  \eM^{a} (x).
\ee
Then, the map
\be \I^{\rm Fock}_{\rm poly} (r):\,\, (\aM(e),\, \eM_{(r)}(g))
\mapsto (\aM^{(r)}(e),\, \eM(g)) \ee
is an isomorphism from the Poisson-Lie algebra of the elementary
observables used in the polymer representation to that used in the
Fock representation. A prototype example of $r_a^{a'} (x,x')$ is
given in Cartesian coordinates by:
\be \label{r} r_a^{a'} (x,x')\, =\, \f{1}{r^3} \, \exp
\f{|x-x'|^2}{2 r^2} \, \delta_a^{a'} \,\,\equiv f_r(x,x')\,
\delta_a^{a'}\ee
with $r >0$, for which $\aM^{(r)} (e)$ is simply the holonomy
around a `thickening' of the edge $e$. For simplicity, we will
restrict ourselves to this specific choice in what follows.

Using isomorphisms $\I^{\rm Fock}_{\rm poly}$, one can pass back
and forth between the polymer and the Fock descriptions.
Specifically, the image of the Fock vacuum can be shown to be the
following element of $\cyl^\star_{\rm Max}$ \cite{mv1,al7}:
\be \label{vac} (V\!\mid \,\, = \,\, \sum_{\alpha,{\vec n}}\, \exp
\left[ -\frac{\hbar}{2} \sum_{IJ} G_{IJ}n_I n_J \right]\,
(F_{\alpha, {\vec n}}\!\mid \, . \ee
where $(F_{\alpha, {\vec n}}|$, called \emph{flux network states},
constitute a basis in $\cyl^\star_{\rm Max}$ and are analogous to
the spin network states in $\cyl^\star$ (see Appendix \ref{a1}).
These states do not have any knowledge of the underlying
Minkowskian geometry. This information is neatly coded in the
matrix $G_{IJ}$ associated with the edges of the graph $\alpha$,
given by:
\be G_{IJ}\ =  \int_{e_I}dt \dot{e}^a_J(t) \int_{e_J}dt'
\dot{e_I}^b (t')\, \int d^3x\, q_{ab}(x)\,  [f_r(x,e_I(t))\,
|\Delta|^{-\frac{1}{2}}\,f(x,e_J(t'))] \ee
where $q_{ab}$ is the flat Euclidean metric and $\Delta$ its
Laplacian. A key insight of Varadarajan \cite{mv1} was to note
that, as in the case of the polymer particle, \emph{one can single
out this state directly in the polymer representation by invoking
Poincar\'e invariance}, without any reference to the Fock space.

Similarly, one can directly locate in $\cyl^\star_{\rm Max}$ all
coherent states, i.e., all eigenstates of the (exponentiated)
annihilation operators. Let us denote by $(C_{(\aM^o,\eM_o)}|$ the
state peaked at classical fields $(\aM^o,\eM_o)$. Given a graph
$\a$, one can show that the restriction of the action of
$(C_{(\aM^o,\eM_o)}|$ to cylindrical functions associated with
$\a$ is fully encoded in a state $C^{(\aM^o,\eM_o)}_{\a}$ in the
Hilbert space $(\H_{\rm Max})_\a$:
\be (C_{(\aM^o, \eM_o)}|\Psi_\alpha\rangle\,  = \, \int_{\Ab_{\rm
Max}} d\mu_o \,[C^{(\aM^o,\eM_o)}_{\alpha}(\bar\aM)]^\star\,
\Psi_\alpha(\bar\aM) \ee
for all cylindrical functions $\Psi_\a$ associated with the graph
$\a$ . The states $C^{(\aM^o,\eM_o)}_{\alpha}(\bar\aM)$ in
$(\H_{\rm Max})_\a$ are referred to as \emph{shadows} of the
element $(C_{(\aM^o, \eM_o)}| \in \cyl^\star$ on graphs $\alpha$.
Note that the set of all shadows captures the full information in
$(C_{(\aM^o, \eM_o)}|$. By analyzing shadows on \emph{sufficiently
refined graphs}, one can introduce criteria to test if a given
element of $\cyl^\star_{\rm Max}$ represents a semi-classical
state \cite{afw,afg}. The states $(C_{(\aM^o, \eM_o)}|$ do satisfy
this criterion and can therefore be regarded as semi-classical in
the polymer framework. Finally, using the isomorphism $\I^{\rm
Fock}_{\rm poly}$ one can check that these states are the images
of the Fock coherent states. To summarize, although the polymer
representation is inequivalent to the Fock, it is possible to
single out and analyze the `correct' semi-classical states of the
quantum Maxwell field directly in the polymer framework
\cite{afg}.

For Maxwell fields, the Fock representation is compatible with
only the Coulomb phase: The vacuum expectation value of
(regularized) Wilson loops goes as the exponential of the
perimeter and one can read-off the Coulomb potential from the
sub-leading term in the exponent. It turns out that one can
translate Wilson's criterion as a condition on the overlap of
certain coherent states defined by the type of loops used in the
original criterion. All these considerations go through also for
the linearized gravitational field in Minkowski space-time
\cite{mv2,agv}. Moreover, the reformulation of the Wilson
criterion provides a means of testing whether candidate
semi-classical states of the \emph{full} theory, approximating
Minkowski space-time and fluctuations thereon, are compatible with
the Coulomb phase ---i.e., if, in a suitable limit, the
gravitational force between two particles will be given by the
Coulomb law \cite{agv}.  Physically, this is a key constraint on
the viability of proposed semi-classical states.

\subsubsection{Quantum geometry}
\label{s9.1.3}

The experience gained from simpler models is currently being used
to construct semi-classical states of quantum geometry peaked at
initial data corresponding to physically interesting space-times.
In particular, are there `preferred' semi-classical states peaked
at such classical space-times, analogous to the coherent states of
photons and gravitons in Minkowski space-time?

The early work \cite{ars,ag} focussed on constructing states which
are peaked at a given spatial triad $E^a_i$ . However, the
mathematical precision was low and, moreover, the analysis ignored
connections altogether. The challenge of constructing states which
are peaked at given values of a set of observables constructed
from both the triads and the connection was taken up in
\cite{cr,toh1,toh2}. In particular, a detailed mathematical
framework developed in the series of papers \cite{toh1,toh2}
focused on observables associated with a given graph: holonomies
of edges of the graph and fluxes of triads across certain surfaces
`dual' to the edges. This work led to states in $\cyl$ which are
sharply peaked at given values of these observables. However, this
set of observables is too small from physical considerations and
these states do not have the `non-local' correlations which are
the hallmark of semi-classical states in Minkowskian physics.
Nonetheless, this analysis introduced a number of mathematical
techniques which continue to be useful in the current
investigations.

As a prelude to current research directions, let us begin by
recasting the construction of the familiar coherent states in a
form that is suitable for generalization. For a harmonic
oscillator (or for free fields in Minkowski space-times) coherent
states can be constructed using heat kernel methods on the
configuration space. In this procedure, one starts by selecting a
suitable, positive function $F$ on the phase space which is
quadratic in momenta. For the harmonic oscillator, this can be
taken to be the simply the kinetic energy, $F= \vec{P}\cdot {\vec
P}$. By rescaling the quantum analog of this function with
suitable constants, one obtains the (negative definite)
\emph{Laplacian} $\Delta$. The associated heat kernel provides a
smoothening operator which maps the generalized eigenstates of the
configuration operator to coherent states. For the oscillator, the
coherent state $C_{\vec{x}_o,0}$, peaked at $\vec{x}=\vec{x}_o$
and $\vec{p}=0$ is given by:
\be C_{(\vec{x}_o, \vec{p}=0)}(x) = [\exp t\Delta]\, \, \delta
(\vec{x},\, \vec{x}_o)\ee
where $t$ determines the width of the Gaussian. ($t$ has physical
dimensions $({\rm length})^2$. The value $t = \hbar/m\omega$
yields the standard coherent states.) A general coherent state
$C_{\vec{x}_o, \vec{p}_o}$ is obtained simply by taking the
analytical continuation of this state with respect to $\vec{x}_o$:
\be C_{\vec{x}_0, \vec{p}_o}(x) = \left([\exp t\Delta] \,\,
\delta(\vec{x},\, \vec{x}_o)\right)_{\vec{x}_o\mapsto \vec{z}_o}
\ee
where $\vec{z}o= \vec{x}_o + (i/\hbar) p_o$. Hall \cite{hall}
generalized this construction for the case when the configuration
space is a compact Lie group. Let us consider the example of a
free particle moving on the group manifold $\SU(2)$ (see section
\ref{s4.1}). We can again use for $F$ the kinetic energy term: $F=
\eta^{ij}p_ip_j$ where as before $\eta^{ij}$ is the Cartan-Killing
metric on $\SU(2)$. One can use the natural isomorphism between
the complexification of and the cotangent bundle over $\SU(2)$ to
label the points in the phase space by elements $g^{{\mathbb C}}$
of ${{\mathbb C}}\SU(2)$. Then, the coherent states
$C_{g^{{\mathbb C}}_o}$ peaked at the point $g^{{\mathbb C}}_o$ of
the phase space is given by:
\be C_{g^{{\mathbb C}}_o} (g)\, =\, \left([\exp t\Delta]\, \,
\delta(g,\, g_o)\right)_{g_o\mapsto g^{{\mathbb C}}_o} \ee
These states are sharply peaked at the phase space point
$g^{{\mathbb C}}_o$ \cite{toh1,toh2}. The generalization of Hall's
procedure to quantum theories of connection on a graph (discussed
in section \ref{s4.2}) is straightforward since now the
configuration space is isomorphic to $[\SU(2)]^n$, where $n$ is
the number of edges of the graph.

For theories of connections in the continuum, one can again follow
the same procedure. For the Maxwell theory in Minkowski space-time
discussed in the section \ref{s9.1.2}, in the polymer picture one
can proceed as follows. First, given a graph $\a$, one can set $F
= G^{IK} p_I p_K$ where $p_I$ denotes the momentum vector in the
cotangent bundle over $\U(1)$ associated with the $I$th edge.
Then, the Laplacian on the Hilbert space $(\H_{\rm Max})_\a$ is
given simply by $\Delta_\a^{\rm Max} = - (\hbar/2)\,\sum_{I,K}\,
[G^{IK} J_IJ_K]$. Interestingly, this family $\Delta_\a^{\rm Max}$
of operators is \emph{consistent} and leads to a negative
definite, self-adjoint operator $\Delta^{\rm Max}$ on the full
Hilbert space $\H_{\rm Max}$. This Laplacian can now be used to
define coherent states. The result is precisely the coherent
states in $\cyl^\star_{\rm Max}$ discussed in section
\ref{s9.1.2}:
\be (C_{\aM^o,\eM_o}| \Psi\rangle = \int_{\Ab} d\mu_o\,
(e^{\Delta^{\rm Max}}\, \delta(\bar{\aM}, \aM_o)|_{\aM_o \mapsto
\aM^o_{{\mathbb C}}})^\star\, \Psi (\bar{\aM}) \ee
for all cylindrical functions $\Psi$, where $\aM^o_{{\mathbb C}}=
\aM^o - i |\Delta|^{-1/2} \eM_o$. In particular $(V|$, the image
of the Fock vacuum in  $\cyl^\star_{\rm Max}$ is obtained by this
procedure simply by setting $\aM_o = \eM_o =0$. These Laplacians
and the corresponding coherent states belong to a general
framework discussed in \cite{cohsttr}.

This procedure can be naturally extended to quantum geometry to
define a candidate semi-classical state $(M|$ corresponding to the
Minkowski space-time, i.e. to the point of the phase space
represented by $(A=0, E=E^o)$ where $E^o$ is a flat triad. Given
any graph $\a$, one can define a Laplacian operator $\Delta_\a$ on
the quantum geometry Hilbert space $\H_\a$:
\be \Delta_\a = - \f{\hbar}{8}\, \sum_{I,K}\, G_{IK} \eta^{ik}
J_i^I J_k^K \ee
Again, this set of operators is consistent and thus defines a
negative definite, self-adjoint operator $\Delta$ on the full
quantum geometry Hilbert space. The desired state $(M| \in
\cyl^\star$ can now be defined using the heat kernel defined by
this Laplacian:
\be (M| \Psi\rangle = \int_{\Ab} d\mu_o\, (e^{\Delta}\,
\delta(\bar{A}, A_o)|_{A_o \mapsto A^o_{{\mathbb C}}})^\star\,
\Psi(\bar{A})  \ee
for all $\Psi \in \cyl$, where $A_o^{{\mathbb C}} = - ib E^o$
where $b$ is a constant with dimensions of inverse length
\cite{al7}. Note however that the state is defined simply by
analogy with the simpler systems. So far, its structure has not
been analyzed in any detail and there is no a priori guarantee
that this is indeed a semi-classical state, i.e., that its shadows
on sufficiently refined graphs are sharply peaked at the point
$(A= 0, E= E_o)$ of the gravitational phase space. (Notion from
statistical geometry \cite{lb} are likely to play an important
role in selecting the appropriate family of graphs.) The `Coulomb
phase criterion' is also yet to be applied.

Thiemann has developed a systematic framework to extend this
procedure to introduce semi-classical states by considering more
general functions $F$, leading to heat kernels based on operators
which are more general than Laplacians \cite{ttcomp}. Thus, rather
powerful tools are now available to explore the semi-classical
regime. However, compelling candidate states are yet to emerge.
Finally, the emphasis in this work is on constructing states which
are peaked at points on the constraint surface of the classical
phase space. These are kinematical states; as simple examples
show, these states will not solve quantum constraints. Indeed,
semi-classical solutions to the quantum constraints would be
peaked at points of the reduced phase space, i.e., roughly, on
equivalence classes of 4-metrics where two are equivalent if they
are related by a diffeomorphism. To make contact with low energy
physics, what we need is states peaked at \emph{individual
classical space-times}, whence it is the kinematical
semi-classical states considered here which are more directly
relevant. The relation between the two is being explored
systematically. The overall picture can be summarized as follows:
i) the kinematical states can be regarded as `gauge fixed
versions' of the semi-classical solutions to constraints; and, ii)
the expectation values and fluctuations of Dirac observables agree
in an appropriate sense on the two sets.

\subsection{Spin foams}
\label{s9.2}

Spin foams can be thought of as histories traced out by `time
evolution' of spin networks and provide a path integral approach
to quantum dynamics. Since an entire review article devoted to
spin foams has appeared recently \cite{ap1}, our discussion will
be very brief.

In the gravitational context, the path integral can play two
roles. First, as in standard quantum field theories, it can be
used to compute `transitions amplitudes'. However outside, say,
perturbation theory about a background space-time, there still
remain unresolved conceptual questions about the physical meaning
of such amplitudes. The second role is `cleaner': as in the
Euclidean approach of Hawking and others, it can be considered as
a device to extract physical states, i.e. solutions to all the
quantum constraint equations. In this role as an \emph{extractor},
it can shed new light on the quantum Hamiltonian constraint and on
the issue of finding a physical inner product on the space of
solutions to all constraints.

The well-defined quantum kinematics of sections \ref{s4} and
\ref{s5} has motivated specific proposals for the definition of
path integrals, often called `state sum models'. Perhaps the most
successful of these is the Barrett-Crane model and its various
modifications. At the classical level, one regards general
relativity as a topological field theory, called the BF theory,
\emph{supplemented with an algebraic constraint}. The BF theory is
itself a 4-dimensional generalization of the 3-dimensional
Chern-Simons theory mentioned in section \ref{s8.2} and has been
investigated in detail in the mathematical physics literature.
However, the role of the additional constraint is very important.
Indeed, BF theory has no local degrees of freedom; it is the extra
constraint that reduces the huge gauge freedom, thereby recovering
the local degrees of freedom of general relativity. The crux of
the problem in quantum gravity is the appropriate incorporation of
this constraint. At the classical level, (modulo issues related to
degenerate configurations) the constrained BF theory is equivalent
to general relativity. To obtain Euclidean general relativity, one
has to start with the BF theory associated with $SO(4)$ while the
Lorentzian theory results if one uses $SO(3,1)$ instead. The
Barrett-Crane model and its extensions are specific proposals to
define quantum geometry based path integrals for the constrained
BF theory in either case.

Fix a 4-manifold $\man$ bounded by two 3-manifolds $M_1$ and
$M_2$. Spin-network states on the two boundaries can be regarded
as `initial' and `final' quantum geometries. One can then consider
histories, i.e., quantum 4-geometries, joining them. Each history
is a spin-foam. Each vertex of the initial spin-network on $M_1$
`evolves' to give a 1-dimensional edge in the spin-foam and each
edge, to give a 2-dimensional face. Consequently, each face
carries a spin label $j$. However, in the course of `evolution'
\textit{new vertices} can appear, making the dynamics non-trivial
and yielding a non-trivial amplitude for an `initial' spin-network
with $n_1$ vertices to evolve into a `final' spin-network with
$n_2$ vertices. For mathematical clarity as well as physical
intuition, it is convenient to group spin-foams associated with
the same 4-dimensional graph but differing from one another in the
labels, such as the spins $j$ carried by faces. Each group is said
to provide a \emph{discretization} of the 4-manifold $\man$.
Physically, a discretization has essentially just the topological
information. The geometrical information ---such as the area
associated with each face--- resides in the labels. This is an
important difference from lattice gauge theories with a background
metric, where a discretization itself determines, e.g., the edge
lengths and hence how refined the lattice is.

A notable development is the discovery that the non-perturbative
path integral, defined by a certain modification of the
Barrett-Crane model, is equivalent to a manageable \textit{group
field theory} (GFT) in the sense specified below \cite{ap2}. The
GFT is a rather simple quantum field theory, defined on four
copies of the underlying group
---$\SL(2,C)$ in the case of Lorentzian gravity and ${\rm
Spin}(4)$ in the case of Euclidean. (Note that these are just
double covers of the Lorentz group and the rotation group of
Euclidean 4-space.) Thus GFTs live in high dimensions. The action
has a `free part' and an interaction term with a coupling constant
$\lambda$. But the free part is non-standard and does not have the
familiar kinetic term, whence the usual non-renormalizability
arguments for higher dimensional, interacting theories do not
apply. In fact, the first key recent result is that \emph{this GFT
is finite order by order in the Feynman perturbation expansion.}
The second key result is $A_{\rm BC}{(n)}= A_{\rm GFT} {(n)}$,
where $A_{\rm BC}{(n)}$ is the modified Barret-crane amplitude
obtained by summing over all geometries (i.e., spin labels $j$)
for a fixed discretization and $A_{\rm GFT}{(n)}$ is the
coefficient of $\lambda^n$ in the Feynman expansion of the GFT.
Together, the two results imply that, in this approach to quantum
gravity, \textit{sum over geometries for a fixed discrete topology
is finite}. This is a highly non-trivial result because, on each
face, the sum over $j$s ranges from zero to infinity; there is no
cut-off.%
\footnote{Perez's Euclidean result has the same `flavor' as the
evidence found by Luscher, Reuter, Percacci, Perini and others
\cite{lr,pp} for non-perturbative renormalizability of
4-dimensional Euclidean quantum general relativity (stemming from
the existence of a non-trivial fixed point).}

However, many open issues remain. First, in the specific proposal
of Perez and others, convergence is achieved at a price: the
integral is dominated by `degenerate' geometries described by by
spin foams where all the spins labelling faces are zero except for
`islands' of higher spin \cite{bcht}. Second, in any of the finite
models, it is not clear if there is a direct physical
interpretation, \emph{in gravitational terms}, of the specific
amplitudes (associated with 2-faces and tetrahedra) that lead to a
suppression of divergences. More importantly, while many of these
developments are very interesting from a mathematical physics
perspective, their significance to quantum gravity is less clear.
Physical issues such as gauge fixing in the path integral are not
fully understood in 3+1 dimensions \cite{bp}. (However, recently
there has been notable progress in 2+1 dimensions \cite{lf,np}.)
Finally, the discrete topology is fixed in most of this work and
issue of summing over all topologies, or a substitute thereof,
remains largely unexplored. However, this is a very active area of
research and the hope is that the current investigations sill soon
yield a sufficient intuition and control on mathematical issues to
enable one to analyze in detail the deeper, physical problems. In
particular, it is likely that a judicious combination of methods
from the canonical treatment of the Hamiltonian constraint and
spin foam models will lead to significant progress in both areas.

\section{Outlook}
\label{s10}

Loop quantum gravity is a non-perturbative, background-independent
approach to the problem of unification of general relativity and
quantum physics. In the last nine sections, we gave a
self-contained account of the core developments in this approach
and then summarized the most important physical applications of
the framework. However, due to space limitation we had to leave
out several interesting developments, particularly at the
forefront of the field. In this section, we will discuss some of
them briefly and outline a few open issues.

\b \emph{Quantum geometry.} As mentioned in section \ref{s1} the
necessity of a quantum theory of geometry was strongly motivated
by the fact that, in general relativity, gravity is coded in
space-time geometry. However, the quantum geometry framework
itself is more general and could be used for background
independent quantization of other theories as well. For example,
in two space-time dimensions, Yang-Mills theory requires only a
background volume element, not a metric. Since the classical
theory is invariant under all volume preserving diffeomorphisms,
it is natural to quantize it in a way that this symmetry is
manifest at every step. Quantum geometry techniques have been used
to carry out this quantization and this construction has certain
advantages over others \cite{yangmills}. Similarly, in the
standard treatments of bosonic string theory, one fixes only a
conformal metric on the world-sheet. In 2 dimensions, the group of
conformal isometries is an infinite dimensional subgroup of the
diffeomorphism group and one can again use the standard techniques
developed in section \ref{s4} to carry out a quantization in which
this symmetry is manifest \cite{asstrings}.

A recent mathematical development is the natural emergence of
quantum groups from quantum geometry considerations
\cite{qgroups}. Suppose for a moment that quantum groups had yet
not been invented and one was trying to extend the construction of
$\cyl$, introduced in section \ref{s4}, to the most general
setting possible, e.g. to obtain mathematically viable
generalizations of quantum gauge theories. Then, one would have
naturally discovered that $\cyl$ can be replaced by a
\emph{non-commutative $C^\star$-algebra which has precisely the
same structure as a quantum group}! This is a fascinating result
which brings out the naturalness of constructions underlying
quantum geometry.

{}From conceptual considerations, an important issue is the
\emph{physical} significance of discreteness of eigenvalues of
geometric operators (see, e.g., \cite{al5}). Recall first that in
the classical theory differential geometry simply provides us with
formulas to compute areas of surfaces and volumes of regions in a
Riemannian manifold. To turn these quantities in to physical
observables of general relativity, one has to define the surfaces
and regions \emph{operationally}, e.g. using matter fields. Once
this is done, one can simply calculate values of these observables
using formulas supplied by differential geometry. The situation is
the same in quantum theory. For instance, the area of the isolated
horizon is a Dirac observable in the classical theory and the
application of the quantum geometry area formula to \emph{this}
surface leads to physical results. In 2+1 dimensions, point
particles have recently been incorporated and physical distance
between them is again a Dirac observable\cite{np}. When used in
this context, the spectrum of the length operator has direct
physical meaning. In all these situations, the operators and their
eigenvalues correspond to the `proper' lengths, areas an volumes
of physical objects, measured in the rest frames. Finally
sometimes questions are raised about compatibility between
discreteness of these eigenvalues and Lorentz invariance. There is
no tension whatsoever \cite{rovelli}: it suffices to recall that
discreteness of eigenvalues of the angular momentum operator
$\hat{J}_z$ of non-relativistic quantum mechanics is perfectly
compatible with the rotational invariance of that theory.

\b  \emph{Quantum Einstein's equations.} The challenge of quantum
dynamics in the full theory is to find solutions to the quantum
constraint equations and endow these physical states with the
structure of an appropriate Hilbert space. The general consensus
in the loop quantum gravity community is that while the situation
is well-understood for Gauss and diffeomorphism constraints, it is
very far from being definitive for the scalar (i.e., the
Hamiltonian) constraint. It \emph{is} non-trivial that
well-defined candidate operators representing the scalar
constraint exist on the space $\H_{\diff}$ of solutions to the
Gauss and diffeomorphism constraints. However as section
\ref{s6.3} shows there is a host of ambiguities and none of the
candidate operators has been shown to lead to a `sufficient number
of' semi-classical states in 3+1 dimensions. A second important
open issue is to find restrictions on matter fields and their
couplings to gravity for which this non-perturbative quantization
can be carried out to a satisfactory conclusion. In the
renormalization group approach, for example, the situation is as
follows. There is significant evidence for a non-trivial fixed
point for pure gravity in 4 dimensions \cite{lr} but when matter
sources are included it continues to exist only when the matter
content and couplings are suitably restricted. For scalar fields
in particular, Percacci and Perini \cite{pp} have found that
polynomial couplings (beyond the quadratic term in the action) are
ruled out, an intriguing result that may `explain' the triviality
of such theories in Minkowski space-times. Are there similar
constraints coming from loop quantum gravity?

To address these core issues, at least four different approaches
are being followed. The first, and the closest to ideas discussed
in section \ref{s6.3} is the `Master constraint program' recently
introduced by Thiemann \cite{tt7}. The idea here is to avoid using
an infinite number of constraints $\C(N)$, each smeared by a lapse
function $N$. Instead, one squares the integrand $C(x)$ itself in
an appropriate sense and then integrates it on the 3-manifold $M$.
In simple examples, this procedure leads to physically viable
quantum theories \cite{master}. In the gravitational case,
however, the procedure does not seem to remove any of the
ambiguities. Rather, its principal strength lies in its potential
to resolve the difficult issue of finding the physically
appropriate scalar product on physical states. The general
philosophy is similar to that advocated by John Klauder
\cite{klauder} over the years in his very interesting approach
based on coherent states. However, there are two key differences.
First, Klauder seeks solutions to constraints in the original,
kinematical Hilbert space rather than in a larger space such as
$\cyl^\star$. Consequently, when zero is in the continuous part of
the spectrum of constraint operators his physical states are only
approximately annihilated by the constraints. Second, in Klauder's
proposal \emph{all} constraints are to be imposed in this manner.
In loop quantum gravity, this does not seem to be feasible for
several important technical reasons; one needs to first solve the
Gauss and the diffeomorphism constraint and work on the Hilbert
space $\H_{\diff}$. Indeed, to our knowledge, the proposal has not
been implemented in sufficient detail to know if the original
strategy can be employed to solve the diffeomorphism constraint
rigorously, even by itself. But the program has a key advantage
that, since it is based on coherent states, the semi-classical
sector can be readily located. A cross-fertilization of this
program and loop quantum gravity is likely to be fruitful in the
analysis of low energy physics.

A second approach to quantum scalar constraint is due to Gambini,
Pullin and their collaborators \cite{bggp}. It builds on their
earlier extensive work \cite{gp1} on the interplay between quantum
gravity and knot theory. The more recent developments use the
relatively new invariants of \emph{intersecting} knots discovered
by Vassiliev. This is a novel approach which furthermore has a
potential of enhancing the relation between topological field
theories and quantum gravity. As our knowledge of invariants of
intersecting knots deepens, this approach is likely to provide
increasingly significant insights. In particular, it has the
potential of leading to a formulation of quantum gravity which
does not refer even to a background manifold (see footnote 5). The
third approach comes from spin-foam models \cite{ap1} discussed
briefly in section \ref{s9.2}. Here, amplitudes used in the path
integrals can be used to restrict the choice of the scalar
constraint operator in the canonical theory. This is a promising
direction and the detailed analysis of restrictions is already in
progress in 2+1 dimensions \cite{np}. In the fourth approach, also
due to Gambini and Pullin, one first constructs consistent
discrete theories at the classical level and then quantizes them
\cite{discrete}. In this program, there are no constraints; they
are solved to find lapse and shift fields. It has already been
applied successfully to gauge theories and certain cosmological
models. An added bonus here is that one can revive a certain
proposal made by Page and Wootters to address the difficult issues
of interpretation of quantum mechanics which become especially
acute in quantum cosmology, and more generally in the absence of a
background physical geometry.

\b  \emph{Applications.} As we saw in sections \ref{s7} and
\ref{s8}, loop quantum gravity has resolved some of the
long-standing physical problems of quantum gravity. As in other
approaches to black hole entropy \cite{gth,sc2,dgw,sc3}, concrete
progress could be made because the constructions do not require
detailed knowledge of how quantum dynamics is implemented in the
\emph{full} quantum theory. Recently, the first law of black hole
mechanics has been extended to fully dynamical situations
\cite{ak}. Its form suggests that the entropy is given by the area
of the dynamical horizon. Can the quantum entropy calculation be
extended to these non-equilibrium situations? This may even
provide an input to non-equilibrium statistical mechanics where
the notion of entropy is still rather poorly understood.

In quantum cosmology, there is ongoing work on obtaining
`effective field equations' which incorporate quantum corrections
\cite{mb8,bd,abw}. Quantum geometry effects significantly modify
the effective field equations which in turn leads to new physics
in the early universe. In particular, not only is the initial
singularity resolved but the (Belinski-Khalatnikov-Lifschitz type)
chaotic behavior predicted by classical general relativity and
supergravity also disappears! As explained \cite{bd}, this is to
be expected on rather general grounds if the underlying geometry
exhibits quantum discreteness because even in the classical theory
chaos disappears if the theory is truncated at any smallest,
non-zero volume. There are also less drastic but interesting
modifications of the inflationary scenario with potentially
observable consequences \cite{mb8,cobe}. While the technical steps
used in these analyses of effective equations are not as clean as
those of section \ref{s7}, it is encouraging that loop quantum
cosmology is already yielding some phenomenological results.

As explained in section \ref{s9.1}, a frontier area of research is
contact with low energy physics. Here, a number of fascinating
challenges appear to be within reach. Fock states have been
isolated in the polymer framework \cite{mv1,al7,als} and elements
of quantum field theory on quantum geometry have been introduced
\cite{st2}. These developments lead to concrete questions. For
example, in quantum field theory in flat space-times, the
Hamiltonian and other operators are regularized through normal
ordering. For quantum field theory on quantum geometry, on the
other hand, the Hamiltonians are expected to be manifestly finite
(see, e.g., Appendix \ref{a1}). Can one then show that, in a
suitable approximation, normal ordered operators in the Minkowski
continuum arise naturally from these finite operators? Can one
`explain' why Hadamard states of quantum field theory in curved
space-times are special? These issues also provide valuable hints
for construction of viable semi-classical states of quantum
geometry. The final and much more difficult challenge is to
`explain' why perturbative quantum general relativity fails if the
theory exists non-perturbatively. As explained in section
\ref{s1}, heuristically the failure can be traced back to the
insistence that the continuum space-time geometry is a good
approximation even below the Planck scale. But a more detailed
answer is needed. Is it because, as recent developments in
Euclidean quantum gravity indicate \cite{lr,pp}, the
renormalization group  a non-trivial fixed point?

Finally, there is the issue of unification. At a kinematical
level, there is already an unification because the quantum
configuration space of general relativity is the same as in gauge
theories which govern the strong and electro-weak interactions.
But the non-trivial issue is that of dynamics. We will conclude
with a speculation. One possibility is to use the `emergent
phenomena' scenario where new degrees of freedom or particles,
which were not in the initial Lagrangian, emerge when one
considers excitations of a non-trivial vacuum. For example, one
can begin with solids and arrive at phonons; start with
superfluids and find rotons; consider superconductors and discover
cooper pairs. In loop quantum gravity, the micro-state
representing Minkowski space-time will have a highly non-trivial
Planck scale structure. The basic entities are 1-dimensional and
polymer like. Even in absence of a detailed theory, one can tell
that the fluctuations of these 1-dimensional entities will
correspond not only to gravitons but also to other particles,
including a spin-1 particle, a scalar and an anti-symmetric
tensor. These `emergent states' are likely to play an important
role in Minkowskian physics derived from loop quantum gravity. A
detailed study of these excitations may well lead on to
interesting dynamics that includes not only gravity but also a
select family of non-gravitational fields.

\section*{Acknowledgments:}

Discussions with a large number of colleagues have added clarity
and precision to this review. Among them, we would especially like
to thank J. Baez, J. Barrett, M. Bojowald, D. Buchholz, A.
Corichi, A. Ghosh, S. Fairhurst, C. Fleischhack, K. Fredenhagen,
L. Freidel, R. Gambini, J. Hartle, G. Horowitz, J. Jain, T.
Jacobson, K. Krasnov, D. L\"ust, D. Maison, P. Majumdar, D.
Marolf, J. Mour\~ao, H. Nicolai, K. Noui, A.  Oko\l\'ow, R.
Penrose, A. Perez, J. Pullin, C. Rovelli, J. Samuel, H. Sahlmann,
D. Sudarsky, T. Thiemann, C. van den Broeck, M. Varadarajan, J.
Wisniewski, J. Willis, W. Unruh, R. Wald and J.A. Zapata. This
work was supported in part by the NSF grant PHY 0090091, the KBN
grant 2 P03B 12724, the Alexander von Humboldt Foundation and the
Eberly research funds of The Pennsylvania State University.

\vfill\break
\appendix

\section{Inclusion of Matter fields:\\ The Einstein-Maxwell theory}
\label{a1}

In section \ref{s5}, to bring out the main ideas we simplified the
discussion of dynamics by ignoring matter fields. Inclusion of
these fields does not require a major modification of the
underlying framework. In this appendix we will illustrate the
procedure using Einstein-Maxwell theory.

\subsection{Classical framework}
\label{a1.1}

The point of departure for canonical quantization is again a
Hamiltonian framework. One can easily repeat the procedure used in
section \ref{s2} by carrying out a 3+1 decomposition also of the
Maxwell action. The phase space now consists of 2 pairs of
canonically conjugate fields $(A,P)$ describing geometry and
$(\aM, \eM)$ describing the Maxwell field, where $\aM$ is our
Maxwell vector potential and $\eM$, our Maxwell electric field. As
usual the only non-vanishing Poisson bracket in the Maxwell sector
is
\be \{\aM_a(x),\eM^b(y)\}\ =\ \delta_a^b\, \delta(x,y).\ee
As in the geometrical sector, the basic configuration variables
will be taken to be holonomies $\aM(e) := \exp -i\int_e\, \aM$.
However, because the Maxwell gauge group $\U(1)$ is Abelian, it
turns out that the electric field $\eM^a$ can be smeared either
along 2-surfaces (as was done for the gravitational $P^a$ in
section \ref{s4.3.1}), or directly in three dimensions. It is more
convenient to use three dimensional smearing and set $\eM(g) :=
\int_M d^3x\, g_a(x) \eM^a(x)$ for all test 1-forms $g_a$ on $M$.
The Poisson bracket between these elementary variables is given
by:
\be \{{\aM(e)},\, {\eM(g)}\}\ =\ -i(\int_e g)\, \aM(e) \, .
\label{maxpb}\ee
Thus the Poisson algebra of elementary variables is closed as
needed.%
\footnote{The physical dimensions of the Maxwell variables are the
same as those of their gravitational analogs. Thus, $[\aM ] =
L^{-1}$ and $[\eM] = ML^{-1}$. The magnetic potential $\tilde\aM$
and electric field $\tilde{\bf E}$ of classical electrodynamics
are given by $\tilde\aM = {\bf e}\, \aM$ and $\tilde{\bf E} =
(1/{\bf e})\, \eM$.  In quantum electrodynamics, the holonomy is
generally written as $\exp (-i{\bf e}/\hbar)\int \aM^\prime$.
Therefore, the vector potential $\aM^\prime$ used there is given
by $\aM^\prime = (\hbar/{\bf e}) \aM$.}

As in section \ref{s2}, one can obtain the Hamiltonian through a
Legendre transform. As expected, the total Hamiltonian density
$h_{\rm EM}$ is a sum of constraints:
\be h_{\rm EM}\ =\ N(C+\CM)+ N^a(C_a+ \CM_{a})+ \omega_t^i\, G_i +
\aM_t\, \GM, \ee
where the lapse $N$ and shift $N^a$ are the same as in the
gravitational sector (see(\ref{h})); $\aM_t$ is a freely
specifiable function, the Lagrange multiplier for the Maxwell
Gauss constraint $\GM = D_a \eM^a$; and $\CM$ and $\CM_{a}$, are
functionals of $P,\aM,\eM$, representing the Maxwell energy and
momentum density, respectively. Specifically, the electromagnetic
contribution to the scalar constraint is
\be \label{maxham} {\C}^{\rm Max} (N) \ =\ \frac{1}{8\pi}\int
d^3x\, N(x)\, \frac{q_{ab}(x)}{\sqrt{\det q(x)}}\, {\Big(
}\eM^a(x) \eM^b(x) + \bM^a(x) \bM^b(x) { \Big) }, \ee
and
\be {\bM}^{a}\ =\
\frac{1}{2}\eta^{abc}(\partial_a\aM_b-\partial_b\aM_a)
 \ee
is the magnetic vector density. In comparison with the Hamiltonian
of the Maxwell field in Minkowski space-time, the presence of the
inverse square root of $q$ may seem surprising. Note however that
the electric and magnetic fields naturally carry density weight
one, whence this factor is quite essential. In Minkowski space the
background metric is implicitly used to remove the density weight.

\subsection{Quantum kinematics}
\label{a1.2}

One can just use the procedure of section \ref{s4.3} to carry out
quantization using $G = \SU(2)\times\U(1)$, where $\SU(2)$ refers
to geometry and $\U(1)$ to the Maxwell field. The Kinematical
Hilbert space of the Einstein-Maxwell theory is given by:
\be \H_{\rm EM}\ =\ \H\otimes \H_{\rm Max}\ee
where $\H$ is the Hilbert space of states of the quantum geometry
of section \ref{s4.3.2} and $\H_{\rm Max}$ is the corresponding
Hilbert space for the case $G = \U(1)$.

Since we discussed the structure of $\H$ and of the operators
thereon in detail in sections \ref{s4.3} and \ref{s5}, let us
focus just on the Maxwell sector. Convenient orthonormal basis
states $F_{\a, \vec{n}}$, called \emph{flux networks}, in $\H_{\rm
Max}$ can be constructed as follows. Given a graph $\a$, assign an
orientation to the edges $(e_1, \ldots e_N)$, label them by
integers $(n_1, \ldots n_N)$ and set
\be F_{\a,\vec{n}}(\aM)\ =\ [{\aM}(e_1)]^{n_1} \ldots
[{\aM}(e_n)]^{n_N} \ee
Note that if the orientation of an edge $e_I$ is reversed, the
state is unchanged if $n_I$ is replaced by $-n_I$.

The Poisson bracket relation (\ref{maxpb}) leads to the definition
of the smeared electric operator $\hat\eM(g)$:
\be \heM(g)\Psi\ =\ i\hbar\, \{\eM(g),\Psi\}\ee
capturing the expectation that $\hat\eM(x)$ should be represented
by $i \hbar \delta/\delta {\aM(x)}$. On the flux network states,
the action reduces to:
\be \label{elec} \heM (g) F_{\g,\vec{n}}\ =\
-\hbar\,\left(\sum_{I}n_I\int_{e_I}g \right)\,F_{\g,\vec{n}}.\ee
If the support of $g$ has non-trivial intersection just with a
single edge $e_I$ of $\a$, then the flux network $F_{\a, \vec{n}}$
is an eigenstate of $\heM (g)$ and the eigenvalue just measures
$n_I$, the `electric flux carried by the oriented edge $e_I$'.
Thus the electric flux is quantized and the each edge of the flux
network $F_{\a, {\vec n}}$ can be thought of as carrying an
integral multiple of the fundamental quantum.

\subsection{The quantum  constraints}
\label{a1.3}

As noted in section \ref{a1.1}, the Einstein-Maxwell theory again
has a set of three first class constraints. The action of the
Gauss constraint for the group $\SU(2)\times \U(1)$ naturally
factors on $\H_{\rm EM} = \H \times \H_{\rm Max}$:
$\hat{\C}_{G}^{\rm EM} = \hat{\C}_G \otimes \hat{\C}_G^{\rm Max}$,
where $\hat{\C}_G^{\rm Max}$ is the Gauss constraint operator on
the quantum geometry Hilbert space $\H$ and $\hat{\GM}$ that on
the Maxwell Hilbert space. Imposition of this constraint selects
the \emph{gauge invariant sub-space} of $\H_{\rm EM}$. The gauge
invariant subspace of $\H$ was obtained in section \ref{s6.1}. On
the Maxwell Hilbert space $\H_{\rm Max}$, the constraint simply
restricts the flux network states as follows: at each vertex the
sum of the labels $n_I$ assigned to the incoming edges is equal to
the sum of the labels assigned to the outgoing edges. Note that
the solution space is a \emph{sub-space} on $\H_{\rm EM}$.

The diffeomorphism constraint $\int d^3x\, N^a(C_a+ \CM_{a})$ is
also straightforward to impose in the exponentiated version. The
general procedure is the same as that of section \ref{s6.2}.
Again, the solutions lie in the dual $\Cyl^\star_{\rm EM}$ of
$\Cyl_{\rm EM} = [\cyl\otimes \cyl_{\rm Max}]$ where $\cyl_{\rm
Max}$ is the space of the cylindrical functions of $\U(1)$
connections.

Finally, we have to impose the scalar constraint. Regularization
of the Einstein part $\hat{\C}(N)$ of the constraint was discussed
in detail in section \ref{s6.3}. Here we will focus just on the
Maxwell part $\hat{\C}^{\rm Max}(N)$. We have organized the
discussion so that it will serve a dual purpose. On the one hand,
it will provide us the Maxwell part of the total Hamiltonian
constraint that must be imposed to select the physical states of
the Einstein-Maxwell theory. For this purpose, we will construct
an operator which is well-defined on the (gauge and)
diffeomorphism invariant sector $(\cyl^\star_{\rm EM})_{\rm diff}$
of $\cyl^\star_{\rm EM}$. On the other hand, in the framework of
field theory in a given classical space-time, $\C^{\rm Max}(N)$
can also be regarded as the \emph{physical Hamiltonian} of the
Maxwell field. Therefore, it is natural to ask if one can
construct from $\C^{\rm Max}(N)$ a well-defined operator which
will act on $(\Psi|_{\rm geo} \otimes \mathcal{F}_{\rm Max}$,
where $(\Psi|_{\rm geo}$ is a given semi-classical state of
quantum geometry and $\mathcal{F}_{\rm Max}$ the Fock-space of
photons on this geometry. We will show that this is also possible.
The result will be a Hamiltonian governing the dynamics of a test
quantum Maxwell field on a fixed, semi-classical quantum geometry.

\subsubsection{Regularization of the 3-geometry part in
$\C^{\rm Max}(N)$}
\label{a1.3.1}

 In contrast to the Maxwell parts $\C_G^{\rm Max}$ and $\C_{\Diff}$
of the Gauss and the vector constraints, the Maxwell part $\C^{\rm
Max}(N)$ of the scalar constraint contains a coefficient
$q_{ab}/\sqrt{q}$ that explicitly depends on geometry.  We will
first `regularize' this term, i.e. express it using  variables
which have direct operator analogs on the quantum geometry Hilbert
space. This discussion will bring out the role played by quantum
geometry in regulating the quantum matter Hamiltonians. For
simplicity, we will work with just the electric term; by
inspection all equations of this sub-section continue to hold if
the electric fields are replaced by magnetic.

Consider then the term
\be \label{E2} \int_M d^3x\, N(x) \,  \frac{q_{ab}(x)}{\sqrt{\det
q(x)}}\, \eM^a(x)\eM^b(x). \ee
We can express the metric $q_{ab}$ using (a slight generalization
of) the expression (\ref{e}) for the orthonormal co-frame $e^i_a$,
\be e^i_a(x)\ =\ \frac{2}{k\g}\{A^i_a(x),\, V_{\cal R}\} \ee
where ${\cal R}$ is an arbitrary open neighborhood of $x$, and
$V_{\cal R}$ is its volume with respect to $q_{ab}$. The Poisson
bracket is independent of ${\cal R}$ and for our regularization
purposes, it is convenient to choose it to be the ball ${\cal
R}_\epsilon$ of coordinate volume $\epsilon^3$, centered at $x$.
Denote the geometric volume of this ball (with respect to
$q_{ab}$) by $V(x,\epsilon)$. Approximating $\sqrt{\det q}$ by
${V(x,\epsilon)}/{\epsilon^3}$ it is easy to verify:
\be \label{geom} \frac{q_{ab}(x)}{\sqrt{\det q(x)}}\ =\
\frac{16}{k^2\g^2}\, \lim_{\epsilon\rightarrow 0}\epsilon^3\,\,
\{A^i_a(x),\sqrt{V(x,\epsilon)}\}
\{A^i_b(x),\sqrt{V(x,\epsilon)}\}\ee

Now, in the quantum Maxwell theory in Minkowski space-time, the
electric field becomes an operator valued distribution whence  the
product of electric fields at the same point, such as the one in
(\ref{E2}), is ill-defined. Therefore, with an eye towards
quantization, let us point-split the product by introducing a
two-point smearing function $\chi_{\epsilon}(x,y)$:
\be \chi_\epsilon(x,y)\ =\ \cases{1, &if $y\in {{\cal
R}_\epsilon}$\cr 0, &otherwise}.\ee
Then, we obtain:
\ba  \label{Heps} \int d^3x\, N(x)\, \frac{q_{ab}}{\sqrt{\det q}}
\eM^a(x) \eM^b(x)\, =  \frac{16}{k^2\g^2}&&\,
\lim_{\epsilon\rightarrow 0} \,\, \int_M d^3x N(x) \,\int_M d^3y\,
\eM^a(x)\eM^b(y)\, \times
\nonumber\\
&& \chi_\epsilon(x,y)\, \{A^i_a(x),\sqrt{V(x,\epsilon)}\}
\{A^i_a(y),\sqrt{V(y,\epsilon)}\}\,\, := \lim_{\epsilon\rightarrow
0}\, \hat{\C}^\epsilon_{\rm elec} \ea
Note that the point splitting procedure requires us to set
$\eM^a(x) = (1/\epsilon^3)\, \int_{\cal R_\epsilon} d^3y\, \eM^a
(y)$ but the factor $1/\epsilon^3$ in the denominator is cancelled
by the factor $\epsilon^3$ in the geometric term (\ref{geom}). We
will see that, thanks to point splitting, this regulated classical
version has a well-defined operator analog in the quantum theory.
Had we worked in Minkowski space-time, the geometric term
$q_{ab}/\sqrt{q}$ would simply be a smooth field on $M$. Then, the
$1/\epsilon^3$ factor required in the point-splitting procedure
would have remained and led to a divergence in the limit as
$\epsilon \mapsto 0$, i.e., when the regulator is removed. This
divergence is now avoided because the quantum geometry operator
corresponding to $\{A_a^i(x),\, \sqrt{V(x,\epsilon)}\}$ has a
well-defined limit as $\epsilon$ tends to infinity. \emph{In this
precise sense, the quantum nature of geometry provides a natural
regulator for matter Hamiltonians} \cite{ttrev}.

\subsubsection{Quantization of the electric part of
$\C^{\rm Max}(N)$}
\label{a1.3.2}

We now wish to find the quantum analog of the expression on the
right side of (\ref{Heps}). Following a strategy introduced by
Thiemann \cite{tt6}, we will proceed in two steps. In the first,
we replace the classical electric field by the corresponding
operators and in the second we do the same for geometric fields.%
\footnote{It is also possible to quantize simultaneously the
electric and geometric fields in the constraint. However then,
subtleties arise in the choice of the holonomies replacing the
$\SU(2)$ connection $A$.}

Consider a flux network state $F_{\g,\vec{n}}$ in the Maxwell
Hilbert space. Then, using the action (\ref{elec}) of smeared
electric field operators, we immediately obtain the action of the
regulated operator $\hat{\C}^\epsilon_{\rm elec}(N)$ (see Eq.
(\ref{maxham} ):
\be \hat{\C}^\epsilon_{\rm elec}(N)\, F_{\g,{\vec n}}\ =\,
{\CM}^\epsilon_{\rm elec}\, F_{\g,{\vec n}} \label{eigen}\ee
where the eigenvalue $\CM^\epsilon_{\rm elec}$ is given by
\ba {\CM}^\epsilon_{\rm elec}(N) \, = \, \frac{2\hbar^2}{k^2\g^2
\pi}\, &&\int_{e_I} dt \int_{e_J} dt'\,\, \sum_{I,J}\, N(e_I(t))\,
\chi_\epsilon(e_I(t),e_J(t')) n_I n_J \, \times \nonumber\\
&&\{A_a^i(e_I(t))\dot{e}_I^a(t),\, \sqrt{V(e_I(t),\epsilon)}\}\,
\{A_b^i(e_J(t'))\dot{e}_J^b(t'),\, \sqrt{V(e_J(t'),\epsilon)}
 \}\,.
 \ea
The second step is now facilitated because the gravitational
 connection $A$ appears only through its along edges of the graph.
By dividing the edges into segments of coordinate length
$\epsilon'$, replacing the integrals by sums of holonomies and
taking the limit $\epsilon^\prime \mapsto 0$ first, followed by
the limit $\epsilon \mapsto 0$, one obtains the action of the
electric part of $\C^{\rm Max}(N)$ on a state $\Psi_\g \otimes
F_{\g, {\vec n}}$ in the full Hilbert space $\H_{\rm EM}$:
\ba  \hat{\C}_{\rm elec}(N)\, [\Psi_\g \otimes F_{\g,\vec{n}}]\ =\
\frac{4}{k^2\g^2 \pi}&& \, \sum_v N(v)
\sum_{I,J}n_In_J\, \, \times \nonumber\\
&&\Tr\Big(\hA(e^{-1}_I)[\hA(e_I),\,\sqrt[4]{\hq_v}]\,
\hA(e_J^{-1})[\hA(e_J), \sqrt[4]{\hq_v}]\Big)\, [\Psi_\g \otimes
F_{\g,\vec{n}}] \label{elecop}\ea
where $v$ ranges over the vertices of the graph $\g$ and, given
$v$, $I,J$ run over the labels of the edges intersecting $v$. Note
that this action preserves each sub-space $(\H_{\rm EM})_{\g}$ of
$\H_{\rm EM}$; it does \emph{not} require us to extend the graph
$\g$.  Finally, it is straightforward to check that $\hat{\C}_{\rm
elec}$ admits a self-adjoint extension to $\H_{\rm EM}$.

\subsubsection{Quantization of the magnetic part}
\label{a1.3.3}

The starting point is again the expression (\ref{Heps}), but with
$\eM^a$ replacing $\bM^a$. We need to define operator analogs of
the magnetic field. The strategy is the same as that used in
section \ref{s6.3} for curvature $F_{ab}$ of the $\SU(2)$
connection of quantum geometry: approximate the dual ${\bf
F}_{ab}$ of the magnetic field $\bM^a$ by holonomies around small
closed loops. For this purpose, as in section \ref{s6.3.1}, we
again cover $M$ with cells $\Box$ (possibly with arbitrary shape)
and, in every cell, introduce edges $s_\Box^I$ and loops
$\alpha_\Box^{IJ}$. Let us label this structure by $\T$. The idea
now is to replace the double integral in the left hand side of
(\ref{Heps}) by a generalized Riemann sum in which the
gravitational connection $A^i_a$ are approximated by holonomies
along the edges $s_\Box^I$ and  the magnetic field is approximated
by the holonomies along $\alpha_\Box^{IJ}$. We are then led to
define the approximate expression of the magnetic part of $\C^{\rm
max}(N)$ as:
\ba \C_{\rm mag}^{(\epsilon,\T)}(N)\ :=\ T_{IJI'J'}
\sum_{\Box,\Box'} N(x_{\Box})\, \chi_{\epsilon}(x_\Box,x_{\Box'})
&& \, \Tr\Big( \aM(\alpha_\Box^{I})\, A((s_\Box^{J})^{-1})
\{A(s_\triangle^{J}),\, \sqrt{V_\R} \} \times\nonumber\\
&& \aM(\alpha_{\Box'}^{I'})\, A(({s}_{\Box'}^{J'})^{-1})\,
\{A(s_{\Box'}^{J'}),\, \sqrt{V_\R'} \} \Big)\label{Hb} \ea
which converges to the magnetic part of $\C^{\rm Max}(N)$ as we
shrink $\T$ and take $\epsilon$ to zero. (Here $\T^{IJI'J'}$ are
constants determined by the geometry of $\T$.) It is now
straightforward to pass to the regulated operator $\hCM_{\rm
mag}^{(\epsilon,\T)}(N)$. While this operator is well defined in
$\cyl_{\rm EM}$, as in the gravitational case, its limit as
$\epsilon \mapsto 0$ fails to be well-defined on $\H_{\rm EM}$.
Therefore, to define the constraint operator, as in section
\ref{s6.3.2}, we pass to $(\cyl^\star_{\rm EM})_{\rm diff}$. Given
$(\Psi|\in(\cyl^\star_{\rm EM})_{\rm diff}$, the procedure used in
section \ref{s6.3.2} leads to the following well-defined action:
\ba [(\Psi|\hat{C}_{\rm mag}(N)] && |\Psi_\g F_{\g,{\vec
n}}\rangle\, =\,
-\hbar^{-2}\, T^{IJI'J'}\, (\Psi|\, \sum_v N(v) \times \nonumber\\
&& \Tr \Big( \aM(\alpha_v^{I}) \hA((s_v^J)^{-1}) \,
[\hA(s_v^J),\sqrt[4]{\hq_v}]\,\,
\aM(\alpha_v^{I'})\hA((s_v^{J'})^{-1})\, [\hA(s_v^{J'}),\,
\sqrt[4]{\hq_{v'}}] \Big)\, |\Psi_\g\rangle \label{magop} \ea
where for every vertex $v$ of the graph $\g$,  $s_v^I$ and
$\alpha_v^I$ are the edges and loops of $\T$ originating at $v$.
Recall from section \ref{s6.3} that the geometric structure of
$\T$ is such the edges $s_v^I$ themselves do not appear int he
final result; the graph changes only through the loops
$\alpha_v^I$. But the geometric operators that appear on the right
hand side do not refer to the loops $\alpha_v^I$. Therefore, while
the action of $\hat{\C}_{\rm mag}(N)$ does add new edges to the
graph, the spin labels of these edges vanish; only the Maxwell
flux quantum numbers ${\vec n}$ are non-trivial on these edges.

\subsubsection{summary}

Collecting the results of the last two sections, solutions to the
quantum scalar constraint are elements $(\Psi|$ of
$(\cyl^\star_{\rm EM})_{\rm diff}$ satisfying
$$(\Psi| [\hat{\C}(N) + \hat{\C}_{\rm ele}(N) + \hat{\C}_{\rm mag}(N)] =0 $$
where the action of the geometrical part $\hat{\C}(N)$ is as in
section \ref{s6.3} and the electric and magnetic Maxwell operators
are given by (\ref{elecop}) and (\ref{magop}).

Finally, as mentioned in the beginning of section \ref{a1.3},
$\C^{\rm Max}(N)$ is also the Hamiltonian of the Maxwell field
propagating on a fixed, static background (where the 4-metric is
determined completely by the 3-metric $q_{ab}$ and the lapse $N$.)
Can we use this operator to `derive', \emph{in a suitable
approximation}, the quantum theory of Maxwell fields on static
space-times? Let us use for $(\Psi|$ the tensor product
$(\Psi|_{\rm geo}\otimes (\Psi|_{\rm Max}$ where $(\Psi|_{\rm
geo}$ is a quantum geometry state peaked at a static space-time
and $(\Psi|_{\rm Max}$ is (the image in $\cyl^\star_{\rm Max}$ of)
a Fock-state of photons associated with the static background.
Note that these states are \emph{not} diffeomorphism invariant.
However, we can exploit the availability of a background metric
and use in place of $\hat{\C}_{\rm mag}(N)$ the regulated operator
$\hCM^{(\epsilon, \T)}(N)$, where the area of the loops
$\alpha^I_\Box$ is given by the minimum non-zero eigenvalue of the
area operator. The resulting Maxwell Hamiltonian $\CM (N)$ has a
well-defined action on $\cyl^\star_{EM}$. Therefore, we can
analyze the evolution of the resulting state and compare it with
the standard evolution in the Fock space. An important viability
criterion for this strategy to work is that the geometry part of
the state does not change appreciably under the action of $\C^{\rm
Max}(N)$. To analyze whether this condition is met, we can expand
$(\Psi|_{\rm geo}$ in terms of spin-network states
$(s_{\g,\vec{j}, \vec{I}}|$. It is easy to check that the action
of the geometric operators in $\hCM(N)$ leaves $\g$ and the spin
labels $\vec{j}$ of \emph{each} of these spin-network component
invariant; only the intertwiners change. Therefore, it is
plausible that $(\Psi|_{\rm geo}$ does not change appreciably.
However, so far it is not obvious that any of the candidate
semi-classical states proposed to date satisfy this condition;
this issue is being investigated. Thus, considerations involving
matter fields provide  detailed, quantitative criteria for
viability of candidate semi-classical states of quantum geometry.

\section{List of Symbols}
\label{a2}

\begin{tabular}{ll}

$a,b, ...$  & spatial indices for tensor fields on the 3-manifold $M$\\
$\alpha,\beta, ...$ &  space-time indices in sections \ref{s2} and \ref{s8.1}\\
$\alpha,\beta, ...$ &  labels for graphs on $M$ in rest of the sections\\
$A_a^i$     & a connection 1-form on $M$\\
$A(e)$      & holonomy along an edge $e$ defined by a connection $A$\\
$\A$        & space of smooth connections on $M$ for a given gauge group $G$\\
$\bA$       & a generalized connection\\
$\bA(e)$    & holonomy along an edge $e$ defined by a generalized connection $\bA$\\
$\hA(e)$    & ---corresponding quantum operator\\
$\Ab$       & quantum configuration space (of generalized connections)\\
$A_S$       & area of a 2-surface (without boundary) $S$ \\
$\hA_S$     & ---corresponding quantum  operator\\
$\aM_a$     & Maxwell vector potential \\
$\aM(e)$    & ---corresponding holonomy along an edge $e$\\
$\bM^a$     & Maxwell magnetic (vector density) field \\
$\Comp$     & the set of complex numbers\\
$\C_{\rm Diff}(\vec{N})$ &  diffeomorphism constraint smeared with ${N^a}$\\
$\hat{\C}_{\rm Diff}(\vec{N})$ & ---corresponding quantum operator\\
$\C(N)$     & scalar constraint smeared with $N$\\
$\hat{\C}(N)$ & ---corresponding quantum operator\\
$\Gauss(\Lambda)$ &  Gauss constraint smeared with $\Lambda^i$\\
$\hat{\Gauss}(\Lambda)$ & ---corresponding quantum operator\\
$C^{(n)}$   & a differentiability class\\
$\cyl$      & algebra of cylindrical functions on $\A$\\
$\cyl_\alpha$ &  algebra of the cylindrical functions defined by a graph $\alpha$\\
$\cyl^\star$& space of linear functionals on $\Cyl$\\
$\cyl^\star_\diff$ &  the image of $\cyl$ under the diffeomorphism averaging map\\
$\Diff$     & group of certain diffeomorphisms of $M$ (defined in Section \ref{s4.3.5})\\
$e$         & a closed-piecewise analytic edge (defined in section \ref{s4.3.1})\\
$E_i^a$     & triads with density weight one, defining the Riemannian geometry on $\M$\\
$\epsilon^{i}_{jk}$ &  structure constants of $\su(2)$ (of a general $\ggot$ in section \ref{s4})\\
$\eta_{ij}$ & the Killing form on $\su(2)$ (on a general Lie algebra $\ggot$ in section \ref{s4})\\
$\eta^{abc}$& metric independent, totally skew pseudo tensor density of weight 1 on $\M$\\
$\eta_{abc}$& metric independent, totally skew pseudo tensor density of weight -1 on $\M$\\
$\eta$      & diffeomorphism averaging map (defined in section \ref{s6.2})\\
$F^i_{ab}$  &  curvature of $A^i_a$\\
$G$         &  a compact Lie group\\
$\ggot$     &  ---its Lie algebra\\
$G$         &  Newton's constant\\
$\gamma$    &  Barbero-Immirzi parameter\\

\end{tabular}
\vfill\break

\begin{tabular}{ll}

$\H$        &  kinematical Hilbert space of quantum geometry\\
$\H_\alpha$ &  subspace of $\H$ defined by cylindrical functions compatible with graph $\alpha$ \\
$\H'_\alpha$&  subspace of $\H_\alpha$ used in the spin-network decomposition of $\H$\\
$i,j, ...$  & internal indices for $\so(3) = \su(2)$  (in section 4, for a general $\ggot$)\\
$I,J, ...$  & 4-dimensional internal indices in section \ref{s2} \\
$I,J, ...$  & labels (e.g. for edges, punctures, etc) in sections \ref{s4}-\ref{s9}\\
$\I_E$      & map from the space of connections on a graph with $n$ edges into $G^n$\\
$\I_V$      & map from the space of gauge transformations on a graph with $m$ vertices into $G^m$\\
$\hJ_i^{(v,e)}$ &  operator on $\cyl_\alpha$ associated to an edge $e$ and a  vertex $v$ of $\alpha$\\
$k$         &  $8\pi$ times Newton's constant \\
$\kappa$    &  surface gravity of isolated horizons\\
$\kappa(S,e)$& a constant ($0, \pm -1$) assigned to a surface $S$ an edge $e$ intersecting it \\
$\lp$       & Planck length\\
$L^2$       & space of square integrable functions\\
$\M$        &  3-dimensional (`spatial') manifold (generally assumed to be compact)\\
$\man$      &  4-dimensional space-time manifold\\
$\Nat$      &  the set of natural numbers\\
$P^i_a$     &  momentum canonically conjugate to $A^i_a$\\
$P(S,f)$    &  flux across a two surface $S$ of $P^a_i$ smeared with a test field $f^i$\\
$\hP(S,f)$  &  quantum operator corresponding to $P(S,f)$\\
$\eM^a$     &  Momentum conjugate to the Maxwell connection $\aM_a$\\
$\eM(g)$    &  Maxwell momentum smeared against a test field $g_a$\\
$\heM(g)$   &  ---corresponding quantum operator\\
$q_{ab}$    &   positive definite metric on $M$\\
$\hq_{v, \a}$&  the quantum operator representing determinant of $q_{ab}(v)$,restricted to  $\cyl_\alpha$\\
$\Real$     &  the set of real numbers\\
$S$         & A closed-piecewise analytic sub-manifold of $M$ (defined in section \ref{s4.3.1})\\
$\Sigma^i_{ab}$& Hodge-dual of the gravitational momentum $P^a_i$ ($\Sigma_{ab}^i = \eta_{abc}\eta^{ij} E^c_j$)\\
$\Tr$       &  trace\\
$V_{\R}$    &  the volume of a region $\R$ defined by $q_{ab}$\\
$\hV_{\R}$  &  ---corresponding quantum volume  \\

\end{tabular}


\begin{thebibliography}{99}

\bibitem[ ]{books}\textsl{Books and recent reviews:}
\bibitem{creutz} Creutz M 1983 Quarks, gluons and lattices
(Cambridge UP, Cambridge)
\bibitem{aa1} Ashtekar A 1991 \textit{Lectures on non-perturbative
canonical gravity} Notes prepared in collaboration with R. S. Tate
(World Scientific, Singapore)
\bibitem{jbbook} Baez J and Muniain J P 1994 \textit{Gauge fields,
knots and gravity} (World Scientific, Singapore)
\bibitem{gp1} Gambini R and Pullin J 1996 \textit{Loops, knots, gauge
theories and quantum gravity}  (Cambridge UP, Cambridge)
\bibitem{sc1} Carlip  S 1998 \textit{Quantum gravity in 2+1 dimensions}
(Cambrige UP, Cambridge)
\bibitem{crrev} Rovelli C 1998 Loop quantum gravity
\textit{Living Rev. Rel} \textbf{1} 1
\bibitem{mb:book} Bojowald M. 2001 \textit{Quantum Geometry and Symmetry}
(Saker-Verlag, Aachen)
\bibitem{aarev} Ashtekar A 2002 Quantum geometry and gravity: Recent
Advances \textit{General Relativity and Gravitation}  ed  Bishop N
T and  Maharaj S D (World Scientific, Singalore),
\texttt{gr-qc/0112038}
\bibitem{ttrev} Thiemann T 2002 Lectures on loop quantum
gravity, to appear in \emph{Aspects of quantum gravity: {}From
theory to experimental search}  Lecture Notes in Physics
(Springer-Verlag, Berlin 2003), \texttt{gr-qc/0210094}\\
 2004 \emph{Modern canonical quantum general relativity}, (Cambridge
University Press, Cambridge, in press), \texttt{gr-qc/0110034}
\bibitem{ap1} Perez  A 2003 Spin foam models for quantum
gravity  \textit{Class. Quant. Grav.} \textbf{20} R43--R104
\bibitem{crbook} C. Rovelli 2004 \emph{Quantum Gravity} (Cambridge
University Press, Cambridge, in press)\\

\bibitem[ ]{class} \textsl{Classical theory:}
\bibitem{aa2} Ashtekar  A 1986  New variables for classical and quantum
gravity  \textit{Phys. Rev. Lett.} \textbf{57}  2244--2247
\bibitem{aa3} Ashtekar A (1987) New Hamiltonian formulation of
general relativity  \textit{Phys. Rev.} \textbf{D36}  1587--1602
\bibitem{lagrangian} Samuel J 1987 A Lagrangian basis for
Ashtekar's reformulation of canonical gravity \textit{Pramana-J.
Phys.} 24 \textbf{28} {L429-L432}\\
Jacobson T and Smolin L 1987 The left handed spin connection as a
variable for canonical greavity \textit{Phys. Lett.} \textbf{B196}
39-42
\bibitem{brst} Ashtekar A, Mazur P and Torre C G 1987 BRST
structure of general relativity in terms of new variables
\textit{Phys. Rev} \textbf{D36} 2955-2962
\bibitem{art} Ashtekar  A,  Romano  J D and Tate  R S 1989 New variables
for gravity: Inclusion of matter  \textit{Phys. Rev.} \textbf{D40}
2572--2587
\bibitem{sugra} Jacobson T 1988 New variables for canonical
supergravity, \textit{Class. Quant. Grav.} \textbf{5} 923--935\\
Matschull H J and Nicolai H  1994 Canonical Quantum Supergravity
in three-dimensions \textit{Nucl. Phys.} \textbf{B411} 609--646
\bibitem{abj} Ashtekar A, Balachandran A P and Jo S 1989 The CP problem
in quantum gravity, \textit{Int. J. Mod. Phys.} \textbf{A4}
1493---1514
\bibitem{jp} Pleba\'nski J F 1977
On the separation of Einsteinian substructures \textit{J. Math.
Phys.} \textbf{18} 2511
\bibitem{cdjm} Capovilla R, Dell J and Jacobson T 1989 General
relativity without a metric  \textit{Phys. Rev. Lett.} \textbf{63}
2325--2328\\
Capovilla R, Dell J, Jacobson T and Mason L 1991 Selfdual two
forms and gravity \textit{Class. Quant. Grav.} \textbf{8} 41--57
\bibitem{holst}Holst S 1996 Barbero's Hamiltonian derived from a
generalized Hilbert-Palatini action \textit{Phys. Rev.}
\textbf{D53} 5966-5969\\
Wisniewski J 2001 Symplectic structure from Holst action,
pre-print
\bibitem{js} Samuel J 2000 Is Barbero's Hamiltonian formulation a
gauge theory of Lorentzian gravity? \textit{ Class. Quant. Grav.}
\textbf{17} L141-L148
\bibitem{dr} Robinson D C 1995 A Lagrangian formalism for the
Einstein-Yang-Mills equations \textit{J. Math. Phys.} \textbf{36}
3733-42\\
Lewandowski J, Okol'ow A 2000 2-Form Gravity of the Lorentzian
Signature \textit{ Class. Quant. Grav.} \textbf{17} L47-L51
\bibitem{Alex} Alexandrov S,  Livine E R 2003
SU(2) Loop Quantum Gravity seen from Covariant Theory
\textit{Phys. Rev.} \textbf{D67} 044009
\bibitem{fk} Freidel L and Krasnov K 1999  BF description of higher
dimensional gravity theories \textit{Adv. Theor. Math. Phys.}
\textbf{3} 1289--1324
\bibitem{deg1} Jacobson T, Romano J D 1992 Degenerate Extensions Of General
Relativity \textit{Class. Quant. Grav.}  \textbf{9}  L119--L124
\bibitem{deg2} Matschull H J 1996 Causal structure and diffeomorphisms in
Ashtekar's gravity \textit{Class. Quant. Grav.} \textbf{13}
765--782
\bibitem{deg3} Jacobson T 1996 (1+1) sector of (3+1) gravity \textit{Class.
Quant. Grav.} \textbf{13} L111--L116, Erratum-ibid. 1996
\textit{Class. Quant. Grav.} \textbf{13} 3269
\bibitem{deg4} Lewandowski J and Wisniewski J 1997 (2+1) sector of (3+1) gravity
\textit{Class. Quant. Grav.} \textbf{14} 775--782
\bibitem{deg5} Lewandowski J and Wisniewski J 1999 Degenerate sectors of the
Ashtekar gravity \textit{Class. Quant. Grav.} \textbf{16}
3057--3069
\bibitem{nlg} Penrose R 1976 Non-linear graviton and curved twistor theory
\textit{Gen. Rel.\& Grav.} \textbf{7} 31-52
\bibitem{hspace} Ko M, Ludvigsen M, Newman E T and Tod P 1981 The theory of
$\H$ space \textit{Phys. Rep.}  \textbf{71} 51--139\\

\bibitem[ ]{loop} \textsl{Connections and loops:}
\bibitem{rs1} Rovelli  C and Smolin  L 1988 Knot theory and quantum
gravity \textit{Phys. Rev. Lett.} \textbf{61} 1155-1158
\bibitem{rs2} Rovelli C and Smolin L 1990 Loop representation for quantum
general relativity  \textit{Nucl. Phys.} \textbf{B331} 80--152
\bibitem{ahrss} Ashtekar  A, Husain  V, Rovelli  C, Samuel J, and Smolin L
1989 2+1 quantum gravity as a toy model for the 3+1 theory
\textit{Class. Quant. Grav.} \textbf{6}  L185--L193
\bibitem{ars1} Ashtekar  A,  Rovelli C, and Smolin  L 1991 Gravitons and
loops  \textit{Phys. Rev.} \textbf{D44}  1740--1755
\bibitem{ls1} Smolin L 1992  Recent developments in
non-perturbative quantum gravity \textit{Quantum Gravity and
Csomology}  ed Mercader J,  Sol\'a J and  Verdaguer R (World
Scientific, Singapore)
\bibitem{gls} Goldberg J N, Lewandowski J,  Stornaiolo C 1992
Degeneracy in loop variables \textit{Commun. Math. Phys.}
\textbf{148} 377--402\\

\bibitem[ ]{qg1} \textsl{Background independent quantization of theories of
connections:}
\bibitem{ai} Ashtekar  A and Isham  C J 1992 Representation of the
holonomy algebras of gravity and non-Abelian gauge theories
\textit{Class. Quant. Grav.} \textbf{9}  1433--1467
\bibitem{al2} Ashtekar  A and Lewandowski  J 1994 Representation
theory of analytic holonomy algebras, in \textit{Knots and Quantum
Gravity}  ed Baez J C (Oxford U.\ Press, Oxford)
\bibitem{jb1} Baez  J C  1994  Generalized measures in gauge theory
\textit{Lett. Math. Phys.} \textbf{31}  213--223
\bibitem{jl-sn} Lewandowski J 1994 Topological Measure and
Graph-Differential Geometry on the Quotient Space of Connections
\textit{Int. J. Mod. Phys.} \textbf{D3}  207--210
\bibitem{charap} Ashtekar A, Lewandowski L, Marolf D,
Mour\~ao J and Thiemann T 1995 A manifestly gauge invariant
approach to quantum gauge theories, in \textit{Geometry of
Constrained Dynamical Systems}  ed Charap J M (Cambridge U.\
Press, Cambridge)60--72
\bibitem{mm} Marolf D and Mour\~ao J 1995
On the support of the Ashtekar-Lewandowski measure \textit{Commun.
Math. Phys.} \textbf{170}  583-606
\bibitem{al3} Ashtekar A and Lewandowski  J 1995
Projective techniques and functional integration  \textit{Jour.
Math. Phys.} \textbf{36}  2170--2191
\bibitem{cohsttr} Ashtekar A, Lewandowski L, Marolf D,
Mour\~ao J and Thiemann T 1996 Coherent State Transforms for
Spaces of Connections \textit{Jour. Funct. Analysis} \textbf{135}
519-551
\bibitem{bs1} Baez  J C and Sawin  S 1997 Functional integration
on spaces of connections  \textit{Jour. Funct. Analysis}
\textbf{150}  1--27
\bibitem{acz} Ashtekar A, Corichi A and Zapata J A 1998 Quantum
theory of geometry: III. Non-commutativity of Riemannian
structures  \textit{ Class. Quant. Grav.} \textbf{15} 2955--2972
\bibitem{mmt1} Marolf D,  Mour\~ao J,  Thiemann T 1997 The Status of
Diffeomorphism Superselection in Euclidean 2+1 Gravity \textit{J.
Math. Phys.} \textbf{38}  4730-4740
\bibitem{mtv} Mour\~ao J M,  Thiemann T,  Velhinho J M 1999
Physical Properties of Quantum Field Theory Measures \textit{J.
Math. Phys.} \textbf{40}  2337--2353
\bibitem{jv}  Velhinho J M 2002 A groupoid  approach to spaces of
generalized connections \textit{J. Geom. Phys.} \textbf{41}
166--180
\bibitem{cf2}  Fleischhack C 2000 Stratification of the Generalized
Gauge Orbit Space  \textit{Commun. Math. Phys.} \textbf{214}
607--649
\bibitem{cf1} Fleischhack C  2003
On the Gribov Problem for Generalized Connections \textit{Commun.
Math. Phys.} \textbf{234} 423--454
\bibitem{cf3} Fleischhack C 2003 Hyphs and the Ashtekar-Lewandowski
Measure \textit{J. Geom. Phys.} \textbf{45}  231--251
\bibitem{hs} Sahlmann H 2002 When Do Measures on the Space of Connections
Support the Triad Operators of Loop Quantum Gravity?
\textit{Preprint} gr-qc/0207112\\
Sahlmann H 2002  Some Comments on the Representation Theory of the
Algebra Underlying Loop Quantum Gravity \textit{Preprint}
gr-qc/0207111
\bibitem{ol} Oko{\l}{\'o}w A,  Lewandowski  J  2003
Diffeomorphism covariant representations of the holonomy-flux
star-algebra \textit{Class. Quant. Grav.} \textbf{20} 3543--3568
\bibitem{st1} Sahlmann  H and Thiemann  T  2003 On the superselection
theory of the Weyl algebra for diffeomorphism invariant quantum
gauge theories \textit{Preprint} gr-qc/0302090\\
Sahlmann  H and Thiemann T 2003 Irreducibility of the
Ashtekar-Isham-Lewandowski representation \textit{Preprint}
gr-qc/0303074
\bibitem{lost} Lewandowski  J,  Oko\l\'ow  A, Sahlmann H, Thiemann T
2003 Uniqueness of the diffeomorphism invariant state on the
quantum holonomy-flux algebra  \textit{Preprint}\\

\bibitem[ ]{qg2} \textsl{Spin networks:}
\bibitem{rp} Penrose  R 1971 Angular momentum: an approach to combinatorial
space-time  \textit{Quantum Theory and Beyond}  ed Bastin  T
(Cambridge University Press)
\bibitem{rs3} Rovelli  C and Smolin  L 1995 Spin networks and quantum
gravity  \textit{Phys. Rev.} \textbf{D52}  5743--5759
\bibitem{jb2} Baez  J C  1996 Spin networks in non-perturbative quantum
gravity, in \textit{The Interface of Knots and Physics } ed
Kauffman  L (American Mathematical Society, Providence) pp.
167--203\\
 Baez  J C 1996 Spin networks in gauge theory  \textit{Adv. Math.}{\bf
117}  253--272
\bibitem{tt-spin} Thiemann T 1998 The inverse loop transform
\textit{J. Math. Phys.} \textbf{39} 1236--1248
\bibitem{bs2} Baez J C and Sawin  S 1998 Diffeomorphism--invariant spin network
states  \textit{Jour. Funct. Analysis} \textbf{158}  253--266
\bibitem{cf4}  C Fleischhack 2003 Proof of a Conjecture by Lewandowski
and Thiemann \textit{Preprint} math-ph/0304002\\

\bibitem[ ]{qg3} \textsl{Geometric operators and their properties}
\bibitem{rs4} Rovelli C and Smolin  L 1995 Discreteness of area and volume
in quantum gravity  \textit{Nucl. Phys.} \textbf{B442}  593--622;
Erratum: \textit{Nucl. Phys.} \textbf{B456}  753
\bibitem{al4} Ashtekar A and Lewandowski  L 1995 Differential
geometry on the space of connections using projective techniques
\textit{Jour. Geo. \& \ Phys.} \textbf{17}  191--230
\bibitem{rl1} Loll  R 1995 The volume operator in discretized quantum
gravity \textit{Phys. Rev. Lett.} \textbf{75} 3048--3051
\bibitem{rl1} Loll  R 1995  Spectrum of the Volume Operator in Quantum
Gravity   \textit{Nucl. Phys.} \textbf{B460} 143--154
\bibitem{flr} Frittelli  S, Lehner  L and Rovelli  C 1996
\textit{Class. Quant. Grav.} \textbf{13}  2921--2932
\bibitem{rdp} De Pietri R,  Rovelli  C 1996
\textit{Phys. Rev.} \textbf{D54}  2664--2690
\bibitem{jl1} Lewandowski  J. 1997  Volume and Quantizations
\textit{Class. Quant. Grav.} \textbf{14} 71--76
\bibitem{al5} Ashtekar  A and Lewandowski  J 1997 Quantum theory of
geometry I: Area operators  \textit{Class. Quant. Grav.}
\textbf{14}  A55--A81
\bibitem{al6} Ashtekar  A and Lewandowski  J 1997
Quantum theory of geometry II: Volume Operators  \textit{Adv.
Theo. Math. Phys.} \textbf{1}  388--429
\bibitem{rl2} Loll  R 1997  Simplifying the spectral analysis of the
volume operator   \textit{Nucl. Phys.} \textbf{B500}  405--420
\bibitem{rl3} Loll  R 1997 Further results on geometric operators in quantum
gravity  \textit{Class. Quant. Grav.} \textbf{14} 1725--1741
\bibitem{dp1} De Pietri  R 1997 Spin Networks and Recoupling in Loop
Quantum Gravity  \textit{Nucl. Phys. Suppl} \textbf{57}  251-254
\bibitem{dp2} De Pietri  R 1997 On the relation between the connection
and the loop representation of quantum gravity \textit{Class.
Quant. Grav.} \textbf{14}  53--70
\bibitem{tt1} Thiemann  T 1998 Closed formula for the matrix elements
of the volume operator in canonical quantum gravity \textit{J.
Math. Phys.} \textbf{39}  3347-3371
\bibitem{tt2} Thiemann  T 1998 A length operator for canonical quantum
gravity  \textit{Jour. Math. Phys.} \textbf{39}  3372--3392
\bibitem{sm} Major S A 1999 Operators for quantized directions
\textit{Class. Quant. Grav.} \textbf{16} 3859-3877\\

\bibitem[ ]{bi}\textsl{Barbero-Immirzi ambiguity}
\bibitem{fb} Barbero F 1996 Real Ashtekar variables for Lorentzian
signature space-times  \textit{Phys. Rev.} \textbf{D51} 5507--5510
\bibitem{gi} Immirzi  G 1997 Quantum gravity and Regge calculus
\textit{Nucl. Phys. Proc. Suppl.} \textbf{57}  65--72
\bibitem{rt1} Rovelli  C. and Thiemann  T. (1998) The Immirizi parameter in
quantum general relativity  \textit{Phys. Rev.} \textbf{D57}
1009--1014
\bibitem{gop} Gambini  R, Obregon  O and Pullin  J 1999
Yang-Mills analogs of the Immirzi ambiguity   \textit{Phys. Rev.}
\textbf{D59} 047505\\

\bibitem[ ]{qd1} \textsl{Quantum Einstein's equation I}
\bibitem{kk} Kucha$\check{\rm r}$  K 1993 Canonical Quantum Gravity
\textit{General Relativity and Gravitation 1992} ed Gleiser R J,
Kozameh C N, Moreschi O M (Institute of Physics Publishing)
119--150
\bibitem{atu} Ashtekar  A, Tate R S and Uggla C 1993
Minisuperspaces: Observables and quantization  \textit{Int. J.
Phys.} \textbf{D2} 15--50\\
 Ashtekar  A, Tate R S and Uggla C 1993
Minisuperspaces: Symmetries and quantization \textit{ Misner
Festschrift} ed Hu B L et al (Cambridge U. P., Cambridge)
\bibitem{at} Ashtekar  A and Tate R S 1994 An algebraic extension
of Dirac quantization: Examples  \textit{Jour. Math. Phys.}
\textbf{35} 6434--6470
\bibitem{dm} Marolf  D 1995 Refined algebraic quantization:
Systems with a single constraint  \ \textit{Preprint}
gr-qc/9508015
\bibitem{almmt1} Ashtekar  A, Lewandowski  J, Marolf  D, Mour\~ao  J
and Thiemann  T 1995 Quantization of diffeomorphism invariant
theories of connections with local degrees of freedom
\textit{Jour. Math. Phys.} \textbf{36} 6456--6493
\bibitem{mmt2} Marolf  D, Mour\~ao J and Thiemann  T 1997
The status of diffeomorphism super-selection in Euclidean 2+1
gravity  \textit{J. Math. Phys.} \textbf{38} 4730-4740
\bibitem{gm} Guilini  N and Marolf  D 1999 On the Generality of
Refined Algebraic Quantization \textit{Class. Quant.
Grav.}\textbf{16}
2479--2488\\
A uniqueness theorem for constraint quantization  \textit{Class.
Quant. Grav.} \textbf{16} 2489--2505
\bibitem{lt} Lewandowski  J and Thiemann  T 1999 Diffeomorphism invariant
quantum field theories of connections in terms of webs  \textit{
Class. Quant. Grav.} \textbf{16}  2299--2322
\bibitem{cz} Corichi  A and Zapata J A 1997  On diffeomorphism
invariance for lattice theories  \textit{Nucl. Phys.} \textbf{
B493}  475--490\\

\bibitem[ ]{qd2}\textsl{Quantum Einstein's equation II}
\bibitem{rs5} Rovelli  C and Smolin  L 1994 The physical Hamiltonian in
nonperturbative quantum gravity  \textit{Phys. Rev. Lett.}
\textbf{72}  446--449
\bibitem{tt3} Thiemann  T 1996 Anomaly-free formulation of non-perturbative,
four-dimensional Lorentzian quantum gravity  \textit{Phys. Lett.}
\textbf{B380}  257--264
\bibitem{tt4} Thiemann  T 1998 Quantum spin dynamics
(QSD)  \textit{Class. Quant. Grav.}  \textbf{15} 839--873
\bibitem{tt5} Thiemann  T. 1998 QSD III: Quantum constraint algebra and
physical scalar product in quantum general relativity
\textit{Class. Quant. Grav.}  \textbf{15}  1207--1247
\bibitem{tt6} Thiemann  T 1998 QSD V: Quantum gravity as the
natural regulator of matter quantum field theories  \textit{Class.
Quant. Grav.} \textbf{15}  1281--1314
\bibitem{tt7} Thiemann T 2001 Quantum Spin Dynamics (QSD) : VII. Symplectic
Structures and Continuum Lattice Formulations of Gauge Field
Theories \textit{Class. Quant. Grav.} \textbf{18} 3293--3338
\bibitem{glmp} Gambini  R, Lewandowski  J, Marolf D and Pullin  J 1998 On
the consistency of the constraint algebra in spin network quantum
gravity  \textit{Int. J. Mod. Phys.}  \textbf{D7}  97--109
\bibitem{lm} Lewandowski  J and Marolf  D 1998 Loop constraints: A habitat and
their algebra  \textit{Int. J. Mod. Phys.} \textbf{D7}  299--330
\bibitem{gr} Gaul  G and Rovelli  C 2001 Generalized hamiltonian
constraint operator in loop quantum gravity and its simplest
Euclidean matrix elements \textit{Class. Quant. Grav.} \textbf{18}
1593--1624\\

\bibitem[ ]{bb}\textsl{Big-bang:}
\bibitem{hk} Kodama H 1988 Specialization of Ashtekar's formalism
to Bianchi cosmology  \textit{Prog. Theor. Phys.}  \textbf{80}
1024--1040\\
Kodama H 1990 Holomorphic wavefunction of the universe
\textit{Phys. Rev.}  \textbf{D42}  2548--2565
\bibitem{mb1} Bojowald  M 2001  Absence of singularity in loop
quantum cosmology  \textit{Phys. Rev. Lett.} \textbf{86}
5227--5230
\bibitem{mb2} Bojowald  M 2001 Inverse scale factor in isotropic
quantum geometry \textit{Phys.\ Rev.} \textbf{D64} 084018
\bibitem{mb3} Bojowald  M 2001 Dynamical initial conditions in quantum
cosmology  \textit{Phys. Rev. Lett.} {\bf 87}  121301
\bibitem{mb4} Bojowald  M 2001 Loop Quantum Cosmology IV: Discrete Time
Evolution \textit{Class. Quant. Grav.} {\bf 18} 1071--1088 \
\bibitem{mb5} Bojowald  M 2001 Loop quantum cosmology III:
Wheeler-DeWitt operators  {Class. Quant. Grav.} \textbf{18} 1055--1070 \\
Bojowald  M 2002 The Semiclassical Limit of Loop Quantum Cosmology
\textit{Class. Quant. Grav.} \textbf{18} L109--L116
\bibitem{mb6} Bojowald  M 2001 The Inverse Scale Factor in Isotropic Quantum
Geometry  \textit{Phys.Rev.} \textbf{D64} 084018
\bibitem{mb7} Bojowald  M 2002 Isotropic Loop Quantum Cosmology
\textit{Class.Quant.Grav.} \textbf{19}  2717--2742
\bibitem{mbh} Bojowald  M, Hinterleitner F 2002
 Isotropic Loop Quantum Cosmology with Matter
\textit{Phys.Rev.} \textbf{D66} 104003
\bibitem{mbv} Bojowald  M and  Vandersloot K 2003
Loop Quantum Cosmology, Boundary Proposals, and Inflation
\textit{Phys.Rev.} \textbf{D67} 124023
\bibitem{mb9} Bojowald M 2003 Homogeneous Loop Quantum Cosmology
\textit{Class.Quant.Grav.} \textbf{20} 2595--2615
\bibitem{mb10} Bojowald M 2003  Initial Conditions for a Universe
\textit{Gen. Rel. Grav.} \textbf{35}, 1877-1883  (First Prize in
the Gravity Research Foundation Essay Contest)
\bibitem{abl:qc} Ashtekar A, Bojowald M, Lewandowski J  2003
Mathematical structure of loop quantum cosmology \textit{Adv.
Theor. Math. Phys.} \textbf{7}  233--268
\bibitem{bdv} Bojowald M, Date G and Vandersloot K 2004
Homogeneous Loop Quantum Cosmology: The Role of the Spin
Connection \textit{Class. Quant. Grav.} textbf{21} 1253-1278
\bibitem{abw} Ashtekar A, Bojowald M and Willis J 2004
Quantum corrections to Friedmann equations in loop quantum
cosmology (in preparation)\\

\bibitem[ ]{bh} \textsl{Black holes:}
\bibitem{jb} Bekenstein  J D 1973 Black holes and entropy
\textit{Phys. Rev.} \textbf{D7} 2333--2346 \\
(1974) Generalized second law of thermodynamics in black hole
physics  \textit{Phys. Rev.} \textbf{D9} 3292--3300\\
Bekenstein  J D and Meisels  A 1977  Einstein A and B coefficients
for a black hole  \textit{Phys. Rev.} {\bf D15} 2775--2781
\bibitem{bch} Bardeen  J W, Carter  B and Hawking  S W 1973 The four
laws of black hole mechanics  \textit{Commun. Math. Phys.}
\textbf{31} 161-170
\bibitem{swh} Hawking  S W 1975 Particle creation by black holes
\textit{Commun. Math. Phys.} \textbf{43} 199--220
\bibitem{gh} Gibbons G and Hawking  S W 1977 Cosmological event
horizons, thermodynamics, and particle creation  \textit{Phys.
Rev.}\textbf{D15} 2738--2751
\bibitem{earlyentropy1}  Smolin  L 1995 Linking topological quantum
field theory and nonperturbative quantum gravity  \textit{J. Math.
Phys.} \textbf{36}  6417--6455
\bibitem{earlyentropy2} Barreira  M, Carfora M and Rovelli C 1996
Physics with non-perturbative quantum gravity: radiation from a
quantum black hole  \textit{Gen. Rel. Grav.} \textbf{28}  1293--1299\\
Rovelli  C 1996 Black hole entropy from loop quantum gravity
\textit{Phys. Rev. Lett.} \textbf{14}  3288--3291\\
Rovelli  C 1996 Loop quantum gravity and black hole physics
\textit{Helv. Phys. Acta.} \textbf{69}  582--611
\bibitem{earlyentropy3} Krasnov  K 1997 Geometrical entropy from loop
quantum gravity  \textit{Phys. Rev.} \textbf{D55}  3505--3513\\
Krasnov  K 1998  On statistical mechanics of Schwarzschild black
holes  \textit{Gen. Rel. Grav.} \textbf{30}  53--68
\bibitem{jkm} Jacobson T, Kang G and Myers R C 1994 On black hole entropy
{\textit Phys. Rev.} {\textbf D49} 6587--6598
\bibitem{iw} Wald  R. (1993) Black hole entropy is Noether charge
\textit{Phys.\ Rev.} \textbf{D48}  3427--3431\\
Iyer  V and Wald  R 1994 Some properties of Noether charge and a
proposal for dynamical black hole entropy  \textit{Phys.\ Rev.}
\textbf{D50}  846--864
\bibitem{abck} Ashtekar  A, Baez  J C, Corichi  A, and Krasnov K
1998 Quantum geometry and black hole entropy  (1998) \textit{Phys.
Rev. Lett.} \textbf{80} 904--907
\bibitem{ack} Ashtekar A, Corichi A and Krasnov  K 1999 Isolated
horizons: the classical phase space  \textit{Adv. Theor. Math.
Phys.} \textbf{3}  418--471
\bibitem{abk} Ashtekar  A, Baez  J C   and Krasnov  K 2000 Quantum
geometry of isolated horizons and black hole entropy  \textit{Adv.
Theo. Math. Phys.}  \textbf{4}  1--95
\bibitem{km} Kaul  R K  and Majumdar  P 2000  Logarithmic
corrections to the Bekenstein-Hawking entropy  \textit{Phys. Rev.
Lett.} \textbf{84}  5255--5257
\bibitem{abf} Ashtekar  A, Beetle C and Fairhurst S 1999
Isolated horizons: a generalization of black hole mechanics
\textit{Class.\ Quantum Grav.} \textbf{16} L1--L7\\
Ashtekar  A, Beetle C and Fairhurst S 2000 Mechanics of isolated
horizons \textit{Class.\ Quantum Grav.} \textbf{17} 253--298
\bibitem{jl2}Lewandowski J 2000 Space-times admitting isolated
horizons  \textit{Class. Quantum Grav.} \textbf{17} L53--L59
\bibitem{afk} Ashtekar  A,  Fairhurst  S and Krishnan  B 2000 Isolated
Horizons: Hamiltonian Evolution and the First Law  \textit{Phys. \
Rev.} \textbf{D62} 104025.
\bibitem{abl2} Ashtekar  A, Beetle  C and Lewandowski  J
2001 Mechanics of rotating isolated horizons  \textit{Phy.\ Rev.\
} \textbf{D64}  044016
\bibitem{abl1} Ashtekar  A, Beetle  C and Lewandowski J 2002
Geometry of generic isolated horizons  \textit{Class. Quantum
Grav.} \textbf{19}  1195--1225
\bibitem{acs} Ashtekar  A, Corichi  A and Sudarski D 2003
Non-Minimally Coupled Scalar Fields and Isolated Horizons
\textit{Class. Quantum Grav.} \textbf{20} 3413-3425
\bibitem{ac} Ashtekar  A and Corichi  A 2003  Non-minimal
couplings, quantum geometry and black hole entropy 20
\textit{Class. Quantum Grav.} \textbf{20} 4473-4484
\bibitem{aepv} Ashtekar A, Engle J, Pawlowski, T and van der
Broeck C 2004 Multipole moments of isolated horizons
\textit{Class. Quant. Grav.} \textbf{21} 2549-2570
\bibitem{aa4} Ashtekar  A 2003 Black hole entropy: Inclusion of
distortion and angular momentum,
http://www.phys.psu.edu/events/index.html?event\_id=517\&event\_type=17\\
Ashtekar A, Engle J and van der Broeck C 2004 (in preparation)
\bibitem{jw} Wheeler  J A   1992  \emph{It} from \emph{bit}
\textit{Sakharov Memorial Lectures on Physics, Vol 2}  ed Keldysh
L and  Feinberg V (Nova Science, Moscow)
\bibitem{gm} Ghosh A and Mitra P 2004 A bound on the log correction to
the black hole area law \textit{pre-print}
\texttt{gr-qc/0401070}\\

\bibitem[ ]{sc}\textsl{Low energy physics:}
\bibitem{ars} Ashtekar  A,  Rovelli  C and Smolin  L 1992 Weaving a
classical geometry with quantum threads  \textit{Phy. Rev. Lett.}
\textbf{69}  237--240
\bibitem{ag} Arnsdorf  M, Gupta  S 2000 Loop quantum gravity on
non-compact spaces   \textit{Nucl. Phys.} \textbf{ B577} 529-546
\bibitem{cr} Corichi  A,  Reyes  J M   2001
A Gaussian Weave for Kinematical Loop Quantum Gravity \textit{Int.
J. Mod. Phys.} \textbf{D10} 325-338
\bibitem{hall} Hall B C 1994 The Segal-Bergmann coherent state
transform for compact Lie Lie groups \textit{J. Funct. Anal.}
\textbf{122} 103-151
\bibitem{toh1}Sahlmann  H, Thiemann T and Winkler O 2001 Coherent
states for canonical quantum general relativity and the infinite
tensor product extension  \textit{Nucl. Phys.} \textbf{B606}
401-440
\bibitem{toh2} Thiemann  T 2001 Gauge field theory coherent states
(GCS): I. general properties  \textit{Class.\ Quant.\
Grav.} {\bf 18} 2025-2064\\
T. Thiemann, O. Winkler 2001 Gauge Field Theory Coherent States
(GCS) : II. Peakedness Properties  \textit{Class. Quant. Grav.}
\textbf{18} 2561--2636\\
Thiemann T,  Winkler O 2001 Gauge Field Theory Coherent States
(GCS) : III. Ehrenfest Theorems \textit{Class. Quant. Grav.}
\textbf{18} 4629-4682\\
Thiemann T,  Winkler O 2001  Gauge Field Theory Coherent States
(GCS) : IV. Infinite Tensor Product  and Thermodynamical Limit
\textit{Class. Quant. Grav.} \textbf{18}  4997-5054
\bibitem{st2} Hanno Sahlmann, Thomas Thiemann 2002 Towards the QFT on Curved
Spacetime Limit of QGR. I: A General Scheme \textit{Preprint}
\texttt{gr-qc/0207030}\\
Sahlmann H Thomas Thiemann 2002  Towards the QFT on Curved
Spacetime Limit of QGR. II: A Concrete Implementation
\textit{Preprint} \texttt{gr-qc/0207031}
\bibitem{ttcomp} Thiemann T 2002 Complexifier Coherent States for Quantum
General Relativity \ \textit{Preprint} gr-qc/0206037
\bibitem{lb} Bombelli L 2002 Statistical geometry of random weave states
\textit{Proceedings of the Ninth Marcel Grossmann Meeting on
General Relativity} ed  Gurzadyan VG, Jantzen RT, and  Ruffini R
(World Scientific, Singapore) \texttt{gr-qc/0101080}
\bibitem{ai2} Ashtekar A and Isham C J 1992 \textit{Phys.\ Lett.}
{\bf B274} 393-398
\bibitem{mv1} Varadarajan  M 2000 Fock representations from U(1) holonomy
algebras  \textit{Phys. Rev.} \textbf{D61} 104001\\
Varadarajan  M. (2001) M. Photons from quantized electric flux
representations  \textit{Phys. Rev.} \textbf{D64} 104003
\bibitem{al7} Ashtekar  A  and  Lewandowski  J 2001 Relation between
polymer and Fock excitations  \textit{Class. Quant. Grav.}
\textbf{18}  L117--L127
\bibitem{jv}  Velhinho  J M 2002
Invariance properties of induced Fock measures for U(1) holonomies
\textit{Commun. Math. Phys.} \textbf{227}  541--550
\bibitem{afw} Ashtekar  A, Fairhurst  S and Willis  J 2003
Quantum gravity, shadow states, and quantum mechanics
\textit{Class. Quantum Grav.} \textbf{20}, 1031-1061
\bibitem{afg} Ashtekar A, Fairhurst S and Ghosh A 2004 (in
preparation)
\bibitem{mv2} Varadarajan M. 2002 Gravitons from a loop representation
of linearised gravity \textit{Phys. Rev.} \textbf{D66} 024017
\bibitem{agv} Ashtekar A, Ghosh A and Van Den Broeck C 2004
(in preparation)
\bibitem{blm} Bobienski M, Lewandowski J, Mroczek M
2002 A 2-Surface Quantization of the Lorentzian Gravity
\textit{Proceedings of the Ninth Marcel Grossmann Meeting on
General Relativity} ed  Gurzadyan VG, Jantzen RT, and  Ruffini R
(World Scientific, Singapore) \texttt{gr-qc/0101069}
\bibitem{als} Ashtekar  A, Lewandowski J and Sahlmann  H 2003
Polymer and Fock representations of a scalar field  \textit{Class.
Quant. Grav.} \textbf{20} L11--L21\\

\bibitem[ ]{sf} \textsl{Spin foams and finiteness:}
\bibitem{jb3} Baez  J C 2000 An introduction to spin foam models
of quantum gravity and BF theory  \textit{Lect. Notes Phys.}
\textbf{543} 25--94
\bibitem{jb4} Baez J C 1998 Spin foam models \textit{Class.
Quant. Grav.} {\bf 15}  1827--1858
\bibitem{mr1} Reisenberger  M P 1997 A lattice worldsheet sum for
 Reisenberger  M P 1997  4-d Euclidean general relativity \
\textit{Preprint} gr-qc/9711052\\
Reisenberger  M P 1999 On relativistic spin network vertices
\textit{J. Math. Phys.} \textbf{40} 2046--2054
\bibitem{mr2} Reisenberger  M P and Rovelli  C 2001 Spacetime as a
Feynman diagram: the connection formulation  \textit{Class. Quant.
Grav.} \textbf{18} 121--140\\
Reisenberger  M P and Rovelli 2002 Spacetime states and covariant
quantum theory \textit{ Phys. Rev.} \textbf{D65} 125016
\bibitem{bc} Barrett  J W and Crane  L 1998 {Relativistic spin networks and
quantum gravity}  \textit{J. Math. Phys.} \textbf{39}  3296--3302\\
Barrett  J W and Crane L  2000 {A Lorentzian signature model for
quantum general relativity}  \textit{Class. Quant. Grav.}
\textbf{17}  3101--3118
\bibitem{ap2} Perez  A 2001 {Finiteness of a spinfoam model for Euclidean
quantum general  relativity}  \textit{Nucl. Phys.} \textbf{B599}
427--434 \\
Perez  A  and Rovelli  C 2001 {Spin foam model for Lorentzian
general relativity}  \textit{Phys. Rev.} \textbf{D63}  041501\\
Crane   L, Perez  A and Rovelli  C 2001 {Perturbative finiteness
in spin-foam quantum gravity}  \textit{Phys. Rev.
Lett.} \textbf{87}  181301\\
Crane   L, Perez  A and Rovelli  C 2001 3+1 spinfoam model of
quantum gravity with spacelike and timelike components
\textit{Phys. Rev.} \textbf{D64} 064002
\bibitem{bcht} Baez J, Christensen D, Halford T R, Tsang DC 2002
Spin Foam Models of Riemannian Quantum Gravity \textit{Class.
Quant. Grav.} \textbf{19}4627-4648
\bibitem{gpspinfoam} Gambini R and Pullin J 2002
A finite spin-foam-based theory of three and four dimensional
quantum gravity \textit{Phys. Rev.} \textbf{D66} 024020
\bibitem{lr} Lauscher  O and Reuter  M 2002 Is quantum Einstein gravity
non-perturbatively renormalizable?  \textit{Class. Quant. Grav.}
\textbf{19}  483--492
\bibitem{pp} Percacci R and Perini D 2003 Asymptotic safety of gravity
coupled to matter \textit{Phys. Rev.} \textbf{D68}044018\\
Perini D 2003 Gravity and matter with asymptotic safety
\textit{pre-print} \texttt{hep-th/0305053}
\bibitem{bp} Bojowald M and Perez A 2003 Spin foam quantization and
anomalies \texttt{gr-qc/0303026}
\bibitem{lf} Freidel L  and Louapre D 2003 Diffeomorphisms and spin
foam models, \textit{Nucl.Phys.} \textbf{B662} 279-298\\
Freidel L  and Louapre D 2003 Non-perturbative summation over 3D
discrete topologies \textit{Phys. Rev.} \textbf{D68} 104004
\texttt{hep-th/0211026}
\bibitem{np} Perez A and Noui K 2004 Three dimensional loop quantum
gravity: coupling to point particles \textit{pre-print}
\texttt{gr-qc/0402111}\\
 2004 Three dimensional loop quantum
gravity: physical scalar product and spin foam models
\textit{pre-print} \texttt{ gr-qc/0402110}\\

\bibitem[ ]{sf} \textsl{Outlook:}
\bibitem{yangmills} Ashtekar  A, Lewandowski  J, Marolf  D, Mour\~ao  J
and Thiemann  T 1997 {\rm SU(N)} Quantum Yang-Mills theory in two
dimensions: A complete solution \textit{J. Math. Phys.}
\textbf{38}, 5453-5482
\bibitem{asstrings} Ashtekar A and Sahlmann H 2004 (in
preparation)
\bibitem{qgroups} Okolw A 204 Representations of quantum geometry
\textit{Ph.D. thesis}, University of Warsaw
\bibitem{rovelli} Rovelli C and Speziale S 2003  Reconcile Planck-scale
discreteness and the Lorentz-Fitzgerald contraction \textit{Phys.
Rev.} \textbf{D67} 064019
\bibitem{tt7} Thiemann T 2003 The Phoenix project: Master constraint
programme for loop quantum gravity  \textit{Preprint}
gr-qc/0305080
\bibitem{master} Dittrich B 2004 Testing the master constraint programme
for loop quantum gravity
http://www.phys.psu.edu/events/index.html?event\_id=850\&event\_type=17
\bibitem{klauder} Klauder J 2003 Affine quantum gravity
\textit{Int. J. Mod. Phys.} \textbf{D12} 1769-1774\\
2004 The utility of coherent states and other mathematical methods
in the foundations of affine quantum gravity, \textit{pre-print}
\texttt{hep-th/0401214}
\bibitem{bggp} Di Bartolo  C, Gambini  R, Griego  J and  Pullin  J 2000
Consistent canonical quantization of general relativity in the
space of Vassiliev knot invariants  \textit{Phys. Rev. Lett.}
\textbf{84}  2314--2317\\
2000 Canonical quantum gravity in the Vassiliev invariants arena:
I. Kinematical structure \textit{Class. Quant. Grav.} \textbf{17}
3211-3238\\
2000 Canonical quantum gravity in the Vassiliev invariants arena:
II. Constraints, habitats and consistency of the constraint
algebra \textit{Class. Quant. Grav.} \textbf{17} 3239-3264
\bibitem{discrete} Gambini R and Pullin J 2003
Canonical quantization of general relativity in discrete
space-times \textit{Phys. Rev. Lett.} \textbf{90} 021301\\
Gambini R, Porto R and Pullin J 2004 Loss of coherence from
discrete quantum gravity  \textit{Class. Quant. Grav.} \textbf{21}
L51-L57\\
Gambini R, Porto R and Pullin J 2004 A relational solution to the
problem of time in quantum mechanics and quantum gravity induces a
fundamental mechanism for quantum decoherence \textit{pre-print}
\texttt{gr-qc/0402118}
\bibitem{gth} Horowitz G. T. 1998 Quantum states of black holes,
in \textit{Black Holes and Relativistic Stars}, ed Wald R M (
University of Chicago Press, Chicago)
\bibitem{sc2} Carlip S 1999 Entropy from Conformal Field Theory at
Killing Horizons  \textit{Class. Quant. Grav.} \textbf{16}
3327--3348
\bibitem{dgw} Dreyer O, Ghosh A and Wisniewski J 2001
Black hole entropy calculations based on symmetries, \textit{
Class. Quant. Grav.}\textbf{18} 1929-1938
\bibitem{sc3} Carlip S 2002 Near-Horizon Conformal Symmetry and Black
Hole Entropy  \textit{Phys. Rev. Lett.} \textbf{88}  241301
\bibitem{ak} Ashtekar A,  Krishnan B 2002
Dynamical Horizons: Energy, Angular Momentum, Fluxes and Balance
Laws, \textit{Phys. Rev. Lett.} \textbf{89} 261101\\
2003 Dynamical horizons and their properties, \textit{Phys. Rev.}
\textbf{D68} 104030
\bibitem{mb8} Bojowald  M  2002  Inflation from Quantum Geometry
\textit{Phys.Rev.Lett.} \textbf{89} 261301
\bibitem{bd} Bojowald M and Date G 2004 Quantum suppression of the
generic chaotic behavior close to cosmological singularities
\textit{Phys. Rev. Lett.} \textbf{92} 071302
\bibitem{cobe} Tsujikawa S, Singh P and Maartens R 2003 Loop
quantum gravity effects on inflation and the CMB,
\textit{pre-print} \texttt{astro-ph/0311015}


\end{thebibliography}
\end{document}